\documentstyle[12pt,fleqn,twoside,hiroko]{report}
\input epsf

\begin{document}
\begin{center}
{\bf \large Dual Higgs Theory for Color Confinement in 
Quantum Chromodynamics
\footnote{This paper is made based on 
Ichie's doctor thesis accepted by Osaka University in 1998.}
} \\
\vspace{0.5cm}
{ Hiroko Ichie
\footnote{
Present address:
Department of Physics, 
Tokyo Institute of Technology,
Meguro, 
Tokyo 152-8551, Japan.} 
and Hideo Suganuma
}  \\
{\it Research Center for Nuclear Physics (RCNP), Osaka University 
Ibaraki, Osaka 567-0047, Japan}
\end{center}
\vspace{0.5cm}
\pagenumbering{roman}
{\small



We study the dual Higgs theory for the confinement mechanism based on 
Quantum Chromodynamics (QCD) in the 't Hooft abelian gauge.
In the abelian gauge, QCD is reduced into an abelian gauge theory
including color-magnetic monopoles, which appear corresponding to the 
nontrivial homotopy group
$\Pi_2($SU$(N_c)/$U$(1)^{N_c-1})$ $= {\bf Z}^{N_c-1}$.
With the two conjectures of 
abelian dominance and monopole condensation, QCD in the abelian 
gauge becomes
the dual Higgs theory, and 
then color confinement can be understood as  one-dimensional 
squeezing of the color-electric flux due to the dual Meissner effect.
In the basis of the dual superconductor picture, confinement phenomena 
are systematically studied using the lattice QCD Monte-Carlo 
simulation, the monopole-current dynamics and the dual 
Ginzburg-Landau (DGL) theory, an infrared effective theory of QCD.

First, we study the origin of abelian dominance for the confinement 
force in the maximally abelian (MA) gauge in terms of the gluon-field 
properties using  the lattice QCD.
In the MA gauge, the gluon-field fluctuation is maximally concentrated 
in the abelian sector.
As the remarkable feature in the MA gauge, the amplitude of the 
off-diagonal gluon is strongly suppressed, and therefore the phase 
variable of the off-diagonal (charged) gluon tends to be random, 
according to the weakness of the constraint from the QCD action.
Using the random-variable approximation for the charged-gluon phase 
variable, we find the perimeter law of the charged-gluon contribution 
to the Wilson loop and show abelian dominance for the string  
tension in the semi-analytical manner.
These theoretical results are also numerically confirmed using the 
lattice QCD simulation.

Second, we study the QCD-monopole appearing in the abelian sector in 
the abelian gauge. The appearance of monopoles is transparently 
formulated in terms of the gauge connection, and is originated from the 
singular nonabelian gauge transformation to realize the abelian gauge.
We investigate the gluon field around the monopole in the lattice QCD. 
The QCD-monopole carries a large fluctuation of the gluon field and 
provides a large abelian action of QCD.
Nevertheless, QCD-monopoles can appear in QCD without large cost of 
the QCD action, due the large cancellation between the abelian and 
off-diagonal parts of the QCD action density around the monopole.
We derive a simple relation between the confinement force and the 
monopole density by idealizing the monopole contribution to the 
Wilson loop.

Third, we study the monopole-current dynamics using the infrared 
monopole-current action defined on a lattice. We adopt  the local 
current action, considering the infrared screening of the 
inter-monopole interaction due to the dual Higgs mechanism.
When the monopole self-energy $\alpha$ is smaller then $\alpha_{c} = 
$ ln$ (2D-1 )$, monopole condensation can be analytically shown, and 
we find this system being in the confinement phase from the Wilson 
loop analysis.
By comparing the lattice QCD with the monopole-current system, the 
QCD vacuum is found to correspond to the monopole-condensed phase in 
the infrared scale.
We consider the derivation of the DGL theory from the monopole 
ensemble, which would be essence of the QCD vacuum in the MA gauge 
because of abelian dominance and monopole dominance.

Finally, we study the QCD phase transition 
at finite temperatures
in the DGL theory.
We formulate the effective potential 
at various temperatures in the imaginary-time formalism.
Thermal effects reduce the QCD-monopole condensate and 
bring a first-order deconfinement phase transition.
We find a large reduction of the self-interaction 
among QCD-monopoles and the glueball mass 
near the critical temperature 
by considering the temperature dependence of the self-interaction.
The string tension is also calculated at finite temperatures. 
We apply  also  the DGL theory  for the bubble formation process in early
Universe and  quark-gluon-plasma (QGP) formation process in the
ultra-relativistic heavy-ion collision.
 


}

\newpage




\pagestyle{plain}

\tableofcontents





\newpage

\pagenumbering{arabic}

\pagestyle{headings}



\chapter{Introduction}

%

Quantum Chromodynamics (QCD) is the fundamental theory of the strong 
interaction, and describes the properties and the 
underlying structure of hadrons in terms of quarks and gluons. 
The  QCD Lagrangian has the local SU($N_c$) symmetry and  is written
by the quark $q$ and the gluon field $A_\mu$ as
\begin{eqnarray}
{\cal L}_{\rm QCD}  =
    - \frac 12{\rm tr}\left( G_{\mu \nu} G^{\mu \nu}\right)
    + \bar{q} \left( i \gamma_\mu D^\mu -  m \right) q,
\end{eqnarray}
where $G_{\mu \nu}$  is the SU($N_c$) field strength
$G_{\mu \nu} \equiv \frac{1}{ie}
[{\hat D_\mu},{\hat D_\nu}] $ and $\hat D_\mu$ is  
covariant-derivative $\hat D_{\mu } = \hat \partial_\mu + ieA_\mu$
 \cite{cheng,Nachtmann,greiner}.
In the chiral limit $m \to 0$ with $N_{f}$ flavor, 
this Lagrangian has also global 
chiral symmetry U($N_{f}$)$_{L}$ $\times$ U($N_{f}$)$_{R}$, 
although U(1)$_A$ is explicitly broken by the U(1)$_A$ anomaly  
at the quantum level 
 \cite{cheng}.
 
Due to the asymptotic freedom, 
the gauge-coupling constant of QCD becomes small 
in the high-energy region
and the perturbative QCD provides a direct and systematic description 
of the QCD system in terms of quarks and gluons. 
On the other hand, in the low-energy region, 
the strong gauge-coupling nature of QCD leads to 
nonperturbative features like color confinement, 
dynamical chiral-symmetry breaking \cite{NGL,higashijima,miransky}
and nontrivial topological effect by instantons \cite{shifman,rajaraman,shuryak}, 
and it is hard to understand them directly from quarks and gluons 
in a perturbative manner. 
Instead of quarks and gluons, 
some collective or composite modes may be relevant degrees of freedom 
for the nonperturbative description 
in the infrared region of QCD. 
As for chiral dynamics,
the pion and the sigma meson play the important role for the low-energy 
QCD, and they are included in the effective theory like the (non-) linear 
sigma model \cite{cheng,bando},
the chiral bag model \cite{thomasA, hosaka}
and the Nambu-Jona-Lasinio model \cite{klevansky,kunihiro},
where these mesons are
described as composite modes of quarks.  
Here, the pion is considered to be the Nambu-Goldstone boson relating to 
spontaneous chiral-symmetry breaking and obeys the 
low-energy theorem and the current algebra \cite{cheng}.
On the other hand, confinement is essentially described by  dynamics of
gluons rather than quarks.
Hence, it is quite desired to extract the relevant collective mode 
from gluon for confinement phenomena.

%

\begin{figure}[tb]
\epsfxsize = 14 cm
\centering \leavevmode
\epsfbox{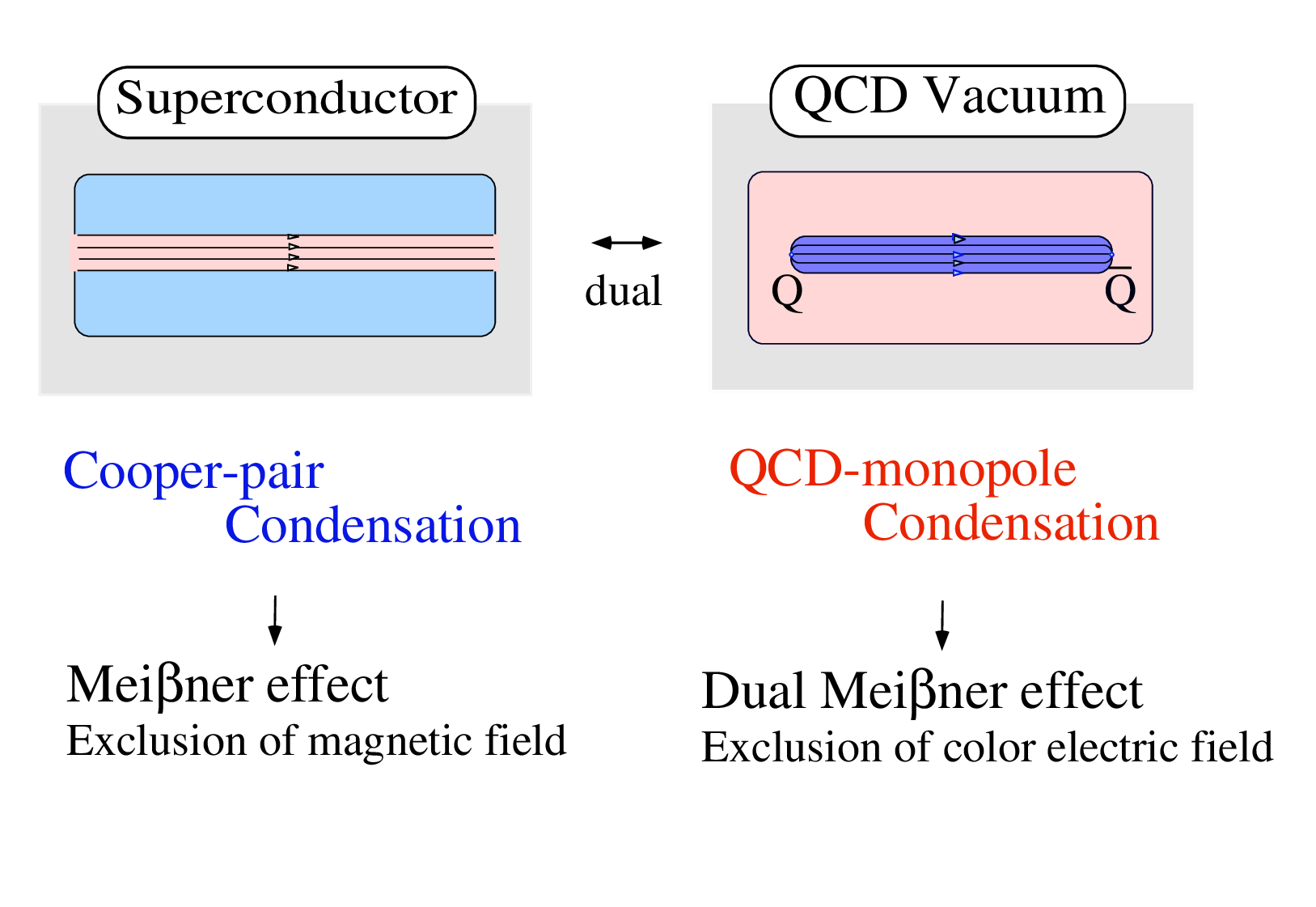}
\vspace{-1.5cm}
\caption{   
Correspondence of the QCD vacuum to the superconductor.
In the QCD vacuum, the color confinement is realized
by monopole condensation.
In the superconductor, the Meissner effect occurs
by Cooper-pair condensation.
}
\label{super}
\vspace{0cm}
\end{figure}

In 1970's, Nambu, 't Hooft and Mandelstam proposed 
an interesting idea that quark confinement can be interpreted 
using the dual version of the 
superconductivity \cite{nambu,thoa,mandelstam} (see Fig.\ref{super}).
In the ordinary superconductor, 
Cooper-pair condensation leads to the Meissner effect, 
and the magnetic flux is excluded or squeezed like a 
quasi-one-dimensional tube as the Abrikosov vortex, 
where the magnetic flux is quantized topologically. 
On the other hand, 
from the Regge trajectory of hadrons and the lattice QCD, 
the confinement force between the color-electric charge is 
characterized by the universal physical quantity of the string tension, 
and is brought by one-dimensional squeezing of the 
color-electric flux \cite{haymaker} in the QCD vacuum. 
Hence, the QCD vacuum can be regarded as the dual version 
of the superconductor based on above similarities 
on the low-dimensionalization of the quantized flux between charges. 
In this dual-superconductor picture for the QCD vacuum, 
as the result of condensation of color-magnetic monopoles, which is the 
dual  version of the electric charge,
the squeezing of the color-electric flux between quarks 
is realized by the dual Meissner effect.
However, there are two following large gaps between QCD and the dual 
superconductor picture.
\begin{enumerate}
\item This picture is based on the abelian gauge theory subject to the 
Maxwell-type equations, where electro-magnetic duality is manifest, 
while QCD is a nonabelian gauge theory.
\item The dual-superconductor scenario  requires condensation of 
(color-) magnetic \\ 
monopoles as key concept, while QCD does not have such a monopole as 
the elementary degrees of freedom.
\end{enumerate}
As the connection between QCD and the dual superconductor scenario, 't Hooft
proposed concept of the abelian gauge fixing \cite{thooft}, the partial gauge 
fixing which only remains abelian gauge degrees of freedom in QCD.
By definition, the abelian gauge fixing reduces QCD into an abelian gauge
theory, where the off-diagonal element of the gluon field behaves as a 
charged matter field
and
provides a color-electric current in terms of the residual abelian gauge 
symmetry.
As a remarkable fact in the abelian gauge, color-magnetic monopole appears 
as topological object corresponding to 
nontrivial homotopy group $\Pi_2( {\rm SU}(N_c)/$U$(1)^{N_c-1}) =
{\bf Z}^{N_c-1}_\infty$.
Thus, by the abelian gauge fixing, QCD is reduced into an abelian
gauge theory including both the electric current $j_\mu$ and 
the magnetic current $k_\mu$, which is expected to provide a theoretical 
basis of the monopole-condensation scheme for the confinement 
mechanism.

For irrelevance of the off-diagonal gluons,
Ezawa and Iwazaki assumed abelian dominance that 
the only abelian gauge fields with monopoles 
would be essential for the description of nonperturbative 
phenomena in the low-energy region of QCD, 
and showed a possibility of monopole condensation in an infrared scale by 
investigating ``energy-entropy balance'' on the monopole current \cite{ezawa} 
in a similar way to 
the Kosterlitz-Thouless transition in the 1+2 dimensional superconductivity
 \cite{kosterlitz}.
Ezawa and Iwazaki formulated the dual London theory as an infrared 
effective theory of QCD, and later Maedan and Suzuki reformulated it as the
dual Ginzburg-Landau (DGL) theory in 1988 \cite{maedan}.

Furthermore, such  abelian dominance and monopole condensation 
have been investigated using 
the lattice QCD simulation in the maximally abelian (MA)
gauge \cite{kronfeld,schierholz,yotsuyanagi,hioki}.
The MA gauge is the abelian gauge where the diagonal component of gluon 
is maximized by the gauge transformation.
In the MA gauge, physical information of the gauge configuration is 
concentrated into the diagonal components as much as possible.
The lattice QCD studies indicate {\it abelian dominance} 
that the string tension \cite{yotsuyanagi,hioki,ichiead} and 
chiral condensate  \cite{miyamura,woloshyn} are almost described only 
by abelian variables in the MA gauge.
In the lattice QCD, 
{\it monopole dominance} is also observed such that 
only the monopole part in the abelian variable 
contributes to the nonperturbative QCD in the MA gauge \cite{miyamura,bali}.
Thus, the lattice QCD simulations show strong evidence on the dual 
Higgs theory for the nonperturbative QCD in the MA gauge \cite{diacomo,poly}.

As the result,
DGL theory is expected as a reliable infrared effective theory 
of QCD.
Recently, RCNP group studied the DGL theory \cite{maedan,suganuma}, and 
derived the simple
formula for the string tension 
and
also pointed out the relevant role of monopole condensation to 
chiral symmetry breaking by solving the Schwinger-Dyson equation with the 
nonperturbative gluon propagator \cite{suganuma,sasaki,umisedo}. 
This abelian dominance for 
chiral-symmetry breaking is confirmed by Miyamura and Woloshyn in more 
rigid framework of the lattice QCD \cite{miyamura,woloshyn}.
Considering the fact that instanton needs the nonabelian component for 
existence, RCNP group pointed out the correlation between instantons and
monopoles in the abelian dominant system in terms of the remaining large
off-diagonal component around the monopole, which is the topological 
defect. The evidence on the strong correlation between  instantons and 
monopoles have been observed also in the lattice QCD 
calculation \cite{markum} and in 
the analytical demonstration using the Polyakov-like 
gauge \cite{suganuma2,suganuma3} and the MA 
gauge \cite{fukushima,brower}.
Thus, the monopole seems to play the essential role for the 
nonperturbative QCD like confinement and chiral symmetry breaking
and topology \cite{diacomo,poly,ssasaki}.

A question arises if the color is confined in the QCD vacuum by 
monopole condensation all the time? 
As the superconducting state at low temperature is changed into the normal phase
at high temperature,
the QCD vacuum would also change
from the confinement phase to the quark-gluon-plasma (QGP) phase,
where the quark and gluon can move freely.
This phase transition is called QCD phase transition,
and becomes one of the most important subject
related to various fields such as
the early Universe and relativistic heavy-ion collision.
According to the big bang scenario, which is a successful model for 
cosmology, the quark gluon plasma phase at high temperature
is changed into the hadron 
phase  10$^{-6}$ second after the big bang.
At the Brookhaven National Laboratory in the United States,
some physicists are trying to form the quark gluon plasma as a
new matter by colliding high energy heavy-ions using RHIC. 
RCNP group also has studied the QCD vacuum at finite temperature
in terms of confinement-deconfinement phase transition and chiral phase transition
using the DGL theory.

In this way, 
the study of color confinement phenomena based on the 
dual Higgs picture is divided into two categories in terms of the
method of approach.

\begin{enumerate}

\item Study with the lattice QCD, which is an useful method
for the direct calculation of the QCD partition functional
$Z_{\rm QCD}$ \cite{rothe,creutz}.

\item Study with the infrared effective theory, 
the DGL theory \cite{maedan,suganuma},
which consists of the essential degrees of the freedom for
infrared phenomena.

\end{enumerate}

In the lattice formalism, space-time coordinate is discretized 
with the lattice spacing $a$, 
and the theory is described by the link variable 
$U_\mu(s) = e^{iaeA_\mu(s)}$, which corresponds to the line integral 
$P\exp\{i\int_s^{s+\hat \mu} dx_\mu eA_\mu(x)\}$ along the link.
The lattice QCD partition functional $Z$
in the Euclidean metric is given as 
\begin{eqnarray}
Z = \int dU_\mu e^{- \beta \hat S[U_\mu]},
\end{eqnarray} 
where $\beta \equiv \frac{2N_c}{e^2(a)}$ is the control parameter 
related to the lattice spacing $a$   \cite{rothe} (Appendix A.1).
Here, $e(a)$ is the QCD running coupling constant.
The standard lattice action is given as 
\begin{eqnarray}
\hat S = \sum_{s,\mu > \nu}[1- \frac{1}{2N_c} {\rm tr} \{
\Box_{\mu\nu}(s) +\Box ^\dagger_{\mu\nu}(s)  \}],
\label{eq:QCDaction}
\end{eqnarray}
where $\Box_{\mu\nu}(s)$ is plaquette defined as
\begin{eqnarray}
 \Box_{\mu\nu}(s) \equiv 
U_\mu(s) U_\nu (s+\hat \mu) U ^\dagger_\mu(s+ \hat \nu) U_\nu^\dagger(s).
\end{eqnarray}
In the limit $a \rightarrow 0$, i.e. $\beta \rightarrow \infty$, 
and the lattice action $\beta \hat S$ 
becomes the QCD action $S_{\rm QCD}$ in the continuum theory.
The gauge configuration of QCD is generated on the lattice 
using the Monte Carlo method (Appendix A.2). 
In the lattice QCD, 
the abelian gauge fixing can be performed 
after the generation of gauge configurations, 
and the abelian link variable is extracted by neglecting the 
off-diagonal part, which is called as the abelian projection. 
In the lattice formalism, the abelian link variable can be 
separated numerically into the photon part and the monopole part 
corresponding to the separation of the electric current 
and the monopole current as will be shown in Chapter \ref{sec:OMDMAGMP}. 
The dual-superconductor scenario for confinement 
has been examined in the lattice QCD by measurements 
of the abelian variable, monopole and so on in the abelian gauge.
 

On the other hand, the DGL theory is derived from the gluon sector of QCD,
and is composed of the dual gauge field ${\cal B}_\mu$ and
the monopole field $\chi$ in the pure gauge case. 
The Lagrangian is expressed as
\begin{eqnarray}
{\cal L}_{DGL}
={\rm tr}{\hat {\cal L}} \nonumber 
\end{eqnarray}
\begin{eqnarray}
\hat {\cal L}=-{1 \over 2}
(\partial _\mu  {\cal B}_\nu -\partial _\nu {\cal B}_\mu )^2 +
[\hat{D}_\mu, \chi]^{\dag}[\hat{D}_\mu, \chi]
-\lambda ( \chi^{\dag} \chi -v^2)^2,
\label{eq:dgllag}
\end{eqnarray}
where $\hat{D}_\mu=\hat{\partial}_\mu +ig{\cal B}_\mu$ is the dual covariant derivative.
Imposing the abelian gauge fixing on QCD,
the monopole appears as the line-like object in 4-dimensional space.
The monopole field is derived by summing  all the paths of the monopole
trajectories and monopole-field interaction is introduced taking the lattice 
result
``monopole condensation'' into the consideration.
Here, the off-diagonal component is neglected 
due to the lattice QCD result
``abelian dominance'', and 
the dual gauge field ${\cal B}_{\mu}$ is used instead of the abelian gauge field
$A_\mu$ adopting the Zwanziger formalism in order to describe gluon dynamics
without the singularity in the gauge field.

The DGL theory provides  an useful method of studying
the various confinement phenomena
such as inter-quark potential, hadron flux-tube system and 
the phase transition,
since it gives not only just the numerical results 
on the various quantities
but also 
their reasons. This is largely different from the lattice simulation.
The quark-antiquark potential arises from the strong correlation between
two quarks in the infrared region, 
which is revealed by the
DGL gluon propagator with 
double pole.
The hadron flux is constructed by a massive dual gauge field, whose mass is 
obtained by the dual Higgs mechanism of  monopole condensation.
However, in the process of the construction of DGL theory,
abelian dominance and  monopole condensation is assumed,
and  its origin is not clear.
Namely, we can not answer
the question what feature of the monopole degrees of freedom is important for
such confinement phenomena  or where  the effect of nonabelian feature 
appears.   


\begin{figure}[tb]
\centering \leavevmode
\epsfxsize = 14 cm
\epsfbox{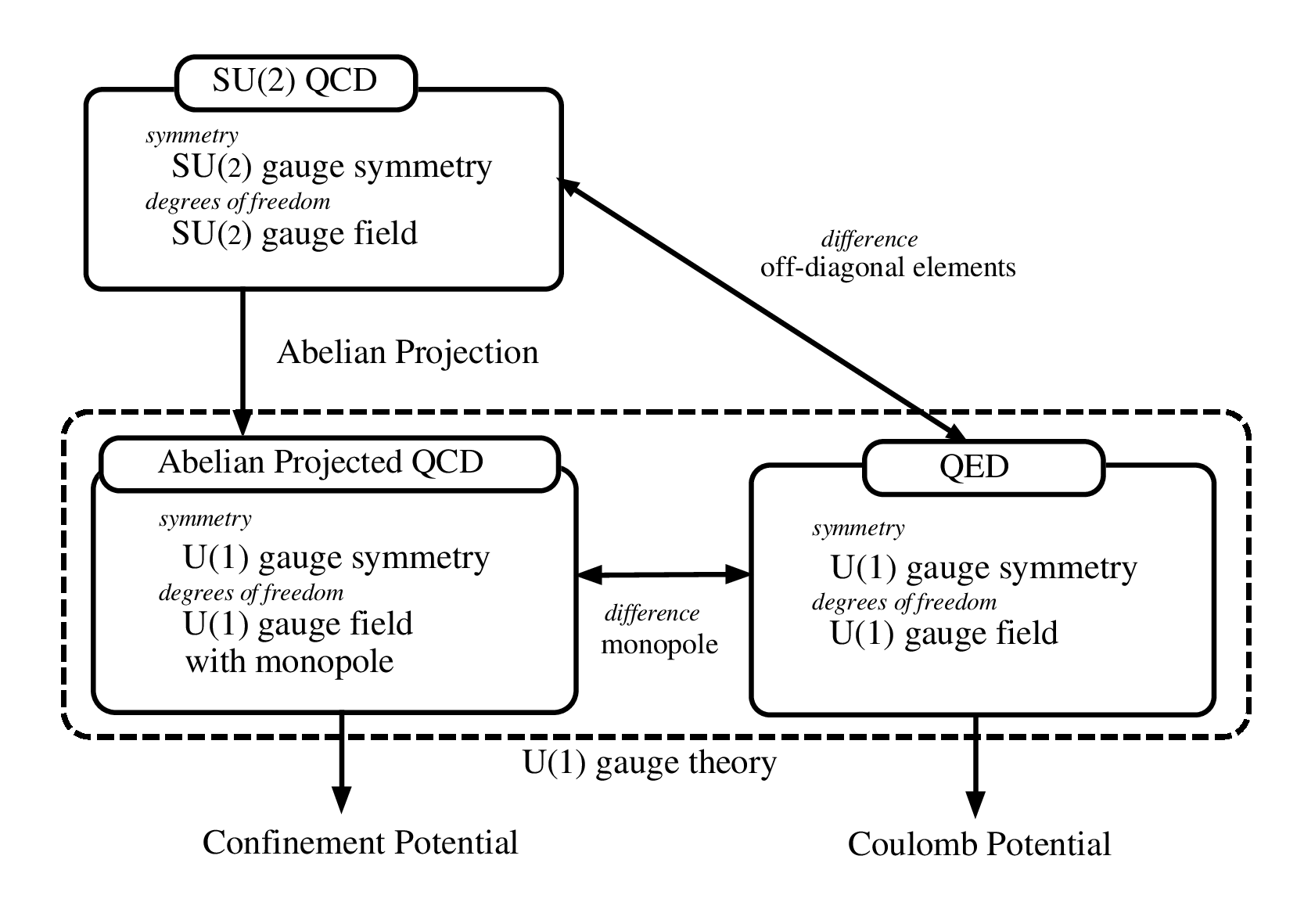}
\caption{    
Comparison among QCD,  abelian projected QCD (AP-QCD) and QED in terms of 
the gauge symmetry and fundamental degrees of freedom. ($N_c$ = 2)
}
\label{apqcd}
\end{figure}

In this paper, we try to understand the confinement mechanism based on the
dual Higgs mechanism using both methods, the lattice QCD simulation and
the DGL theory.
In the first half of this paper, we study the origin of abelian 
dominance and
monopole condensation in terms of the gluon configuration 
using the lattice QCD, and in the second half we apply the DGL theory to
the confinement phenomena such as QCD phase transition and multi-hadron
flux tube system. 

The point at issue in  the first half is the following three subjects.
\begin{enumerate}
\item  What is the origin of abelian dominance for infrared quantities
like  the string tension in the MA gauge?
\item Why does monopole appear in  abelian projected QCD(AP-QCD),
although AP-QCD is an abelian gauge theory like QED?
\item 
What is the role of monopoles  for the confinement phenomena?
What is the role of the  off-diagonal component of gluon  in  QCD in 
the MA gauge?
\end{enumerate}

In the MA gauge, 
AP-QCD neglecting 
the off-diagonal gluon component almost reproduces 
essence of the nonperturbative QCD, 
although AP-QCD is an abelian gauge theory like QED.
One may speculate that the strong-coupling nature 
leads to the similarity between AP-QCD and QCD, 
because the gauge coupling $e$ in AP-QCD  \cite{kondo} is 
the same as that in QCD in the lattice simulation. 
However, the strong-coupling nature would not be enough 
to explain the nonperturbative feature, 
because, if monopoles are eliminated from AP-QCD, 
nonperturbative features are lost in 
the remaining system called as photon part, 
although the gauge coupling $e$ is same as that in QCD.
For further understanding,
let us compare the theoretical structure of QCD, 
AP-QCD and QED in terms of the gauge symmetry 
and the fundamental degrees of freedom as shown in Fig.1.2. 
As for the interaction, the linear confinement potential arises 
both
in QCD and in AP-QCD, 
while only the Coulomb potential appears in QED.
On the symmetry, QCD has a nonabelian gauge symmetry,
while both AP-QCD and QED have abelian gauge symmetry. 
The obvious difference between QCD and QED is  existence of  
off-diagonal gluons in QCD. 
On the other hand, the difference between AP-QCD and QED is 
existence of the monopole, since the magnetic monopole does not 
exist in QED because of the Bianchi identity. 
This indicates the close relation between monopoles and off-diagonal 
gluons.
In particular, off-diagonal gluon components play a crucial role 
for  existence of the monopole \cite{ichiemp}, 
and the monopole itself is expected to play an alternative role of 
off-diagonal gluons for  confinement.

In Chapter \ref{sec:AGFRGS},
we review the abelian gauge fixing 
in line with 't~Hooft 
to discuss the confinement phenomena 
in terms of the monopoles based on the dual superconductor picture.
We show the gauge invariance condition in the abelian gauge.
As the abelian gauge fixing, we introduce the 
maximally abelian gauge, which is considered to be 
the best abelian gauge for the
infrared physics according to recent lattice QCD studies.
In addition, we generalize the maximally abelian gauge fixing.

In Chapter \ref{sec:OADMA}, 
we investigate the origin of abelian dominance in the MA gauge
in the SU(2) gauge theory.
We introduce the abelian projection rate,
which is defined as the overlapping factor  between SU(2) link variable and
abelian link variable, and investigate  abelian dominance for the abelian
link variable. 
We study abelian dominance for the Wilson loops
in terms of by approximating the off-diagonal angle variable as a random 
variable.
Using the U(1)$_3$ Landau gauge, we study the abelian gauge field and 
abelian field strength directly in the MA gauge. 
Then, we compare the abelian gauge field in the SU(2) Landau gauge,
where the gauge field is fixed most continuously and study the
feature of the MA gauge in terms of the gluon field fluctuation.

In Chapter \ref{sec:QCDMAG}, 
we investigate the mechanism of the appearance of monopole 
in AP-QCD 
both in the continuum theory and in the lattice QCD formalism. 
We show the appearance of monopole using the covariant derivative
in the abelian sector of QCD, clarifying the role of the off-diagonal 
components of QCD. 

In Chapter \ref{sec:LFFM},
we investigate the relations between 
monopoles and gauge-field fluctuations in the MA gauge,
measuring the probability distribution of gluons around the monopole.  
We show the distribution of the action density on the SU(2), U(1)$_{3}$ and 
off-diagonal part around the monopoles and consider the
appearance of the monopole  in terms of the action density.

In Chapter \ref{sec:OMDMAGMP},  
we study extraction of the monopole degrees of freedom from U(1)$_{3}$ gauge theory. 
The abelian gauge field is decomposed into the monopole and photon 
parts. We investigate the properties on the magnetic and electric currents,
action density, field variable itself in both parts.
We show the scaling properties on the monopole current and the Dirac sheet
and obtain the good scaling on variables related to the Dirac sheet.
Furthermore, 
we obtain the simple relation between the string tension and
monopole current density. 

In Chapter \ref{sec:MCD}, we study
monopole condensation in the QCD vacuum by comparing to the 
monopole-current system.
We first generate the monopole-current system on the lattice using 
a simple monopole current action, and
study the role of monopole current to  confinement.

In the second part of this paper, 
making the best use of the DGL theory,
we investigate the QCD phase transition using effective potential formalism,
and multi-flux-tube system,
which cannot be studied by the lattice QCD simulation.

In Chapter \ref{sec:DGLT}, 
we first review the derivation of the DGL theory starting 
from the QCD Lagrangian.
We derive the monopole field by summing all of the monopole trajectories, and
the dual gauge field is introduced using the Zwanziger formalism
in order to describe the gluon dynamics 
without the singularities originated from the monopole.
Then, using the DGL theory, we show the dual Meissner effect 
by monopole condensation and 
the structure of the flux tube as the dual version of the Abrikosov 
vortex in the superconductor.
We also demonstrate the quark confinement potential using the
DGL gluon propagator including nonperturbative effect.
Finally, we discuss the asymptotic behavior in terms of the DGL theory.  

In Chapter \ref{sec:QCDPTFT}, we consider the QCD vacuum at finite 
temperature in the DGL theory.
We formulate the effective potential at various temperatures by 
introducing the quadratic source term, which is a new useful method to 
obtain the effective potential in the negative-curvature region.
We find the thermal effects reduce the monopole condensate and bring a 
first-order phase transition. 

In Chapter \ref{sec:AQGP}, we apply the DGL theory to the multi-flux tube system.
We formulate this system by regarding it as the system of two flux tubes 
penetrating through a two-dimensional sphere surface.
We find the multi-flux-tube configuration becomes uniform above some 
critical flux-tube density.



\chapter{Abelian Gauge Fixing}
\label{sec:AGFRGS}

In  infrared QCD, there appear interesting nonperturbative phenomena
such as color confinement and chiral symmetry breaking due to 
the strong coupling nature.
At the same time, however, the large gauge-coupling constant
leads to breaking down of the perturbative technique.
As the result, it becomes difficult to treat the infrared phenomena 
analytically.
We have to use an effective model, which includes 
essence. Otherwise, we can perform the
partition functional of QCD directly using the huge supercomputer.
Historically,  the dual superconductor picture 
was proposed by Nambu and 't Hooft more than 20 years ago
to understand the confinement mechanism.
The QCD vacuum is regarded as a dual version of the superconductor and
color confinement is understood as the exclusion of the 
color-electric field by monopole condensation.
Later, 't Hooft and Ezawa-Iwazaki showed that 
QCD is reduced into the U(1) gauge theory with monopoles
by the abelian gauge fixing.
If ``abelian dominance'' and ``monopole condensation'' occurs in the 
QCD vacuum,
 color confinement would be realized through the dual Higgs mechanism.
Recently, these assumptions are supposed by the 
Monte Carlo simulation of the lattice QCD.

In this chapter, 
we study the abelian gauge fixing in QCD
in terms of the residual gauge symmetry.
%
%
%
In the abelian gauge, the SU($N_c$) gauge theory is reduced into the 
U(1)$^{ N_c-1}$ gauge theory  \cite{thooft,suganuma2}, and 
confinement phenomena can be studied by the dual abelian Higgs theory. 
In this gauge, 
the diagonal gluon component remains to be a U(1)$^{ N_c-1}$ 
gauge field, and
the off-diagonal gluon component behaves as a charged matter field
and provides the color electric current $j_\mu$
in terms of the residual U(1)$^{ N_c-1}$ gauge symmetry. 
In the abelian gauge, color-magnetic monopoles also appear as 
topological defects provided by the singular gauge transformation. 
Thus, QCD in the abelian gauge is 
reduced into an abelian gauge theory including both the electric and 
monopole currents, which is described by  
the extended Maxwell equation with the magnetic current.
For simplicity, we concentrate ourselves on the $N_c=2$ case hereafter.

\section[Residual Symmetry and Gauge Invariance Condition]
{Residual Symmetry and Gauge Invariance Condition in the Abelian Gauge}
\label{subsec:GICAG}

The abelian gauge fixing,
the partial gauge fixing which remains the abelian gauge symmetry, 
is realized by the diagonalization of a 
suitable SU($N_c$)-gauge dependent variable 
as $\Phi[A_\mu(x)] \in su(N_c) $
by the SU($N_c$) gauge transformation.
In the abelian gauge, $\Phi[A_\mu(x)]$ plays the 
similar role of the Higgs field on the determination of the gauge 
fixing.

For an hermite operator $\Phi[A_\mu(x)]$ which 
obeys the adjoint transformation,
$\Phi(x)$ is transformed as
\begin{eqnarray}
\Phi(x) = \Phi^a T^a 
\rightarrow \Phi^\Omega(x) & =  & \Omega(x) \Phi(x) \Omega^{\dagger}(x) 
\equiv \vec H \cdot \vec \Phi_{diag}(x)  \nonumber \\ 
&  = & {\rm diag}(\lambda^1(x), \cdots, \lambda^{N_c}(x)),
\end{eqnarray}
using a suitable gauge function 
$\Omega(x) = {\rm exp} \{ i \xi^a (x) T^a \} \in$ SU($N_{c}$).
Here, each diagonal component $\lambda^{i}$ 
($i$=1,$\cdots$,$N_c$) is to be real for the 
hermite operator $\Phi[A_\mu(x)]$.
In the abelian gauge, the SU($N_{c}$) gauge symmetry is reduced into 
the U(1)$^{N_{c}-1}$ gauge symmetry 
corresponding to the gauge-fixing ambiguity. 
The operator $\Phi(x)$ is diagonalized to 
$\vec H \cdot \vec \Phi_{diag}(x)$ also 
by the gauge function $\Omega^\omega (x) \equiv \omega(x)\Omega(x)$ with 
$\omega(x) = {\rm exp}(- i \, \vec H \cdot  \vec \varphi(x)  ) \in$
U(1)$^{N_{c}-1}$, 
\begin{eqnarray}
\Phi(x) \rightarrow  \Omega^\omega(x) \Phi(x)
\Omega^{\omega\dagger}(x) = \omega(x) \vec  H \cdot \vec \Phi_{diag}(x)
\omega^{\dagger}(x) = \vec H \cdot  \vec \Phi_{diag}(x),
\end{eqnarray}
and therefore U(1)$^{N_{c}-1}$ abelian gauge symmetry remains 
in the abelian gauge.

In the abelian gauge, there also remains the global Weyl symmetry as a 
``relic'' of the nonabelian theory \cite{suganuma2,suganuma4}.
Here, the Weyl symmetry corresponds to the subgroup of SU($N_{\rm c}$)
relating to the permutation of the basis in the fundamental representation.
Then, the Weyl group is expressed as 
the permutation group {\bf P}$_{N_c}$ including $N_c!$ elements.
For simplicity, let us consider the $N_c=2$ case.  
For SU(2) QCD, the Weyl symmetry corresponds to the 
interchange of the SU(2)-quark color, 
$| + \rangle \equiv $
$({}^1_0)$ and $| - \rangle \equiv $ $({}^0_1)$, in the fundamental 
representation.
The global Weyl transformation is expressed by the global gauge function,
\begin{eqnarray}
W  & =  & 
e^{ i \{  \frac{\tau_1}{2} \cos \alpha + \frac{\tau_2}{2}\sin \alpha 
 \} \pi }  
=
i \{ \tau_1 \cos \alpha + \tau_2 \sin \alpha \}  \nonumber \\
 & = & i \left( 
{\matrix{
0  & e^{-i \alpha}  \cr
e^{i \alpha} & 0 \cr
}}
\right)
\hspace{0.2cm} \in  {\bf P}_2 \subset \mbox{SU}(2)
\label{eq:weyl}
\end{eqnarray}
with an arbitrary constant $\alpha \in {\bf R}$. 
By the global Weyl transformation $W$, 
the SU(2)-quark color is interchanged 
as $W | + \rangle = i e^{i \alpha} | - \rangle $ and 
$W | - \rangle = i e^{-i \alpha} | + \rangle $ except 
for the global phase factor.
This global Weyl symmetry remains in the abelian gauge, 
because the operator $\Phi(x)$ is also diagonalized by using
$\Omega^W(x) \equiv W \Omega(x)$, 
\begin{eqnarray}
\Phi(x) \rightarrow  \Omega^{W}(x) \Phi(x)
\Omega^{W{\dagger}}(x) = W \Phi_{diag}(x) \frac{\tau_3}{2} W ^{\dagger}
= - \Phi_{diag}(x) \frac{\tau_3}{2}.
\end{eqnarray}
Here, the sign of $\Phi_{diag}(x)$, or the order of 
the diagonal component $\lambda^{i}(x)$, is globally changed 
by the Weyl transformation. 
It is noted that 
the sign of the U(1)$_3$ gauge field 
${\cal A}_\mu \equiv A^3_\mu \frac{\tau_3}{2}$ is globally changed
under the Weyl transformation,
\begin{eqnarray}
{\cal A}_\mu \rightarrow {\cal A}_\mu^{W} = W  A^3_\mu \frac{\tau_3}{2} W^\dagger = 
- A^3_\mu \frac{\tau_3}{2} = -{\cal A}_\mu.
\end{eqnarray}
Therefore, all the sign of the abelian field strength,
electric and  magnetic charges are also globally changed:
\begin{eqnarray}
{\cal F}_{\mu\nu}  & \equiv & F_{\mu\nu}\frac{\tau_3}{2}  \rightarrow   
{\cal F}_{\mu\nu}^{W} = W  {\cal F}_{\mu\nu}  W^\dagger 
= - {\cal F}_{\mu\nu},  \nonumber \\
j_\mu & \equiv & \partial^\alpha {\cal F}_{\alpha\mu}   
\rightarrow   j_\mu^{W}=-j_\mu, \nonumber \\
k_\mu & \equiv & \partial^\alpha {}^*{\cal F}_{\alpha\mu}   
\rightarrow   k_\mu^{W}=-k_\mu.
\end{eqnarray}
In the abelian gauge, the absolute signs of the electric 
and the magnetic charges are settled, only when the Weyl symmetry 
is fixed by the additional condition.
When $\Phi[A_{\mu}(x)]$ obeys the adjoint-type gauge transformation 
like the nonabelian Higgs field, the global Weyl symmetry can be 
easily fixed by imposing the additional gauge-fixing condition 
as $\Phi_{\it diag} (x) \ge 0$ for SU(2), or 
the ordering condition of the diagonal components $\lambda^{i} $ 
in $\vec H \cdot \vec \Phi_{\it diag}$ 
as $\lambda^{1} \ge..\ge \lambda^{N_c}$ for the SU($N_{c}$) case. 
As for the appearance of monopoles in the abelian gauge, 
the global Weyl symmetry ${\bf P}_{N_c}$ is not relevant, because 
the nontriviality of the homotopy group is 
not affected by the global Weyl symmetry. 
However, the definition of the magnetic monopole charge, 
which is expressed by the nontrivial dual root 
of ${\rm SU}(N_c)_{\rm dual}$ \cite{ezawa}, 
is globally changed by the Weyl transformation.

Now, we consider the abelian gauge fixing 
in terms of the coset space of the fixed gauge symmetry.
The abelian gauge fixing is a sort of the partial 
gauge fixing which reduces the gauge group 
$G \equiv $SU($N_{c})_{\rm local}$ of the system 
into its subgroup $H \equiv 
$U(1) $^{N_{c}-1}_{\rm local} (\times {\rm P}_{N_c}^{\rm global})$ 
including the maximally torus subgroup of $G$. 
In other words, the abelian gauge fixing freezes the gauge symmetry 
relating to the coset space $G/H$, 
and hence the representative gauge function $\Omega$ which brings 
the abelian gauge belongs to the coset space $G/H$: $\Omega \in G/H$. 
In fact, $\Omega \in G/H$ is uniquely determined 
without the ambiguity on the residual symmetry $H$, if 
the additional condition on $H$ is imposed for $\Omega$.

However, such a partial gauge fixing makes the total gauge 
invariance unclear. Here, let us consider 
the SU($N_c$) gauge-invariance condition on the 
operator defined in the abelian gauge \cite{suganuma4}.
To begin with, 
we investigate the gauge-transformation property of the 
gauge function $\Omega \in G/H$ which brings the abelian 
gauge (see Fig.\ref{gfig2}). 
For simplicity, the operator $\Phi $ to be diagonalized is assumed to 
obey 
the adjoint gauge transformation as $\Phi \to \Phi ^g=g\Phi 
g^\dagger$ with $^\forall g\in G$.
After the gauge transformation by $^\forall g\in G$, 
$\Omega ^g\in {G / H}$ is defined so as to diagonalize $\Phi ^g$ as
$\Omega ^g\Phi ^g(\Omega ^g)^\dagger 
=\Phi 
_{diag}$, 
and hence the gauge function 
$\Omega ^g\in {G / H}$ which realizes the abelian gauge is 
transformed as
\begin{eqnarray}
\Omega \rightarrow \Omega ^g=h[g]\Omega g^\dagger
\label{eq:function}
\end{eqnarray}
under arbitrary SU($N_c$) gauge transformation by $g\in G$.
Here, $h[g] \in H$ is chosen so as to make $\Omega^g$ belong to  
{\it G/H}, i.e., $\Omega^g \in G/H$. 
(If the additional condition on $H$ is imposed 
to specify $\Omega \in G/H$, $\Omega g^\dagger$ 
does not satisfy it in general.) 
This is similar to the argument on the hidden local 
symmetry \cite{bando} in the nonlinear representation.
In general, the gauge function $\Omega \in G/H$ is transformed 
nonlinearly by the gauge function $g$ due to $h[g]\in H$.
Thus, the gauge-transformation property 
on the gauge function $\Omega \in G/H$ 
becomes nontrivial in the partial gauge fixing.

Owing to the nontrivial transformation (\ref{eq:function}) of $\Omega 
\in$$G/H$, any operator $O^{\Omega}$ defined in the abelian gauge is 
found to be transformed as $O^{\Omega} \to (O^{\Omega})^{h[g]}$ by
the SU($N_c$) gauge transformation of ${}^\forall g \in G$.
We demonstrate this for the gluon field $A_\mu ^\Omega \equiv 
\Omega 
(A_\mu +{i \over e}\partial _\mu )\Omega ^\dagger$ in the  abelian 
gauge.
By the gauge transformation of $^\forall g\in G$, $A_\mu ^\Omega $ 
is 
transformed as
\begin{eqnarray}
A_\mu ^\Omega \to (A_\mu ^g)^{\Omega ^g}=A_\mu ^{\Omega ^gg}=A_\mu 
^{h[g]\Omega } = (A_{\mu}^\Omega)^{h[g]}
 =h[g](A_\mu ^\Omega +{i \over e}\partial _\mu )h^\dagger[g].
\end{eqnarray}
Here, we have used 
\begin{eqnarray}
(A_\mu ^{g_{1}})^{g_{2}}
& = &  g_{2}( A_\mu ^{g_{1}} +{i \over e}\partial _\mu ) g_{2}^\dagger
=
( g_{2} g_{1}) (A_\mu  +{i \over e}\partial _\mu ) ( g_{2} 
g_{1})^\dagger = (A_\mu)^{ g_{2} g_{1}} 
\end{eqnarray} 
for the successive gauge transformation by 
$g_{1}$, $g_{2} \in {\rm SU}(N_c)$.
Similarly, the operator $O^\Omega $ defined in the abelian gauge is 
transformed by
$^\forall g\in G$ as
\begin{eqnarray}
O^\Omega \to (O^g)^{\Omega ^g} &  = & \Omega ^gO^g\Omega 
^{g\dagger} = h[g]\Omega g^\dagger  \cdot gOg^\dagger \cdot g\Omega 
^\dagger h^\dagger[g]  \nonumber \\
 & = & h[g]\Omega O\Omega ^\dagger h^\dagger[g]=h[g]O^\Omega 
h^\dagger[g]=(O^\Omega )^{h[g]},
\end{eqnarray}
as shown in Fig.\ref{gfig2}.
Here, $O$ is assumed to obey the adjoint transformation as 
$O^g = gOg^\dagger$ 
for simplicity.

\begin{figure}[bt]
\centering \leavevmode
{\tt    \setlength{\unitlength}{0.92pt}
\begin{picture}(214,352)
\thinlines    \put(70,22){{{\large $h[g]$} $\in$ {\large $H$}}}
              \put(20,81){{\large $\Omega $}}
              \put(70,150){{{\large $g$} $\in$ {\large $G$}}}
              \put(150,271){{{\large $\Omega^g $} $ \equiv $ {\large $h[g] 
\Omega 
              g^{-1}$} $\in$ {\large $G/H$} }}
              \put(20,296){{{\large $\Omega$} $\in$ {\large 
              $G/H$}}}
              \put(70,349){{{\large $g$} $\in$ {\large $G$}}}
              \put(210,82){{\large $\Omega^g $} $\equiv$ {\large $h[g] 
\Omega 
              g^{-1}$}}
              \put(5,9){{\large $O^\Omega$}}
              \put(190,9){{\Large $(O^{\Omega})^{h[g]}$}}
              \put(190,141){{\Large $O^g$ }}
              \put(5,141){{\Large $O$ }}
              \put(5,219){{\Large $\Phi_{\rm diag}$}}
              \put(190,339){{\Large $\Phi^g$}}
              \put(5,339){{\Large $\Phi $}}
              \put(36,12){\vector(1,0){141}}
              \put(196,126){\vector(0,-1){95}}
              \put(10,125){\vector(0,-1){95}}
              \put(36,141){\vector(1,0){141}}
              \put(10,216){{}}
              \put(10,221){{}}
              \put(177,323){\vector(-3,-2){125}}
              \put(10,325){\vector(0,-1){85}}
              \put(36,340){\vector(1,0){141}}
\end{picture}}
\caption{  
The gauge transformation property of $\Phi$ and
gauge function $\Omega \in$ $G/H$.
(a) After the gauge transformation by $^\forall g\in G$,  
the operator $\Phi^g$ is diagonalized by the gauge function   $\Omega 
^g=h[g]\Omega g^\dagger$
$\in$ $G/H$.
(b) The gauge transformation property of the operator $O^\Omega$
defined in the abelian gauge.
If $O^\Omega$ is $H$-invariant,
$O^\Omega$ is 
found to be invariant under the whole gauge transformation of $G$.
}
\label{gfig2}
\vspace{0cm}
\end{figure}

Thus, arbitrary SU($N_c$) gauge transformation by $g\in G$ is mapped into 
the partial gauge transformation by $h[g]\in H$ for the operator 
$O^\Omega $ defined in the abelian gauge, and $O^\Omega $ transforms 
nonlinearly as $O^\Omega \to (O^\Omega )^{h[g]}$ by the SU($N_c$) gauge 
transformation $g$.
If the operator $O^\Omega $ is $H$-invariant, one gets 
$(O^\Omega)^{h[g]}=O^\Omega $ for any $h[g] \in H$, 
and hence $O^\Omega$ is also $G$-invariant 
or total SU($N_c$) gauge invariant, 
because $O^\Omega $ is transformed into $(O^\Omega)^{h[g]}=O^\Omega$ 
by $^\forall g\in G$. 
Thus, we find a useful criterion on the SU($N_c$) gauge 
invariance of the operator defined in the abelian gauge \cite{suganuma4}:
If the operator $O^\Omega$ defined in the abelian gauge is $H$-invariant,
$O^\Omega$ is 
also invariant under the whole gauge transformation of $G$.

Here, let us consider the application of this criterion 
to the effective theory of QCD in the abelian gauge, 
the dual Ginzburg-Landau (DGL) theory \cite{maedan,suganuma} (see Chapter
\ref{sec:DGLT}). 
In the DGL theory, 
the local U(1)$^{N_c-1}$ and the global Weyl symmetries remain, 
and the dual gauge field ${\cal B}_{\mu}$ and the monopole field 
$\chi_\alpha$ [$\alpha$=1, $\cdots$, $\frac12 N_{c}(N_{c}-1)$]
are the relevant modes for infrared physics. 
Although ${\cal B}_{\mu}$ is invariant under the local transformation of 
${\rm U(1)}^{N_c-1} \subset {\rm SU}(N_c)$, 
${\cal B}_{\mu} \equiv \vec B_{\mu} \cdot \vec H$ is variant 
under the global Weyl transformation, 
and therefore $B_{\mu}$ is SU($N_c$)-gauge dependent object 
and does not appear in the real world alone.
As for the monopole field, 
there exists one Weyl-invariant combination of 
the monopole field fluctuation, 
$\tilde \chi \equiv \sum_{\alpha}  \tilde \chi_{\alpha}$ \cite{ichieB},
which is also {\rm U(1)}$^{N_c-1}$-invariant. 
Therefore, the monopole fluctuation $\tilde \chi$ 
is completely residual-gauge invariant in the abelian gauge, 
so that $\tilde \chi$ is SU($N_c$)-gauge invariant 
and is expected to appear as a scalar glueball with $J^{PC}=0^{++}$, 
like the Higgs particle in the standard model.

\section{Maximally Abelian Gauge}

The abelian gauge has some arbitrariness
corresponding to the choice of the operator $\Phi$ to be diagonalized. 
As the typical abelian gauge, 
the maximally abelian (MA) gauge, the Polyakov gauge and the F12 gauge  
have been tested on the dual superconductor scenario for the 
nonperturbative QCD.
Recent lattice QCD studies show that 
infrared phenomena such as confinement properties and chiral symmetry
breaking are almost reproduced in the MA 
gauge \cite{kronfeld,schierholz,yotsuyanagi,hioki,miyamura,woloshyn,bali}. 
In the SU(2) lattice formalism, the MA gauge is defined so as to maximize
\begin{eqnarray}
R_{\rm MA}[U_\mu]  & \equiv &
\sum_{s,\mu} {\rm tr} \{  U_\mu(s) \tau_3 U^{\dagger}_\mu(s) \tau_3 \}  
\nonumber \\
   & = & 
2\sum_{s,\mu}\{
U^0_\mu(s)^2+U^3_\mu(s)^2-U^1_\mu(s)^2-U^2_\mu(s)^2 \} \nonumber \\
   & = & 
2\sum_{s,\mu} \left[
1 - 2\{ U^1_\mu(s)^2+U^2_\mu(s)^2 \} \right]
\end{eqnarray}
by the SU(2) gauge transformation (Appendix B).
Here, the link variable 
$U_{\mu}(s) \equiv U_{\mu}^0(s) + i \tau^a U_{\mu}^a (s) \in$ SU(2) with
$U_{\mu}^0(s)$, $U_{\mu}^a(s) \in$ {\bf R} relates to the (continuum) 
gluon field $A_\mu \equiv A_\mu^a T^a \in$ $su$(2) as $U_{\mu}(s) = 
e^{iaeA_{\mu}(s)}$, where $e$ denotes the QCD gauge coupling and $a$ the 
lattice spacing.   
In the MA gauge, the absolute value of off-diagonal components, 
$U_\mu^1(s)$ and $U_\mu^2(s)$, are
forced to be small.
In  the continuum limit $a \rightarrow 0$,
the link variable reads 
$U_\mu(s) = e^{iaeA_\mu(s)} = 1+iae A_{\mu}(s) + O(a^2)$,
and hence the MA gauge is found to minimize the functional, 
\begin{eqnarray}
 R_{ch}[A_\mu] \equiv \frac12 e^2 \int d^4 x \{ A_\mu^1(x)^2 + 
A_\mu^2(x)^2 \}
= e^2 \int d^4 x A_\mu^+(x) A_\mu^-(x),
\end{eqnarray}
with $A_\mu^\pm (x) \equiv {1 \over {\sqrt 2}}
  \{ A_\mu^1(x) \pm 
i A_\mu^2(x) \}$.
Thus, in the MA gauge, the off-diagonal gluon component is globally forced 
to be small by the gauge transformation, and hence the QCD system 
is expected to be describable only by its diagonal part
approximately.

The MA gauge is a sort of the abelian gauge which diagonalizes the 
hermite operator
\begin{eqnarray}
\Phi[U_{\mu}(s)] \equiv \sum _{\mu,\pm } U_{\pm \mu}(s) \tau_3 U^{\dagger}_{\pm 
\mu}(s).
\end{eqnarray}
Here, we use the convenient notation $U_{-\mu}(s) \equiv
U^{\dagger}_\mu(s- \hat \mu)$ in this paper.
In the continuum limit, the condition of the MA gauge becomes 
$\displaystyle \sum_\mu$$(i\partial_\mu \pm e A^3_{\mu} ) A^{\pm }_\mu = 
0$.
This condition can be regarded as the maximal decoupling condition 
between the abelian gauge sector and the charged gluon sector.

In the MA gauge, $\Phi(s)$ is diagonalized as 
$\Phi_{\rm MA}(s) = \Phi_{diag}(s) \tau_{3}$ with  $\Phi_{diag}(s) \in$ 
{\bf R}, and there remain the local U(1)$_3$ symmetry and the  global Weyl 
symmetry \cite{suganuma4}.
As a remarkable fact, $\Phi(s)$ does not obey the adjoint 
transformation  in the MA gauge, and  
the sign of $\Phi_{diag}(s)$ 
is not changed by the  Weyl transformation by $W $ in 
Eq.(\ref{eq:weyl}), 
\begin{eqnarray}
\lefteqn{\Phi_{\rm MA}(s) = \Phi_{diag}(s) \tau_3 } \nonumber \\ 
  & \rightarrow &   
\Phi_{\rm MA}^W (s)  =  
\sum _{\mu,\pm }  W  U_{\pm \mu}(s)  W ^{\dagger}
\tau_3  W  U^{\dagger}_{\pm \mu}(s)
 W^{\dagger} \nonumber \\
&  & = 
- \sum _{\mu,\pm } W  U_{\pm \mu}(s)   
\tau_3  U^{\dagger}_{\pm \mu}(s)
 W ^{\dagger}
= - W \Phi_{diag}(s) \tau_3 W^\dagger  = \Phi_{diag}(s) \tau_3. 
\end{eqnarray}
Thus, the Weyl symmetry is not fixed in the MA gauge by the simple 
ordering condition as $ \Phi_{diag} \ge 0$, unlike the adjoint case.  
We find the gauge invariance condition on the operator $O^\Omega$
defined in the MA gauge:
if $O^\Omega$ is invariant both under the local U$(1)^{N_c-1}$ 
gauge transformation and the global Weyl transformation,
$O^\Omega$ is also invariant under 
the SU($N_c$) gauge transformation.

In the continuum SU($N_c$) QCD, 
it is more fundamental and convenient to define the MA gauge fixing 
by way of the SU($N_c$)-covariant derivative operator 
${\hat D_\mu} \equiv {\hat \partial_\mu} + ie A_\mu $, 
where $\hat \partial_\mu$ is the derivative operator 
satisfying $[\hat \partial_\mu, f(x)]= \partial_\mu f(x)$. 
The MA gauge is defined so as to make 
SU($N_{c}$)-gauge connection $\hat D _{\mu}= \hat \partial_{\mu} + ie 
A_{\mu}^{a} T^{a}$ close to U(1)${}^{N_{c}-1}$-gauge connection 
$\hat D _{\mu}^{\rm Abel}= \hat \partial_{\mu} + ie \vec A_{\mu} \cdot 
\vec H$ by minimizing 
\begin{eqnarray}
 R_{\rm ch} \equiv \int d^4 x \; {\rm tr}[{\hat D_\mu}, \vec 
H]^\dagger
[{\hat D_\mu}, \vec H]
= e^{2} \int d^{4}x \; {\rm tr}[A_\mu, \vec H]^{\dagger}[A_\mu, \vec H] 
\nonumber \\
= e^{2} \int d^4 x \; \sum_{\alpha, \beta}C_\mu^{\alpha*}C_\mu^{\beta}
\vec \alpha \cdot \vec \beta {\rm tr} (E_\alpha^{\dagger} E_\beta)
= \frac{e^{2}}{2} \int d^{4}x \sum_{\alpha = 1}^{N_c(N_{c}-1)}
|C_{\mu}^{\alpha}|^{2}, 
\label{eq:rch}
\end{eqnarray}
which expresses the total amount of the off-diagonal gluon component. 
Here, we have used the Cartan decomposition,
$\displaystyle
A_{\mu} \equiv A_\mu^{a} T^{a} = \vec A_{\mu}  \cdot \vec H +
\sum_{\alpha = 1}^{N_c(N_{c}-1)} C_{\mu} ^{\alpha} E^{\alpha}$; 
$\vec H \equiv  (T_3, T_8, \cdots, T_{N_c^2-1})$ is the Cartan 
subalgebra, and $E^{\alpha}(\alpha=1,2,\cdots,N_{c}^{2}-N_c)$ 
denotes the raising or lowering operator.
In this definition with $\hat D _{\mu}$, the gauge transformation 
property of $ R_{\rm ch}$ becomes quite clear, 
because the SU($N_c$) covariant derivative ${\hat D_\mu}$
obeys the simple adjoint gauge transformation, 
${\hat D_\mu} \rightarrow 
\Omega {\hat D_\mu} \Omega^{\dagger}$,
with the SU($N_c$) gauge function $\Omega \in$ SU($N_c$).
By the SU($N_c$) gauge transformation, $R_{\rm ch}$ is transformed as
 \begin{eqnarray}
R_{\rm ch} \rightarrow R_{\rm ch}^{\Omega} =
 \int d^{4} x \; {\rm tr}
 \left( [\Omega \hat D_{\mu} \Omega^{\dagger}, \vec H]^{\dagger}
[\Omega \hat D_{\mu} \Omega^{\dagger}, \vec H] \right) \nonumber \\
=     \int d^{4} x \; {\rm tr} 
 \left( [\hat D_{\mu}, \Omega^{\dagger}\vec H \Omega]^{\dagger}
[\hat D_{\mu}, \Omega^{\dagger}\vec H \Omega]\right),
\end{eqnarray} 
and hence
the residual symmetry corresponding to the invariance of $R_{\rm ch}$ 
is found to be U(1)$^{N_{c}-1}_{\rm local} \times P^{N_{c}}_{\rm 
global} \subset $SU($N_{c})_{\rm local}$, where  $P^{N_{c}}_{\rm 
global}$ denotes the global Weyl group relating to the permutation of 
the $N_{c}$ basis in the fundamental representation.
In fact, one finds $\omega^{\dagger} \vec H \omega = \vec H$ for 
$\omega = e^{-i \vec \varphi(x) \cdot \vec H} \in$   
U(1)$^{N_{c}-1}_{\rm local}$, and the global Weyl transformation by 
$W \in$ $P^{N_{c}}_{\rm global}$ only exchanges the permutation of the 
nontrivial root $\vec \alpha_{j}$ and never changes $R_{\rm ch}$.
In the MA gauge, arbitrary gauge transformation by ${}^\forall \Omega \in 
G$ is to increase $R_{\rm ch}$ as $R_{\rm ch}^{\Omega} \ge R_{\rm ch}$.
Considering arbitrary infinitesimal gauge transformation $\Omega = 
e^{i \varepsilon} \simeq 1 + i \varepsilon $ with ${}^\forall \varepsilon 
\in$$su$($N_{c}$), one finds $\Omega^{\dagger} \vec H \Omega \simeq \vec 
H + i[\vec H, \varepsilon]$ and
\begin{eqnarray}
R^{\Omega}_{\rm ch} \simeq  R_{\rm ch} + 2i\int d^{4}x 
{\rm tr} \left( [\hat D_{\mu}, [\vec H, \varepsilon]] [\hat D_{\mu}, \vec H] 
\right)
\nonumber \\
= R_{\rm ch} + 2i\int d^{4}x 
{\rm tr}\left( \varepsilon
[\vec H, [\hat D_{\mu},[\hat D_{\mu},\vec H] ]] \right).
\end{eqnarray}
For the $N_{c}$=2 case, 
the MA gauge extremum-condition of $R^{\Omega}_{\rm ch}$ on 
${}^\forall \varepsilon \in $$su$(2)
provides 
\begin{eqnarray}
[\tau_3, [\hat D_{\mu}, [\hat D_{\mu},\tau_3] ]] = 0,
\end{eqnarray}
which leads to $\sum_{\mu}(i \partial_{\mu} \pm eA^{3}_{\mu}) 
A_{\mu}^{\pm} = 0$.
Thus, the operator $\Phi$ to be diagonalized in the MA gauge is found 
to be
\begin{eqnarray}
\Phi[A_{\mu}] = [\hat D_{\mu}, [\hat D_\mu, \tau_3]] 
\end{eqnarray}
in the continuum theory.
Here, $ \Phi[A_{\mu}]$ is hermite as $\Phi^\dagger [A_{\mu}]
=\Phi[A_{\mu}]$ because of $\hat D_{\mu}^\dagger = -\hat 
D_{\mu}$,
and hence the diagonal elements of $\Phi[A_{\mu}]$ should be real.

In the commutator form, the diagonal part of the variable 
$\hat O[A_\mu(x)]$ is expressed as \cite{ichiemp}
\begin{eqnarray} 
\hat O^{\vec H} = \hat O - [\vec H, [\vec H, \hat O]]. 
\label{eq:APdif}
\end{eqnarray}
For the covariant derivative operator, one finds 
\begin{eqnarray} 
\hat D_\mu^{\vec H} = \hat D_\mu - [\vec H, [\vec H, \hat D_\mu]] 
=\hat \partial_\mu+ie \vec A_\mu(x) \cdot \vec H 
\end{eqnarray}
with $A_\mu(x)=\vec A_\mu(x) \cdot \vec H + C_\mu^\alpha(x)E^\alpha$. 
Then, the abelian projection, 
$\hat D_\mu \rightarrow \hat D_\mu^{\vec H}$, is expressed 
by the simple replacement as 
$A_\mu(x) \in su(N_c) \rightarrow {\cal A}_\mu(x) \equiv
\vec A_\mu(x) \cdot \vec H$ $\in$ $u(1)^{N_{c}-1}$. 

\section{Generalization of the Maximally Abelian Gauge}

In the MA gauge, $R_{\rm ch}[A_\mu]$ in Eq.(\ref{eq:rch}) is 
forced to be reduced by the MA gauge transformation 
$\Omega_{\rm MA}(x) \in G/H$ \cite{ichiead},
and therefore the gluon field $A_\mu(x)$ 
is maximally arranged in the diagonal direction 
$\vec H$ in the internal SU$(N_c)$ color space. 
In the definition of the MA gauge, $\vec H$ is the 
specific color-direction, 
since $\vec H$ explicitly appears in the MA gauge-fixing 
condition with $R_{\rm ch}[A_\mu]$. 
On this point of view, the MA gauge can be called as the 
``maximally diagonal gauge''. 
However, for the extraction of the abelian gauge theory from the 
nonabelian theory, 
we need not take the specific direction as $\vec H$ in the internal 
color-space, although the system becomes transparent 
when the specific color-direction as $\vec H$ is introduced 
on the maximal arrangement of the gluon field $A_\mu(x)$. 

In this section, we consider the generalization of the 
framework of the MA gauge and the abelian projection, 
without explicit use of the specific direction $\vec H$
in the internal color-space 
on the gauge fixing \cite{ichiemp}.
Instead of the special color-direction $\vec H$,
we introduce the 
``Cartan frame field'' 
$\vec \phi(x) \equiv (\phi_1(x), \phi_2(x), \cdots, \phi_{N_c-1}(x))$,
where $\phi_i(x) \equiv \phi^a_i(x) T^a$ $(\phi_i^a(x) \in {\bf R})$
commutes each other as $[\phi_i(x), \phi_j(x)]=0$, and satisfy the
orthonormality condition
2tr$(\phi_i(x) \phi_j(x))=\sum_{a=1}^{N_c-1} \phi^a_i(x) \phi^a_j(x) 
= \delta_{ij}$. 
At each point $x_\mu$, $\vec \phi(x)$ forms the Cartan sub-algebra, and 
can be expressed as
\begin{eqnarray}
\vec \phi(x) = \Omega_C^\dagger(x) \vec H \Omega_C(x)
\end{eqnarray} 
using $\Omega_C(x)$ $\in$ $G/H$.
For the fixed Cartan frame field $\vec \phi(x)$, 
we define the generalized maximally abelian (GMA) gauge 
so as to minimize the functional 
\begin{eqnarray}
R_{\phi {\rm ch}}[A_\mu] \equiv \int d^4 x 
{\rm tr}[{\hat D_\mu}, \vec \phi(x)]^\dagger 
[{\hat D_\mu}, \vec \phi(x)]
\end{eqnarray}
by the SU$(N_c)$ gauge transformation.
Here, the Cartan frame field $\vec \phi(x)$ is defined at each $x_\mu$
independent of the gluon field like $\vec H$, and never changes under the 
SU($N_c$) gauge transformation.
For the special case of $\vec \phi(x)=\vec H$, 
the GMA gauge returns to the usual MA gauge. 
In the GMA gauge, the SU($N_c$) covariant derivative $\hat D_\mu$ 
is maximally arranged to be ``parallel'' 
to the $\vec \phi(x)$-direction in the internal color-space 
using the SU($N_c$) gauge transformation.

In the GMA gauge, the gauge symmetry 
is reduced from SU($N_c$) 
into ${\rm U}(1)_\phi^{N_c-1}$, and
the generalized AP-QCD leads to the monopole 
in the similar manner to the MA gauge. 
In the GMA gauge,  the remaining U(1)$^{N_c-1}_{\phi}$ gauge symmetry 
corresponds to the invariance of $R_{\phi {\rm ch}}[A_\mu]$ 
under the U(1)$^{N_c-1}_{\phi}$ gauge transformation by 
\begin{eqnarray}
\omega_\phi(x) \equiv e^{i \vec \phi(x) \cdot 
\vec \chi(x)} \in {\rm U}(1)_\phi^{N_c-1},
\qquad  \vec \chi(x) \in {\bf R}^{N_c-1}.
\end{eqnarray}
In fact, using
$\omega_{\phi}^\dagger(x) \vec \phi(x) \omega_{\phi}(x) 
= \vec \phi(x)$, U(1)$^{N_c-1}_\phi$ invariance of $R_{\phi{\rm ch}}[A_\mu]$
is easily confirmed as
\begin{eqnarray}
(R_{\phi{\rm ch}}[A_\mu])^\omega
&  = & \int d^4x {\rm tr}
[\omega(x)_{\phi}\hat D_\mu \omega^{\dagger}_{\phi}(x), \vec \phi(x)]^\dagger
[\omega(x)_{\phi} \hat D_\mu \omega^{\dagger}_{\phi}(x), \vec \phi(x)]   
\\
&  = & \int d^4x {\rm tr}
[\hat D_\mu , \omega_{\phi}^\dagger (x) \vec \phi(x) \omega_{\phi}(x)]^\dagger
[\hat D_\mu , \omega_{\phi}^\dagger (x) \vec \phi(x) \omega_{\phi}(x)] 
  =  R_{\phi{\rm ch}}[A_\mu]. \nonumber 
\end{eqnarray}
There also remains the global Weyl symmetry ${\bf P}_{N_{c}}$
similarly in the usual MA gauge,
although the gauge function takes a complicated from.

Here, we consider the generalized abelian projection 
to $\vec \phi(x)$-direction. 
Similar to the ``diagonal part'' in Eq.(\ref{eq:APdif}), 
we define the ``$\vec \phi(x)$-projection'' of the operator $\hat O(x)$ as 
\begin{eqnarray} 
\hat O^{\phi}(x) = \hat O(x) - [\vec \phi(x), [\vec \phi(x), \hat 
O(x)]],
\end{eqnarray}
using the commutation relation.
For the SU($N_c$) covariant derivative operator $\hat D_\mu \equiv \hat 
\partial_\mu + ie A_\mu$, its $\vec \phi(x)$-projection is defined as 
\begin{eqnarray} 
\hat D_\mu^\phi \equiv \hat D_\mu -[\vec \phi(x), [\vec \phi(x), 
\hat D_\mu]]= \hat \partial_\mu + ie {\cal A}_\mu^\phi(x) + 
[\vec \phi(x), \partial_\mu \vec \phi(x)]
\end{eqnarray}
with ${\cal A}_{\mu}^\phi(x) \equiv \vec A_{\mu}^\phi(x) \cdot \vec 
\phi = 2 {\rm tr}(\vec \phi(x) A_{\mu}(x)) 
\cdot \vec \phi(x)$.
Here, the nontrivial term $[\vec \phi(s), \partial_\mu \vec \phi(x)]$ 
appears in $\hat D_\mu^\phi$ owing to the $x$-dependence of the Cartan-frame
field $\vec \phi(x).$
The U(1)$^{N_c-1}_\phi$ gauge field is defined as the difference 
between $\hat D^\phi_\mu $ and $\hat \partial_\mu$,
\begin{eqnarray} 
\tilde {\cal A}^\phi_\mu(x) \equiv \frac{1}{ie}(\hat D^\phi_\mu - \hat 
\partial_\mu) = {\cal A}_\mu^\phi(x) + \frac{1}{ie}[\vec \phi(x), \partial_\mu 
\vec \phi(x)] \hspace{.5cm} \in \hspace{.5cm} su(N_c).
\end{eqnarray}
Here, $\tilde {\cal A}_\mu^\phi(x)$ includes both the $\vec 
\phi(x)$-component 
${\cal A}_\mu^\phi(x) = 2 {\rm tr}( A_\mu(x) \vec \phi(x)) \cdot \vec \phi(x)$ 
and the non-$\vec \phi(x)$-component  $\frac{1}{ie}[\vec \phi(x), \partial_\mu 
\vec \phi(x)]$, because $[\vec \phi(x), \partial_\mu \vec \phi(x)]$ 
does not include $\vec \phi(x)$-component as
${\rm tr} \left( \phi_i(x) [\vec \phi(x), \partial_\mu \vec \phi(x)] 
\right)=0$. 
Here,  $\vec A_\mu^\phi(x)$ is the image of 
$\tilde {\cal A}_\mu^\phi(x)$ mapped into the U(1)$_\phi^{N_c-1}$-manifold. 
The  generalized abelian projection for the variable $O[A_\mu(x)]$ 
is defined via the two successive mapping, 
$O[A_\mu(x)] \rightarrow O[\tilde {\cal A}_\mu^\phi(x)] 
 \rightarrow \vec O_{AP} \equiv 
 2{\rm tr}(\vec \phi(x) O[\tilde {\cal A}_\mu^\phi(x)])$, 
after the GMA gauge fixing.

Under the U$(1)_\phi^{N_c-1}$ abelian gauge transformation by 
$\omega_\phi(x)=e^{i \vec \phi(x) \cdot \vec \chi(x)} \in 
{\rm U}(1)_\phi^{N_c-1}$, 
$\tilde {\cal A}_\mu^\phi(x)$ or $\vec A_\mu^\phi(x)$ 
behaves as the U(1)$^{N_c-1}_\phi$ abelian gauge field,
\begin{eqnarray}
\tilde {\cal A}^\phi_\mu(x) \rightarrow (\tilde {\cal A}^\phi_\mu(x))^\omega
= \tilde {\cal A}_\mu^\phi(x) + \frac{1}{e} \partial_\mu \vec \chi_\mu(x) \cdot
\vec \phi(x).
\end{eqnarray}
The abelian field-strength matrix is defined as 
\begin{eqnarray} 
\tilde {\cal F}_{\mu\nu}^\phi(x) & \equiv & 
\frac{1}{ie} \left( [\hat D_\mu^\phi, \hat D_\nu^\phi]
-[\hat \partial_\mu, \hat \partial_\nu] \right) \nonumber \\
& = & \partial_\mu \tilde {\cal A}^\phi_\nu(x) - \partial_\nu \tilde 
{\cal A}^\phi_\mu(x) + ie[\tilde {\cal A}_\mu^\phi(x), \tilde {\cal A}_\nu^\phi(x)],
\end{eqnarray}
which generally includes the non-$\vec \phi$-component as well as 
$\tilde {\cal A}_\mu^\phi(x)$. 
The $\vec \phi$-component of $\tilde {\cal F}_{\mu\nu}^\phi(x)$ is 
the image of $\tilde {\cal F}_{\mu\nu}^\phi(x)$ projected into the 
U(1)$_\phi^{N_c-1}$ gauge manifold, and is 
observed as the ``real abelian field-strength'' in the 
abelian-projected gauge theory. 
The explicit form of $\vec F_{\mu\nu}^\phi(x)$ is derived as 
\begin{eqnarray}
\vec F_{\mu\nu}^\phi(x)  & \equiv  & 2 {\rm tr}
\left( \tilde {\cal F}_{\mu\nu}^\phi(x) \vec \phi(x) \right)  \nonumber \\
& = & \partial_\mu \vec A_\nu^\phi(x) - 
      \partial_\nu \vec A_\mu^\phi(x) +
      \frac{4}{ie} {\rm tr} (\vec \phi(x) [\partial_\mu \phi_{i}(x), 
      \partial_\nu \phi_{i}(x)]) \\
& = & \partial_\mu \vec A_\nu^\phi(x) - 
      \partial_\nu \vec A_\mu^\phi(x) + \frac{2}{e} f_{abc}
      \vec \phi^a \partial_\mu \phi_{i}^b \partial_\nu \phi_{i}^c, 
\end{eqnarray}
where the last term breaks the abelian Bianchi identity and 
provides the monopole current. 
The magnetic monopole current is derived as 
\begin{eqnarray} 
\vec k_\mu^\phi(x) \equiv \partial^{\alpha*} \vec F_{\alpha\mu}^\phi(x) 
= -\frac{1}{e}  \varepsilon_{\mu\alpha\beta\gamma}  f_{abc}
\partial^\alpha \vec \phi^a(x) \partial^\beta \phi_{i}^b(x) 
\partial^\gamma \phi_{i}^c(x), 
\end{eqnarray}
which is the topological current induced by $\vec \phi$. 
Hence, the monopole appears from the center of the hedgehog 
configuration of $\vec \phi$ as shown in Fig.\ref{Higgs} 
in the SU(2) case.

Next, we investigate the properties of the GMA gauge function $\Omega_{\rm 
GMA}(x)$,
which brings the GMA gauge.
Here, $\Omega_{\rm GMA}(x)$ is a complicated function of $A_\mu(x)$ and is 
expressed by an element of the coset space 
$G/H= {\rm SU}(N_c)/ \{ {\rm U}(1)^{N_c-1}_\phi \times
{\rm Weyl} \} $ as the representative element 
because of the residual gauge symmetry.
For instance, we impose here
\begin{eqnarray}
{\rm tr}(\Omega_{\rm GMA}(x) \vec \phi(x)) = \vec 0
\end{eqnarray}
for the selection of $\Omega_{\rm GMA}$ $\in$ $G/H$. 
Similarly to the MA gauge function \cite{ichiead}, 
$\Omega_{\rm GMA}[A_\mu]$ obeys
the nonlinear transformation as
\begin{eqnarray}
\Omega_{\rm GMA}(x) \in G/H \rightarrow (\Omega_{\rm GMA}(x))^V =
d^V(x)\Omega_{\rm GMA}(x)V^\dagger(x) \,\, \in \,\, G/H 
\end{eqnarray}
by the SU($N_c$) gauge transformation with $V(x)$ $\in$ $G$.
Here, $d^V(x) \in  H
\equiv$ U(1)$^{N_{c}-1}_\phi$ $\times$ Weyl 
appears to keep $(\Omega_{\rm GMA})^V$ belonging to $G/H$.
Therefore, the gluon field 
$A^{\rm GMA}_\mu = \Omega_{\rm GMA}(A_\mu+\frac{1}{ie}
\partial_\mu)
\Omega^\dagger_{\rm GMA} $ $\in$ $g$ in the GMA gauge is transformed as
\begin{eqnarray}
A_\mu^{\rm GMA} \rightarrow (A_\mu^{\rm GMA})^V  & = &
\Omega^V_{\rm GMA}(x)
(A_\mu^V+\frac{1}{ie}\partial_\mu) \Omega_{\rm GMA}^{V\dagger}(x) 
\nonumber \\
& = & d^V(x) (A_\mu^{\rm GMA}+\frac{1}{ie}\partial_\mu) d^{V\dagger}(x) = 
(A_\mu^{\rm GMA})^{d^V}
\label{eq:subabelian}
\end{eqnarray}
by the SU($N_c$) gauge transformation.
As a remarkable feature, the SU$(N_c)$ gauge transformation by $V(x)$ 
$\in$ $G$ is mapped as 
the abelian  sub-gauge transformation by $d^V(x)$ $\in$ $H$  in the GMA gauge:
$(A_\mu^{\rm GMA})^V = (A_\mu^{\rm GMA})^{d^V}$. 
In particular, for the residual gauge transformation by $\omega(x) = 
e^{i \vec \phi(x) \cdot \vec \chi(x)}$ $\in$
$H$, we find $d^\omega(x) = \omega(x)$ to keep the representative-element 
condition
tr$(\Omega^\omega_{\rm GMA}(x) 
\vec \phi(x)) = \vec 0$ imposed above, 
and then $A_\mu^{\rm GMA}$ obeys the ordinary $H$-gauge 
transformation
\begin{eqnarray}
A_\mu^{\rm GMA}(x) \rightarrow (A_\mu^{\rm GMA}(x))^\omega = 
\omega(x) (A_\mu^{\rm GMA}+\frac{1}{ie}\partial_\mu) 
\omega^{\dagger}(x).
\end{eqnarray}
For the arbitrary variable $\hat O[A_\mu^{\rm GMA}]\equiv \hat 
O[A_\mu^{\Omega_{\rm GMA}}]$ defined in the GMA gauge, 
we find $\hat O[A_\mu^{\rm GMA}]^V =
\hat O[A_\mu^{\rm GMA}]^{d^V} $ with $d^V$ $\in$ $H$ from 
Eq.(\ref{eq:subabelian}),
and hence we get a similar criterion on the SU($N_c$) gauge invariance:
if $\hat O[A_\mu]$ is $H$-invariant as 
$\hat O[A_\mu^{\rm GMA}]^\omega = \hat O[A_\mu]$ for ${}^\forall \omega$ 
$\in$ $H$, $\hat O[A_\mu^{\rm GMA}]$ is also $G$-invariant, because of 
$\hat O[A_\mu^{\rm GMA}]^V = 
\hat O[A_\mu^{\rm GMA}]^{d^V} = \hat O[A_\mu^{\rm GMA}]$
for ${}^\forall V$ $\in$ $G$.


The correspondence between $\Omega _{\rm GMA}$ and 
$\Omega _{\rm MA}$ is 
straightforward.  
Using  
$\Omega _C(x) \in {\rm SU}(N_{c})$ satisfying 
$\vec \phi(x)=\Omega _C^\dagger(x) \vec H \Omega_C(x)$,
$\Omega _{\rm GMA}$ is expressed as 
\begin{eqnarray}
\Omega^{\rm GMA}(x)=\Omega _C^\dagger(x)\Omega ^{\rm MA}(x).
\end{eqnarray}
Then, for regular $\vec \phi (x)$, $\Omega _C(x)$ becomes
regular, and the singularity of $\Omega _{\rm MA}$ 
is directly mapped to that of $\Omega _{\rm GMA}$. 
However, if singular $\vec \phi(x)$ is used, 
the singularity of $\Omega _{\rm MA}$ can be mapped in $\vec \phi(x)$
 or $\Omega_C(x)$
instead of $\Omega _{\rm GMA}$.
In this case, the gluon field $A_\mu^{\rm GMA}$ is kept to be regular, and 
the Cartan frame field
 $\vec \phi(x)$ includes the multi-valuedness or the 
singularity, which leads to the monopole. 

\chapter{Origin of Abelian Dominance in the MA gauge}
\label{sec:OADMA}

In the abelian gauge, the diagonal and the off-diagonal gluons 
play different  roles in terms of the residual abelian gauge symmetry: 
the diagonal gluon behaves as the abelian gauge field,
while off-diagonal gluons behave as charged matter 
fields \cite{kronfeld}.
Under the U(1)$_3$ gauge transformation by   
$\omega = {\rm exp}(- i \varphi \frac{\tau_3}{2}) \in$ U(1)$_3$,
one finds
\begin{eqnarray}
A_\mu^3 & \rightarrow & (A_\mu^{\omega})^3 = A_\mu^3  + \frac{1}{e} 
\partial_\mu 
\varphi \\ 
A_\mu^{\pm} & \rightarrow & (A_\mu^{\omega})^\pm = A_\mu^{\pm} e^{ \pm 
i\varphi}
\end{eqnarray} 
with $A^\pm_\mu = \frac{1}{\sqrt{2}} (A^1_\mu \pm i A^2_\mu)$.
The abelian projection is simply defined as the 
replacement of the gluon field $A_\mu = A_\mu^a {\tau^a \over 2} \in $ 
$su(2)$ by the diagonal part ${\cal A}_\mu \equiv  A_\mu^3 {\tau^3 \over 2} 
\in u(1)_3 \subset su(2)$.
 
We call
``abelian dominance for an operator $\hat O[A_\mu]$'', when the
expectation value $\langle \hat O \rangle$ is almost unchanged by 
the  abelian projection $A_{\mu} \to {\cal A}_{\mu}$ as 
$\langle \hat O[A_\mu] \rangle \simeq \langle \hat O[{\cal A}_\mu] 
\rangle_{\rm A.G.}$, when 
$\langle$ $\rangle_{\rm A.G.}$ denotes the expectation value in the 
abelian gauge. Ordinary abelian dominance is observed for the long-distance
physics in the MA gauge, and this would be physically interpreted as the 
effective-mass generation of the off-diagonal gluon 
induced by the MA gauge fixing \cite{amemiya,suganuma1}.

In the lattice formalism, the SU(2) link-variable $U_\mu(s)$ can be 
factorized as 
\begin{eqnarray}
U_\mu(s) & = & 
M_\mu(s)  u_\mu(s)   \hspace{2cm} \in G \nonumber \\
M_\mu(s) & = & 
{\rm exp} \left(
i \{  \tau_1 \theta^1_\mu(s)+ \tau_2 \theta^2_\mu(s)  \} 
\right) \hspace{1cm} 
\in  G/H, \nonumber \\
u_\mu(s) & = & {\rm exp} \left( {i \tau^3 \theta^3_\mu (s) } \right) 
\hspace{3cm} \in  H
\label{eq:para}
\end{eqnarray}
with respect to the Cartan decomposition of $G = G/H$ $\times$ $H$ into 
$G/H=$SU(2)/U(1)$_3$ 
and $H=$U(1)$_3$.
Here, the abelian link variable, 
\begin{eqnarray}
u_\mu(s) = e^{i \tau^3  \theta^3_\mu (s) } =
 \left( {\matrix{
e^{i\theta^3_\mu  (s)}  &          0
\cr
0                     &  e^{-i\theta^3_\mu (s)} \cr
}} \right)  \hspace{1cm}  \in {\rm U(1)}_3 \subset {\rm SU(2)}, 
\end{eqnarray}
plays the similar role as the SU(2)-link variable 
$U_\mu(s) \in {\rm SU(2)}$ 
in terms of the residual U(1)$_3$ gauge symmetry 
in the abelian gauge, and $\theta^3_\mu(s) \in (-\pi, \pi]$ 
corresponds to the diagonal component of the gluon 
in the continuum limit. 
On the other hand, the off-diagonal factor 
$M_\mu(s) \in $SU(2)/U(1)$_3$ is expressed as 
\begin{eqnarray}
 M_\mu(s)  & = & 
{\rm exp} \left(
{i \{  \tau_1 \theta^1_\mu(s)+ \tau_2 \theta^2_\mu(s)  \} }
\right)
\nonumber \\
& = & 
\left( 
{\matrix{
{\rm cos}{\theta_\mu(s)} & -{\rm sin}{\theta_\mu(s)}e^{-i\chi_\mu(s)} \cr
{\rm sin}{\theta_\mu(s)}e^{i\chi_\mu(s)} & {\rm cos}{\theta_\mu(s)} 
}}
 \right) 
\label{divided} 
 \\ & = &
\left( {\matrix{
\sqrt{1-|c_\mu(s)|^2} & -c_\mu^*(s) \cr
c_\mu (s) & \sqrt{1-|c_\mu(s)|^2}
}} \right)  \nonumber 
\end{eqnarray}
with  
$\theta_\mu (s) \equiv {\rm mod}_\frac{\pi}{2} \sqrt{ (\theta^1_\mu )^2 + 
(\theta^2_\mu)^2} \in [0, \frac{\pi}{2}]$ and $\chi_\mu (s) \in ( - \pi, \pi]$.
Near the continuum limit, the off-diagonal elements of $M_\mu(s)$ 
correspond to the off-diagonal gluon components. 
Under 
the residual U(1)$_3$ gauge transformation by 
$\omega(s) = e^{-i \varphi(s) \frac{\tau_3}{2}}\in$ U(1)$_3$, 
$u_\mu(s)$ and $M_\mu(s)$ are transformed as
\begin{eqnarray}
u_\mu(s) &\rightarrow & 
u^\omega _\mu(s) = \omega(s) u_\mu(s) \omega^\dagger(s+ \hat \mu 
)  \hspace{1cm}  \in H\\
M_\mu(s) &\rightarrow &
M^\omega_\mu(s) = \omega(s) M_\mu(s) \omega^\dagger(s) \hspace{1cm} \in G/H
\end{eqnarray}
so as to keep $M_\mu^\omega(s)$ belonging to $G/H$.
Accordingly, 
$\theta^3_\mu(s)$ and $c_\mu(s) \in {\bf C}$ are transformed as 
\begin{eqnarray}
\theta^3_\mu(s) &\rightarrow & 
\theta^{3 \omega}_\mu(s)=
{\rm mod}_{2\pi} [ \theta^3_\mu(s) + \{ \varphi(s+ \hat \mu)-\varphi(s) \} /2 ]
\\
c_\mu(s) &\rightarrow & 
c_\mu^{\omega}(s) = c_\mu(s)e^{i\varphi(s)}.
\end{eqnarray}
Thus, on the residual U(1)$_3$ gauge symmetry, 
$u_\mu(s)$ behaves as the U(1)$_3$ lattice gauge field, and 
$\theta^3_\mu(s)$ behaves as the U(1)$_3$ gauge field 
in the continuum limit. 
On the other hand, 
$M_\mu(s)$ and $c_\mu(s)$ behave as the charged matter field 
in terms of the residual U(1)$_3$ gauge symmetry, which is 
similar to the charged weak boson $W_\mu^{\pm}$ in the standard model.

In this parameterization
(\ref{eq:para}), there are two U(1)-structures embedded in SU(2) 
corresponding to $e^{i \theta^3_\mu }$ and $e^{ i  \tilde { \chi}_\mu}$. 
To clarify this structure, we reparametrize the SU(2) link variable as
\begin{eqnarray}
U_\mu(s) & = &
 \left( {\matrix{
{\rm cos}{\theta_\mu}e^{i \theta^3_\mu } &
 -{\rm sin}{\theta_\mu}e^{-i {\tilde \chi}_\mu} \cr
{\rm sin}{\theta_\mu}e^{i {\tilde \chi}_\mu} & 
{\rm cos}{\theta_\mu}e^{-i \theta^3_\mu }
 }} \right), 
\label{eq:repara}
\end{eqnarray}
or equivalently
\begin{eqnarray}
U^0_\mu & = & \cos \theta _\mu \cos \theta^3_\mu, \hspace{1cm} 
U^1_\mu  =  \sin \theta _\mu \sin \tilde \chi_\mu, \nonumber \\
U^3_\mu & = & \cos \theta _\mu \sin \theta^3_\mu,  \hspace{1cm} 
U^2_\mu  =  \sin \theta _\mu \cos \tilde \chi_\mu,
\end{eqnarray}
with $\tilde \chi_\mu \equiv \chi_\mu +\theta^3_\mu$.
The range of the angle variable can be redefined as 
$0 \le \theta_\mu \le \frac{\pi}{2}$
 and $-\pi < \theta^3_\mu,  \tilde \chi_\mu \le \pi$.
Here, $(U^0_\mu,U^1_\mu,U^2_\mu,U^3_\mu)$ forms an element of the 
3-dimensional 
hyper-sphere $S^3 \simeq $SU(2), because of
$(U_\mu^0)^2 +(U_\mu^1)^2 + (U_\mu^2)^2 + (U_\mu^3)^2  = 1$.
 For a fixed $\theta_\mu$, both 
$(U^0_\mu,U^3_\mu)$ and
$(U^1_\mu,U^2_\mu)$ form the two $S^1 \simeq $U(1) subgroups embedded in 
$S^3$ in
a symmetric manner.
From the parametrization in Eq.(\ref{eq:repara}), 
the SU(2) measure can be easily found as 
\begin{eqnarray}
\int d U_\mu & \equiv & 
\int d U_\mu^0 U_\mu^1 U_\mu^2 U_\mu^3  \,\,\, \delta \,(  \,\sum_{a = 0}^3  
\,( \, U^a_\mu )^2 -1 ) \nonumber \\
& = &\frac{1}{2 \pi^2} 
\int_{0}^{\frac{\pi}{2}} d \theta_\mu \sin  \theta_\mu \cos \theta_\mu 
\int_{-\pi}^{ \pi} d \tilde \chi_\mu 
\int_{-\pi}^{ \pi} d \theta^3_\mu.
\label{eq:measure}
\end{eqnarray}
In the lattice formalism,
the abelian projection is defined by replacing the SU(2) link variable 
$U_\mu(s) \in$ SU(2) by the abelian link variable $u_\mu(s) \in$ U(1)$_3$.

\section[Microscopic Abelian Dominance]{Microscopic Abelian Dominance in the MA gauge}

In the MA gauge, the off-diagonal gluon component is strongly suppressed, and the 
SU(2) link variable is expected to be U(1)$_3$-like as 
$U_\mu(s) \simeq u_\mu(s)$ in the relevant gauge configuration.
In the quantitative argument, this can be expected as 
$\langle U_{\mu}(s) u_{\mu}^{\dagger}(s) \rangle_{\rm {MA}} \simeq 1$,
where $\langle \: \rangle_{\rm {MA}}$ denotes the expectation value in 
the MA gauge.
In order to estimate the difference between $U_{\mu}(s)$ and 
$u_{\mu}(s)$,
we introduce the ``abelian projection rate'' $R_{\rm {Abel}}$ 
\cite{ichiead,poulis,ichie6},
which is 
defined as the overlapping factor as 
\begin{eqnarray}
R_{\rm Abel} (s,\mu) \equiv \frac12 {\rm Re} \; {\rm tr} \{ U_{\mu}(s) 
u_{\mu}^{\dagger}(s) \} = \frac12 {\rm Re} \; {\rm tr} M_{\mu}(s) = \cos 
\theta_\mu(s).
\end{eqnarray}
This definition of $R_{\rm {Abel}}$ is inspired by the ordinary 
``distance'' between two 
matrices $A, B \in {\rm GL}(N,{\bf C})$ defined as
$d^{2}(A,B) \equiv \frac12 {\rm tr} \{ (A-B)^{\dagger} (A-B) 
\}$ \cite{georgi}, 
which 
leads to 
$d^{2}(A,B) = 2-{\rm Re} \;{\rm tr}(AB^{\dagger})$ for $A,B \in$SU(2).
The similarity between $U_\mu(s)$ and $u_\mu(s)$ 
can be quantitatively measured 
in terms of the ``distance'' between them.
For instance, if $\cos \theta _\mu (s) =1$,
the SU(2) link variable becomes completely abelian as
\begin{eqnarray}
U_\mu(s) =
 \left( {\matrix{
e^{i\theta^3_\mu}  &          0
\cr
0                     &  e^{-i\theta^3_\mu} \cr
}} \right), \nonumber
\end{eqnarray}
while, if $\cos \theta _\mu(s) =0$, it becomes completely off-diagonal as 
\begin{eqnarray}
U_\mu(s)  =
 \left( {\matrix{
0 & -e^{-i \tilde \chi_\mu}  
\cr
e^{i\tilde \chi_\mu} & 0\cr
}} \right). \nonumber
\end{eqnarray}

\begin{figure}[bt]
\epsfxsize = 10 cm
\centering \leavevmode
\epsfbox{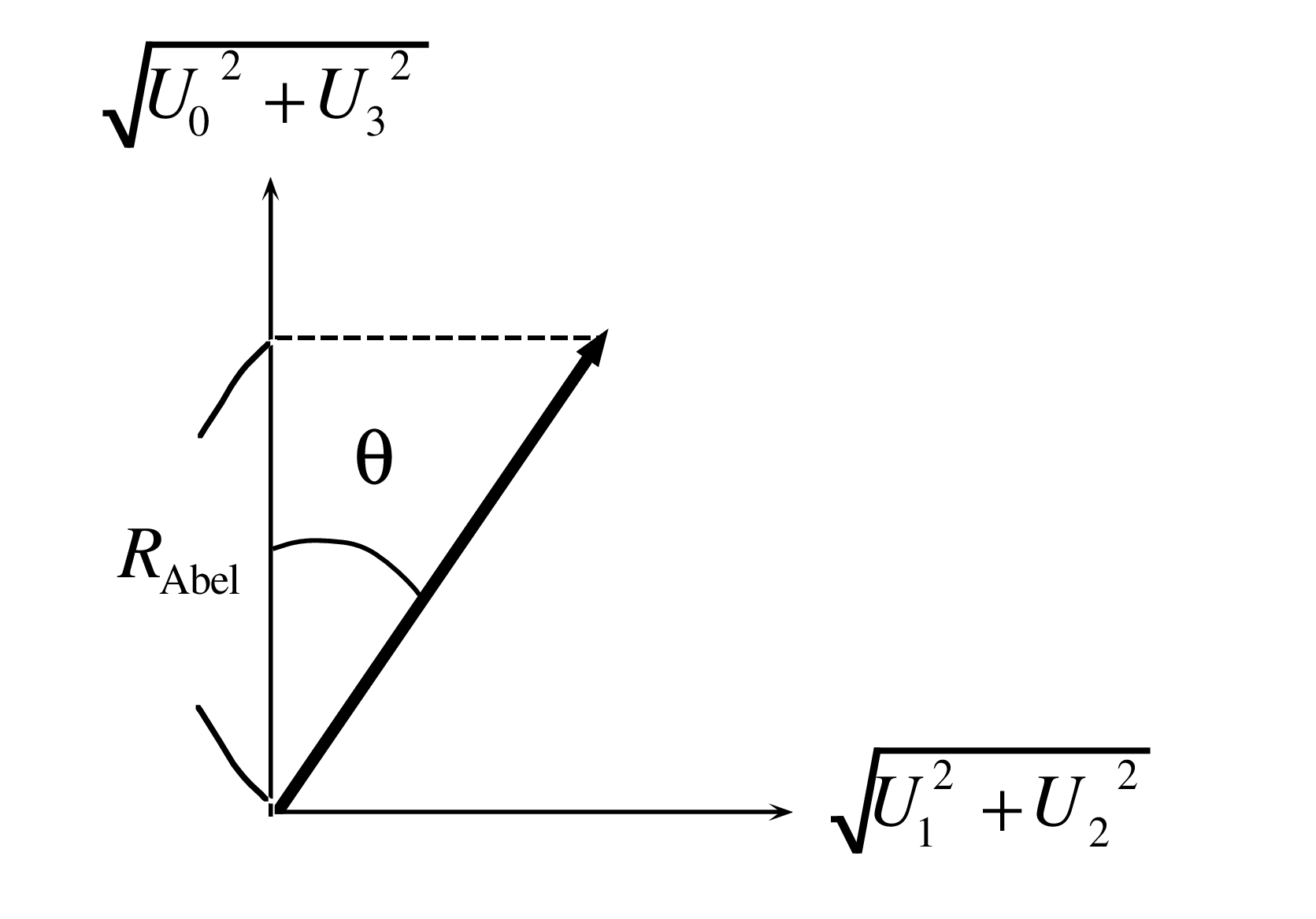}
\vspace{0cm}
\caption{
Geometrical explanation of the abelian projection rate $R_{\rm Abel}$
in terms of $U \equiv U_0 + iU_a \tau_a$.
The deviation from the vertical axis indicates the
magnitude of the off-diagonal component.
This description will be used in Figs \ref{arrow5} and \ref{arrow6}.
}
\label{arrow4hosoku}
\vspace{0cm}
\end{figure}

\begin{figure}[p]
\epsfxsize = 13 cm
\centering \leavevmode
\epsfbox{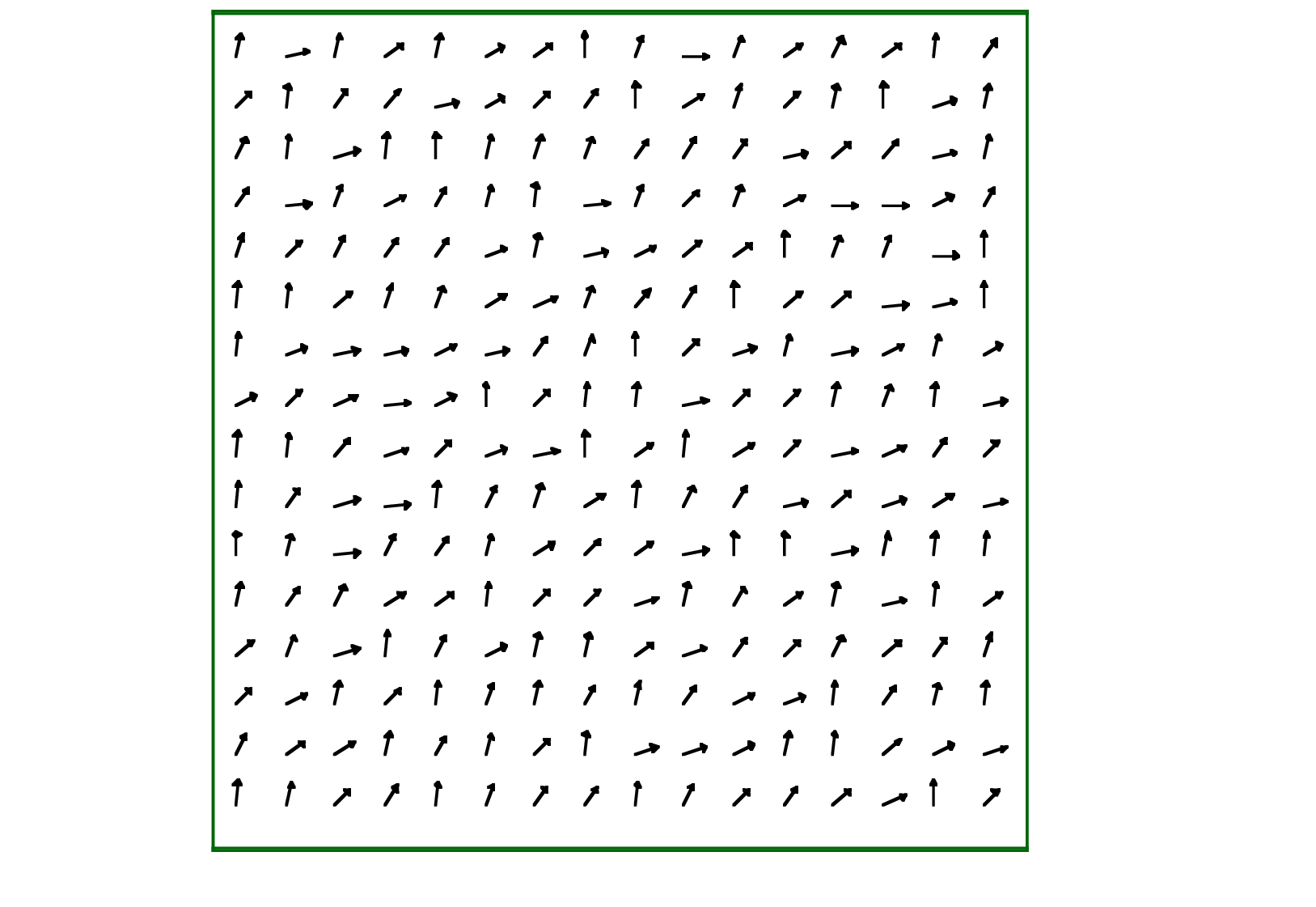}
\vspace{0cm}
\caption{
Local abelian rate $R_{\rm Abel} =  \cos \theta_\mu(s)$ at 
$\beta = $2.4 on $16^4$ lattice without 
gauge fixing. 
The meaning of the arrow is shown in Fig.\ref{arrow4hosoku}.
}
\label{arrow5}
\vspace{1cm}
\epsfxsize = 13 cm
\centering \leavevmode
\epsfbox{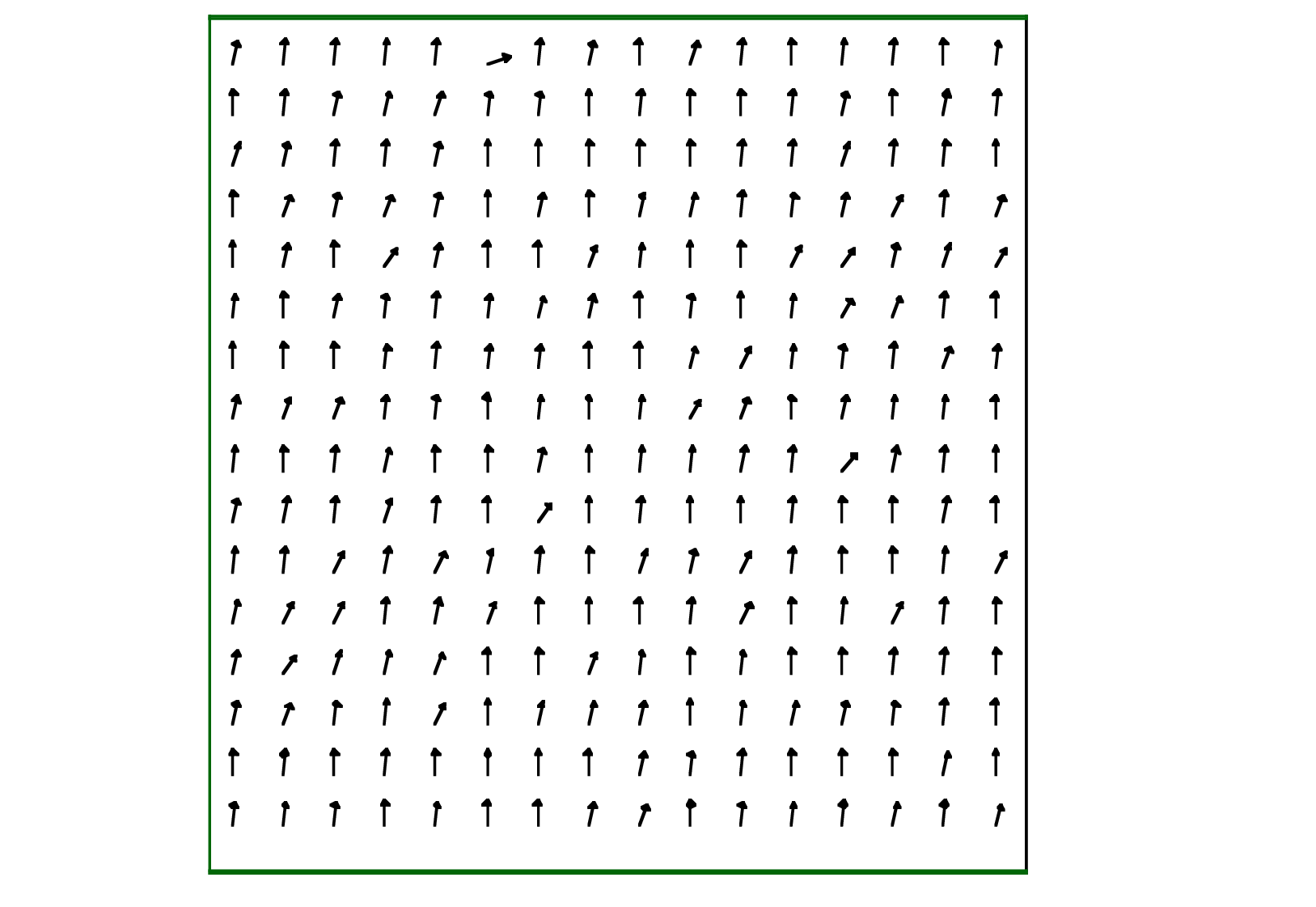}
\vspace{0cm}
\caption{
Local abelian rate $R_{\rm Abel} =  \cos \theta_\mu(s)$ at 
$\beta = $2.4 on $16^4$ lattice in the MA gauge. 
The meaning of the arrow is shown in Fig.\ref{arrow4hosoku}.
}
\label{arrow6}
\vspace{0cm}
\end{figure}

\begin{figure}[p]
\epsfxsize = 11 cm
\centering \leavevmode
\epsfbox{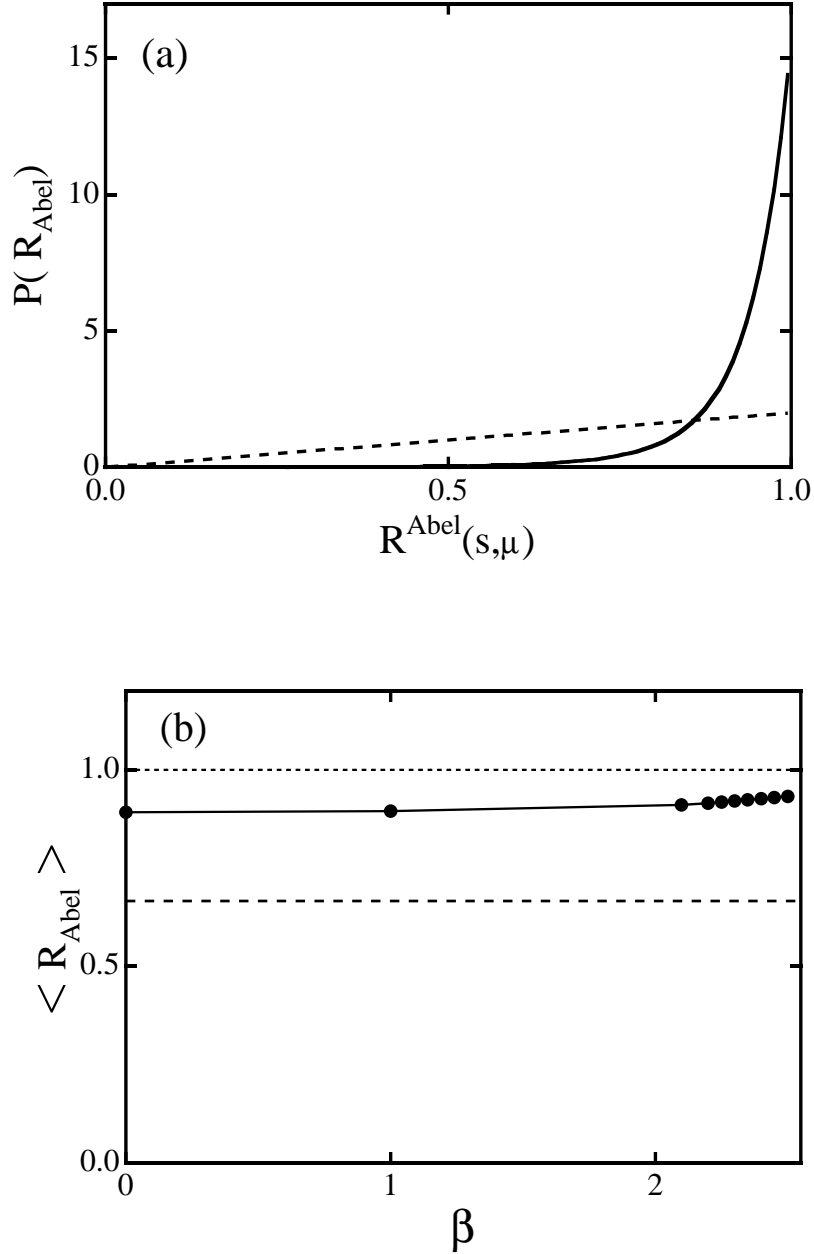}
\vspace{0cm}
\caption{
(a) The probability distribution $P(R_{\rm Abel})$
of the abelian projection rate $R_{\rm Abel} $
at $\beta =2.4$ on the $16^4$ lattice from 40 gauge configurations.
The solid curve denotes $P(R_{\rm Abel})$ in the MA gauge,
and  the dashed line denotes $P(R_{\rm Abel})$
without gauge fixing.
(b) The average of the abelian projection rate $\langle  R_{\rm Abel} 
\rangle$ in the MA gauge
as the function of $\beta$.
For comparison, we plot also $\langle  R_{\rm Abel} \rangle$ without gauge 
fixing. }
\label{gfig3}
\vspace{0cm}
\end{figure}

We show in Fig.\ref{arrow5} and Fig.\ref{arrow6} the spatial distribution of the abelian projection rate
$R_{\rm Abel} = \cos \theta$
as an arrow $(\sin \theta, \cos \theta)$.
In the MA gauge, most of all SU(2) link variables become U(1)$_{3}$-like.
We also show in Fig.\ref{gfig3}(a) the probability distribution  $P(R_{\rm 
Abel})$  
of the abelian projection rate 
$R_{\rm Abel} (s,\hat \mu) $
in the MA gauge. 
Here, $\langle  R_{\rm Abel} \rangle_{\beta=0}$  
in the strong coupling limit ($\beta = 0$) \cite{poulis,ichie6} is analytically 
calculable as  
\begin{eqnarray}
\langle  R_{\rm Abel} \rangle _{\beta=0}   & = & 
\langle \cos \theta_\mu (s) \rangle _{\beta=0} = 
\frac{  \int d U_\mu(s) \cos \theta_\mu(s)}
 {  \int d U_\mu(s)}   \nonumber \\
& = &
\frac{\int_{0}^{\frac{\pi}{2}} d \theta_\mu(s) 
\sin \theta_\mu (s) \cos^2 \theta_\mu(s) }
 { \int_{0}^{\frac{\pi}{2}} d \theta_\mu (s) 
\sin \theta_\mu(s) \cos \theta_\mu(s)  } =
\frac23,
\end{eqnarray}
using Eq.(\ref{eq:measure}).
In the MA gauge, $R_{\rm Abel}$ approaches to unity as shown in  Fig.\ref{gfig3}(a).
The off-diagonal component of the SU(2) link variable 
is forced to be reduced.
As a typical example, one obtains   
$\langle  R_{\rm Abel} \rangle_{\rm MA} \simeq$  0.926  on $16^4$ lattice
with $\beta = 2.4$.
In Fig.\ref{gfig3}(b), we show  the abelian projection rate 
$\langle  R_{\rm Abel} \rangle_{\rm MA}$ as the function of $\beta$.
For larger $\beta$, 
$\langle \cos \theta_\mu(s) \rangle_{\rm MA}$ becomes slightly larger.
Without gauge fixing, the average  
$\langle  R_{\rm Abel} \rangle$ is found to be about $2 \over 3$ 
without dependence on $\beta $.
In the continuum limit in the MA gauge,
$U_\mu^1(s) $ and  $U_\mu^2(s)$ become at most
$O(a)$,  and  therefore  $\langle  R_{\rm Abel} \rangle_{\rm MA}$
approaches to unity as  
$\langle  R_{\rm Abel} \rangle_{\rm MA} = 1 + O(a^2)$
due to the trivial dominance 
of $U_\mu^0(s)$, which differs from abelian dominance in the physical 
sense.
The remarkable feature of the MA gauge is  the high abelian 
projection rate as $\langle  R_{\rm Abel} \rangle_{\rm MA} \simeq 1$ in the 
whole region of $\beta$. In fact, we find  
$\langle  R_{\rm Abel} \rangle_{\rm MA} \simeq 
0.88 $ even for the strong coupling limit $\beta=0$, where the 
original link variable $U_\mu$ is completely random.
Thus, abelian dominance for the link variable $U_{\mu}$ is observed at 
any scale in the MA gauge.

To understand the origin of the high abelian projection rate  
as $\langle  R_{\rm Abel} \rangle_{\rm MA} \simeq 1$, 
we estimate the lower bound of 
$\langle  R_{\rm Abel} \rangle_{\rm MA} $ in the MA gauge using the 
statistical 
consideration.
The MA gauge maximizes
\begin{eqnarray}
R_{\rm MA}[U_\mu]   \equiv 
\sum_{s,\mu} {\rm tr} \{  U_\mu(s) \tau_3 U^{\dagger}_\mu(s) \tau_3 \}  
={\rm tr}(\tau_3  \sum_{s,\mu} \hat \phi_\mu(s)),
\end{eqnarray}
where $\hat  \phi_\mu(s) \equiv 
U_{\mu}(s) \tau_3 U^{\dagger}_{\mu}(s)$
is an  $su$(2) element satisfying $\hat \phi^2_{\mu} = 1$.
Denoting $\hat  \phi_\mu(s) =\hat  \phi^a_\mu(s) \tau^a $,
we parameterize the 
3-dimensional unit vectors
$\vec \phi_\mu \equiv (\hat  \phi_\mu^1,\hat  
\phi_\mu^2,\hat  \phi_\mu^3) \in S^2$
$(\mu=1,2,3,4)$ as 
$\vec \phi_\mu =
(\sin 2 \theta_{\mu} \cos \chi_{\mu}, 
 \sin2 \theta _{\mu} \sin \chi_{\mu},\cos 2\theta_{\mu})$
using Eqs.(\ref{eq:para}) and (\ref{divided}).
The MA gauge  maximizes the third component $\hat \phi_\mu^3$
using the gauge transformation.
Under the local gauge transformation by $V(s) \equiv 1 + \{ V(s_0)-1 \} 
\delta_{ss_0} \in$ SU(2), $\hat \phi_\mu(s_0)$ is transformed as the unitary 
transformation,
\begin{eqnarray}
\hat \phi_\mu(s_0) \rightarrow \hat \phi'_\mu(s_0) \equiv 
V(s_0) \hat \phi_\mu(s_0)  V^{-1}(s_0), 
\label{eq:rot}
\end{eqnarray}
which leads to a simple rotation of the unit vectors $\vec \phi_\mu$.
In the MA gauge, 
$\displaystyle \sum_{s,\mu}$$\vec \phi_\mu$ is 
``polarized''
along the
positive third direction.
On the 4-dimension lattice with $N$ sites, 4$N$ unit vectors 
$\vec \phi_\mu(s)$ are  maximally polarized by 
$N$ gauge functions $V(s) $ in the MA gauge.
Then,  $\langle R_{\rm Abel} \rangle_{\rm MA}$ 
 is expressed as the maximal
``polarization rate'' of 4$N$ unit vectors 
$\vec \phi_\mu$ 
by suitable $N$ gauge functions $V(s)$.
On the average, this estimation of $\langle R_{\rm Abel} \rangle_{\rm MA}$
is approximately given by the estimation of the maximal 
polarization rate of 
4 unit vectors $\vec \phi_{\mu}$ by a suitable rotation with $V \in$ 
SU(2).
The lower bound of $\langle R_{\rm Abel} \rangle_{\rm MA}$ is obtained 
from the 
strong-coupling system with $\beta=0$, where link variables $U_\mu(s)$ 
are completely random. Accordingly, $\vec \phi_\mu$ can 
be regarded as 
random unit vectors on $S^2$.
The maximal ``polarization'' of 4 unit vectors $\vec \phi_\mu$
is realized by the rotation which moves  
 $\vec \phi  \equiv
$$\displaystyle \sum_{\mu=1}^4$$
\vec \phi_\mu /|
$$\displaystyle \sum_{\mu=1}^4$$
 \vec \phi_\mu|$  to the unit vector $\vec \phi^R \equiv (0,0,1)$ 
 in third direction.
Here, $\cos 2\theta_\mu^R$ after the rotation is 
identical to the inner product between $\vec \phi_\mu$ and $\vec \phi$, 
because of $\vec \phi \cdot \vec \phi_\mu = 
\vec \phi^R \cdot \vec \phi_\mu^R =
 (\hat \phi_\mu^R)^3= \cos 2\theta_\mu^R$.
Then, we estimate $\langle R_{\rm Abel} \rangle_{\rm MA} 
= \langle \cos \theta_\mu \rangle_{\rm MA}$ at $\beta=0$ as
\begin{eqnarray}
\lefteqn{ \langle \cos \theta_\mu \rangle_{\rm MA}^{\beta=0} 
\simeq  \! \{ \prod_{\mu=1}^{4} \int d U_\mu \}
\left(
\frac14 \sum_{\mu=1}^4 \cos  \theta_\mu^R 
\right) }  \nonumber \\
 &  =  & \{ \prod_{\mu=1}^{4} 
\frac{1}{\pi}
\int_0^{\frac{\pi}{2}}  d\theta_\mu \cos \theta_\mu \sin \theta_\mu 
\int_{-\pi}^{\pi}           d  \chi_\mu 
\}
\left(
\frac14 \sum_{\mu=1}^4 \cos \{ \frac12 \cos^{-1}(
\vec \phi \cdot 
\vec \phi_\mu  )  \}  \right).
\label{eq:lower}
\end{eqnarray}
Using this estimation (\ref{eq:lower}), 
we obtain  
$\langle  R_{\rm Abel} \rangle_{\rm MA} \simeq $ 0.844, 
which is close to the lattice result
$\langle R_{\rm Abel} \rangle \simeq $ 0.88  in the strong coupling 
limit ($\beta=0$).
Such a high abelian rate 
$\langle  R_{\rm Abel} \rangle_{\rm MA}$ in the MA gauge 
would provide a microscopic basis of abelian dominance for 
the infrared physics.

%
%

%
%

\section[Abelian Dominance for Confinement Force]{
Abelian Dominance for Confinement Force in the MA Gauge}

In this section, 
we study the origin of abelian dominance on  the string tension
as the confinement force in 
a semi-analytical manner,
considering the relation with  { microscopic abelian dominance}
on the link variable \cite{ichiead,ichie6}.

In the MA gauge, the {\rm diagonal element} $\cos \theta_\mu(s)$
in $M_\mu(s)$ is maximized by the gauge transformation
as large as possible. For instance, the abelian projection rate
is almost unity as 
$R_{\rm Abel}=\langle\cos \theta_\mu(s)\rangle_{\rm MA}\simeq 0.93$
at $\beta=2.4$. Then, the {off-diagonal element
$e^{i\chi_\mu(s)}\sin\theta_\mu(s)$ is forced to take a small value
in the MA gauge} due to the factor $\sin \theta_\mu(s)$, 
and therefore the approximate treatment
on the off-diagonal element would be allowed in the MA gauge.
Moreover, the {angle variable $\chi_\mu(s)$ is not
constrained} by the MA gauge-fixing condition at all,
and {tends to take a random value} besides the residual
${\rm U(1)}_3$ gauge degrees of freedom.
Hence, $\chi_\mu(s)$ can be regarded as a {random angle variable}
on the treatment of $M_\mu(s)$ in the MA gauge 
in a good approximation.

Let us consider the Wilson loop
$ \displaystyle \langle W_C[U_\mu(s)]\rangle \equiv
\langle{\rm tr}\prod_C U_\mu(s)\rangle
=\langle{\rm tr}\prod_C\{M_\mu(s)u_\mu(s)\}\rangle$
in the MA gauge.
In calculating $\langle W_C[U_\mu(s)]\rangle$,
the expectation value of $e^{i\chi_\mu(s)}$
in $M_\mu(s)$ vanishes as
\begin{eqnarray}
\langle e^{i\chi_\mu(s)}\rangle
\simeq \int_0^{2\pi} d\chi_\mu(s)\exp\{i\chi_\mu(s)\}=0,
\end{eqnarray}
when $\chi_\mu(s)$ behaves as a {random angle variable.}
Then, within the random-variable approximation for $\chi_\mu(s)$, 
the {off-diagonal factor} $M_\mu(s)$ appearing in
$\langle W_C[U_\mu(s)]\rangle$ is simply reduced as a $c$-number factor,
$
M_\mu(s) \rightarrow \cos \theta_\mu(s) \ {\bf 1},
$
and therefore the SU(2) link variable $U_\mu(s)$ in the Wilson loop
$\langle W_C[U_\mu(s)]\rangle$
is simplified as a {diagonal matrix,}
\begin{eqnarray}
U_\mu(s)\equiv M_\mu(s)u_\mu(s)
\rightarrow
\cos \theta_\mu(s) u_\mu(s).
\end{eqnarray}

Then, for the $I \times J$ rectangular $C$, the Wilson loop
$W_C[U_\mu(s)]$ in the MA gauge is approximated as
\begin{eqnarray}
\langle W_C[U_\mu(s)]\rangle &\equiv&
\langle{\rm tr}\prod_{i=1}^L U_{\mu_i}(s_i)\rangle
\simeq
%
%
\langle\prod_{i=1}^L \cos \theta_{\mu_i}(s_i) \cdot
{\rm tr} \prod_{j=1}^L u_{\mu_j}(s_j)\rangle_{\rm MA} \cr
&\simeq&
\langle\exp\{\sum_{i=1}^L \ln (\cos \theta_{\mu_i}(s_i))\}\rangle_{\rm MA}
\ \langle W_C[u_\mu(s)]\rangle_{\rm MA} \cr
&\simeq&
\exp\{L \langle \ln (\cos \theta_\mu(s)) \rangle_{\rm MA} \}
\ \langle W_C[u_\mu(s)]\rangle_{\rm MA},
\label{eq:wilson}
\end{eqnarray}
where $L\equiv 2(I+J)$ denotes the perimeter length and
$ \displaystyle W_C[u_\mu(s)]\equiv {\rm tr}\prod_{i=1}^L u_{\mu_i}(s_i)$
the abelian Wilson loop.
Here, we have replaced
$ \displaystyle \sum_{i=1}^L \ln \{\cos(\theta_{\mu_i}(s_i)\}$
by its average
$L \langle \ln \{\cos \theta_\mu(s)\} \rangle_{\rm MA}$
{in a statistical sense}, and 
such a statistical treatment becomes more accurate for larger $I,J$ 
and becomes exact for infinite $I,J$.

In this way, we derive a simple estimation as
\begin{eqnarray}
W_C^{\rm off}\equiv
\langle W_C[U_\mu(s)]\rangle/\langle W_C[u_\mu(s)]\rangle_{\rm MA}
\simeq \exp\{L\langle \ln(\cos \theta_\mu(s))\rangle_{\rm MA}\}
\label{eq:offw}
\end{eqnarray}
for the {contribution of the off-diagonal
gluon element to the Wilson loop}.
From this analysis, the contribution of off-diagonal gluons 
to the Wilson loop is expected to obey the {perimeter law}
in the MA gauge for large loops, where the statistical
treatment would be accurate.

\begin{figure}[p]
\epsfxsize = 11 cm
\centering \leavevmode
\epsfbox{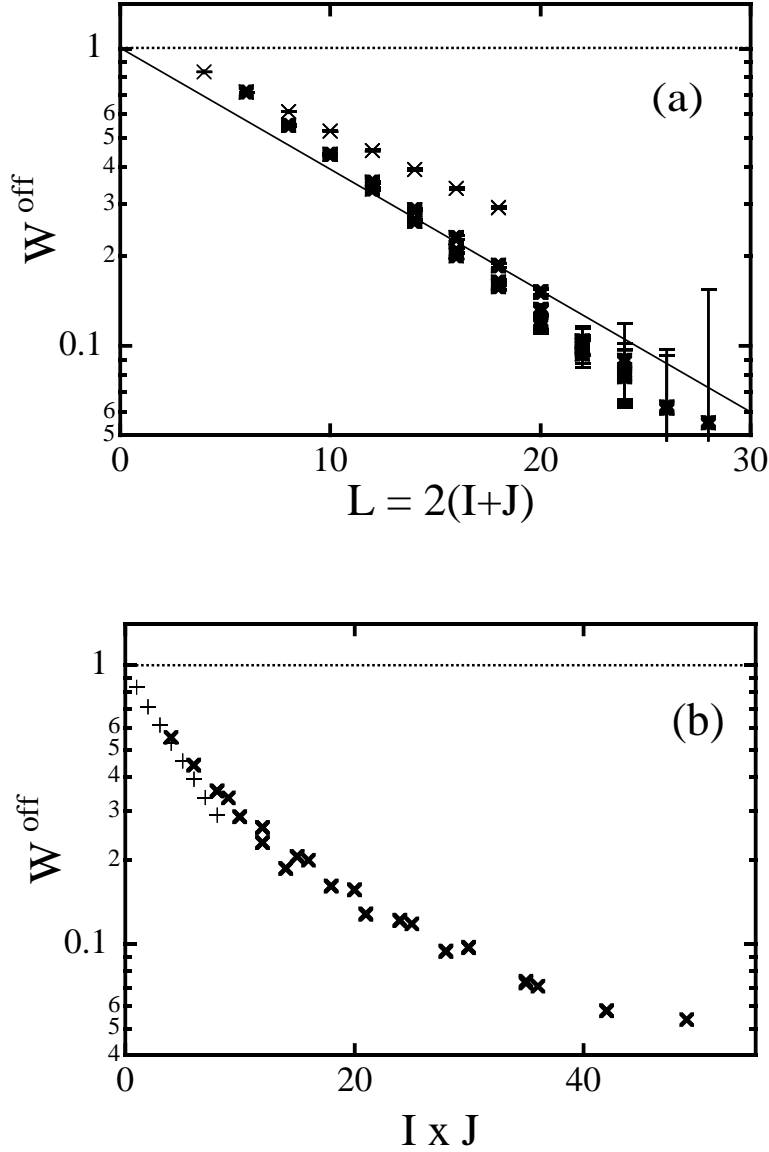}
\vspace{0cm}
\caption{    
The off-diagonal gluon contribution on the Wilson loop,
$W^{\rm off} 
 \equiv  \frac{  \langle  W_C[U_\mu(s)]  \rangle }
                          {  \langle  W_C[u_\mu(s)]  \rangle }$,
as the function of the perimeter length $L\equiv 2 (I + J)$
in the MA gauge on $16^4$ lattice with $\beta = 2.4$.
The thick line denotes the theoretical estimation 
in Eq.(\ref{eq:offw})
with the microscopic input 
$\langle \ln \{\cos \theta_\mu(s)\}\rangle_{\rm MA} \simeq -0.082$ 
at $\beta=2.4$.
The data of the Wilson loop with $I=1$ or $J=1$ are distinguished by 
the thin cross. 
}
\label{gfig4}
\vspace{0cm}
\end{figure}

\begin{figure}[bt]
\epsfxsize = 11 cm
\centering \leavevmode
\epsfbox{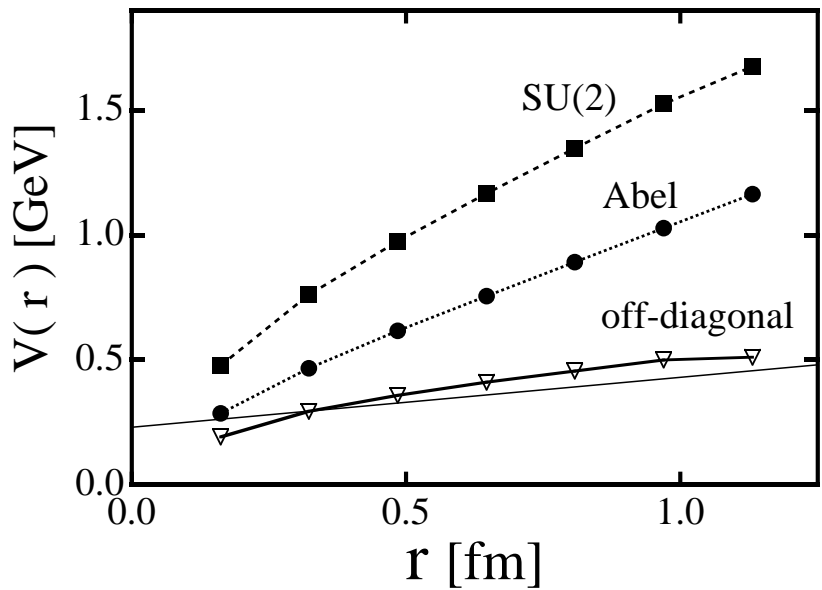}
\vspace{0cm}
\caption{    
The inter-quark potential $V(r)$ as the function of
the inter-quark distance $r$.
The lattice data are obtained from the Wilson loop in the MA gauge
on $16^4$ lattice with $\beta = 2.4$ and $T=7$.   
The square, the circle and the rhombus denote the full SU(2), 
the abelian
and the off-diagonal contribution of the static potential,
respectively.
The thin line denotes the theoretical estimation
in Eq.(\ref{eq:voff}).
Here, the lattice spacing $a$ is determined so as to produce $\sigma = 0.89 $ GeV/fm.
Due to the artificial finite-size effect of the Wilson loop,
the off-diagonal contribution $V^{\rm off}$ gets a slight linear part.}
\label{gfig4c}
\vspace{0cm}
\end{figure}

Now, we study the behavior of the off-diagonal contribution \\
$W_C^{\rm off}\equiv
\langle W_C[U_\mu(s)]\rangle/\langle W_C[u_\mu(s)]\rangle_{\rm MA}$
in the MA gauge using the lattice QCD, considering the theoretical 
estimation Eq.(\ref{eq:offw}).
As shown in Fig.\ref{gfig4}, we find 
that $W_C^{\rm off}$ seems to obey the
{perimeter law} for the Wilson loop with $I,J \ge 2$
in the MA gauge in the lattice QCD simulation with $\beta = 2.4$ and 
$16^4$.
We find also that the behavior on $W_C^{\rm off}$
as the function of $L$ is well reproduced
by the above analytical estimation with  {microscopic information}
on the diagonal factor $\cos\theta_\mu(s)$ as
$\langle \ln \{\cos\theta_\mu(s)\} \rangle_{\rm MA}\simeq -0.082$
for $\beta=2.4$.
Thus, the off-diagonal contribution $W_C^{\rm off}$
to the Wilson loop obeys
the perimeter law in the MA gauge, and therefore the
{abelian Wilson loop} $\langle W_C[u_\mu(s)]\rangle_{\rm MA}$
 should obey the
{area law} as well as
the SU(2) Wilson loop $W_C[U_\mu(s)]$.
From Eq.(\ref{eq:offw}),
the off-diagonal contribution to the string tension vanishes as
\begin{eqnarray}
\Delta\sigma & \equiv & 
\sigma_{\rm SU(2)}-\sigma_{\rm Abel}  \\ \nonumber 
&\equiv&
-\lim_{R,T \rightarrow \infty}
{1 \over RT}\ln \langle W_{R \times T}[U_\mu(s)]\rangle
+\lim_{R,T \rightarrow \infty}
{1 \over RT}\ln \langle W_{R \times T}[u_\mu(s)]\rangle_{\rm MA}
\cr
&\simeq&
-2 \langle \ln \{\cos\theta_\mu(s)\} \rangle_{\rm MA}
\lim_{R,T \rightarrow \infty} {R+T \over RT}=0.
\label{eq:dsig}
\end{eqnarray}
Thus, {abelian dominance for the string tension},
$\sigma_{\rm SU(2)}=\sigma_{\rm Abel}$,
can be proved in the MA gauge by
replacing the off-diagonal angle variable $\chi_\mu(s)$
as a random variable.

The analytical relation in Eq.(\ref{eq:offw}) indicates also that the finite
size effect on $R$ and $T$ in the Wilson loop leads to the deviation 
between the SU(2) string tension $\sigma_{\rm SU(2)}$ and the
abelian string tension $\sigma_{\rm Abel}$ as 
$\sigma_{\rm SU(2)} > \sigma_{\rm Abel}$
in the actual lattice QCD simulations.
Here, we consider this deviation 
$\Delta \sigma \equiv \sigma_{\rm SU(2)} - \sigma_{\rm Abel}$
in some detail.
Similar to the SU(2) inter-quark potential $V_{\rm SU(2)}(r)$ from
$\langle \, W_{\rm SU(2)}
 \,  \rangle \equiv
\langle \, W[U_\mu(s)] \,  \rangle$, we define the abelian inter-quark
potential $V_{\rm Abel}(r)$ and the off-diagonal contribution 
$V_{\rm off}(r)$ of the potential from
$\langle \, W_{\rm Abel} \,  \rangle \equiv
\langle \, W[u_\mu(s)] \,  \rangle$ and $W_{\rm off}$, respectively,
\begin{eqnarray}
V_{\rm SU(2)}(r) & \equiv  & -\frac{1}{Ta} {\rm ln}  \, \langle \, W_{\rm SU(2)}
({R \times T}) \,  \rangle, \nonumber \\ 
V_{\rm Abel}(r)& \equiv  & -\frac{1}{Ta} {\rm ln}  \, \langle  \,  W_{\rm Abel}
({R \times T})  \, \rangle, \nonumber \\
V_{\rm off}(r) & \equiv  & -\frac{1}{Ta} {\rm ln}   \,  W_{\rm off}
({R \times T})   
 =   -\frac{1}{Ta} {\rm ln}\,  
 \frac{ \langle \, W_{\rm SU(2)}
({R \times T}) \,  \rangle }{
 \langle  \,  W_{\rm Abel}
({R \times T})  \, \rangle } \nonumber \\
& = & V_{\rm SU(2)}(r)   - V_{\rm Abel}(r),
\end{eqnarray}
where $r \equiv Ra$ denotes the inter-quark distance in the physical unit.
We show in Fig.\ref{gfig4c} $V_{\rm SU(2)}(r)$,  $V_{\rm Abel}(r)$ and $V_{\rm off}(r)$
extracted from the Wilson loop with $T = 7$ in the lattice QCD simulation
with $\beta = 2.4$ and $16^4$.
As shown in Fig.\ref{gfig4c}, the lattice result for $V_{\rm off}(r)$ seems to be  
reproduced by the theoretical estimation obtained from Eq.(\ref{eq:offw}),
\begin{eqnarray}
V_{\rm off}(r) = V_{\rm SU(2)}(r) - V_{\rm Abel}(r) \simeq - \frac{2(R+T)}{Ta} 
\langle \ln(\cos \theta_\mu(s))\rangle_{\rm MA}
\label{eq:voff}
\end{eqnarray}
using the microscopic information of  
$\langle \ln(\cos \theta_\mu(s))\rangle_{\rm MA} = -0.082 $ at 
$\beta = 2.4$.
From the slope of $V_{\rm off}(r)$ in Eq.(\ref{eq:voff}), we
can estimate 
$\Delta\sigma  \equiv 
\sigma_{\rm SU(2)}-\sigma_{\rm Abel}$
in the physical unit as
\begin{eqnarray}
\Delta\sigma  \equiv  \sigma_{\rm SU(2)}-\sigma_{\rm Abel}    & \simeq   &
-2 
\langle  \, \ln \, ( \, \cos \theta_\mu(s) \,  ) \,  \rangle_{\rm MA} 
\,  \frac{1}{Ta^2}
\nonumber  \\ 
& = & - 
\langle \ln( 1- \sin ^2 \theta_{\mu}(s) ) \rangle_{\rm MA}  \frac{1}{Ta^2}.
\end{eqnarray}
In the MA gauge, $ \sin ^2 \theta_\mu(s) $ takes a small value
and can be treated in a perturbation manner so that one finds
\begin{eqnarray}
\Delta \sigma
  \simeq 
\langle \sin ^2 \theta_\mu(s)  \rangle_{\rm MA}  \frac{1}{Ta^2}
= \langle (U^{1}_{\mu}(s))^2 + (U^{2}_{\mu2}(s))^2  \rangle_{\rm MA} \frac{1}{Ta^2}.
\end{eqnarray}
Near the continuum limit $a \simeq 0$, we find $U_{\mu}^a \simeq ae 
A_{\mu}^a /2$ ($a$=1,2,3) from $U_{\mu} = e^{iae A_{\mu}^a 
{\tau^a}/{2}}$, and
 then we derive the relation between $\Delta\sigma $ and the  
off-diagonal gluon in the MA gauge as
\begin{eqnarray}
\Delta\sigma 
 \simeq  
\frac{1}{4T} 
\langle (eA_{\mu}^1)^2  + (eA_{\mu}^2) ^2 \rangle_{\rm MA } =
\frac{a}{4t} \langle (eA_{\mu}^1)^2  + (eA_{\mu}^2) ^2 \rangle_{\rm MA},
\label{eq:delsig}
\end{eqnarray}
where $t \equiv Ta$ is the temporal length of the Wilson loop in the 
physical unit.
In Eq.(\ref{eq:delsig}), 
$\langle (eA_{\mu}^1)^2  + (eA_{\mu}^2) ^2 \rangle_{\rm MA }$ is the 
off-diagonal gluon-field fluctuation, and is strongly suppressed in the 
MA gauge by its definition. It would be interesting to note that
microscopic abelian dominance or the suppression of off-diagonal gluons in 
the MA gauge is directly connected to reduction of the deviation 
$\Delta\sigma$  in Eq.(\ref{eq:delsig}).
Since $\langle (eA_{\mu}^1)^2  + (eA_{\mu}^2) ^2 \rangle_{\rm MA }$ is a 
local continuum quantity, it is to be independent on both $a$ and $t$.
Hence, the deviation $\Delta\sigma $ between the SU(2) string tension 
$\sigma_{\rm SU(2)}$ and the abelian string tension $\sigma_{\rm Abel}$ 
can be removed by taking the large Wilson loop as 
$t \rightarrow \infty $ or the small mesh as $a \rightarrow 0 $ with fixed $t$.

\section{Gluon Field in the MA Gauge with the U(1)$_3$ Landau Gauge}

In the MA gauge, 
the linear confinement potential can be almost reproduced
only by the abelian degrees 
of freedom, which is called as
abelian dominance on the string tension.
In this section, 
we study the probability distribution of
the gauge field such as the abelian gauge field, the abelian field strength,
the off-diagonal gluon in the MA gauge. 
 
From the abelian angle variable $\theta^3_\mu(s)$,
the abelian field strength ${\bar \theta_{\mu\nu}}(s)$ is defined as
\begin{eqnarray}
{\bar \theta_{\mu\nu}} (s) \equiv {\rm mod }_{2\pi} 
 (\partial \wedge \theta^3)_{\mu\nu}(s) \in (-\pi,\pi],
\label{eq:abelFS}
\end{eqnarray}
where
$(\partial \wedge \theta^3)_{\mu\nu}(s)$ is expressed as
\begin{eqnarray}
(\partial \wedge \theta^{\rm 3})_{\mu\nu}(s) \equiv
\theta_{\mu\nu}(s) \equiv
 \theta^{\rm 3}_\mu(s) +  \theta^{\rm 3}_\nu (s+ \hat \mu) -
\theta^{\rm 3}_\mu(s+\hat \nu) - \theta^{\rm 3}_\nu(s). 
\end{eqnarray}
The abelian field strength ${\bar \theta_{\mu\nu}}(s)$ is related to 
the abelian plaquette as 
\begin{eqnarray}
 \Box^{\rm Abel}_{\mu\nu}(s) \equiv 
u_\mu(s) u_\nu (s+\hat \mu) u ^\dagger_\mu(s+\hat \nu) u_\nu^\dagger(s) = 
e^{i \bar \theta_{\mu\nu}(s)}.
\end{eqnarray}
In terms of the U(1)$_3$ gauge symmetry remaining
in the abelian gauge,
the abelian field strength
${\bar \theta_{\mu\nu}}(s)$ is a U(1)$_3$ gauge-invariant quantity
because of the gauge invariance of the abelian plaquette, while both
$\theta^3_\mu(s)$ and  $(\partial \wedge \theta^3)_{\mu\nu}(s)$ are U(1)$_3$
gauge variant. 


\begin{figure}[p]
\epsfxsize = 13 cm
\centering \leavevmode
\epsfbox{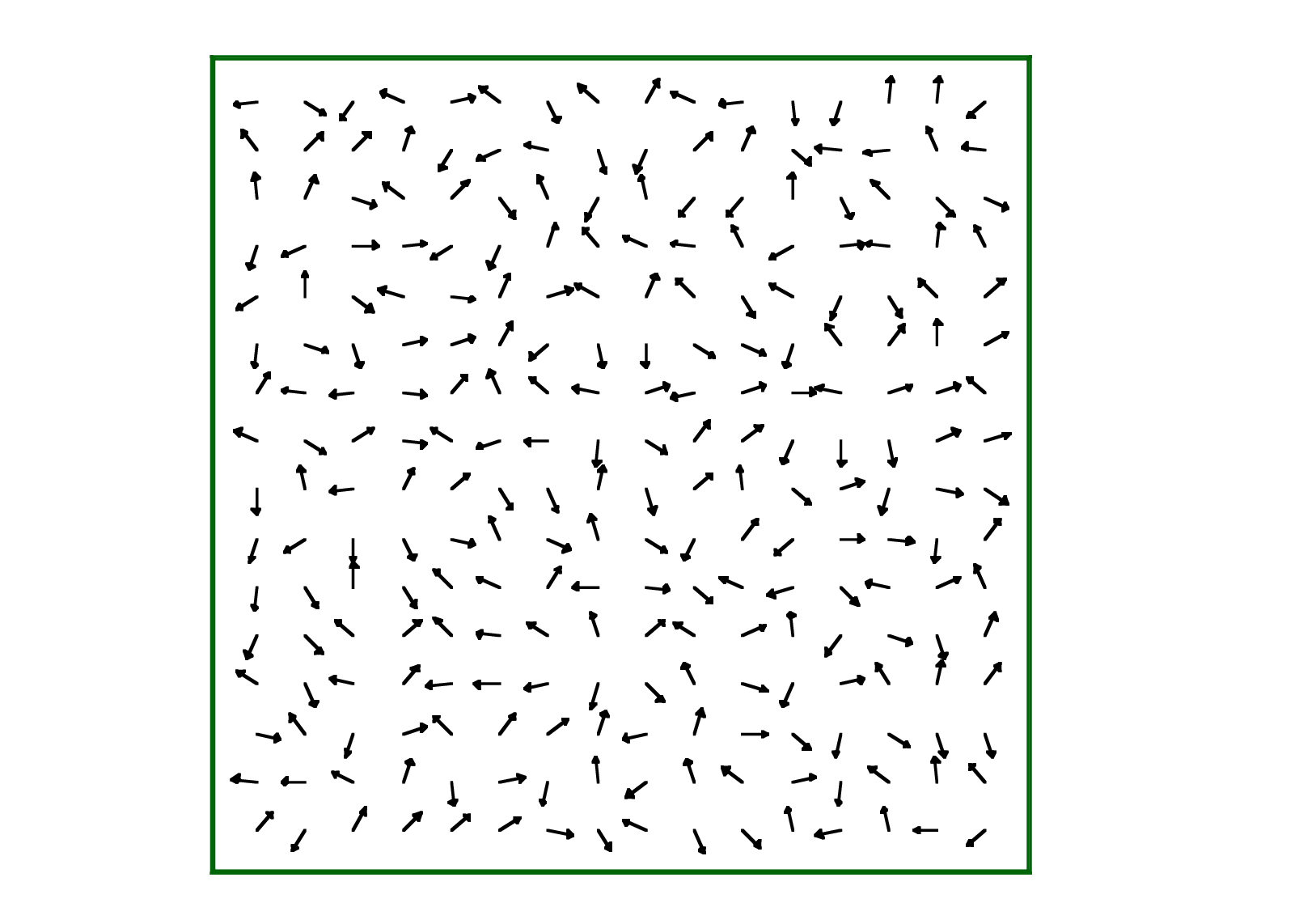}
\vspace{0cm}
\caption{    
The abelian angle variable $\theta_\mu^3(s)$ without U(1)$_3$ Landau gauge 
fixing
in the MA gauge at $\beta=2.4$.
}
\label{arrow4}
\epsfxsize = 13 cm
\centering \leavevmode
\epsfbox{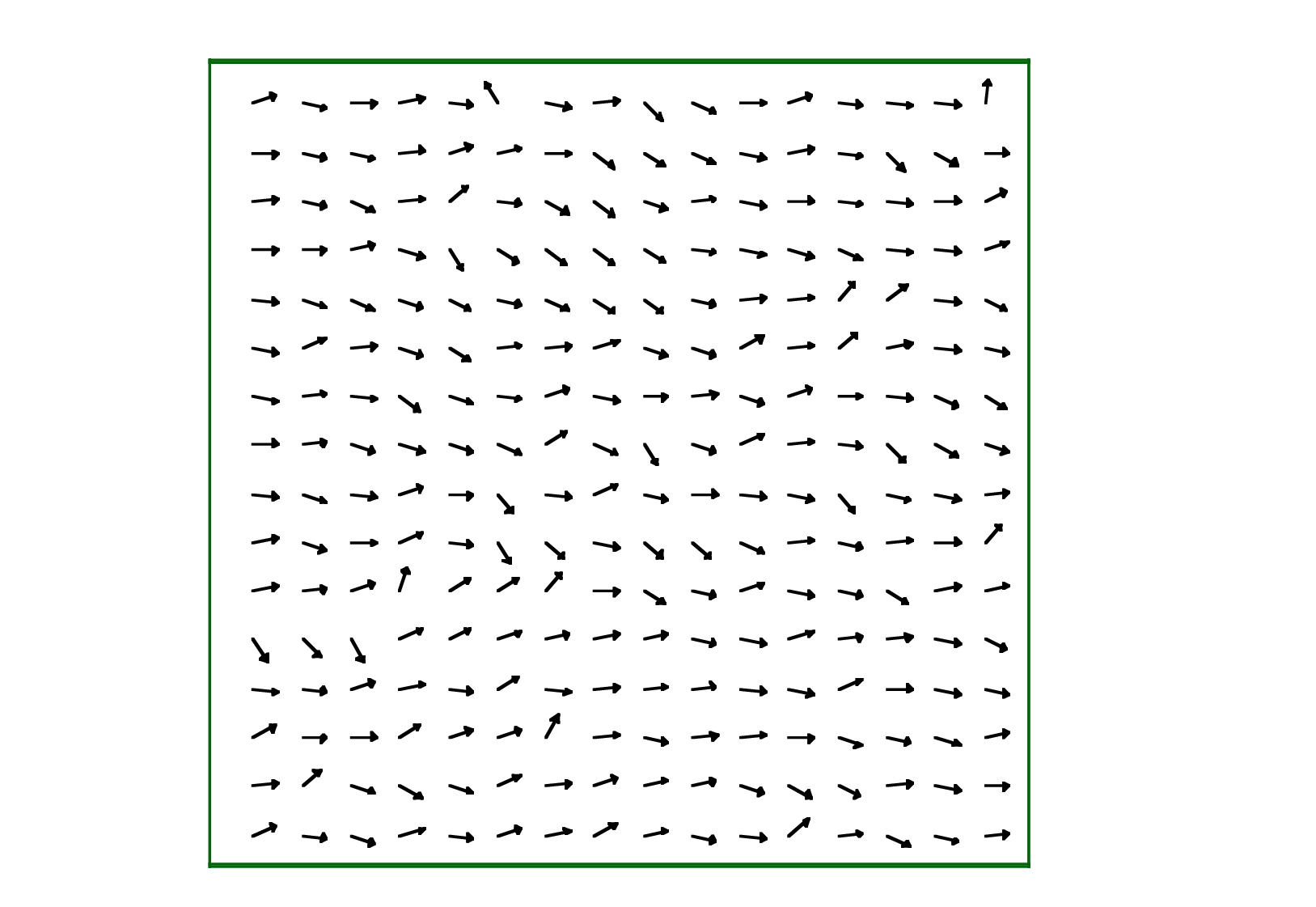}
\vspace{0cm}
\caption{  
The abelian angle variable $\theta_\mu^3(s)$ in the MA gauge
with U(1)$_3$ Landau gauge fixing.
}
\label{arrow1}
\vspace{0cm}
\end{figure}

To investigate the features of U(1)$_3$ gauge-variant quantities like
the abelian angle variable $\theta^3_\mu(s)$,
it is necessary to fix the residual U(1)$_3$ gauge degrees of freedom
in addition to the abelian gauge fixing.  
In this paper, we introduce U(1)$_3$ lattice Landau gauge 
\cite{mandula}, where
the gluon field is mostly continuous under the constraint of the MA gauge 
condition,  and 
the lattice field can be compared with continuum field variable more directly.


\begin{figure}[p]
\epsfxsize = 10 cm
\centering \leavevmode
\epsfbox{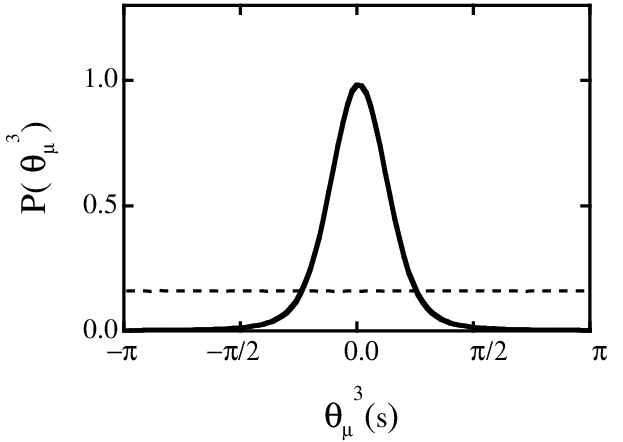}
\vspace{0cm}
\caption{    The probability distribution $P(\theta_\mu^3)$ of the abelian 
angle variable
$\theta^3_\mu(s) \in (-\pi,\pi]$ in the MA gauge on $16^4$ lattice with 
$\beta=2.4$.
The solid curve denotes $P(\theta_\mu^3)$
in the U(1)$_3$ Landau gauge fixing after the MA gauge fixing, and 
the dashed line denotes $P(\theta^3_\mu)$ without the U(1)$_3$ gauge 
fixing.
Without the U(1)$_3$ gauge fixing, the  angle variable
$\theta^3_\mu(s)$ is randomly distributed, 
while $\theta^3_\mu(s)$ has a Gaussian-type peak around $\theta^3_\mu (s)= 
0$
in the U(1)$_3$ Landau gauge.
}
\label{gfig5a}
\vspace{1.5cm}
\epsfxsize = 10 cm
\centering \leavevmode
\epsfbox{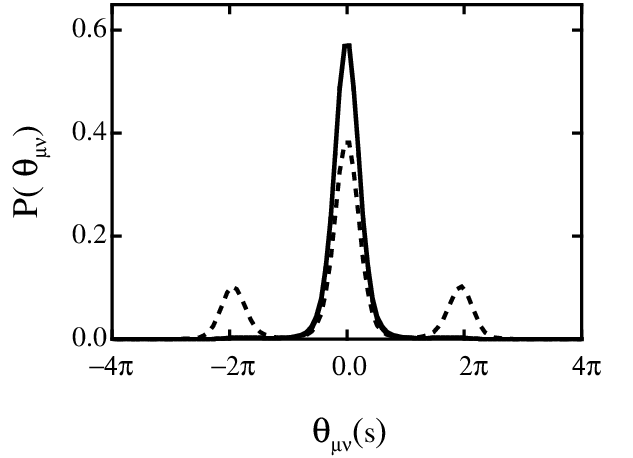}
\vspace{0cm}
\caption{  
The probability distributions $P(\theta_{\mu\nu})$ of the two form 
$\theta_{\mu\nu} \equiv (\partial \wedge \theta)_{\mu\nu}(s)$
 in the MA gauge with and without U(1)$_3$-Landau gauge fixing,
 which are denoted by the solid and dashed curves,
 respectively.
 In the U(1)$_3$-Landau gauge, $P(\theta_{\mu\nu})$  has single peak around 
 $\theta_{\mu\nu}=0$.
}
\label{gfig5b}
\vspace{0cm}
\end{figure}

The U(1)$_3$ lattice Landau gauge is defined by maximizing
\begin{eqnarray}
R_L[U_\mu] \equiv \displaystyle \sum_{s,\mu} {\rm Re} \,\, u_\mu(s)
\end{eqnarray}
using the residual U(1)$_3$ gauge transformation.
In the U(1)$_3$ Landau gauge, the abelian angle variable
$\theta^3_\mu(s)$ is suppressed as small as possible.
In the continuum limit $a \rightarrow 0$, the abelian gauge field
$A_\mu^3(x)$ satisfies the ordinary Lorentz-gauge condition as in QED,
$\partial_\mu A_\mu^3=0$. 

We show in Figs \ref{arrow4} and \ref{arrow1} 
the configuration of the abelian angle variable $\theta^{3}_\mu(s)$ 
before and after the U(1)$_3$ Landau gauge fixing
in the MA gauge at $\beta=2.4$.
The magnitude of the angle variable is found to become small and 
continuous in the U(1)$_3$ Landau gauge.
We  show also in Fig.\ref{gfig5a} the probability distribution 
$P(\theta_\mu^3)$ of the 
abelian angle variable $\theta^3_\mu(s) \in (-\pi,\pi]$ 
in the MA gauge with and without 
U(1)$_3$ Landau gauge fixing.
We find that the whole shape of the distribution seems Gaussian-type peak 
around $\theta_\mu^3 = 0$
in the U(1)$_3$ Landau gauge,
while $\theta^3_\mu(s) $ is not settled 
without the U(1)$_3$ gauge fixing.

\begin{figure}[tb]
\vspace{0cm}
\epsfxsize = 11 cm
\centering \leavevmode
\epsfbox{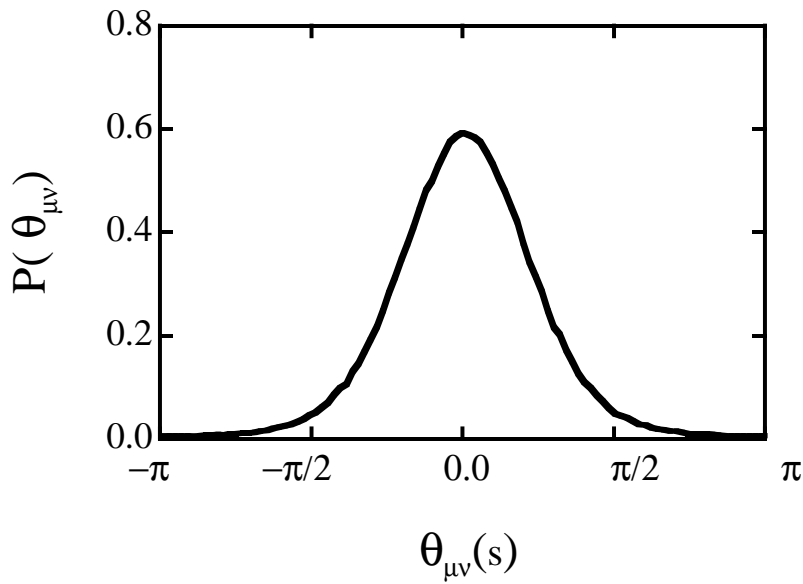}
\vspace{0cm}
\caption{   
The probability distribution $P(\bar \theta_{\mu\nu})$ of the 
abelian field strength $\bar \theta_{\mu\nu}(s)$, which is 
U(1)$_3$ gauge invariant.
}
\label{gfig5c}
\vspace{0cm}
\end{figure}
\begin{figure}[tb]
\vspace{0cm}
\epsfxsize = 11 cm
\centering \leavevmode
\epsfbox{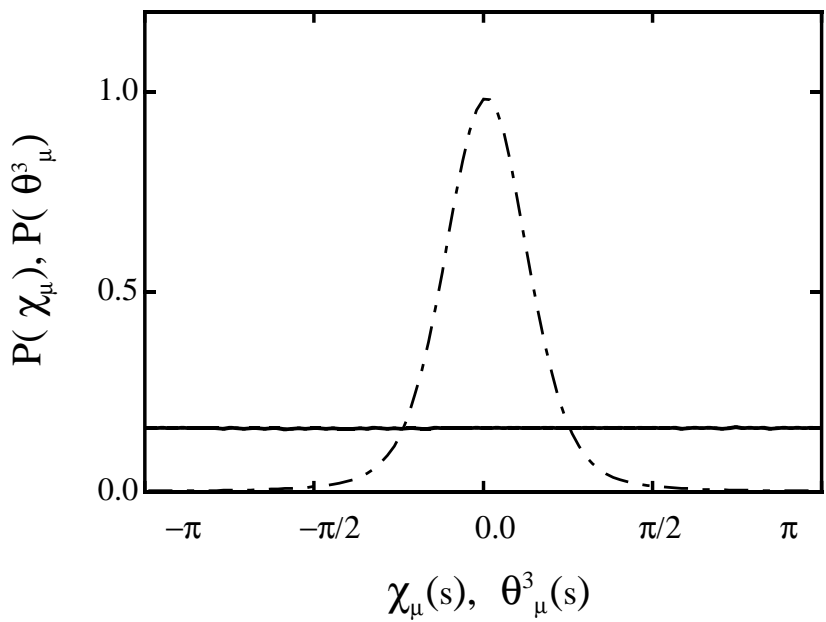}
\vspace{0cm}
\caption{   
The probability distributions $P(\chi_\mu)$ (solid line)
and $P(\theta^3_\mu)$ (dash-dotted curve)
in the MA gauge with the U$(1)_{3}$ Landau gauge
at $\beta =2.4$ on the $16^4$ lattice from 40 gauge configurations.}
\label{gfig4d}
\vspace{0cm}
\end{figure}

We show in Fig.\ref{gfig5b} the probability distribution 
$P(\theta_{\mu\nu})$ of the 
two form $\theta_{\mu\nu}(s) \equiv (\partial \wedge \theta^{\rm 
3})_{\mu\nu}
(s) $ of the abelian field $\theta^3_\mu(s)$.
Without the gauge fixing, there appear three peaks
around $\theta_{\mu\nu}(s)=$ $-2\pi$, $0$, $2 \pi$.
In the MA gauge,
because of $U_\mu \simeq u_\mu$, not only the SU(2) action
$\displaystyle  \hat S \equiv  S/\beta = \sum_{s, \mu > \nu}[1- \frac12
{\rm tr} \Box_{\mu\nu}(s) ]$ 
but also the abelian action 
$\displaystyle  \hat S^{\rm Abel} \equiv \sum_{s, \mu > \nu}[1- \frac12
{\rm tr} \Box^{\rm Abel}_{\mu\nu}(s)]$
is suppressed by the action factor $e^{-\beta \hat S}$
in the partition functional.
Since the abelian action is written by 
$\displaystyle \sum_{s, \mu > \nu}[1- \cos( \theta_{\mu\nu})]$, 
$P(\theta_{\mu\nu})$ has peaks around  $\cos( \theta_{\mu\nu}) =1$.
As shown in Fig.\ref{gfig5b},
most of $\theta_{\mu\nu}(s)$ 
distribute  around  $\theta_{\mu\nu}(s)=0$  in the U(1)$_3$ Landau gauge,
because the abelian gauge field is
mostly continuous in the U(1)$_3$ Landau gauge. 
On the other hand, the probability distribution $P(\bar \theta_{\mu\nu})$
of the abelian field strength 
$\bar \theta_{\mu\nu} \in (-\pi,\pi]$
is U(1)$_3$  gauge invariant.
The whole shape of 
$P(\bar \theta_{\mu\nu})$ is Gaussian-type, as shown in Fig.\ref{gfig5c}.

Finally in this section, we investigate the off-diagonal phase variable 
$\chi_{\mu}(s)$
in Eq.(\ref{divided}).
We show in Fig.\ref{gfig4d}(a) the probability distributions $P(\chi_{\mu})$
and $P(\theta^3_{\mu})$ at $\beta=2.4$ in the MA gauge with U(1)$_3$ 
Landau gauge.
Unlike $P(\theta^3_{\mu})$, $P(\chi_{\mu})$ is flat distribution
without any structure.
This property on the off-diagonal element would
lead to the validity 
of the random-variable approximation for $\chi_\mu(s)$, which has been 
used for
estimation of the Wilson loop in Eq.(\ref{eq:wilson}).


\section{Randomness of Off-diagonal Gluon Phase and
 Abelian Dominance}

In this section, 
we reconsider the origin of abelian dominance in the 
MA gauge in terms of the properties of the off-diagonal element
\begin{eqnarray}
c_\mu(s) \equiv e^{i \chi_{\mu}(s)} \sin \theta_{\mu}(s)
\end{eqnarray}
in $M_{\mu}(s)$ in the link variable $U_\mu(s)$, 
considering the validity 
of the random-variable approximation for $\chi_\mu(s)$ 
in the MA gauge
with U(1)$_3$ Landau gauge \cite{ichiead}.
 In this treatment, the contribution of the off-diagonal element in 
the link variable $U_{\mu}(s)$ is completely dropped off, and its 
effect indirectly  remains as the appearance of the $c$-number
factor $\cos  \theta_{\mu}(s)$ in the link variable. Such a reduction 
of the contribution of the off-diagonal elements is brought by the two 
relevant features on the two local variables, $\theta_{\mu}(s)$
and  $\chi_{\mu}(s)$, in the MA gauge.
One is  microscopic abelian dominance as 
$\langle \cos \theta_{\mu}(s) \rangle_{\rm MA} \simeq 1$ in the MA 
gauge, and 
the other is the randomness of the off-diagonal phase variable $\chi_{\mu}(s)$.  
\begin{enumerate}
\item In the MA gauge,  microscopic abelian dominance holds as 
$\langle \cos \theta_{\mu}(s) \rangle_{\rm MA}$ $\simeq 1$, and the absolute value of the 
off-diagonal element 
$|c_{\mu}(s)| = |\sin  \theta_{\mu}(s) |$ is strongly reduced. Such a 
tendency becomes more significant as $\beta$ increases.
\item The off-diagonal angle variable  $\chi_{\mu}(s)$ is not constrained by the 
MA gauge-fixing condition at all, and tends to be a random variable.
In fact, $\chi_{\mu}(s)$ is affected only by the QCD action factor $e^{- 
\beta \hat S_{\rm QCD}}$ in the QCD generating functional, but the effect 
of the action to $\chi_{\mu}(s)$ is quite weaken due to the small factor 
$\sin  \theta_{\mu}(s)$ in the MA gauge.
The randomness of $\chi_{\mu}(s)$ tends to vanish the contribution of 
the off-diagonal elements.
\end{enumerate}

Here, the randomness of the off-diagonal angle-variable
$\chi_\mu(s)$ is closely related to 
microscopic abelian dominance.
In fact, the randomness of $\chi_{\mu}(s)$ is controlled 
only by the action factor $e^{- 
\beta \hat S_{\rm QCD}}$ in the QCD generating functional, however the effect 
of the action to $\chi_{\mu}(s)$ is quite weaken due to the small factor 
$\sin  \theta_{\mu}(s)$ in the MA gauge, because $\chi_{\mu}(s)$ always 
accompanies $\sin \theta_{\mu}(s)$  in the link variable $U_{\mu}(s)$.
Near the strong-coupling limit $\beta \simeq 0$,
the action factor $e^{-\beta \hat S_{\rm QCD}}$ brings almost no 
constraint on  $\chi_{\mu}(s)$ in the MA gauge.
The independence of $\chi_{\mu}(s)$ from the action factor is enhanced 
by the small factor $\sin \theta_{\mu}(s)$
accompanying $\chi_{\mu}(s)$.
Hence, 
$\chi_\mu(s)$ behaves as a random angle-variable almost exactly, and
the contribution of the off-diagonal element is expected 
to disappear in the strong-coupling region.
As $\beta$ increases, the action factor 
$e^{-\beta \hat S_{\rm QCD}}$ becomes relevant and will reduce the 
randomness of 
$\chi_{\mu}(s)$ to some extent.
Near the continuum limit $\beta \rightarrow \infty$, however, the 
factor $\sin \theta_{\mu}(s)$ tends to approach 0 in the MA gauge as shown in 
Fig.\ref{gfig3}(b), and hence such a constraint on $\chi_{\mu}(s)$ from the 
action is largely reduced, and the strong  randomness of
$\chi_{\mu}(s)$ is expected to hold there.
Moreover, the reduction of the absolute value $| c_{\mu}(s)| = | \sin 
\theta_{\mu}(s)|$ itself
further reduces the 
contribution of the off-diagonal element $|c_{\mu}(s)|$ in the MA gauge.

\begin{figure}[bt]
\epsfxsize = 11 cm
\centering \leavevmode
\epsfbox{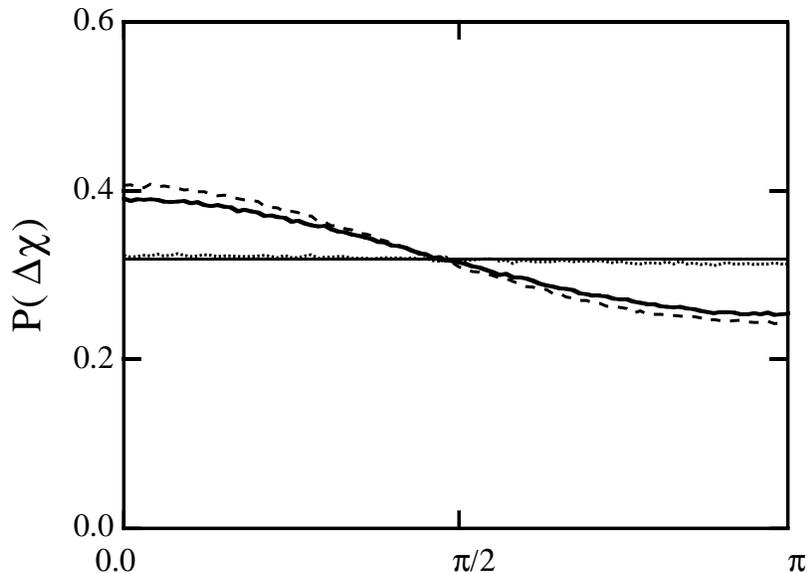}
\vspace{0cm}
\caption{    
The probability distribution $P(\Delta \chi)$
of the correlation $\Delta \chi \equiv {\rm 
mod}_{\pi}(|\chi_\mu(s)-\chi_\mu(s+ \hat \nu)|)$ in the same gauge 
at $\beta$ = 0 (thin line), 1.0 (dotted curve), 2.4 (solid curve), 
3.0 (dashed curve).
}
\label{gfig4e}
\vspace{0cm}
\end{figure}

Now, we examine the randomness of $\chi_\mu(s)$
using the lattice QCD simulation.
We calculate 
the correlation 
between $\chi_\mu(s)$ and $\chi_\mu(s+ \hat \nu)$
in the MA gauge with the U(1)$_3$-Landau gauge.
If  $\chi_{\mu}(s)$ is an exact random angle variable, 
no correlation is 
observed between $\chi_\mu(s)$ and $\chi_\mu(s+ \hat \nu)$.
We show in Fig.\ref{gfig4e}
the probability distribution $P(\Delta \chi)$
of the correlation
\begin{eqnarray}
\Delta \chi(s) \equiv d(\chi_\mu(s),\chi_\mu(s+ \hat \nu)) \equiv 
{\rm mod}_{\pi}|\chi_\mu(s)-\chi_\mu(s+ \hat \nu)| \in [0,\pi],
\end{eqnarray}
which is the difference between two neighboring angle variables,
at $\beta$=0, 1.0, 2.4, 3.0.
In the strong-coupling limit $\beta=0$,
$\chi_\mu(s)$ is a completely random variable, and
there is no correlation between neighboring $\chi_\mu$.
In the strong-coupling region as $\beta \le 1.0$,
almost no correlation is observed between neighboring $\chi_\mu$,
which suggests the strong randomness of $\chi_\mu(s)$.
As a remarkable feature, the correlation between
neighboring $\chi_\mu$ seems weak
even in the weak-coupling region as $\beta \ge 2.4$,
where the action factor $e^{-\beta \hat S_{\rm QCD}}$ becomes dominant
and remaining variables $\theta_\mu^3(s)$ and $\theta_\mu(s)$ behave as 
continuous variables with small difference between their neighbors as
$\Delta \theta_\mu^3 \simeq 0$ and $\Delta \theta_\mu \simeq 0$.
Such a weak correlation of neighboring $\chi_\mu$ would be originated 
from the reduction of the accompanying factor $\sin \theta_{\mu}(s)$ in the MA gauge.
Moreover, in the weak-coupling region,
the smallness of $\sin \theta_\mu(s)$ makes $c_\mu(s)$
more irrelevant in the MA gauge, which permits some approximation on $\chi_\mu(s)$.
Thus, the random-variable approximation for $\chi_\mu(s)$
would provide a good approximation
in the whole region of $\beta$ in the MA gauge.
To conclude, the origin of abelian dominance for confinement in the MA 
gauge is stemming from the strong randomness of the off-diagonal angle 
variable $\chi_\mu(s)$ and the strong reduction of the off-diagonal 
amplitude $|\sin \theta_\mu(s)|$ as the result of the MA gauge fixing.

\section{Comparison with SU(2) Landau Gauge}

In this section, we study the feature of the MA gauge in terms of the 
concentration of the gluon field fluctuation into the U(1)$_3$ sector by 
comparison with the SU(2) Landau gauge \cite{mandula}.
In the MA gauge, off-diagonal gluon components are forced to be small by 
the gauge transformation. 
Instead,  the gluon field fluctuation is maximally concentrated into the 
abelian sector, and monopoles appear in the abelian sector as the result 
of the large fluctuation of the abelian field component.
For the qualitative argument on the share of the gluon fluctuation into 
each component, we measure $\langle U_\mu^a(s) \rangle$ ($a$ =1,2,3) and 
\begin{eqnarray} 
R_{\rm diag}(s) & \equiv  & 
\frac{U^{3}_\mu(s)^2}{ U^{1}_\mu(s)^2+U^{2}_\mu(s)^2+U^{3}_\mu(s)^2}. 
\end{eqnarray}

In the MA gauge with the U(1)$_3$ Landau gauge, we find a strong 
concentration of the gluon field fluctuation into the abelian sector as 
$\langle U_\mu^1(s)\rangle =\langle U_\mu^2(s)\rangle \ll 
\langle U_\mu^3(s)\rangle $ and $R_{\rm diag} \gg \frac13;$ 
as a typical example, we find at $\beta = 2.4$
$\langle U_\mu^1(s)\rangle =\langle U_\mu^2(s)\rangle \simeq 0.067 $,
$\langle U_\mu^3(s)\rangle \simeq 0.43$ and $R_{\rm diag} \simeq 0.68$.
%
%

For comparison,  we consider the SU(2)  lattice Landau 
gauge \cite{mandula}
defined 
by maximizing
\begin{eqnarray}
R_L[U_\mu] \equiv \sum_{s,\mu}  {\rm tr} U_\mu(s) = 2 \sum_{s,\mu}  
U^0_\mu(s),
\end{eqnarray}
where all the lattice gluon components
fields become mostly continuous owing to the 
suppression of their fluctuation around $U_\mu(s) = 1$. 
In the continuum limit, 
this gauge fixing condition  coincides the ordinary SU(2) 
Landau gauge condition
$\partial_\mu A_\mu = 0$.
In the  lattice SU(2) Landau  gauge,
one finds $R_{\rm diag} = \frac13$ and 
$\langle U_\mu^1(s)\rangle =\langle U_\mu^2(s)\rangle  =  
\langle U_\mu^3(s)\rangle \ll 1$, for instance 
$\langle U_\mu^a(s)\rangle =$ 0.076 at $\beta =2.4$, so that all the 
gluon components are forced to be small equally.
%
%

\begin{figure}[p]
\epsfxsize = 11 cm
\centering \leavevmode
\epsfbox{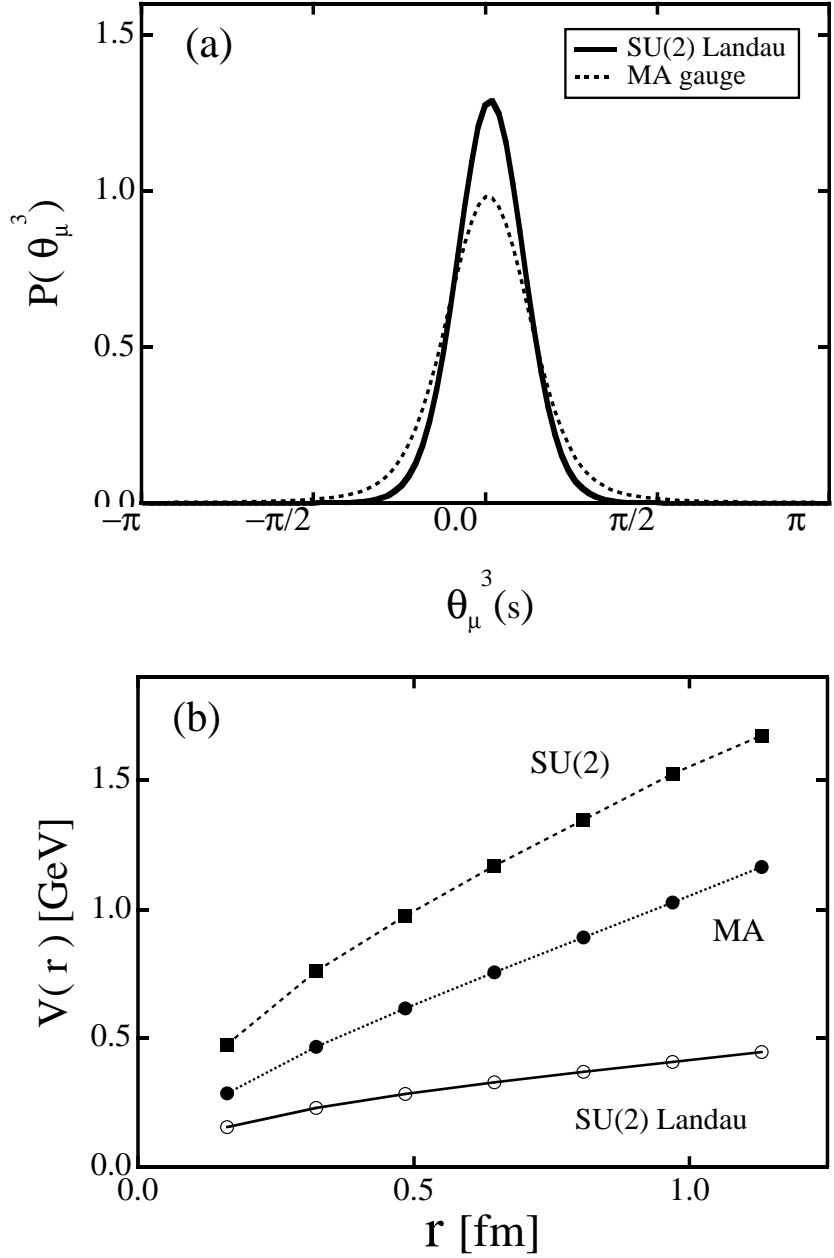}
\vspace{0cm}
\caption{   
Comparison between the MA gauge and the SU(2) Landau gauge
with global SU(2)/(U(1) $\times$ Weyl) fixed.
(a) The probability distribution $P(\theta^3_\mu)$ of the 
lattice angle variable 
$\theta^3_\mu(s)$
in the SU(2) Landau gauge (solid curve) and the MA gauge (dotted curve )
with $\beta=2.4$ on $16^4$ lattice.
The large fluctuation on the link variable $\theta^3_\mu(s)$ disappears
in the SU(2) Landau gauge. 
The abelian part of the interquark potential $V(r)$ as the function of 
interquark distance $r$ in the SU(2) Landau gauge  and in the MA gauge on 
$16^4$ lattice with $\beta=2.4$. 
For comparison, the full SU(2) result  is added.
}
\label{gfig6}
\vspace{0cm}
\end{figure}

In the SU(2) Landau gauge, the local symmetry of SU(2) is fixed, and only 
the global SU(2) symmetry remains,
because $\displaystyle R_L [U_\mu] \equiv \sum_{s,\mu}  {\rm tr} U_\mu(s)$ 
is invariant by 
any global gauge transformation.
In order to compare with the MA gauge, we fix SU(2)/(U(1)$_3$ $\times$ 
Weyl)
in this global SU(2) symmetry  by the additional condition so as to 
maximize 
$\displaystyle R_{\rm MA}[U_\mu] = \sum_{s,\mu} {\rm tr}(U_\mu(s) \tau_3 
U_\mu^\dagger(s) \tau_3)$ 
by the remaining global SU(2) gauge transformation.
Here, the SU(2) Landau gauge with SU(2)$_{\rm global}$/(U(1)$_{3\rm 
global} 
\times$ Weyl) fixing is regarded as a kind of the abelian gauge. 
Then, we can extract the abelian variable $\theta_\mu^3(s)$ 
 even for this SU(2) Landau gauge.
Figure \ref{gfig6}(a) shows the probability distribution $P(\theta_\mu^3)$
of the abelian angle variable $\theta_\mu^3(s)$ in the SU(2) 
Landau gauge and in the MA gauge at $\beta = 2.4$.
We show also in Fig.\ref{gfig6}(b) the interquark potential in the abelian 
sector 
evaluated from the abelian Wilson loop in these gauges.
Although the global shape of the distribution $P( \theta_\mu^3)$ in the
SU(2) Landau gauge is similar to that in the MA gauge except for the 
reduction of the large fluctuation apart from $\theta_\mu^3(s)=0$,
the abelian string tension in the SU(2) Landau gauge is much smaller than 
that in the MA gauge \cite{bali}.
Therefore, the large fluctuation ingredient is expected to be responsible 
for the confinement property.

To summarize, in the MA gauge, the field fluctuation is maximally 
concentrated into the abelian sector, and hence large fluctuation 
ingredient appears and the confinement property is almost reproduced 
only by the abelian variable.
Another clear difference between the MA gauge and the SU(2) Landau gauge 
observed on lattice is the density of monopoles
appearing in the abelian sector.
Indeed, the SU(2) Landau gauge includes scarcely monopoles in the 
abelian sector in comparison with the MA gauge,
for instance, the ratio on the monopole density is less than 1/10 at 
$\beta=2.4$.
 This result seems natural because  the SU(2) Landau gauge fixing 
 provides a mostly continuous gluon field, while the monopole arises from 
 a singular-like large fluctuation of the abelian field as will be shown 
 in Chapter \ref{sec:QCDMAG} and \ref{sec:LFFM} in detail.
 In the next chapter, we study features of monopole appearing in 
 the MA gauge in relation with confinement and large field fluctuation
 concentrated into the abelian sector.

  
\chapter{QCD-Monopole in the Abelian Gauge}
\label{sec:QCDMAG}

In the abelian gauge, QCD is reduced into an abelian gauge theory 
with QCD-monopoles, which appear from the hedgehog-like configuration 
corresponding to the nontrivial homotopy group on the nonabelian gauge 
manifold,
$\Pi_2($SU$(N_c)/$U$(1)^{N_c-1})$ $= {\bf Z}^{N_c-1}$.
The relevant role of the QCD-monopole to the infrared phenomena has been
studied by using the infrared effective theory
and the lattice gauge simulation \cite{diacomo,poly}.
In the dual Ginzburg-Landau(DGL) theory, 
the linear static quark potential,
which characterizes quark confinement,
is obtained in the monopole condensed vacuum \cite{suganuma}. 
In addition, chiral symmetry breaking is also
brought from the monopole contribution in the DGL 
theory \cite{suganuma,sasaki}.
The recent lattice QCD studies in the MA gauge  suggest 
monopole condensation   
in the confinement phase in the MA gauge, and show abelian dominance and
monopole dominance for nonperturbative QCD \cite{diacomo,poly}.
Here, monopole dominance means that QCD phenomena are described only by
the monopole part of the abelian variables in the abelian gauge.  
In this chapter, we study appearance of QCD-monopoles
in the abelian sector of QCD 
and clarify the difference between the ordinary
QED and abelian projected QCD (see Fig.1.1).

\section{Appearance of  Monopoles in the SU(2) Singular Gauge 
Transformation}
\label{appearance}

The abelian gauge fixing, which reduces QCD into an abelian gauge
theory, is realized by the diagonalization of a suitable variable
$\Phi[A_\mu(x)]$. 
In the continuum theory of QCD, the continuous field $A_\mu(x)$ can be
taken to be regular everywhere in a suitable gauge as the Landau 
gauge, and then
$\Phi[A_\mu(x)]$ is expected to be a regular function almost everywhere.
In the abelian gauge, however, there appears the singular point,
where the gauge function to diagonalize $\Phi[A_\mu(x)]$ is not 
uniquely determined even for the off-diagonal part, and  such a 
singular point 
leads to the
appearance of the monopole.

\begin{figure}[bt]
\epsfxsize = 11 cm
\centering \leavevmode
\epsfbox{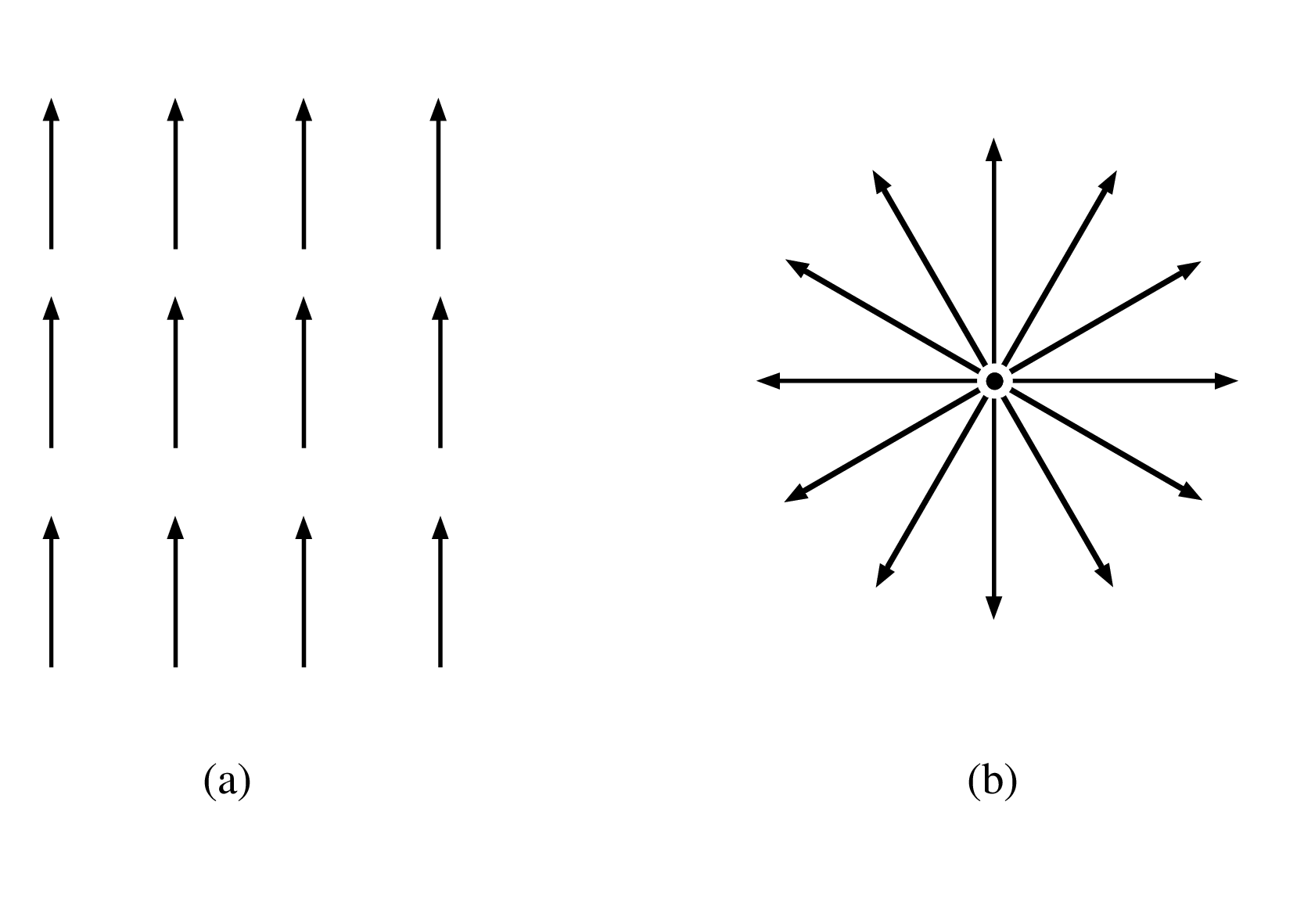}
\vspace{-0.8cm}
\caption{
Topological structure of variable
$\Phi[A_{\mu}(x)]$ 
in the abelian gauge fixing in the SU(2) QCD.
In the abelian gauge, the monopole appears at the singular
point of $\hat \Phi(x) \equiv \Phi/|\Phi|$ with $|\Phi| \equiv 
(\Phi^a\Phi^a)^{1/2}$. 
(a) For the regular (trivial) configuration of $\hat 
\Phi[A_{\mu}(x)]$,
 no monopole appears in the abelian gauge.
(b) For the hedgehog configuration of $\hat 
\Phi[A_{\mu}(x)]$,
the unit-charge monopole appears in the abelian gauge.
}
\label{Higgs}
\end{figure}

Here, let us consider the appearance of QCD-monopoles in the abelian gauge 
in terms of the singularity in the gauge transformation \cite{suganuma}.
For the variable 
$\Phi(x)$ obeying the adjoint transformation,
the monopole appears at the ``degeneracy point'' of 
the diagonal elements of $\vec H \cdot \vec \lambda(x) = 
diag(\lambda^{1}(x),\lambda^{2}(x),\cdots, \lambda^{N_{c}}(x))$
after the abelian gauge fixing: 
$(i,j)$-monopole appears at the point satisfying 
$\lambda^{i}(x)=\lambda^{j}(x)$.
For the $(i,j)$-monopole, the SU(2) subspace relating  to $i$ and $j$
is enough to consider, so that the essential feature of the monopole 
can be understood in the SU(2) case without loss of generality.
Then, we consider the SU(2) case for simplicity. For the SU(2) case,
the diagonalized element of $\Phi(x)$ are given by 
$\lambda = \pm(\Phi_{1}^2 +\Phi_{2}^2 +\Phi_{3}^2)^{1/2}$, and hence 
the ``degeneracy point'' satisfies the condition $\Phi(x)= 0$, 
which is ${\rm SU}(2)$ gauge invariant. 
This gauge-invariant condition $\Phi(x)=0$ can be regarded as 
the singularity condition on $\hat \Phi(x)\equiv \Phi(x)/|\Phi(x)|$ 
with $|\Phi(x)|\equiv (\Phi^a(x)\Phi^a(x))^{1/2}$. 
In fact, the ``degeneracy point'' in the abelian gauge 
appears as the singular point of $\hat \Phi(x)$ like the center 
of the hedgehog configuration as shown in Fig.\ref{Higgs}(b) 
before the abelian gauge fixing.

Since the singular point on $\hat \Phi(x)$ is to satisfy 
three conditions $\Phi^1(x)=\Phi^2(x)=\Phi^3(x)=0$ simultaneously, 
the set of the singular point forms the point-like manifold in ${\bf R}^3$
or the line-like manifold in ${\bf R}^4$.
We investigate the topological nature near the singular point 
$({\bf x}_0,t)$ of $\hat \Phi(x)$ for fixed $t$, i.e., 
$\Phi({\bf x}_0,t)=0$ \cite{suganuma}.
Using the Taylor expansion,
one finds 
\begin{eqnarray}
\Phi({\bf x},t) = \Phi^a({\bf x},t) \frac{\tau^a}{2} 
\simeq \tau^a C^{ab} ({\bf x}-{\bf x}_0)^b,
\end{eqnarray}
with $C^{ab} \equiv \frac12 \partial^{b} \Phi^a({\bf x}_0,t) $.
In the general case, one can expect 
det$C \ne 0$, i.e., det$C > 0$ or
det$C < 0$, and the fiber-bandle $\Phi^a({\bf x})$ can be deformed 
into the (anti-)hedgehog configuration
$\Phi(\tilde {\bf x}) \simeq \pm \tau^a \tilde {\bf x}^a$
around the singular point ${\bf x}_0$
by using the continuous modification on the spatial 
coordinate ${\bf x}^a \rightarrow \tilde {\bf x}^a \equiv 
{\rm sgn}({\rm det}C) \cdot C^{ab} ({\bf x}-{\bf x}_0)^b$.
The linear transformation matrix $C$ can be written 
by a combination of the rotation $R$ and the dilatation 
of each axis $\lambda = diag(\lambda^1,\lambda^2,\lambda^3)$
with $\lambda^i > 0$ as $C = {\rm sgn}( {\rm det}C) R \lambda$. 
Here, topological nature is never changed by such a continuous deformation.
For det$C > 0$, the configuration $\Phi({\bf x})$
can be continuously deformed
into the hedgehog configuration around  ${\bf x}_0$, $\Phi( \tilde {\bf x})
\simeq \tau^a \tilde {\bf x}^a$, while,
for det$C < 0$,  $\Phi({\bf x})$ can be continuously deformed
into the anti-hedgehog configuration, $\Phi( \tilde {\bf x})
\simeq - \tau^a \tilde {\bf x}^a$.
Since det$C=0$ is the exceptionally special case and  det$C < 0$ is 
similar to det$C > 0$, we have only to consider the hedgehog 
configuration.
This hedgehog configuration around the singular point 
of $\hat \Phi(x)$ corresponds to the simplest nontrivial topology of 
the nontrivial homotopy group 
$\Pi_2({\rm SU(2)/U(1)_3})=Z_\infty$, and 
the abelian gauge field has the singularity as the monopole 
appearing from the hedgehog configuration.

Using the polar coordinate $(r,\theta,\varphi)$ of $\tilde {\bf x}$,
the hedgehog configuration is expressed as
\begin{eqnarray}
\Phi & =  & \tau^a \tilde {\bf x}^a= 
   r \sin \theta \cos \varphi \cdot \tau_1 
+  r \sin \theta \sin \varphi \cdot \tau_2
+  r \cos \theta              \cdot \tau_3 
\nonumber \\  
 & = &
r  
\left( {\matrix{
{\rm cos}{\theta}   &
e^{-i\varphi}{\rm sin} \theta \cr
e^{i\varphi} {\rm sin} \theta &
 -{\rm cos}\theta }
} \right),
\end{eqnarray}
and $\Phi$ can be diagonalized by the gauge transformation with
\begin{eqnarray}
 \Omega^H = 
\left( {\matrix{
e^{i\varphi}{\rm cos}{\frac{\theta}{2}}  & {\rm sin}{\frac{\theta}{2}} \cr
-{\rm sin}{\frac{\theta}{2}} & e^{-i\varphi}{\rm cos}{\frac{\theta}{2}}}
} \right),
\label{eq:gauge-function}
\end{eqnarray}
where $\theta$, $\varphi$ denote the polar and the azimuthal angles, 
respectively.
Here, on the $z$-axis ($\theta=0$ or $\theta=\pi$), $\varphi$ is the 
``fake parameter'', and the unique description does not allow the 
$\varphi$-dependence on the $z$-axis. However, at the positive region 
of $z$-axis, $\theta = 0$, $\Omega^H$ depends on $\varphi$ and is 
multi-valued as
\begin{eqnarray}
 \Omega^H = 
\left( 
{
\matrix{
e^{i\varphi} & 0 \cr
0 & e^{-i\varphi}
}
} 
\right). 
\end{eqnarray}
Such a multi-valuedness of $\Omega^H$ leads to the divergence in the 
derivative $\partial_\mu  \Omega^H$ at $\theta = 0$. 
In fact, $\partial_\mu  \Omega^H$  includes the singular part as
 $\cos \frac{\theta}{2}$ 
$(\nabla \varphi)_\varphi = \frac{\cos \frac{\theta}{2}}{r \sin \theta} 
\frac{\partial}{\partial \varphi} \varphi = \frac{1}{r \sin 
\frac{\theta}{2}}$, which  diverges
at $\theta=0$.
By the gauge transformation with $ \Omega^H$,
the variable $\Phi$ becomes
$\Phi^{\Omega} = \Omega  \Phi  \Omega^\dagger = r \tau^3$, and
the gauge field is transformed as
\begin{eqnarray}
A_\mu  \rightarrow  A_\mu^{\Omega} = 
 \Omega  (A_\mu  - \frac{i}{e} \partial_\mu)  \Omega^\dagger.
\end{eqnarray}
For regular $A_\mu$, the first term $ \Omega A_\mu  \Omega^\dagger$
is regular, while 
$A^{\rm sing}_\mu \equiv 
- \frac{i}{e}  \Omega  \partial_\mu  \Omega^\dagger$
is singular and
the monopole appears in the abelian sector originating from the singularity
of $A^{\rm sing}_\mu $
 \cite{suganuma}.
To examine the appearance of the monopole at the origin $\tilde {\bf 
x}=0$, we consider the magnetic flux
$\Phi^{\rm flux}( \theta )$
which penetrates the area inside the closed contour 
$c(r,\theta) \equiv \{ (r,\theta,\varphi)  | 0 \le \varphi< 2 \pi \} $.
One finds that
\begin{eqnarray}
\Phi^{\rm flux}( \theta ) 
& = & \int_c d{\bf x} \cdot {\bf A}^{\rm sing}  
= -\frac{i}{e} \int_c  d{\bf x} \Omega {\nabla}  \Omega^\dagger  
 \nonumber \\ 
& = & -\frac{i}{e} \int^{2\pi}_{0} d \varphi  \Omega 
\frac{\partial}{\partial \varphi}   \Omega^{\dagger} 
=-\frac{4\pi}{e} \cdot \frac{1 + \cos \theta}{2}  \frac{\tau_3}{2},
\end{eqnarray}
which denotes the 
magnetic flux of the monopole with the unit-magnetic charge 
$g = \frac{4 \pi}{e}$ with the Dirac string \cite{suganuma}.
Here, 
the direction of the Dirac string from 
the monopole can be arbitrary changed by the singular U$_{3}(1)$ gauge 
transformation, which can move $e^{i\varphi}$ in $\Omega^H$ from the 
$\tau_{3}$-sector to the off-diagonal sector.
In fact, 
the multi-valuedness of $\Omega$ is not necessary to be fixed 
in $\tau^3$-direction.
Nevertheless, 
the singularity in 
$\Omega \partial_{\mu} \Omega^\dagger$ appears only in the 
$\tau_{3}$-sector, and $\tau_{3}$-direction becomes special in the 
abelian gauge fixing.

The anti-hedgehog configuration of 
$\Phi( \tilde {\bf x})
= - \tau^a \tilde {\bf x}^a$
provides a
monopole with the  opposite magnetic charge, because anti-hedgehog 
configuration is transformed to the hedgehog configuration by the 
Weyl transformation. 
Thus, the only unit-charge magnetic monopole appears
in the general case of det$ C \ne 0$. 
In principle, the multi-charge 
monopole can 
also appear when  det$C =0$, however, the condition is scarcely satisfied
in general,
because this exceptional case is realized only when four conditions
$\Phi^1 = \Phi^2 = \Phi^3 = {\rm det}C = 0$ are simultaneously 
satisfied.  
To summarize, in the abelian gauge,
the unit-charge magnetic monopoles appear from the singular points of 
$\hat \Phi(x),$ however, multi-charge monopoles do not appear in general 
cases.


In this way, by the singular SU(2) gauge transformation,
there appears  the monopole with the Dirac string.
Here, we consider the role of the off-diagonal component in the SU(2)
gauge function $\Omega^H$ to appearance of the monopole, by comparing with the 
U(1)$_3$ gauge transformation.
Let us consider the singular gauge transformation
$\Omega^{\rm U(1)} = e^{i\varphi \tau_3} \in$ U(1)$_{3}$
instead of $ \Omega^H$. 
This U(1)$_{3}$ gauge function 
$\Omega^{\rm U(1)}$ is multi-valued on the whole region of the $z$ axis 
($\theta=0$ and $\theta=\pi$),
and  
$A_{\mu}^{\rm sing} \equiv -\frac{i}{e} \Omega^{\rm U(1)} 
\partial_{\mu} \Omega^{{\rm U(1)}\dagger}$ also has a singularity.
The magnetic flux which penetrates the area inside the 
closed 
contour $c(r,\theta) = \{ r,\theta,\varphi | 0 \le \varphi < 2 \pi \}$ 
is found to be
\begin{eqnarray}
\Phi^{\rm flux}( \theta ) 
=\int_c d{\bf x} \cdot {\bf A}^{\rm sing}   = - \frac{4\pi}{e}  \frac{\tau_3}{2},
\end{eqnarray}
which corresponds to the endless Dirac string along the $z$-axis.
It is noted that the singular U(1)$_3$ gauge transformation can provide 
the endless Dirac string, however, it never creates the monopole.

The monopole is created not by above singular U(1)$_3$ 
gauge transformation but by a singular SU(2) gauge transformation.
Since the multi-valuedness of $ \Omega^H$ is 
originated from 
the $\varphi$-dependence
at $\theta = 0$ or $\theta = \pi$, 
we separate the SU(2) gauge function  (\ref{eq:gauge-function}) as
\begin{eqnarray}
 \Omega = {\rm cos}\frac{\theta}{2} e^{i\varphi \tau_3}+  
( \varphi \mbox{-independent term} ). \nonumber
\end{eqnarray}
At $\theta=0$ or the positive side of $z$
 axis, $ \Omega^H $ coincides with  $\Omega^{\rm U(1)} \equiv e^{i 
 \varphi \tau_3}$ and is multi-valued like $\Omega^{\rm U(1)}$.
Therefore the Dirac string is created at $\theta = 0$ by the gauge
transformation $ \Omega^H$.
 On the other hand, at $\theta = \pi$ or the negative side of $z$-axis,
 $\varphi$-dependent part of $ \Omega$ vanishes due to $\cos 
 \frac{\theta}{2} = 0$, so that the Dirac string never appears in $ \Omega 
 \partial_\mu  \Omega^\dagger$ at $\theta = \pi$.
 Thus, by the SU(2) singular gauge transformation  $ \Omega^H$,
 the Dirac string is generated only on the positive side of the $z$-axis
 and terminates at the origin $r = 0$, and hence the monopole appears at the end 
 of the Dirac string.
Around the origin $ \tilde {\bf x}=0$, the factor $\cos \frac{\theta}{2}$ varies from 
unity to zero continuously 
with the polar angle $\theta$, and this makes the Dirac string 
terminated. 
Such a variation of the norm of the diagonal component  $\cos 
\frac{\theta}{2}
e^{i\varphi}$ cannot be realized in the U(1)$_3$ gauge transformation
with $\Omega^{\rm U(1)}$.
In the SU(2) gauge transformation with $\Omega^H$, the norm of the 
diagonal component can be changed owing to existence of the 
off-diagonal component of $\Omega^H$, and the difference of the 
multi-valuedness between $\theta=0$ and $\theta=\pi$ leads to the 
terminated Dirac string and the monopole. 
In this way, to create the monopole in QCD,
full SU(2) components of the (singular)
gauge transformation is necessary, and therefore one can expect a close 
relation between monopoles and the off-diagonal component of the gluon 
field.


\section{Appearance of Monopoles in the  Connection Formalism}

In this section, we study the appearance of monopoles 
in the abelian sector of QCD in the abelian gauge in detail using
the gauge connection formalism.
In the abelian gauge, 
the monopole or the Dirac string appears as the result of 
the SU($N_{c}$) singular gauge transformation from a 
regular (continuous) gauge configuration. 
For the careful description of the singular gauge transformation, 
we formulate the gauge theory in terms of 
of the gauge connection, described 
by the covariant-derivative operator $\hat D_\mu$
and
$
\hat D_\mu  \equiv \hat \partial_\mu +ieA_\mu (x), 
$
where $\hat \partial_\mu $ is the derivative operator satisfying 
$[\hat \partial_\mu ,f(x)]=\partial_\mu f(x)$.

To begin with, let us consider the system holding the local 
difference of the internal-space coordinate frame. 
We attention the neighbor of the real space-time $x_\mu$,
and denote by $|q(x) \rangle$ the basis of the internal-coordinate frame.
At the neighboring point $x_\mu + \varepsilon_\mu$, we express the 
difference of the internal-coordinate frame as
$
|q(x+\varepsilon) \rangle  =
R_\varepsilon (x) |q(x) \rangle
$
with $R_\varepsilon (x) = e^{ir_\varepsilon(x)}$ $\in G$ being
the ``rotational matrix'' of the internal space.
We require the ``local superposition'' on $r_\varepsilon$ as
$r_{\varepsilon_1+\varepsilon_2}= r_{\varepsilon_1}+r_{\varepsilon_2}$
up to $O(\varepsilon)$, and then we can express 
$r_\varepsilon(x) = -e \varepsilon_\mu A^\mu(x)$ using a
$\varepsilon$-independent local variable $A_\mu(x)$ $\in$ $g$ :
$|q(x+\varepsilon) \rangle = e^{-i e\varepsilon_\mu A^\mu(x) } |q(x) \rangle.$
Then, the ``observed difference'' of the internal space 
coordinate depends on 
the real space-time $x_\mu $, 
the observed difference of the local operator $O(x)$ 
between neighboring points, 
$x_\mu $ and $x_\mu +\varepsilon _\mu$, is given by 
\begin{eqnarray}
&  & \langle q(x+\varepsilon )|O(x+\varepsilon )|q(x+\varepsilon ) \rangle  
 -  \langle q(x)|O(x)|q(x) \rangle  \nonumber \\
& = & 
 \langle q(x)| e^{ie \varepsilon _\mu  A^\mu (x)}O(x+\varepsilon ) 
 e^{-ie \varepsilon _\mu  A^\mu (x)}|q(x) \rangle - \langle q(x)|O(x)|q(x) 
\rangle
\nonumber \\
& \simeq & 
\varepsilon _\mu  \langle q(x)|\{\partial^\mu O(x)
 +ie[A^\mu (x), O(x)]\}|q(x) \rangle 
\nonumber \\
& = &
\varepsilon _\mu  \langle q(x)|\{[\hat \partial^\mu +ieA^\mu (x), 
O(x)]\}|q(x) \rangle 
\equiv
\varepsilon _\mu  \langle q(x)|[\hat D^\mu , O(x)]|q(x) \rangle. 
\end{eqnarray}
Here, one finds natural appearance of the covariant derivative operator, 
$
\hat D_\mu  \equiv \hat \partial_\mu +ieA_\mu (x).
$
The gauge transformation is simply defined by the 
arbitrary internal-space rotation as 
$|q(x) \rangle \rightarrow \Omega (x) |q(x) \rangle $
with $\Omega(x)$ $\in$ $G$,
and therefore the covariant derivative operator
is transformed as 
$
\hat D_\mu \rightarrow \hat D^\Omega_\mu  =\Omega (x)\hat D_\mu \Omega^\dagger 
(x)$
with $\Omega(x)$ $\in$ $G$,
which is consistent with 
$A_\mu \rightarrow A^\Omega_\mu =\Omega (A_\mu -\frac{i}{e}\partial_\mu )\Omega 
^\dagger$. 

In the general system including singularities such as the Dirac string,
the gauge field and the field strength are defined as
the difference between the gauge connection and the derivative 
connection,
\begin{eqnarray}
A_\mu & \equiv & \frac{1}{ie}({\hat D_\mu}-{\hat \partial_\mu}) \\
G_{\mu\nu} & \equiv & \frac{1}{ie}
([{\hat D_\mu},{\hat D_\nu}]-[\hat \partial_\mu,\hat \partial_\nu]).
\end{eqnarray}
This expression of $G_{\mu\nu}$ is returned to the standard definition 
$G_{\mu\nu} 
= \frac{1}{ie} [{\hat D_\mu},{\hat D_\nu}] 
= \partial_\mu A_\nu -\partial_\nu A_\mu + ie[A_\mu, A_\nu]$
in the regular system.
By the general gauge transformation with the gauge function $\Omega $, 
the field strength $G_{\mu\nu}$ is transformed as
\begin{eqnarray}
G_{\mu\nu} \rightarrow G^\Omega_{\mu\nu} & = &  
\Omega G_{\mu\nu} \Omega^{\dagger} 
 =  \frac{1}{ie} ([\hat D^\Omega_\mu ,\hat D^\Omega_\nu ]
-\Omega [\hat \partial_\mu , \hat \partial_\nu ]\Omega {}^\dagger) 
\nonumber \\
& = & \partial_\mu A^\Omega_\nu -\partial_\nu A^\Omega_\mu 
+ ie[A^\Omega_\mu, A^\Omega_\nu] +\frac{i}{e}
(\Omega[\hat \partial_\mu, \hat \partial_\nu] \Omega^{\dagger} 
-[\hat \partial_\mu, \hat \partial_\nu]) 
\nonumber \\
& = & (\partial_\mu A^\Omega_\nu -\partial_\nu A^\Omega_\mu )
+ ie[A^\Omega_\mu, A^\Omega_\nu] +\frac{i}{e}
\Omega[\partial_\mu, \partial_\nu] \Omega^{\dagger}
\nonumber \\ 
& \equiv  & G^{\rm linear}_{\mu\nu} 
+  G^{\rm bilinear}_{\mu\nu} + G^{\rm sing}_{\mu\nu}.
\label{eq:field-strength}
\end{eqnarray}
The last term remains only for the singular gauge transformation
on $\Omega^H$ and $\Omega^{\rm U(1)}$, 
and can provide the Dirac string.

\begin{figure}[bt]
\centering \leavevmode
\vspace{0cm}

{\tt    \setlength{\unitlength}{0.92pt}
\begin{picture}(400,250)

\thinlines    \put(130,180){\vector(1,2){28}}
              \put(130,180){\vector(2,1){52}}
              \put(130,180){\vector(-1,2){28}}
              \put(130,180){\vector(2,-1){52}}
              \put(130,180){\vector(-1,-2){28}}
              \put(130,180){\vector(-2,-1){52}}
              \put(130,180){\vector(1,-2){28}}
              \put(130,180){\vector(-2,1){52}}
              
              \put(130,180){\vector(-1,0){60}}
              \put(130,180){\vector(1,0){60}}
              \put(130,180){\vector(0,-1){60}}

\thinlines    \put(400,180){\line(1,2){28}}
              \put(400,180){\line(2,1){52}}
              \put(400,180){\line(-1,2){28}}
              \put(400,180){\line(2,-1){52}}
              \put(400,180){\line(-1,-2){28}}
              \put(400,180){\line(-2,-1){52}}
              \put(400,180){\line(1,-2){28}}
              \put(400,180){\line(-2,1){52}}
              
              \put(400,180){\line(-1,0){60}}
              \put(400,180){\line(1,0){60}}
              \put(400,180){\line(0,1){60}}
              \put(400,180){\line(0,-1){60}}

\thinlines    \put(414,208){\vector(-1,-2){0}}
              \put(427,166.5){\vector(-2,1){0}}
              \put(386,208){\vector(1,-2){0}}
              \put(427,193.5){\vector(-2,-1){0}}
              \put(386,152){\vector(1,2){0}}
              \put(373,166.5){\vector(2,1){0}}
              \put(414,152){\vector(-1,2){0}}
              \put(373,193.5){\vector(2,-1){0}}
              
              \put(368,180){\vector(1,0){0}}
              \put(433,180){\vector(-1,0){0}}
              \put(400,213){\vector(0,-1){0}}
              \put(400,147){\vector(0,1){0}}

              \put(217,180){\Large $ + $}
              \put(300,180){\Large $ + $}
              \put(260,240){\vector(0,1){0}}

\thinlines    \put(130,210){\line(1,1){12}}
              \put(130,210){\line(-1,1){12}}

\thinlines    \put(260,225){\line(1,-1){12}}
              \put(260,225){\line(-1,-1){12}}

              \linethickness{1.mm} 
              \put(260,180){\vector(0,1){60}}
              \put(130,240){\vector(0,-1){60}}
\put(10,80){{\large $G^\Omega_{\mu\nu}  
 = 
 \hspace{8mm}
 \underbrace{
   \partial_\mu A^\Omega_\nu  - \partial_\nu A^\Omega_\mu
 \hspace{8mm}
   +   \hspace{3mm} \frac{i}{e}
 \Omega[\partial_\mu, \partial_\nu]  \Omega^{\dagger}
 \hspace{3mm} }   + \hspace{8mm} ie[A^\Omega_\mu, A^\Omega_\nu]  $}
}

 \linethickness{0.4mm}  

\put(200,55){\vector(0,-1){28}}

\put(120,5){\large Abelian Projected QCD}

\end{picture}
}
\caption{   
Appearance of monopoles in abelian projected QCD(AP-QCD).
After the abelian gauge fixing, monopole with the Dirac string 
appears
from 
$G^{\rm linear}_{\mu\nu} $ 
in Eq.(\ref{eq:field-strength})
and the ``anti-Dirac string'' appears in the singular part  
$G^{\rm sing}_{\mu\nu}$. 
The off-diagonal contribution $G_{\mu\nu}^{\rm bilinear} = 
ie[A_{\mu},A_{\nu}]$ forms the anti-monopole configuration and 
compensates to the singularity of the other parts.
As the result, the monopole without the Dirac string appears
in the abelian field strength 
${\cal F}_{\mu\nu} $
in AP-QCD. 
}
\label{gfig7}
\vspace{0cm}
\end{figure}

Figure \ref{gfig7}
shows the SU(2) field 
strength $G_{\mu\nu}^{\rm linear}, 
G_{\mu\nu}^{\rm bilinear}$ and $ G_{\mu\nu}^{\rm sing}$ 
in the abelian gauge provided 
by $\Omega^H$ in Eq.(\ref{eq:gauge-function}).
The linear term $G^{\rm linear}_{\mu\nu} \equiv
(\partial_\mu A^\Omega_\nu -\partial_\nu A^\Omega_\mu ) $
includes in the abelian sector the singular gauge configuration of
{\it the monopole with the Dirac string}, which
supplies the magnetic flux from infinity.
Since each component satisfies the Bianchi identity 
$\partial^\alpha {}^* G^{ \rm linear}_{\alpha\mu} = \partial^\alpha {}^*
(\partial \wedge A^\Omega )_{\alpha\mu}=0$,
the abelian magnetic flux is conserved.
The abelian part of $ G^{\rm bilinear}_{\mu\nu}
\equiv ie[A^\Omega_\mu, A^\Omega_\nu]$, $ (G^{\rm bilinear}_{\mu\nu})^3
= -e(A_{\mu}^1 A_{\nu}^2-  A_{\nu}^1 A_{\mu}^2)$, 
includes the effect  
of off-diagonal components, and it is dropped by the abelian projection.
The last term
$G^{\rm sing}_{\mu\nu} \equiv 
\frac{i}{e}
\Omega[\partial_\mu, \partial_\nu]  \Omega^{\dagger}$
appears from the singularity of the gauge function $ \Omega$, and 
it plays the important role of the appearance of the magnetic monopole 
in the abelian sector.

First, we consider the singular part $G^{\rm sing}_{\mu\nu}$.
In general, $G^{\rm sing}_{\mu\nu}$ disappears in the regular point 
in $\Omega$. 
It is to be noted that $G^{\rm sing}_{\mu\nu}$ is found to be diagonal 
from the direct calculation  with  $\Omega^H$ in Eq.(\ref{eq:gauge-function}),
\begin{eqnarray}
G^{\rm sing}_{\mu\nu}  & \equiv & 
\frac{i}{e} \Omega^H[\partial_\mu, \partial_\nu] \Omega^{H\dagger} \nonumber 
 =   \frac{i}{e}(g_{\mu1} g_{\nu2} - g_{\mu2} g_{\nu1})
{\rm cos} ^2 \frac{\theta}{2} e^{i \varphi \tau_3} [\partial_1, \partial_2]
e^{-i \varphi \tau_3} \nonumber  \\
& = & \frac{1}{e} (g_{\mu1} g_{\nu2} - g_{\mu2} g_{\nu1})
\frac{1+ {\rm cos} \theta}{2} [\partial_1, \partial_2]  \varphi
\cdot \tau_3  
\nonumber \\ & = &
\frac{4 \pi}{e} (g_{\mu1} g_{\nu2} - g_{\mu2} g_{\nu1}
)
\theta(x_3)  \delta(x_1) \delta(x_2) \cdot \frac{\tau_3}{2},
\label{eq:monopole}
\end{eqnarray}
where we have used relations,
\begin{eqnarray}
[\partial_1, \partial_2]\varphi = - 2 \pi \delta(x_1) \delta(x_2), 
\hspace{0.4cm}
 \frac{1+ {\rm cos} \theta}{2} \delta(x_1) 
\delta(x_2)
= \theta(x_3) \delta(x_1) \delta(x_2). 
\end{eqnarray}
The off-diagonal component of $ \Omega ^H [\partial_\mu, \partial_\nu] 
\Omega^{H\dagger}$ disappears, since the 
singularity appears only from $\varphi$-dependent term. 
As a remarkable fact,
the last expression  in Eq.(\ref{eq:monopole})  
shows the terminated Dirac string,
which is placed along the positive $z$-axis with the end at the origin.
Hence, in the abelian part of the SU(2) field strength,
$G^{\rm sing}_{\mu\nu}$ leads to the breaking of the U(1)$_3$ Bianchi identity,
\begin{eqnarray}
k_\mu  & = & \partial^\alpha {}^* G^{\rm sing}_{\alpha\mu} 
= \frac12  \varepsilon_{\alpha\mu}{}^{\beta\gamma}
\partial^\alpha G^{\rm sing}_{\beta\gamma}
= \frac{4\pi}{e}  
 \varepsilon_{\alpha\mu12}\partial^\alpha \{ \delta(x_1)\delta(x_2)\theta(x_3) 
\}
\frac{\tau_3}{2}  \nonumber \\
 & = & 
 \frac{4\pi}{e} g_{\mu0} \delta(x_1)\delta(x_2)\delta(x_3) 
\frac{\tau_3}{2},
\label{eq:kk}
\end{eqnarray}
which is the expression for the static monopole
with the magnetic charge $g = \frac{4\pi}{e}$  at the origin.
Thus, the magnetic current $k_\mu$ is induced in the abelian sector
by the singular gauge transformation with $\Omega^H$ and {\it the Dirac 
condition $eg = 4 \pi$ is automatically derived} in this 
gauge-connection formalism.


In the covariant manner, 
$G^{\rm sing}_{\mu\nu}$ is expressed as
$G^{\rm sing}_{\mu\nu} = \frac{1}{n \cdot \partial } {}^{*}(n \wedge 
k)_{\mu\nu}$ 
using the monopole current $k_\mu$ in Eq.(\ref{eq:kk}) and a
constant 
4-vector $n_\mu$.
Actually, for the above case, one finds for $n_{\mu} = g_{\mu3}$
\begin{eqnarray}
\frac{1}{n \cdot \partial} {}^{*}(n \wedge k)_{\mu\nu}
\label{eq:k}
& = & 
\int dx^{'}_3 \langle x_3|\frac{1}{n \cdot \partial}|x^{'}_3 \rangle 
\varepsilon_{\mu\nu30} n^3
\frac{4\pi}{e} \delta (x_1) \delta(x_2) \delta(x^{'}_3) \frac{\tau_3}{2} 
\nonumber  \\
& = & \frac{4 \pi}{e} (g_{\mu1} g_{\nu2} -
g_{\mu2} g_{\nu1}) \theta(x_3)
\delta(x_1) \delta(x_2)  \frac{\tau_3}{2} \nonumber \\ 
& = & 
\frac{i}{e}  \Omega^H[\partial_\mu, \partial_\nu]  \Omega^{H\dagger}
= G^{\rm sing}_{\mu\nu},
\end{eqnarray}
using the relation 
$ \langle x_n|\frac{1}{n \cdot \partial}|x_n^{'} \rangle 
 = \theta(x_n-x_n^{'})$.  

Thus, the last term $G^{\rm sing}_{\mu\nu}$ 
corresponds to the Dirac string terminated at the origin.
Since $G^{\rm linear}_{\mu\nu}$  shows 
the configuration of the monopole together with the Dirac string,
the sum of $G^{\rm linear}_{\mu\nu} + G^{\rm sing}_{\mu\nu}$
provides the gauge configuration of the monopole
without the Dirac string in the abelian sector.
Thus, 
by dropping the off-diagonal gluon element,
$G^{\rm bilinear}_{\mu\nu}$ vanishes and 
the remaining part $(G_{\mu\nu}^{\rm linear} +G_{\mu\nu}^{\rm sing})^3$
describing  abelian projected QCD includes the field 
strength of monopoles.

Next, we consider the role of off-diagonal gluon components for  
appearance of the monopole.
The gluon  field 
 is divided into the regular part 
$ \Omega  A_\mu  \Omega^{\dagger}$
and the singular part
$   -\frac{i}{e}  \Omega  \partial_\mu  \Omega^{\dagger}$.
Since we are interested in the behavior of the singularity,
we neglect the regular part of the gluon field.
Then, $G^{\rm bilinear}_{\mu\nu}$ is written as
\begin{eqnarray}
ie[A^\Omega_\mu, A^\Omega_\nu]  & = &
\frac{1}{ie}[ \Omega \partial_\mu  \Omega^{\dagger},
              \Omega \partial_\nu  \Omega^{\dagger}]  \nonumber  \\
& = & - \frac{1}{ie}\{ (\partial_\mu  \Omega) \partial_\nu  
\Omega^{\dagger}
                 - (\partial_\nu  \Omega) \partial_\mu  \Omega^{\dagger} 
\} 
                 \nonumber \\
& = & -\frac{1}{ie} \{ \partial_\mu ( \Omega  \partial_\nu  
\Omega^{\dagger} )
                 - \partial_\nu ( \Omega  \partial_\mu  \Omega^{\dagger} ) 
\}
  -\frac{i}{e}    \Omega[\partial_\mu, \partial_\nu]  \Omega^{\dagger} 
  \nonumber \\
& = & -(\partial_\mu A^\Omega_\nu -\partial_\nu A^\Omega_\mu) 
 -\frac{i}{e}    \Omega[\partial_\mu, \partial_\nu]  \Omega^{\dagger}, 
\end{eqnarray}
where the last term appears as the breaking of the Maurer-Cartan 
equation.
In the abelian gauge, the singularity 
of the monopole  appearing in  
$G^{\rm linear}_{\mu\nu} + G^{\rm sing}_{\mu\nu}$ is exactly canceled by  that of  
$G^{\rm bilinear}_{\mu\nu}$. 
Thus, in the abelian gauge,  the off-diagonal gluon combination 
$(G_{\mu\nu}^{\rm    bilinear})^3 = -e
\{ \, (A_{\mu}^{\Omega})^1 \, (A_{\nu}^{\Omega})^2 - 
(A_{\nu}^{\Omega})^1  (A_{\mu}^{\Omega})^2 \, \}$ includes the field 
strength of the anti-monopole, and hence
the off-diagonal gluons $(A_{\mu}^{\Omega})^1$ and $(A_{\mu}^{\Omega})^2$
 have to include some singular structure around the monopole.

The abelian projection is defined by
dropping the off-diagonal component of the gluon field $A_\mu$,
\begin{eqnarray}
A_\mu^\Omega \equiv A_{\mu a}^\Omega \frac{\tau^a}{2} \rightarrow 
{\cal A}_\mu \equiv {\rm tr}(A_\mu^\Omega \tau^3) \frac{\tau^3}{2} = 
 (A_\mu^\Omega)^3 \frac{\tau^3}{2}.
\end{eqnarray}
Accordingly, the SU(2) field strength $G_{\mu\nu}^\Omega$ is projected to 
the abelian field strength 
${\cal F}_{\mu\nu} \equiv F_{\mu\nu}  \frac{\tau^3}{2}$,
\begin{eqnarray}
G_{\mu\nu}^\Omega  & \equiv & (G_{\mu\nu}^\Omega)^a \frac{\tau^a}{2}
 =  (\partial_\mu A^\Omega_\nu -\partial_\nu A^\Omega_\mu )
+ ie[A^\Omega_\mu, A^\Omega_\nu] +\frac{i}{e}
\Omega[\partial_\mu, \partial_\nu] \Omega^{\dagger} \nonumber \\
&  \rightarrow   &{\cal F}_{\mu\nu}  = 
\partial_\mu {\cal A}_\nu -\partial_\nu {\cal A}_\mu +\frac{i}{e}
 \Omega[\partial_\mu, \partial_\nu]  \Omega^{\dagger} \nonumber \\
& &  = 
\partial_\mu {\cal A}_\nu -\partial_\nu {\cal A}_\mu - 
{\cal F}_{\mu\nu}^{\rm sing},
\end{eqnarray}
where ${\cal F}_{\mu\nu}^{\rm sing} \equiv  
{F}_{\mu\nu}^{\rm sing}  \frac{\tau_3}{2}  \equiv -
\frac{i}{e} \Omega[\partial_\mu, \partial_\nu]  \Omega^{\dagger}  $ is diagonal
and remains.
Here, the bilinear term 
 $ie[A^\Omega_\mu, A^\Omega_\nu]$ vanishes in  AP-QCD because 
 it is projected to $ie[{\cal A}_\mu, {\cal A}_\nu]=0$ by the abelian 
 projection.
The appearance of ${\cal F}_{\mu\nu}^{\rm sing} $ 
 leads to the breaking of the abelian  Bianchi identity in the U(1)$_3$ sector,
\begin{eqnarray}
\partial^\alpha {}^*{\cal F}_{\alpha\mu}  =
- \partial^\alpha {}^*{\cal F}_{\alpha\mu}^{\rm sing} 
= \partial^\alpha {}^*  \{ \frac{i}{e} \Omega[\partial_\alpha, 
    \partial_\mu]  \Omega^{\dagger} \}
= k_\mu,
\end{eqnarray}
where Eq.(\ref{eq:k}) is used.
Thus, the magnetic current $k_\mu$ is induced into the abelian gauge theory 
through the singularity of the SU(2)  gauge transformation. 

Here, we compare AP-QCD and QCD in terms of the field strength.
The SU($N_{c}$) field strength $G_{\mu\nu}$ is controlled by the QCD action,
$
S_{\rm QCD} = \int d^4x \{ -\frac12 {\rm tr}G_{\mu\nu}G^{\mu\nu}\},
$
so that each component $G_{\mu\nu}^a$ cannot diverge.
On the other hand, the field strength 
${\cal F}_{\mu\nu}$ in AP-QCD is not directly
controlled by $S_{\rm QCD}$, since the QCD action includes
also off-diagonal components.  
It should be noted that the point-like monopole appearing in AP-QCD 
makes the  U(1)$_3$ action 
$
S_{\rm Abel} = \int d^4x \{ -\frac12 {\rm tr}{\cal F}_{\mu\nu}{\cal 
F}^{\mu\nu} \}
$
divergent around the monopole, 
such a divergence  in ${\cal F}$ should cancel exactly with the 
remaining off-diagonal contribution from 
$G^{\rm bilinear}_{\mu\nu}$ to keep the total QCD action finite.
Thus, the appearance of monopoles in AP-QCD is supported by the 
singular contribution of off-diagonal gluons.
In this way, abelian projected QCD includes monopoles generally.


\section{Monopole Current in the Lattice Formalism  }
\label{monolattice}



In this section, we show the extraction of the monopole current 
in the lattice formalism \cite{degrand}. 
The monopole in lattice QCD is defined in the same manner as  
in the continuum theory.

The abelian field strength ${\bar \theta_{\mu\nu}}(s)$ is defined as 
$
{\bar \theta_{\mu\nu}} (s) \equiv {\rm mod }_{2\pi} 
 (\partial \wedge \theta^3)_{\mu\nu}(s) \in (-\pi,\pi]$,
which is U$(1)_{3}$ gauge invariant.
In general, the two form of the abelian angle variable 
 $\theta^3_\mu(s) $ is divided as
\begin{eqnarray}
\theta_{\mu\nu}(s) \equiv (\partial \wedge \theta^3)_{\mu\nu}(s)
= {\bar \theta_{\mu\nu} }(s) + 2 \pi n_{\mu\nu}(s),
\label{eq:twoform}
\end{eqnarray}
where $n_{\mu\nu}(s) \in {\bf Z}$
corresponds to the quantized magnetic flux of 
the ``Dirac string'' penetrating through the plaquette.  
Although $n_{\mu\nu} \ne 0$ provides the infinite magnetic field is 
the continuum limit as $2 \pi n_{\mu\nu}/a$,
the term $2 \pi n_{\mu\nu}(s)$ does not contribute to the abelian
plaquette $\Box^{\rm Abel}_{\mu\nu}(s)$,  
and it is changed by the singular U(1)$_3$ gauge-transformation as
$\theta^3_\mu(s) \rightarrow \theta^3_\mu(s)+ \partial_\mu \varphi(s) $
with $\varphi(s)$ being the azimuthal angle.
Thus,  $2 \pi n_{\mu\nu}$ corresponds to the Dirac string as an 
unphysical object.

\begin{figure}[bt]
\epsfxsize = 11 cm
\centering \leavevmode
\epsfbox{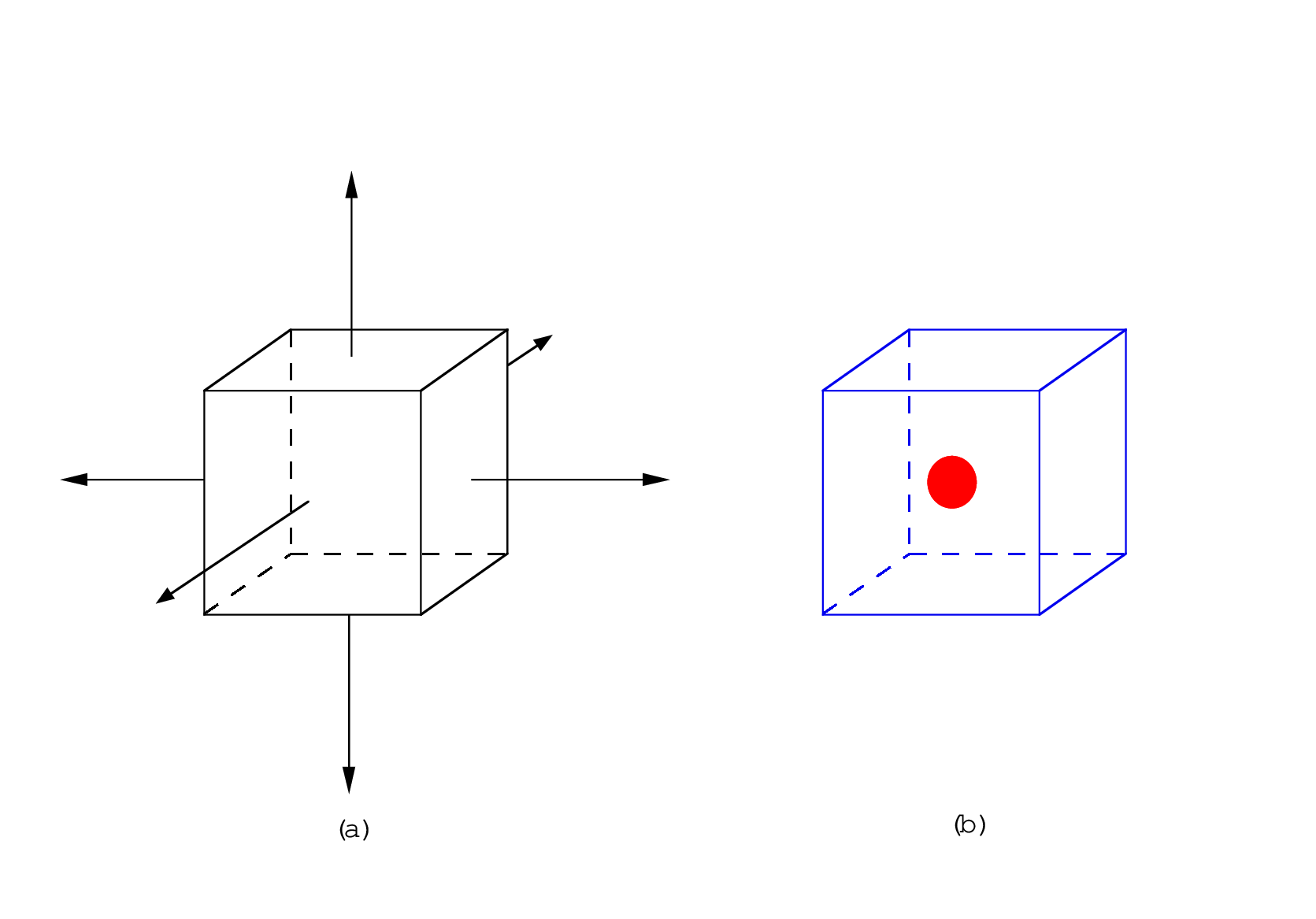}
\vspace{0cm}
\caption{    
(a) The (static) monopole defined on the dual lattice
is equivalent to the total magnetic flux of the Dirac string.
(b) The neighboring links/plaquettes around the dual link.
}
\label{monopoledef}
\vspace{0cm}
\end{figure}

The monopole $k^{lat}_\mu(s)$ is defined on the {\it dual link} as
\cite{degrand},
\begin{eqnarray}
k^{lat}_\mu(s)  \equiv  \frac{1}{2\pi} \partial_\alpha {}^*{\bar 
\theta_{\alpha\mu}} (s)
 = - \partial_\alpha  {}^*n_{\alpha\mu}(s),
 \label{eq:monodif}
\end{eqnarray}
using the abelian field strength $\bar \theta_{\mu\nu}(s)$.
Here,  $k^{lat}_\mu(s) $ is 
defined such that the topological quantization is manifest,
 $k^{lat}_\mu(s) \in {\bf Z}$.
In this definition,  
for instance, one finds    
$k^{lat}_0 =  \frac12 \varepsilon_{ijk} \partial_i n_{jk}$ and
$k^{lat}_i = 0$ ($i=1,2,3$) for the static monopole.
The magnetic charge of the monopole on the dual lattice is
determined
by the total magnetic flux of the Dirac strings
entering
the cube around the monopole (see Fig.\ref{monopoledef}(a).)

\begin{figure}[bt]
\epsfxsize = 9 cm
\centering \leavevmode
\epsfbox{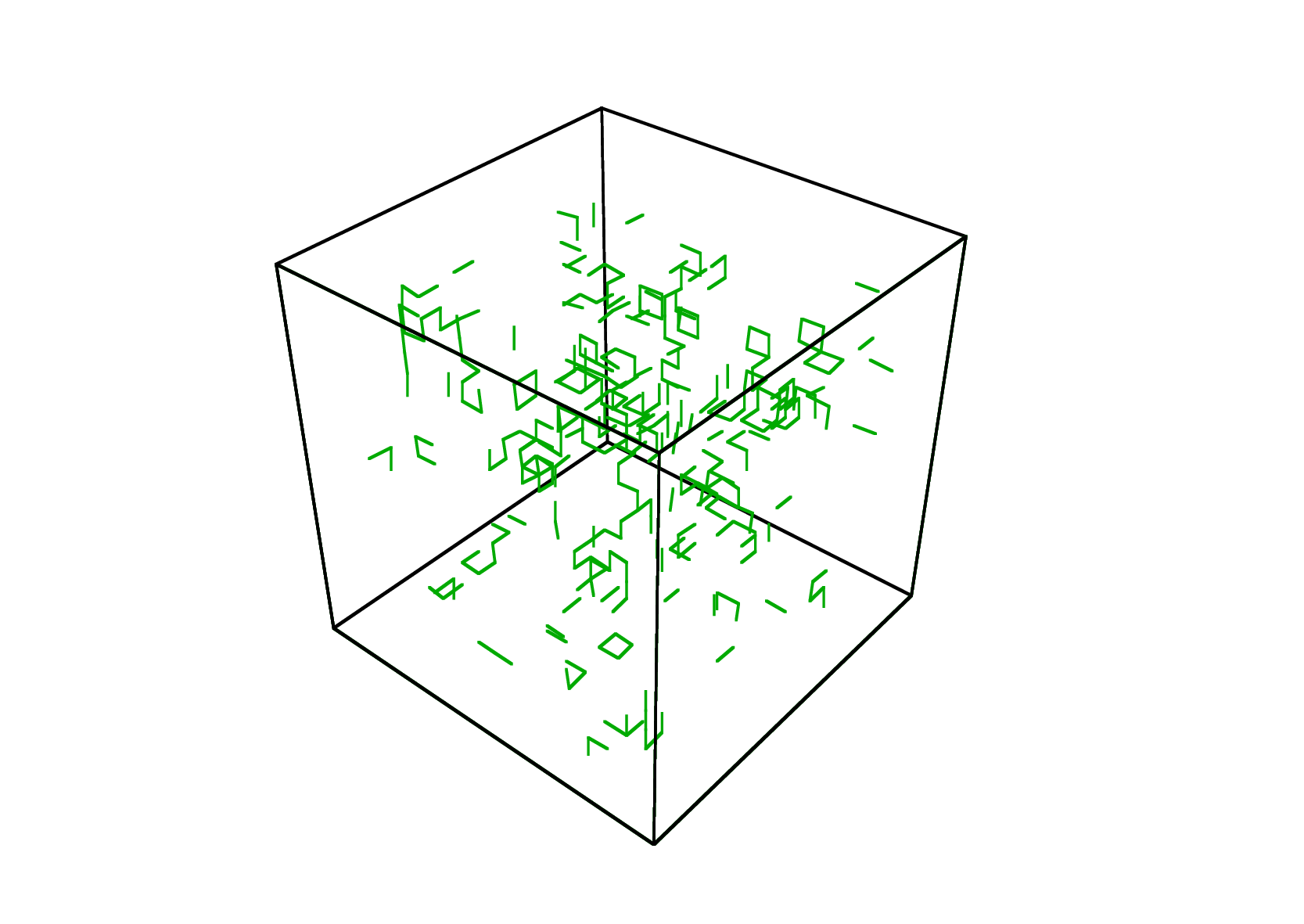}
\vspace{0cm}
\caption{    
The typical example of the 3-dimensional time-slice of the 
monopole current in the MA gauge  in the lattice QCD with 
$\beta$ = 2.4 on $16^4$.
}
\label{monopolecur}
\vspace{0cm}
\end{figure}

We show in Fig.\ref{monopolecur} a typical example of the
monopole current at a time slice in the lattice QCD
at $\beta$ = 2.4  in the maximally abelian (MA) gauge.
In each gauge configuration, the monopole current 
appears as a distinct line-like object, and 
the neighbor of the monopole can be defined 
on the lattice. However, taking the temporal direction into 
account, the monopole current forms a global network covering over 
${\bf R}^4$.
 
Here, we summarize several relevant properties of $k_\mu(s)$.
\begin{enumerate}
\item
The monopole current $k_\mu$ is topologically quantized 
and $k^{lat}_\mu(s)$ takes an integer 
$k^{lat}_\mu(s)$ $\in$ ${\bf Z}$ in the definition of 
Eq.(\ref{eq:monodif}).
As the result, 
$k^{lat}_\mu(s)$ forms 
a line-like object in the space-time ${\bf R}^4$,
since $k^{lat}_\mu$ is a conserved current as $\partial_\mu k_\mu ^{lat} = 0$.
These features of $k_\mu^{lat}(s)$ $\in$ ${\bf Z}$ are quite unique and 
different from the electric current $j_\mu(s)$ $\in$ ${\bf R}$, which can 
spread as a continuous field.  
\item
In the lattice formalism, 
$k^{lat}_\mu \equiv \frac{1}{2\pi} \partial_\alpha ^* \bar 
\theta_{\alpha\mu}$ 
is defined as a three-form on the dual link. 
For the use of the forward derivative, $k^{lat}_\mu(s)$ is to be 
defined on the dual link between 
$s^{\rm dual}_{\pm \mu} \equiv 
s+\frac{\hat x}{2}+\frac{\hat y}{2}+\frac{\hat z}{2}+\frac{\hat t}{2}
\pm \frac{\hat \mu}{2}$.
For instance, $k^{lat}_0(s)$ is placed on the dual link between 
$(s_x+\frac12,s_y+\frac12,s_z+\frac12,s_t)$ and 
$(s_x+\frac12,s_y+\frac12,s_z+\frac12,s_t+1)$.
Thus, the monopole is defined to appear 
at the center of the 3-dimensional cube 
perpendicular to the monopole-current direction 
as shown in Fig.\ref{monopoledef}(b).
\item
Because of 
$k_\mu \equiv \partial_\alpha ^* F_{\alpha\mu}
=-\frac{1}{2} \varepsilon_{\mu\alpha\beta\gamma}
\partial_\alpha F_{\beta\gamma}$, 
$k_\mu$ only affects the perpendicular components to the 
$\hat \mu$-direction
for the ``electric variable'' as $F_{\alpha\beta}$
 in a direct manner.
For instance, the static monopole with $k_0 \ne 0$ creates 
the magnetic field $F_{ij}$ ($i,j$=1,2,3) around it, but 
does not bring the electric field $F_{0i}$.
Hence, in testing the field around the monopole in the next chapter, 
one has to consider the difference between 
such perpendicular components and others. 
\end{enumerate}

We now consider the relationship between the lattice variable
and the field variable in the continuum theory.
The continuous abelian field 
${\cal A}_\mu (x)\equiv A^3_\mu (x) \frac{\tau^3}{2}$ is expressed as 
\begin{eqnarray}
eA^3_\mu \equiv  \theta^3_{\mu} \cdot \frac{2}{a}
\end{eqnarray}
with the gauge coupling constant $e$ and the lattice spacing $a$.
The abelian field strength
${\cal F}_{\mu\nu} (x)\equiv F_{\mu\nu} (x) \frac{\tau^3}{2}$ 
 in the continuum theory is written as
\begin{eqnarray}
eF_{\mu\nu} & \equiv  &  {\rm mod}_{2 \pi} (\theta_{\mu\nu}) 
\cdot \frac{2}{a^2}
=  {\bar \theta_{\mu\nu}} \cdot \frac{2}{a^2}, 
\end{eqnarray}
and $F_{\mu\nu}$ is composed of 
two parts according to the decomposition (\ref{eq:twoform})
\begin{eqnarray}
F_{\mu\nu} & = & (\partial \wedge A^3)_{\mu\nu} - F^{\rm sing}_{\mu\nu}.
\label{eq:twopars}
\end{eqnarray}
Thus, in the SU$(N_c)$-lattice formalism, the difference between the 
field strength $F_{\mu\nu}$ and two-form $(\partial \wedge A)_{\mu\nu}$ 
arises from the periodicity of the angle variable in the compact subgroup 
U(1)$^{N_c-1}$ embedded in SU($N_c$).
Here, the singular Dirac-string part $F^{\rm sing}_{\mu\nu}$ is
directly related to $2 \pi n_{\mu\nu}$ and is 
written by
\begin{eqnarray}
F^{\rm sing}_{\mu\nu} =  2\pi n_{\mu\nu} \cdot \frac{2}{ea^2} = 
\frac{4\pi}{e} 
n_{\mu\nu}  \frac{1}{a^2}.
\end{eqnarray}
Owing to existence of $F^{\rm sing}_{\mu\nu}$ in Eq.(\ref{eq:twopars}),
the monopole current 
$k_{\mu} (x) \equiv k^3_{\mu} (x) \frac{\tau^3}{2}
\equiv \partial_\alpha {}^* F_{\alpha\mu} \frac{\tau^3}{2}$ 
appears in the continuum theory and is written as
\begin{eqnarray}
k^3_{\mu}  =
  k^{lat}_{\mu} \cdot \frac{4\pi}{ea^3}
= -\frac{4\pi}{e}  \partial_\alpha {}^*  n_{\alpha\mu} \frac{1}{a^3},
\end{eqnarray}
where the magnetic-charge unit $g \equiv \frac{4\pi}{e}$
naturally appears in $k_\mu$.


\chapter{Large Field Fluctuation around Monopole}
\label{sec:LFFM}

In this chapter, 
we study the QCD-monopole appearing 
in the abelian gauge in terms of the gluon field 
fluctuation \cite{ichiemp,ichiem}.
For simplicity, we take $N_c=2$.
In the static frame of the QCD-monopole 
with the magnetic charge $g$, 
a spherical ``magnetic field'' is created around the monopole 
in the abelian sector of QCD as 
\begin{eqnarray}
{\bf H}(r) =  \frac{g}{4\pi r^3} {\bf r}
\end{eqnarray}
with ${\bf H}_i\equiv  \varepsilon _{ijk} \partial_j A^3_k$. 
Thus, the QCD-monopole inevitably accompanies 
a large fluctuation of the abelian gluon component $A^3_\mu$ around it. 
As was discussed in the previous chapter, in the abelian gauge, 
the formal action of  abelian projected QCD 
or the abelian part of the QCD action is given by 
$S^{\rm Abel}\equiv -\frac14 \int d^4x
\{ (\partial_\mu A_\nu^3-\partial_\nu A_\mu^3)^2  
-F_{\mu\nu}^{\rm sing}\}$, where 
$-F_{\mu\nu}^{\rm sing}$ appears and 
eliminates the Dirac-string contribution. 
In the abelian part, 
the field energy created around the monopole 
is estimated as the ordinary electro-magnetic energy, 
\begin{eqnarray}
{\cal E}(a) = \int_a^\infty d^3 x \frac12 {\bf H}(r)^2 = 
\frac{g^2}{ 8 \pi a},
\label{monopolemass1}
\end{eqnarray}
where $a$ is an ultraviolet cutoff like a lattice mesh.
As the ``mesh'' $a$ goes to 0, 
the monopole inevitably accompanies an infinitely large 
energy-fluctuation in the abelian part 
and makes $S^{\rm Abel}$ divergent. 

Since there seems no plausible reason to eliminate 
such a divergence via renormalization, 
the monopole seems difficult to appear in the abelian 
gauge theory controlled by $S^{\rm Abel}$.
This is the reason why QED does not have the point-like Dirac 
monopole.
Then, why can the QCD-monopole appear in  abelian projected QCD ? 
To answer it, let us consider the division of the total QCD action 
$S^{\rm QCD}$ into the abelian part $S^{\rm Abel}$ and 
the remaining part $S^{\rm off}\equiv S^{\rm QCD}-S^{\rm Abel}$, which 
is contribution from the off-diagonal gluon component. 
While $S^{\rm QCD}$ and $S^{\rm Abel}$ are positive definite in the 
Euclidean metric, 
$S^{\rm off}$ is not positive definite and can take a negative value.
Then, around the QCD-monopole, 
the abelian action $S^{\rm Abel}$ should be 
partially canceled by the remaining contribution $S^{\rm off}$ from 
the off-diagonal gluon component, so as to keep 
the total QCD action $S^{\rm QCD}$ 
%
%
finite even for $a\rightarrow 
0$. 
Similar cancellation between the gauge field and the Higgs field 
fluctuation is also found around the GUT monopole. 
Thus, we expect large off-diagonal gluon components 
around the QCD-monopole for its existence as well as a 
large field fluctuation in the abelian part.
Based on this analytical consideration, we study the field fluctuation and 
monopoles in the MA gauge using the lattice QCD.

\section{Gluon Field Configuration around Monopoles}

We study the properties of monopole in terms of the gluon configuration 
in the MA gauge. In particular, we investigate the correlation between 
monopoles 
and the abelian angle variable $\theta_\mu^3(s)$ and abelian projection rate
$R_{\rm Abel}$.

\begin{figure}[bt]
\epsfxsize = 13 cm
\centering \leavevmode
\epsfbox{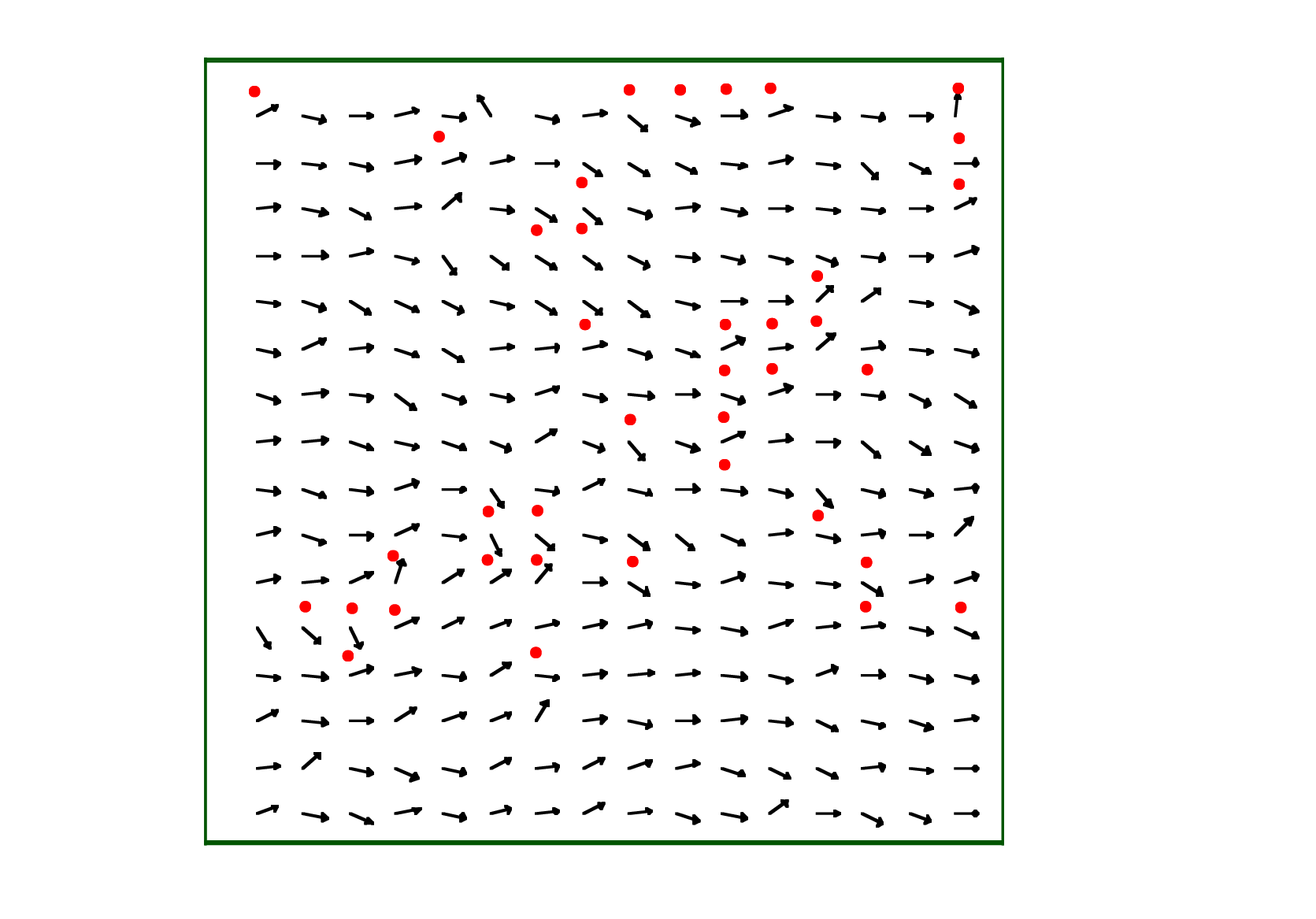}
\vspace{0cm}
\caption{    
The local correlation between the abelian angle variable $\theta_\mu^3(s)$
and the monopole current $k_\mu$ in the MA gauge
with U(1)$_3$ Landau gauge at $\beta=2.4$.
The closed symbol denotes the monopole current on the dual link.
}
\label{arrow1m}
\vspace{0cm}
\end{figure}

First, we show the local correlation between the abelian angle variable and
the monopole using the gauge configuration
in the MA gauge with U(1)$_3$ Landau gauge
in Fig.\ref{arrow1m}.
The closed symbol denotes the monopole current on the dual link.
Around the monopoles, the abelian angle variable tends to fluctuate 
largely.

\begin{figure}[tb]
\epsfxsize = 11 cm
\centering \leavevmode
\epsfbox{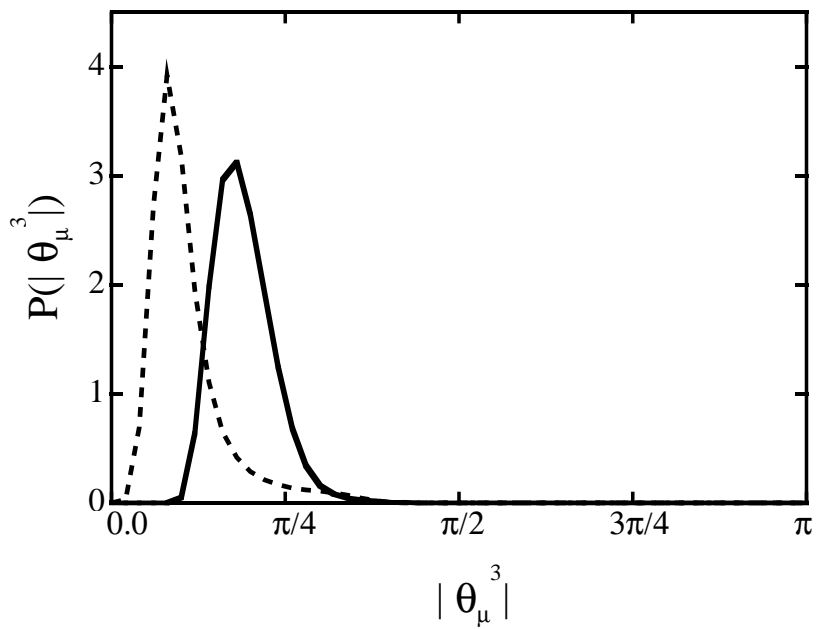}
\vspace{0cm}
\caption{  The solid curve denotes the 
probability distribution $P( |\bar \theta^3|)$ of the 
averaged abelian angle variable $|\bar \theta^3(s, \hat \mu)|$ 
around the monopole current
in the MA gauge with the U(1)$_3$ Landau gauge fixing.
Here, $ |\bar \theta^3(s,\hat \mu)|$ is the average of 
$|\theta_\alpha^3(s)|$ over the 
neighboring links around the dual link,
For comparison, the total distribution $P$ on the whole lattice is 
also added by the dashed curve.
Around the monopole, $ |\bar \theta^3|$ corresponding to the abelian 
gluon component takes a large value.
}
\label{gfig8}
\vspace{0cm}
\end{figure}

Next, we consider the correlation between the field variables 
and the monopole quantitatively in the lattice QCD. 
For this argument, one has to recall the property of the monopole current 
shown in Section \ref{monolattice}. 
In particular, one should note that $k_\mu(s)$ is defined on the dual 
link and only affects the perpendicular components to the $\hat \mu$-direction
 for the electric variable as $F_{\alpha\beta}$ because of
$k_\mu \equiv \partial_\alpha^* F_{\alpha\mu} = -\frac12
\varepsilon_{\mu\alpha\beta\gamma} \partial_\alpha F_{\beta\gamma}$.
Taking account of these properties, 
we study the local correlation between the field variables 
and the monopole current $k_\mu(s)$ 
in the MA gauge with the U(1)$_3$ Landau gauge.
We first measure the average of the abelian angle 
variable $\theta^3_\mu(s)$ over the neighboring links 
around the dual link (see Fig.\ref{monopoledef}(a)),
\begin{eqnarray}
 |\bar \theta^3(s,\hat \mu)| \equiv \frac{1}{12} \sum_{\alpha\beta\gamma} 
\sum_{m,n=0}^{1} 
\frac12 | \varepsilon_{\mu\alpha\beta\gamma}| \cdot 
|\theta^3_\alpha (s+m \hat \beta + n \hat \gamma) |,
\end{eqnarray} 
which only consists of the perpendicular components considering 
the above monopole property.
%
%
Here, the index $\hat \mu$ denotes the direction of the dual link, 
and $|\bar \theta^3(s,\hat \mu)|$ corresponds to the average over the 12 
sides of the 3-dimensional cube perpendicular to the 
$\hat \mu$-direction \cite{ichiemp}.
We show in Fig.\ref{gfig8} the probability distribution $P(|\bar \theta^3|)$ of 
$|\bar \theta^3(s,\hat \mu)|$ in the MA gauge with the U(1)$_{3}$ Landau gauge at 
$\beta = 2.4$.
The solid curve denotes $P(|\bar \theta^3|)$ 
around the monopole current,
while the dashed curve denotes the total distribution 
on the whole lattice. The abelian angle variable $|\theta^3_\mu(s)|$ 
takes a large value around the monopole. 
In other words, the monopole provides the large fluctuation of 
the  abelian gauge field, which would enhance the randomness 
of the abelian link variable.

\begin{figure}[p]
\vspace{3cm}
\epsfxsize = 11 cm
\centering \leavevmode
\epsfbox{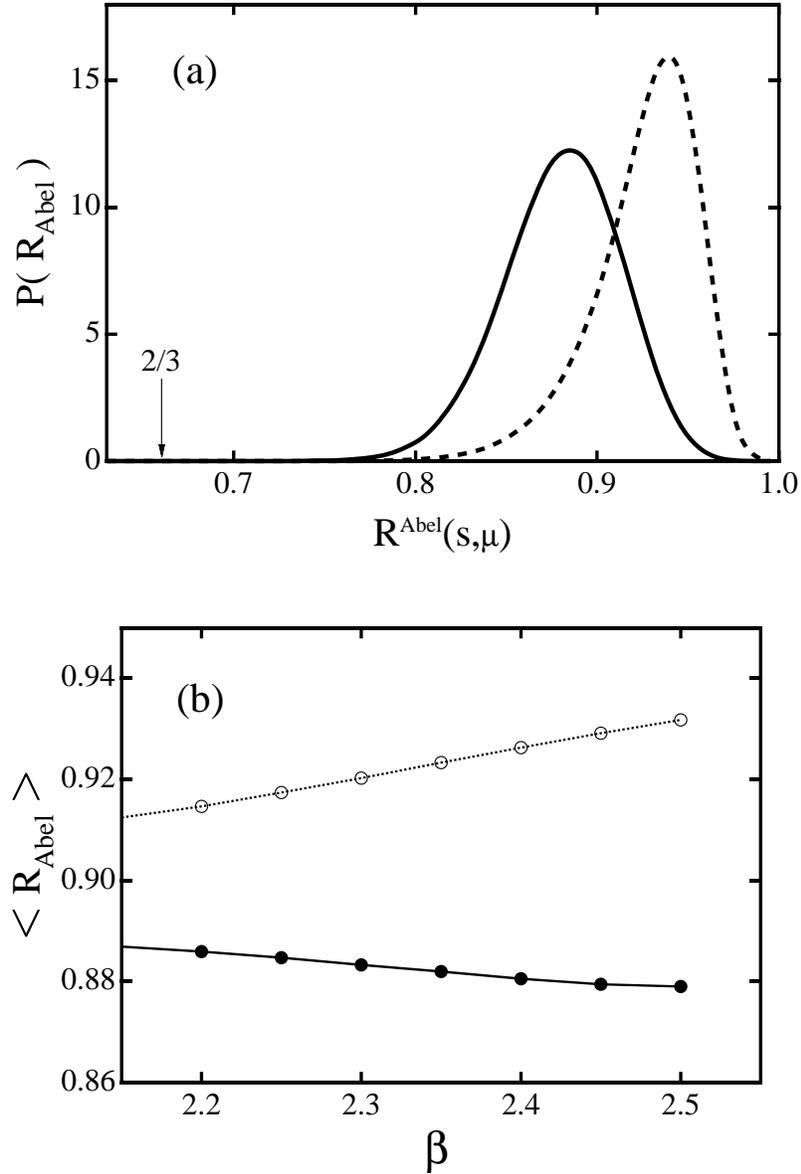}
\vspace{0cm}
\caption{    
(a) The solid curve denotes the 
probability distribution $P(\bar R_{\rm Abel})$
of the averaged abelian projection rate $\bar R_{\rm Abel}(s,\hat \mu)$
 around the monopole current
in the MA gauge in the SU(2) lattice QCD with  $\beta =2.4$ on $16^4$.
For comparison, the total distribution $P$ on the whole lattice is 
also added by the dashed curve.
(b) The solid curve denotes abelian projection rate 
$\langle \bar R_{\rm Abel} \rangle $
around the monopole current in the MA gauge as the function of $\beta$.
The dashed curve denotes 
$\langle \bar R_{\rm Abel} \rangle $ on the whole lattice. 
}
\label{gfig9}
\vspace{0cm}
\end{figure}

Similar to $|\bar \theta^3(s,\hat \mu)|$, 
we measure the average $\bar R_{\rm Abel}$ of the 
abelian projection rate $R_{\rm Abel}(s,\hat \mu)\equiv \cos\theta_\mu(s)$
over the neighboring links 
around the dual link, 
\begin{eqnarray}
\bar R_{\rm Abel} (s,\hat \mu) \equiv \frac{1}{12} \sum_{\alpha\beta\gamma} 
\sum_{m,n=0}^{1} 
\frac12 | \varepsilon_{\mu\alpha\beta\gamma} | \cos \theta_\alpha 
(s+m \hat \beta + n \hat \gamma)
\end{eqnarray} 
in the MA gauge 
to investigate the correlation between off-diagonal gluons and monopoles.
As shown in Fig.\ref{gfig9}(a), 
$\bar R_{\rm Abel}$ around the monopole current becomes smaller
than the total average of $\bar R_{\rm Abel}$ 
and therefore 
{\it the magnitude of the off-diagonal gluon component becomes larger around 
the monopole}.
The $\beta$ dependence of the abelian projection rate $\langle R_{\rm Abel} 
\rangle$
is shown in  Fig.\ref{gfig9}(b).
Although $\langle R_{\rm Abel} \rangle$  on the whole lattice 
approaches to unity as 
$\beta \rightarrow \infty$,  
$\langle R_{\rm Abel} \rangle$ around the monopole is about 0.88
and is not changed even in the large $\beta$ region.
Thus, the monopole provides the large fluctuation both for 
the abelian 
field and for the off-diagonal gluon. 

\section{Plaquette Action Density around Monopoles}


We next study monopoles in terms of the plaquette action density.
We define the SU(2), abelian  and 
``off-diagonal''  plaquette action densities as
\begin{eqnarray}
S^{\rm SU(2)}_{\mu\nu} (s) & \equiv & 1 - \frac12 {\rm tr} 
\Box^{\rm SU(2)}_{\mu\nu}(s),  
\label{eq:abaction1}
\\
S^{\rm Abel}_{\mu\nu} (s)   &  \equiv  &
1 - \frac12 {\rm tr} \Box^{\rm Abel}_{\mu\nu}(s), 
\label{eq:abaction2}
\\
S^{\rm off}_{\mu\nu} (s)  & \equiv  &
 S^{\rm SU(2)}_{\mu\nu}(s) - S^{\rm Abel}_{\mu\nu}(s),
\end{eqnarray}
where $\Box^{\rm SU(2)}_{\mu\nu}(s)$ and 
$\Box^{\rm Abel}_{\mu\nu}(s)$ denote the SU(2) and the abelian plaquette 
variables, respectively; 
\begin{eqnarray}
 \Box^{\rm SU(2)}_{\mu\nu}(s) & \equiv &
U_\mu(s) U_\nu (s+ \hat \mu) U^{\dagger}_\mu(s+\hat \nu)
U^{\dagger}_\nu(s), \\
\Box^{\rm Abel}_{\mu\nu}(s) & \equiv &
u_\mu(s) u_\nu(s + \hat \mu) u^{\dagger}_\mu(s + \hat \nu)
u^{\dagger}_\nu(s).
\end{eqnarray}
Here, all of $S_{\mu\nu}$ are defined as symmetric tensors, 
$S_{\mu\nu}=S_{\nu\mu}$, 
instead of the Lorentz scalar, considering the above property  
of the monopole current. 
In the continuum limit $a \rightarrow 0$, $S^{\rm SU(2)}_{\mu\nu} (s)$ and
$S^{\rm Abel}_{\mu\nu} (s)$ are related to the SU(2) and the abelian 
action densities as 
$S^{\rm SU(2)}_{\mu\nu} (s) \rightarrow \frac14 a^4 e^2  {\rm tr} 
G^2_{\mu\nu}$ and
$S^{\rm Abel}_{\mu\nu} (s) \rightarrow \frac14 a^4 e^2  {\rm tr} {\cal 
F}^2_{\mu\nu}$, 
and then we call $S_{\mu\nu}$ as the action density, 
in spite of the lack of the summation on the Lorentz indices. 
Here, $S^{\rm off}_{\mu\nu}$ corresponds to the contribution of the
 off-diagonal gluon. 
While $S^{\rm SU(2)}_{\mu\nu}$ and $S^{\rm Abel}_{\mu\nu}$
are positive-definite, 
$S^{\rm off}_{\mu\nu}$ is not positive-definite and can take a 
negative value.

In order to examine the correlation between the action densities and 
the monopole current defined on the dual link, we measure the average of 
the action density $ S(s)$ over the neighboring plaquettes 
around the dual link, 
\begin{eqnarray}
\bar S(s,\hat \mu) \equiv
\frac16 \sum_{\alpha\beta\gamma} \sum_{m=0}^1 \frac12 
| \varepsilon_{\mu\alpha\beta\gamma} |
S_{\alpha\beta}(s + m \hat \gamma).
\label{eq:duallink}
\end{eqnarray}
%
%
Here, $\hat \mu$ appearing in $\bar S(s,\hat \mu)$ denotes 
the direction of the dual link, and $\bar S(s,\hat \mu)$ corresponds to 
the average over 6 faces of the 3-dimensional cube perpendicular to 
the $\hat \mu$-direction.

\begin{figure}[p]
\epsfxsize = 11 cm
\centering \leavevmode
\epsfbox{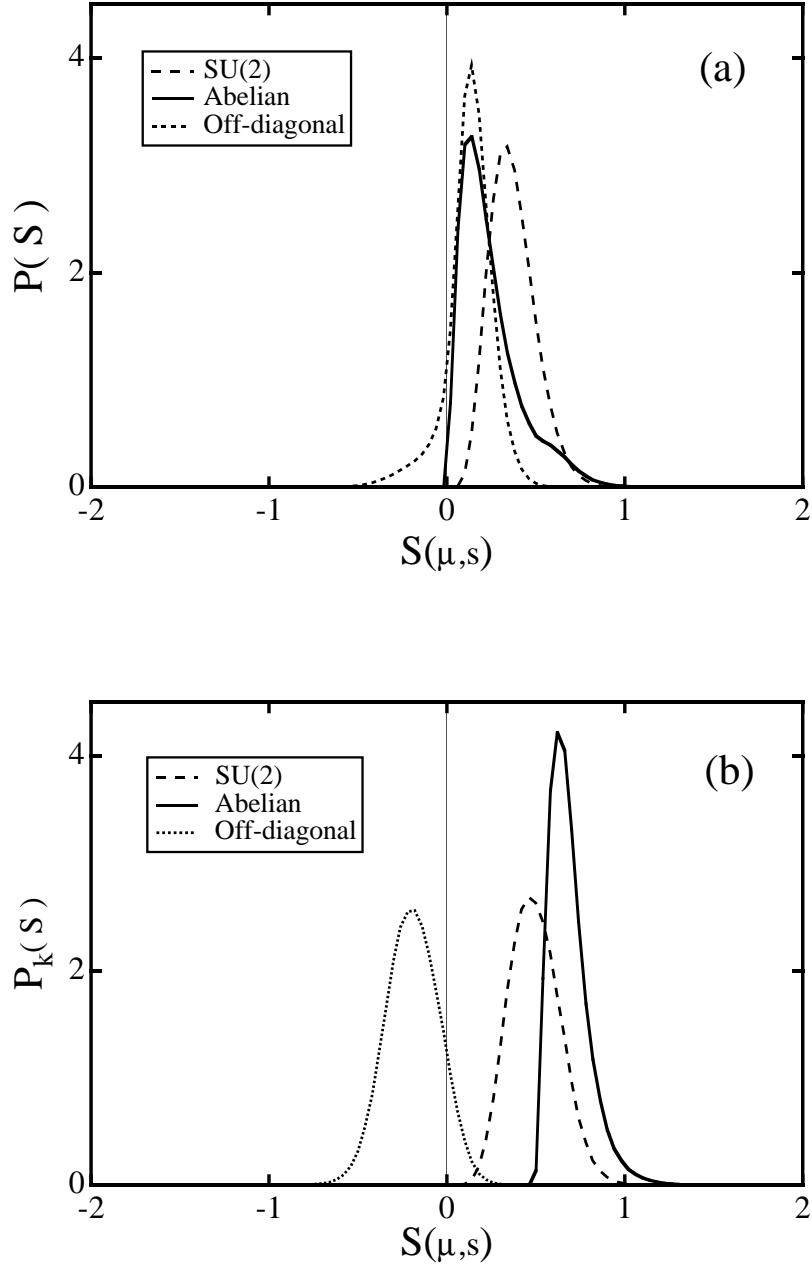}
\vspace{0cm}
\caption{   
(a) The probability distribution $P(\bar S)$ of 
density $\bar S(s,\hat \mu)$ on the whole lattice
in the MA gauge at $\beta =2.4$ on $16^4$ lattice.
(b)The probability distribution $P_k(\bar S)$ 
of the action density $\bar S(s,\hat \mu)$ around the monopole 
current ${k_\mu}$.
The dotted and the solid curves denote
$P(\bar S^{\rm SU(2)})$ and  $P(\bar S^{\rm Abel})$, respectively.
The dashed curve denotes  
$P(\bar S^{\rm off})$ for the off-diagonal part 
$\bar S^{\rm off}$ of the action density. 
}
\label{gfig10}
\vspace{0cm}
\end{figure}

\begin{figure}[bt]
\epsfxsize = 11 cm
\centering \leavevmode
\epsfbox{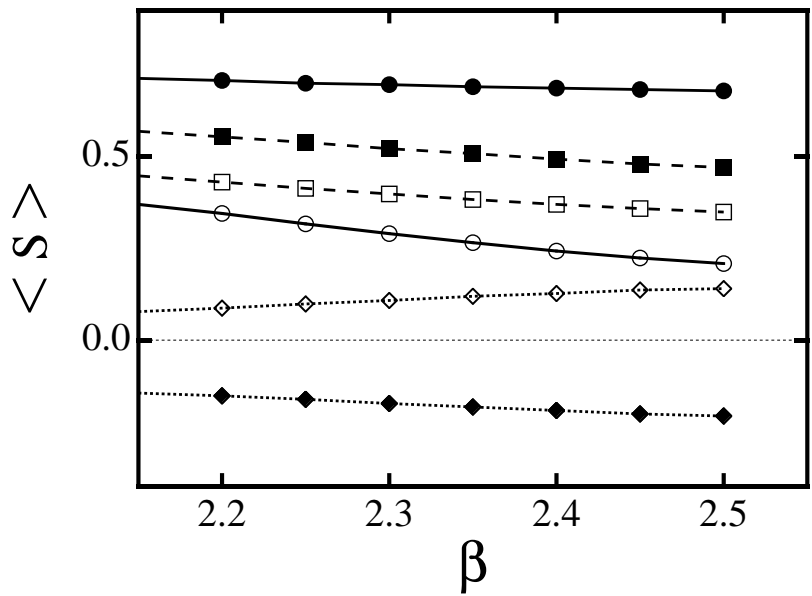}
\vspace{0cm}
\caption{ The action density as the function of $\beta$
in the MA gauge in the SU(2) lattice QCD.
The closed symbols denote the action densities $\langle  S \rangle$
around the monopole current, while the open symbols denote those on 
the whole lattice.
The square, circle and rhombus denote  $\langle S^{\rm SU(2)} \rangle$,
$\langle  S^{\rm Abel} \rangle$ and $\langle S^{\rm off} \rangle$,
respectively.
The monopole  accompanies a large U(1)$_3$ plaquette action,
however, such a large U(1)$_3$ action is 
strongly canceled by the off-diagonal part.
}
\label{gfig10c}
\vspace{0cm}
\end{figure}

We show in Fig.\ref{gfig10}
the probability distribution $P(\bar S)$ of the action densities
$\bar S(s,\hat \mu)$ in the SU(2), 
the abelian and the off-diagonal parts. 
Before the argument around the monopole current, 
we show the action densities 
on the whole lattice in Fig.\ref{gfig10} (a).
On the whole lattice, most 
$\bar S^{\rm off}$ are positive, and both $\bar S^{\rm Abel}$ and 
$\bar S^{\rm off}$ tend to take smaller values than
 $\bar S^{\rm SU(2)} = \bar S^{\rm Abel} +  \bar S^{\rm off}$. 
In other words, $\bar S^{\rm Abel}$ and positive $\bar S^{\rm off}$
additionally contribute to $\bar S^{\rm SU(2)}$.

However,  such a tendency of the action densities is drastically changed 
around the monopole as shown in Fig.\ref{gfig10}(b). 
We find remarkable features of 
the action densities around the monopole as follows.
\begin{enumerate}
\item 
Around monopoles, most $\bar S^{\rm off}$ take negative values,
and $\bar S^{\rm Abel}$ is larger than 
$\bar S^{\rm SU(2)} = \bar S^{\rm Abel} +  \bar S^{\rm off}$. 
\item 
Due to the cancellation between  $\bar S^{\rm Abel}$ and $\bar S^{\rm 
off}$, $\bar S^{\rm SU(2)}$ does not take an extremely 
large value around the monopole. 
\end{enumerate}
Thus, the large abelian action density $S^{\rm Abel}$ around the monopole 
is strongly canceled by the off-diagonal contribution $S^{\rm off}$ to 
keep
the total QCD action $S^{\rm QCD}=S^{\rm Abel}+S^{\rm off}$ small.
Here, different from $S^{\rm SU(2)}$, $S^{\rm Abel}$ itself does not 
control the system directly, and hence there is no severe constraint from 
$S^{\rm Abel}$. However, large $S^{\rm Abel}$ is still not preferable, 
because the large-cancellation requirement between $S^{\rm Abel}$
and $S^{\rm off}$ leads to a strong constraint on the off-diagonal gluon 
and brings the strong reduction of the configuration number.

Around the monopole, the abelian action density $S^{\rm Abel}$ 
takes a large value, and this value can be estimated from a following 
simple calculation.
Without loss of generality, the monopole-current direction is locally 
set to be parallel to the temporal direction as 
$k^{lat}_0 (s)  \equiv  \frac{1}{2\pi} \partial_\alpha {}^*{\bar 
\theta_{\alpha 0}} (s) 
= \pm 1$. 
Here, $k_0^{lat}(s)$ is expressed as the sum of 
six plaquette variables $\bar \theta_{ij}$ ($i,j$=1,2,3) 
around the monopole, because of 
$k_0^{lat}(s)
=-\frac{1}{4\pi}  \varepsilon_{ijk} \partial_i \bar \theta_{jk}(s)
=-\frac{1}{2\pi} \sum_i \sum_{j<k}  \varepsilon_{ijk} 
\{\bar \theta_{jk}(s+\hat i)-\bar \theta_{jk}(s)\}$. 
Hence, the total sum of six $|\bar \theta_{ij}(s)| (i<j) $
is to exceed $2 \pi$ to realize $k_0(s)=\pm 1$.
%
Since large $|\bar \theta_{ij(s)}|$ accompanying large $S^{\rm Abel}$ is 
not preferable,  the magnetic field $|\bar \theta_{ij}|$ around the 
monopole is estimated as 
$| \bar \theta_{ij} | \simeq  2 \pi / 6 = \pi / 3$ on the average, 
using the spherical symmetry of the magnetic field 
in the vicinity of the monopole. 
Accordingly, we estimate as 
$S^{\rm Abel}_{ij} = 1-{\rm  cos}(| \bar \theta_{ij} |)
\simeq 1- \cos \frac{\pi}{3} = \frac12$ 
around the monopole on the average. 
The above argument can be easily generalized to the case with 
arbitrary monopole-current direction. 

Then, existence of monopoles brings a peak 
around $S^{\rm Abel} = \frac12 $ in the distribution $P(S^{\rm Abel})$. 
In fact, the abelian action density $S^{\rm Abel}$ has two ingredients; 
one is nontrivial large fluctuation about $S^{\rm Abel}=1/2$ 
originated from the monopole, and the other is 
remaining small fluctuations, which is expected to vanish 
as $S^{\rm Abel} \rightarrow 0$ as $a \rightarrow 0$.
As shown in Fig.\ref{gfig10c}, the peak originated from
 the monopole is almost $\beta$ independent,
while the other fluctuation becomes small 
for large $\beta$.
At a glance from this result, 
the monopole seems hard to exist at the small mesh $a$, 
since the monopole needs a large abelian action $S^{\rm Abel}$. 
Nevertheless, the monopole can exist in QCD 
even in the large $\beta$ region 
owing to the contribution of the off-diagonal gluon. 
As shown in Fig.\ref{gfig10}(b), 
the off-diagonal part $S^{\rm off}$ of the action density 
around the monopole tends to take a large negative value, and 
strongly cancels with the large abelian action $S^{\rm Abel}$ 
to keep the total SU(2) action $S^{\rm QCD}$ finite.

Here, we consider the angle variable $\tilde \chi_\mu(x)$ of 
the off-diagonal gluons $A_\mu^\pm(x)$ around the monopole. 
In the MA gauge, the amplitude of $A_\mu^\pm(x)$ is strongly reduced, 
and $\tilde \chi_\mu(x)$ can be approximated as a random variable 
on the whole, because $\tilde \chi_\mu(x)$ is free from 
the MA gauge condition entirely and is less constrained from 
the QCD action due to the small $|A_\mu^\pm(x)|$. 
However, around the monopole, 
the off-diagonal gluon $A_\mu^\pm(x)$ inevitably has 
a large amplitude even in the MA gauge to cancel the large 
abelian action density. 
This requirement on the reduction of the total action density 
severely constrains the randomness of the angle variable 
$\tilde \chi_\mu(x)$ of the off-diagonal gluon $A_\mu^\pm(x)$ 
around the monopole. 
As the result, the randomness of $\tilde \chi_\mu(x)$ is weaken, and 
continuity of $\tilde \chi_\mu(x)$ or $A_\mu^\pm(x)$ becomes clear 
in the vicinity of the monopole even in the MA gauge. 
This continuity of $A_\mu^\pm(x)$ around the monopole 
ensures the topological stability 
of the monopole itself as $\Pi_2({\rm SU}(2)/{\rm U}(1))={\bf 
Z}_{\infty}$.

To summarize, existence of the monopole  inevitably accompanies a large 
abelian plaquette action $S^{\rm Abel}$ around it, however, the 
off-diagonal part $S^{\rm off}$ takes a large negative value around the 
monopole and strongly cancels with $S^{\rm Abel}$ to keep $S^{\rm QCD}$ 
not so large. 
Due to this strong cancellation 
between $S^{\rm Abel}$ and $S^{\rm off}$, 
monopoles can appear in the abelian sector in QCD 
without large cost of the QCD action $S^{\rm QCD}$, which 
controls the generating probability of the gluon configuration. 
The extension of the off-diagonal rich region around the monopole
can be interpreted as the effective size or 
the structure of the monopole, because the abelian gauge theory is 
largely modified inside the QCD-monopole like the 
't~Hooft-Polyakov monopole.

Finally, in this section, let us consider the correlation between 
monopoles and instantons \cite{fukushima} in terms of the 
gluon-field fluctuation. 
The instanton is a nontrivial classical solution of 
the Euclidean Yang-Mills theory, 
corresponding to the homotopy group 
$\Pi_3({\rm SU}(N_c)) = {\bf Z}_\infty $ \cite{shifman,rajaraman,shuryak}. 
For the instanton, the SU(2) structure of the gluon field 
is necessary at least. 
In spite of the difference on the topological origin, 
recent studies indicate the strong correlation 
between monopoles and instantons in the QCD 
vacuum in the MA gauge \cite{suganuma3,brower}.
What is the origin of the relation between two different 
topological objects, monopoles and instantons ? 
In the MA gauge, off-diagonal components are forced to be small, 
and the gluon field configuration seems abelian on the whole. 
However, even in the MA gauge, off-diagonal gluons largely remain 
around the QCD-monopole. 
The concentration of off-diagonal gluons around monopoles leads to 
the local correlation between monopoles and instantons: 
instantons appear around the monopole world-line in the MA gauge, 
because instantons need full SU(2) gluon components for existence.




\chapter{
Monopole Projection and Scaling Properties in the MA Gauge}
\label{sec:OMDMAGMP}

\section{Decomposition into  Monopole and Photon Angle-Variables}

Abelian projected QCD is obtained by  neglecting the off-diagonal 
gluon component in the abelian gauge, 
and it includes both the electric current $j_\mu$
and the monopole current $k_\mu$. 
Here, $j_\mu$ is generated by charged gluons, and 
$k_\mu$ is generated by the singular SU(2) 
gauge transformation in the abelian gauge.
In the lattice formalism,
the U(1)$_3$-gauge field 
can be separated into the ``monopole part'' and  the ``photon part''
using the Coulomb propagator.
Here, the monopole part only includes the monopole current $k_\mu$ and 
reproduces 
the purely linear part of the static quark potential \cite{stack}.
On the other hand, the  photon part includes the electric current $j_\mu$  
only, and 
provides the Coulomb potential between the quark and the anti-quark, similarly in 
the ordinary QED. In this section, we extract the monopole contribution from 
 abelian projected QCD using the lattice Coulomb propagator \cite{degrand},
and study the role of the monopole for confinement.

The abelian gauge field $\theta^3_\mu(s)$ can be decomposed 
into the monopole part $\theta^{\rm Mo}_\mu(s)$ and the photon part  
$\theta^{\rm Ph}_\mu(s)$ \cite{miyamura},
\begin{eqnarray}
\theta^{\rm Mo}_\mu (s) & \equiv & 
2 \pi \sum_{s'} \langle s|\partial^{-2} |s'\rangle 
\partial_\alpha  n_{\alpha \mu}(s'), 
\label{eq:monopart} \\
\theta^{\rm Ph}_\mu (s) & \equiv & 
\sum_{s'} \langle s|\partial^{-2} |s'\rangle
 \partial_\alpha \bar \theta _{\alpha \mu}(s'),
\label{eq:photpart} 
\end{eqnarray}
where $\langle s|\partial^{-2} |s'\rangle = -\frac{1}{4 \pi}
\frac{1}{(s-s')^2}$ is the Coulomb propagator in the 4-dimensional
Euclidean space.
As for the U(1)$_3$ gauge invariance, 
$\theta^{\rm Mo}_{\mu}(s)$ is gauge-variant,
since  $2 \pi n_{\mu\nu}$ is gauge-variant.
On the other hand,
$\theta^{\rm Ph}_{\mu}(s)$ is U(1)$_3$ gauge-invariant,
since it is composed of gauge-invariant $\bar \theta_{\mu\nu}$.
In the Landau gauge $\partial_\mu \theta^3_\mu (s) = 0$, one finds 
\begin{eqnarray}
\theta^{\rm Mo}_\mu(s) + \theta^{\rm Ph}_\mu(s) 
= \theta_\mu^3 (s).
\label{eq:decompose}
\end{eqnarray}

Here, we investigate  properties of the monopole and the photon parts
in terms of the electric and the magnetic currents.
The two form of these angle variables are written as
\begin{eqnarray}
(\partial \wedge \theta^{ \rm Mo})_{\mu\nu} 
& =  {\bar \theta^{\rm Mo}_{\mu\nu} } + 2 \pi n_{\mu\nu}^{\rm Mo}  
 \hspace{0.3cm} \mbox{with} 
  \hspace{0.3cm} &
\bar \theta^{\rm Mo}_{\mu\nu} = {\rm mod}_{2 \pi}
(\partial \wedge \theta^{ \rm Mo})_{\mu\nu} \in (-\pi,\pi]   
\nonumber 
\end{eqnarray}
\begin{eqnarray}
(\partial \wedge \theta^{ \rm Ph})_{\mu\nu} 
& =   {\bar \theta^{\rm Ph}_{\mu\nu} } + 2 \pi n_{\mu\nu}^{\rm Ph}
\hspace{0.3cm} \mbox{with}
   \hspace{0.3cm}  &
\bar \theta^{\rm Ph}_{\mu\nu} = {\rm mod}_{2 \pi}
(\partial \wedge \theta^{ \rm Ph})_{\mu\nu}\in (-\pi,\pi],
\nonumber 
\end{eqnarray}
using $n_{\mu\nu}^{\rm Mo}$, $n_{\mu\nu}^{\rm Ph} \in {\bf Z}$.
The monopole current  $k_\mu(s)$
and
the electric current $j_\mu(s)$ in the monopole 
part are defined as
\begin{eqnarray}
k^{\rm Mo}_\mu  \equiv  
\frac{1}{2\pi} \partial_\alpha ^*{\bar \theta_{\alpha\mu}^{\rm Mo}}
= - \partial_\alpha ^* n^{\rm Mo}_{\alpha \mu},
\hspace{1cm} 
j^{\rm Mo}_\mu \equiv \partial_\alpha {\bar \theta^{\rm Mo}_{\alpha\mu}},
\end{eqnarray}
and those in the photon part are defined as
\begin{eqnarray}
k^{\rm Ph}_\mu  \equiv 
\frac{1}{2\pi} \partial_\alpha ^* {\bar \theta_{\alpha\mu}^{\rm Ph}}
= - \partial_\alpha ^* n^{\rm Ph}_{\alpha \mu},
\hspace{1cm} 
j^{\rm Ph}_\mu \equiv \partial_\alpha {\bar \theta^{\rm Ph}_{\alpha\mu}}.
\end{eqnarray}

From the physical point of view, the decomposition of the abelian sector 
into the photon and monopole sectors directly corresponds to the 
separation of the electric current $j_\mu$ and the monopole current $k_\mu$
near the continuum limit.

\begin{figure}[p]
\epsfxsize = 11 cm
\centering \leavevmode
\epsfbox{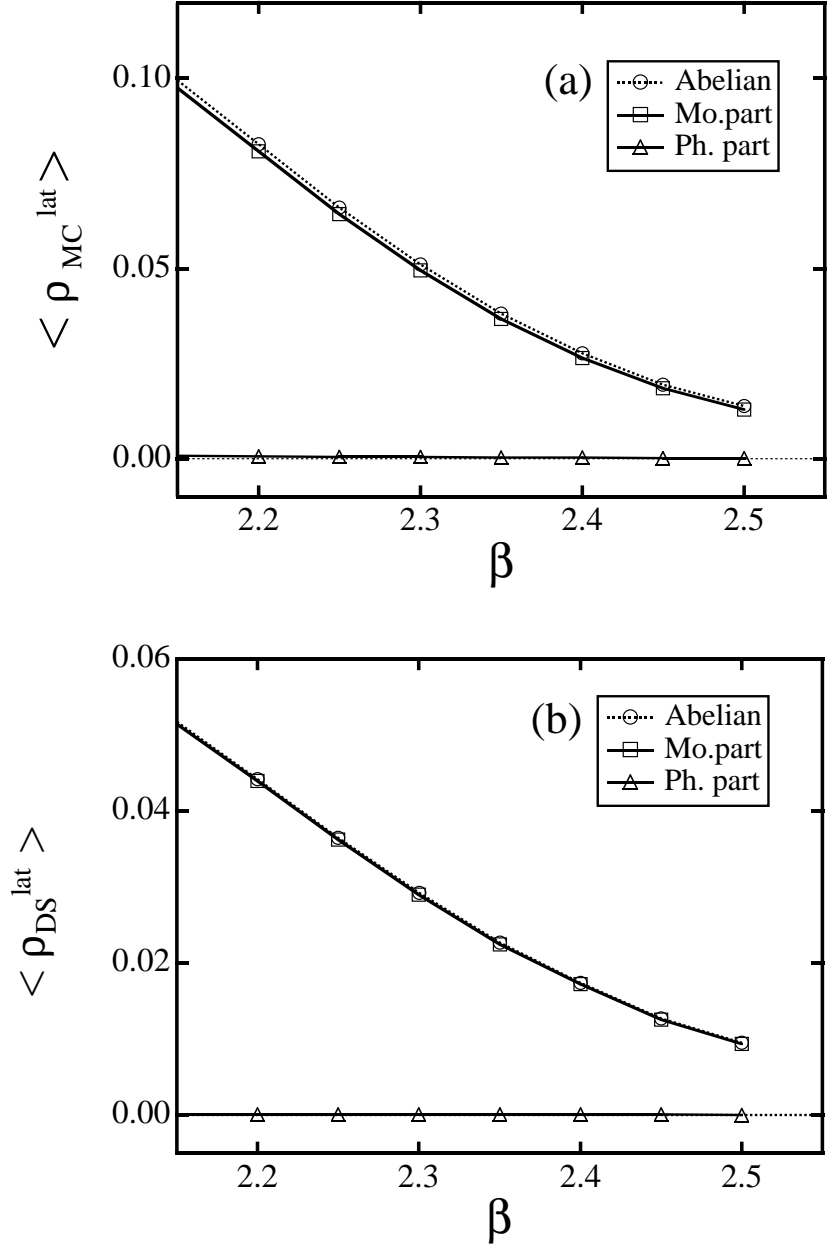}
\vspace{0cm}
\caption{   
(a)  The monopole-current density  $\rho^{lat}_{\rm MC}$ 
in the MA gauge as the function of $\beta$.
The circle, square and triangle denote
$\rho^{lat}_{\rm MC}$ in the abelian, monopole and photon sectors, 
respectively.
(b) The Dirac-sheet density $\rho_{\rm DS}^{lat}$  
in the MA gauge with U(1)$_3$ Landau-gauge as the function of $\beta$.
The circle, square and triangle denote $\rho_{\rm DS}^{lat}$ in the abelian, 
monopole and photon sectors, respectively.
}
\label{gfig11}
\vspace{0cm}
\end{figure}

In the actual lattice simulation, we can observe $k_\mu^{\rm ph} \simeq 0$,
$j_\mu^{\rm ph} = j_\mu$ and $k^{\rm Mo}_\mu \simeq 0$,
$k_\mu^{\rm Mo} = k_\mu$ in the MA gauge 
within a few percent error for large $\beta$.
For instance, we show in Fig.\ref{gfig11}(a) the lattice result of the 
monopole current density $\rho^{lat}_{\rm MC} \equiv 
\sum_{s\mu} |k_\mu(s)|/ \sum_{s\mu} 1$ in the MA gauge.
One fined $\rho^{\rm Abel}_{\rm MC} \simeq \rho^{\rm Mo}_{\rm MC}$ and 
$\rho^{\rm Ph}_{\rm MC} \simeq 0$.

Existence of the monopole accompanies the Dirac string.
We show also the Dirac-sheet density 
$\rho_{DS} \equiv  \sum_{s\mu\nu} |n_\mu(s)|/ \sum_{s\mu\nu} 1$ in the MA 
gauge 
with U(1)$_3$-Landau gauge in  Fig.\ref{gfig11}(b).
Owing to the U(1)$_3$-Landau gauge fixing, the abelian angle variable 
$\theta^3_\mu$ becomes mostly continuous and $|(\partial \wedge 
\theta^3)_{\mu\nu}|$ scarcely exceeds $\pi$,
so that redundant Dirac sheets are eliminated and the correlation between 
$\rho_{\rm MC}^{lat}$ and $\rho_{\rm DS}^{lat}$ appears.
More quantitative argument of $\rho_{\rm MC}^{lat}$ and  $\rho_{\rm DS}^{lat}$
will be done in the physical unit in Section \ref{subsec:SPMCDS}. 

For comparison, we investigate the monopole density $\rho_{\rm MC}^{lat}$ in 
the SU(2) Landau gauge  with SU(2)/U(1)$_3$  fixed, which is a kind of 
the abelian gauge.
 In the SU(2) Landau gauge, each component of the gluon field is 
 maximally smooth, and monopoles scarcely appear because they are 
 generated from the singular-like large fluctuation of the abelian field.
 For instance, $\rho_{\rm MC}^{lat}$ in the SU(2) Landau gauge in less than 
 1/10 of $\rho_{\rm MC}^{lat}$ in the MA gauge at $\beta=2.4$.
 In fact, in the regular SU(2) gauge configuration, the gluon fluctuation 
 is
 shared into each component almost equivalently, and there appears no 
 singularity as the monopole due to the ``wise sharing'' of the fluctuation.
 On the other hand, in the MA gauge, large gluon fluctuations are 
 concentrated into the abelian sector by the suppression of off-diagonal 
 gluon components, and therefore the monopole appears as the singularity 
 or the topological defect from the large field fluctuation in the 
 abelian sector.

\section{Link and Plaquette Variables in Monopole and Photon Sectors}

\begin{figure}[p]
\epsfxsize = 8 cm
\centering \leavevmode
\epsfbox{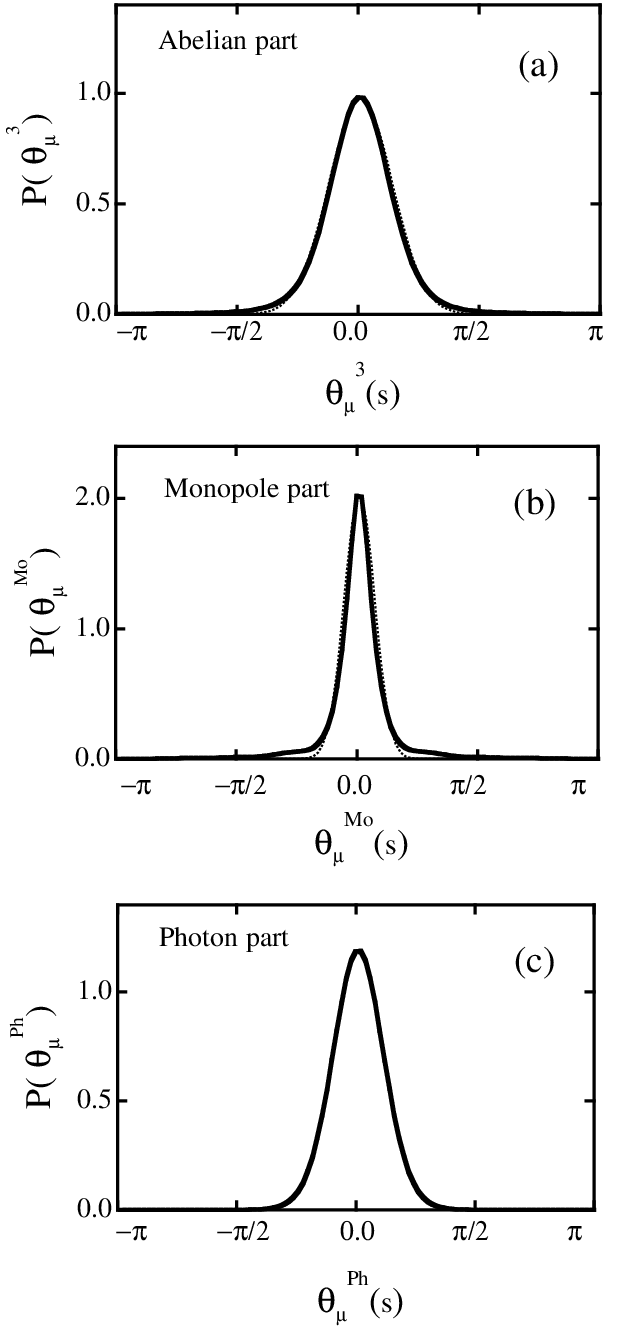}
\vspace{0cm}
\caption{  
The probability distribution $P(\theta_\mu)$  of lattice angle variables
$ \theta^3_\mu(s)$, $ \theta^{\rm Mo}_\mu(s)$,
$ \theta^{\rm Ph}_\mu(s) \in (-\pi,\pi] $ in the MA gauge with 
the U(1)$_3$ Landau gauge fixing at $\beta =2.4$ on $16^4$ lattice.
(a)  
the abelian angle variable
 $P( \theta^3_\mu ) $.
(b)
the monopole angle variable 
$P( \theta^{\rm Mo}_\mu )$.
(c) the photon angle variable 
$P( \theta^{\rm Ph}_\mu )$.
The dotted curves are fitting Gaussian curves in Eq.(\ref{gaussian})
with  $\alpha=$0.167 for  $P( \theta^3_\mu ) $,
$\alpha=$0.038 for $P( \theta^{\rm Mo}_\mu )$ and   $\alpha=$0.114 for 
$P( \theta^{\rm Ph}_\mu )$.
}
\label{gfig12}
\vspace{0cm}
\end{figure}

In general, confinement observed as the area-low reduction of the 
Wilson loop is closely related to the large gluon fluctuation as is 
suggested by the strong coupling QCD.
Hence, from monopole dominance for the confinement force observed on 
the lattice, one may expect large fluctuation of the gluon field in the 
monopole sector and small ones in the photon sector.
This naive expectation on the large fluctuation $\theta_\mu^{\rm Mo}$ 
may be also suggested from the definition of 
$\theta_{\mu}^{\rm Mo}$ and $\theta_{\mu}^{\rm Ph}$
 in Eq.(\ref{eq:decompose}), since the 
monopole angle variable $\theta_{\mu}^{\rm Mo}$ consists of the integer 
part $2\pi n$ $\in$ $2\pi {\bf Z}$, while the photon angle variable 
$\theta^{\rm Ph}_\mu$ consists of the fractional part $\bar \theta_{\mu}$ 
$\in$ $[-\pi,\pi)$.
We measure the lattice angle variables $\theta^{3}_{\mu}(s)$, 
$\theta^{\rm Mo}_{\mu}(s)$ and $\theta^{\rm Ph}_{\mu}(s)$ in the MA 
gauge with U(1)$_{3}$ Landau gauge as shown in Fig.\ref{gfig12}.
On the whole shape of the probability distribution $P(\theta_{\mu})$
of link variable, the distribution of monopole part 
$P(\theta_{\mu}^{\rm Mo})$ is smaller than that of the photon part  
$P(\theta_{\mu}^{\rm Ph})$, and $P(\theta_{\mu}^{3})$ is similar 
to $P(\theta_{\mu}^{\rm Ph})$ rather than $P(\theta_{\mu}^{\rm Mo})$.
These results are in contradiction to the above expectation on the large 
fluctuation of $P(\theta_{\mu}^{\rm Mo})$, and seem to be 
surprising in terms of the confinement properties,
because the string tension in the abelian sector resembles that in the 
monopole 
sector and the string tension in the photon sector is almost zero.

Quantitatively, the whole shape of probability distributions 
$P(\theta_{\mu}^{\rm 3})$, $P(\theta_{\mu}^{\rm Mo})$ and 
$P(\theta_{\mu}^{\rm Ph})$ can be fitted with the Gaussian curve 
around $\theta = 0$ as
\begin{eqnarray}
P(\theta) =  \frac{1}{\sqrt{2 \alpha \pi }} e^{- \frac{1}{2 
\alpha} 
  \theta^2 }
  \label{gaussian}
\end{eqnarray}
with 
$\alpha=$0.167 for  $P( \theta^3_\mu ) $,
$\alpha=$0.038 for $P( \theta^{\rm Mo}_\mu )$ and   $\alpha=$0.114 for 
$P( \theta^{\rm Ph}_\mu )$.  

\begin{figure}[bt]
\epsfxsize = 11 cm
\centering \leavevmode
\epsfbox{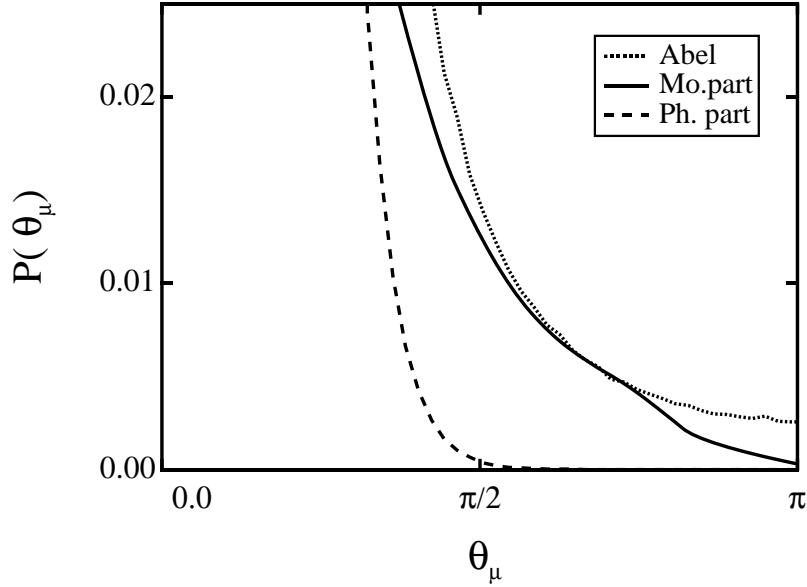}
\vspace{0cm}
\caption{   
The tail of 
$P( \theta^3_\mu )$, $ P( \theta^{\rm Mo}_\mu)$ and
$P(  \theta^{\rm Ph}_\mu)$.
Both $P( \theta^{\rm 3}_\mu )$ and $P( \theta^{\rm Mo}_\mu )$ include
a large fluctuation ingredient as $\theta_\mu > \frac{\pi}{2}$, while
$P( \theta^{\rm Ph}_\mu )$ does not have such a large fluctuation ingredient.
}
\label{gfig12d}
\vspace{0cm}
\end{figure}

Now let us reconsider the fluctuation of the monopole part.
In the U(1)$_{3}$ Landau gauge, the appearance of the Dirac sheet 
with $ n_{\mu\nu} \ne 0 $ is strongly suppressed, because the field is 
forced to be continuous there.
Therefore, almost all of the two form 
$\theta_{\mu\nu}^{\rm Mo} \equiv \partial_{\mu} \theta_{\nu}^{\rm Mo} - \partial_{\nu} 
\theta_{\mu}^{\rm Mo}$ satisfies $- \pi < \theta_{\mu\nu}^{\rm Mo} \le \pi$, i.e., one 
finds $ n_{\mu\nu}^{\rm Mo}(s) = 0$ almost everywhere on the lattice.
This is the reason why the fluctuation of the monopole angle 
variable $\theta^{\rm Mo}_\mu$ becomes small.
However, there appears the relic of the large fluctuation of the 
abelian angle variable in the monopole angle variable,
as shown in Fig.\ref{gfig12d}.
Similar to  $P( \theta^3_\mu ) $, the distribution $P( \theta^{\rm Mo}_\mu )$
of the monopole angle variable has a non-Gaussian large tail even 
for $|\theta^{\rm Mo}| \le \frac{\pi}{2}$, while  $P( \theta^{\rm Ph}_\mu )$
has the Gaussian tail only and mainly distributes within 
$- \frac{\pi}{2} < \theta^{\rm Ph}_\mu < \frac{\pi}{2}$ as shown in 
Fig.\ref{gfig12d}.
Such a large fluctuation of $\theta^{\rm Mo}_\mu$ is responsible for 
the appearance of the monopole, and it 
also would play a relevant role for confinement properties,
because a large reduction of the string tension is observed in the 
lattice QCD by the artificial elimination of the large fluctuation.

\begin{figure}[p]
\epsfxsize = 11 cm
\centering \leavevmode
\epsfbox{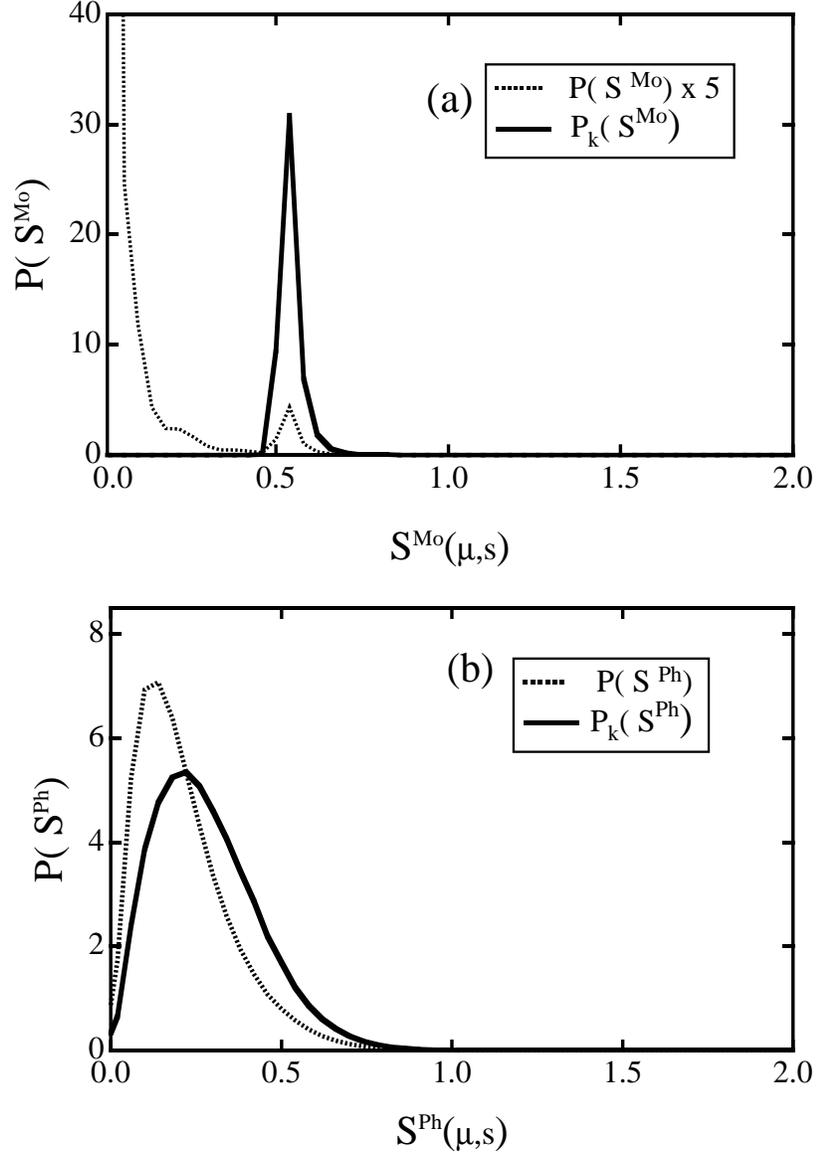}
\vspace{0cm}
\caption{ 
(a) The probability distribution $P(S^{\rm Mo})$ of the monopole part of 
the action density
$S^{\rm Mo} (s,\hat \mu) $ around the monopole. 
The dotted curve denotes $S^{\rm Mo} (s,\hat \mu) $ on the whole lattice.
The value of $S^{\rm Mo} (s,\hat \mu) $ is the average of
the action density $S^{\rm Mo}_{\mu\nu}(s) $
over the neighboring plaquettes around the dual link.
(b) The probability distribution $P(S^{\rm Ph})$ of the photon
part of action density
$S^{\rm Ph} (s,\hat \mu) $ around the monopole.
The dotted curve denotes $S^{\rm Ph} (s,\hat\mu) $ on the whole lattice.
}
\label{gfig13}
\vspace{0cm}
\end{figure}

Next, we consider the plaquette action densities of the  monopole and the photon 
parts,
\begin{eqnarray}
S^{\rm Mo}_{\mu\nu} (s)  & \equiv & 
1 - \frac12 {\rm tr} \Box^{\rm Mo}_{\mu\nu}(s),  \\
S^{\rm Ph}_{\mu\nu} (s)  &  \equiv& 
1 - \frac12 {\rm tr} \Box^{\rm Ph}_{\mu\nu}(s),  
\end{eqnarray}
similar to the SU(2) and abelian action densities in 
Eqs.(\ref{eq:abaction1}) and (\ref{eq:abaction2}).
Here, $\Box^{\rm Mo}_{\mu\nu}(s) $ and $\Box^{\rm Ph}_{\mu\nu}(s)$ are
the plaquette variables in the monopole and photon parts, 
 \begin{eqnarray}
\Box^{\rm Mo}_{\mu\nu}(s) & \equiv &
u^{\rm Mo}_\mu(s) u_\nu^{\rm Mo}(s + \hat \mu) u^{{\rm Mo}\dagger}_\mu(s + 
\hat \nu)
u^{{\rm Mo}\dagger}_\nu(s), \\
\Box^{\rm Ph}_{\mu\nu}(s) & \equiv &
u^{\rm Ph}_\mu(s) u_\nu^{\rm Ph}(s + \hat \mu) u^{{\rm Ph}\dagger}_\mu(s + 
\hat \nu)
u^{{\rm Ph}\dagger}_\nu(s),
\end{eqnarray}
respectively.
To see the monopole effect on the action density $S_{\mu\nu}(s)$, 
we measure the average $\bar S(s,\mu)$ 
over the neighboring plaquettes around the dual link,
\begin{eqnarray}
\bar S(s,\hat \mu) \equiv
\frac16 \sum_{\alpha\beta\gamma} \sum_{m=0}^1 \frac12 
| \varepsilon_{\mu\alpha\beta\gamma} |
S_{\alpha\beta}(s + m \hat \gamma),
\end{eqnarray}
similar to Eq.(\ref{eq:duallink}).
As shown in Fig.\ref{gfig13}, both action densities $P(S^{\rm Mo})$ and
$P(S^{\rm Ph})$ have a peak near $S=0$.
For the monopole parts, 
there are two ingredients in the probability distribution $P(S^{\rm Mo})$
of the monopole action density $S^{\rm Mo}(s,\mu)$:
one is the large fluctuation around 
$S^{\rm Mo}=\frac12$ and the other is the 
small fluctuation near $S^{\rm Mo} \simeq 0$.
The large fluctuation ingredient corresponds to the  
average action-density over six plaquettes of the cube including the monopole 
inside
\begin{eqnarray}
S_{\mu\nu}^{\rm Mo} 
  = 1-\frac12
  {\rm tr} \Box_{\mu\nu}^{\rm Mo}
= 1- \cos \bar \theta_{\mu\nu}^{\rm Mo} \simeq 1 - \cos \frac{2 \pi}{6}  
= \frac12,
\label{geome}
\end{eqnarray}
which is similar to the 
abelian action density $P(S^{\rm Abel})$
around the monopole, as shown in Fig.\ref{gfig10}.
Here, $2\pi$  is the total magnetic flux created by the unite charge 
monopole.
On the other hand, 
the small fluctuation ingredient corresponds to the plaquette apart from
the monopole.
For the photon part, 
the global shape of the probability distribution $P(S^{\rm Ph})$ of the 
photon action density $S^{\rm Ph}(s,\hat \mu)$ resembles $P(S^{\rm Abel})$
as shown in Fig.\ref{gfig10},
however,
$P(S^{\rm Ph})$ does not  have the large fluctuation peak around 
$S^{\rm Ph}=\frac12$,
corresponding to the absence of the monopole unlike $P(S^{\rm Abel})$.
In the distribution $P(S^{\rm Mo})$, the peak position around 
$S^{\rm Mo} = \frac 12 $ is almost $\beta$-independent, since it is 
geometrically determined as Eq.(\ref{geome}).
On the other hand, the peak near $S = 0$ becomes narrow both 
in $P(S^{\rm Mo})$ and in $P(S^{\rm Ph})$ as $\beta$ increases.

\begin{figure}[bt]
\epsfxsize = 11 cm
\centering \leavevmode
\epsfbox{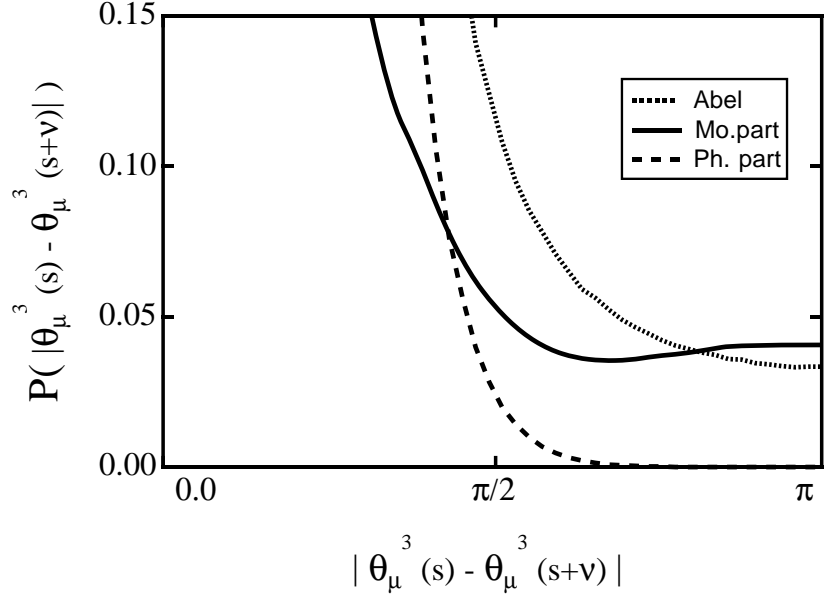}
\vspace{0cm}
\caption{    
The probability distribution of the derivative of the lattice angle
variables.
The dashed, solid and dotted curves denote 
$P(| \theta^3_\mu(s)-\theta^3_\mu(s + \hat \nu) |)$,
$P(| \theta^{\rm Mo}_\mu(s)-\theta^{\rm Mo}_\mu(s + \hat \nu) |)$ and 
$P(| \theta^{\rm Ph}_\mu(s)-\theta^{\rm Ph}_\mu(s + \hat \nu) |)$,
respectively.
}
\label{gfig17}
\vspace{0cm}
\end{figure}



Finally in this section, we consider also the monopole effect on the 
discontinuity of the angle variables.
As the result of the large fluctuation of the monopole plaquette 
variable around %
the monopole as $S^{\rm Mo} \simeq \frac12$, 
there appears the discontinuity of the link variable 
$\theta_{\mu}^{\rm Mo}$.
We measure the correlation neighboring links $u_{\mu}(s)=e^{i\theta(s)}$
and $u_{\mu}(s+ \hat \nu)=e^{i\theta(s + \hat \nu)}$. 
We show in Fig.\ref{gfig17} the angle difference 
$\Delta \equiv |{\rm mod}_{2 \pi} (\theta_{\mu}(s) - 
\theta_{\mu}(s+\hat \nu) ) |$
with $\mu \ne \nu$. The photon angle variable $\theta_\mu^{\rm Ph}(s)$
tends to be continuous with the only small difference between the 
neighboring links as
$\Delta \le \frac{\pi}{2}$.
This tendency on the continuity of $\theta^{\rm Ph}(s)$ becomes clear 
as $\beta$ increases.
On the other hand, similar to the abelian sector, the monopole angle 
variable $\theta^{\rm Mo}_\mu(s)$ includes the discontinuity as
$\Delta \ge \frac{\pi}{2}$. 
Thus, the appearance of point-like monopoles in the MA gauge brings 
the large fluctuation and the discontinuity of the  angle variable 
$\theta^{\rm Mo}_\mu(s)$ around them.



\section{Dual Field Formalism}
\label{dff}

In the presence of magnetic monopoles, the  abelian gauge
field $A_\mu(x)$  inevitably includes the singularity as the Dirac string,
which leads to some difficulties in the field theoretical treatment.
Here, we investigate the description with  
the dual field formalism, which is  useful to describe the monopole
sector in the QCD vacuum.
In the dual formalism, the dual gauge field $B_\mu(x)$ is introduced as 
$F^{\rm dual}_{\mu\nu} = \partial_\mu B_\nu - \partial_\nu B_\mu$ with
$F^{\rm dual}_{\mu\nu} \equiv ^{*} F_{\mu\nu} \equiv
\frac12 \epsilon_{\mu\nu\alpha\beta} F^{\alpha\beta}$, and
the interchange between $A_\mu$ and $B_\mu$ corresponds to the 
electro-magnetic duality transformation, 
$F_{\mu\nu} \leftrightarrow F^{\rm dual}_{\mu\nu}$ or ${\bf H} \leftrightarrow 
{\bf E}$ \cite{suganuma}. 
The monopole sector, which carries 
essence of the 
nonperturbative QCD, includes the magnetic current only and does not 
include the electric current: $k_\mu \ne 0$ and $j_\mu =0$. 
Therefore, the monopole sector is the dual version of the ordinary QED 
system with $j_\mu \ne 0$ and $k_\mu = 0$. 
The dual gauge field $B_\mu(x)$
can be introduced without the singularity like the Dirac string 
owing to the dual Bianchi identity,
$\partial^\mu {}^*(\partial \wedge B)_{\mu\nu} = 
-\partial^\mu F^{\rm dual}_{\mu\nu} = - j_{\nu} = 0$, corresponding to 
the absence of the electric current.

Let us consider the derivation of the dual gauge field $B_\mu(x)$ 
from the monopole current $k_\mu(x)$, taking the dual Landau gauge, 
$ \partial_\mu B^\mu = 0$.
Then, the relation $\partial^\mu  F^{\rm dual}_{\mu\nu} 
= \partial^2 B_\nu - \partial_\nu ( \partial^\mu B_\mu ) = k_\nu$
becomes a simple form $\partial^2 B_\mu = k_\mu$.
Therefore, starting from the monopole current configuration
$k_\mu(x)$,
we derive the dual gauge field $B_\mu(x)$ as
\begin{equation}
B_\mu (x) = \int d^4 y \langle x|\partial^{-2} |y\rangle k_\mu(y) = 
-\frac{1}{4\pi^2}
\int d^4 y \frac{k_\mu(y)}{(x-y)^2}.
\label{eq:dual}
\end{equation}
Using the dual gauge field $B_{\mu}$, the  Wilson loop in the monopole sector 
is expressed as
\begin{eqnarray}
 W^{\rm Mo} &
= &  {\rm exp} \{ {ie\oint dx^\mu A_\mu ^{\rm Mo} } \} 
= {\rm exp}
\{ {ie\int  \, \int d\sigma^{\mu\nu} {}^* F_{\mu\nu}^{\rm Mo} } \} \nonumber \\
& = & {\rm exp}
({-ie \int \, \int
d\sigma^{\mu\nu} {}^* F^{\rm dual}_{\mu\nu}}) 
\nonumber \\  
& = & {\rm exp}
(-ie \int \, \int
d\sigma^{\mu\nu} {}^* (\partial \wedge B)_{\mu\nu}), 
\end{eqnarray}
where the Stokes theorem is applicable because of abelian nature.

Now, we apply this dual field formalism to the monopole sector in the 
lattice QCD in the Euclidean metric.
The lattice dual gauge field
$\theta_\mu^{\rm dual} $ is defined as
$\theta_\mu^{\rm dual} \equiv a e B_\mu /2$
similar to $\theta_{\mu}^3 \equiv  a e A_\mu /2$.
In the dual Landau gauge,  $\theta_\mu^{\rm dual}$ is obtained from 
the monopole current $k_{\mu}$ as
\begin{equation}
\theta_\mu^{\rm dual} (s+ \hat \mu) =
2 \pi \sum_{s'}  \langle s|\partial^{-2} |s'\rangle k^{ lat}_\mu(s'),
\end{equation}
using the lattice Coulomb propagator $ \langle s|\partial^{-2} |s'\rangle
=\langle s|(\partial_\mu
\partial'_\mu)^{-1} | s'\rangle$,
where $\partial_\mu$ and $\partial'_\mu$ denote the forward and 
backward derivatives, respectively:
\begin{equation}
\partial_\mu f(x) \equiv f(s+\hat \mu) -f(s),
\end{equation}  
\begin{equation}
\partial'_\mu f(x) \equiv f(s) -f(s-\hat \mu).
\end{equation}  
The two-form of
$\theta_{\mu\nu}^{\rm dual} $
is defined by
\begin{equation}
\theta_{\mu\nu}^{\rm dual}(s) \equiv
\partial'_\mu \theta^{\rm dual}_\nu(s)
- \partial'_\nu \theta^{\rm dual}_\mu(s) 
= \bar \theta_{\mu\nu}^{\rm dual} + 2 \pi n_{\mu\nu}^{\rm dual},
\end{equation}
where the dual abelian field strength
$\bar \theta_{\mu\nu}^{\rm dual} 
= e a^2 F^{\rm dual}_{\mu\nu}/2$ is defined as
\begin{equation}
\bar \theta_{\mu\nu}^{\rm dual}(s)  \equiv {\rm mod}_{2\pi} (\partial 
\wedge \theta^{\rm dual})_{\mu\nu}(s) \in (-\pi,\pi]
\end{equation}
in the dual gauge invariant manner.

\begin{figure}[bt]
\epsfxsize = 11 cm
\centering \leavevmode
\epsfbox{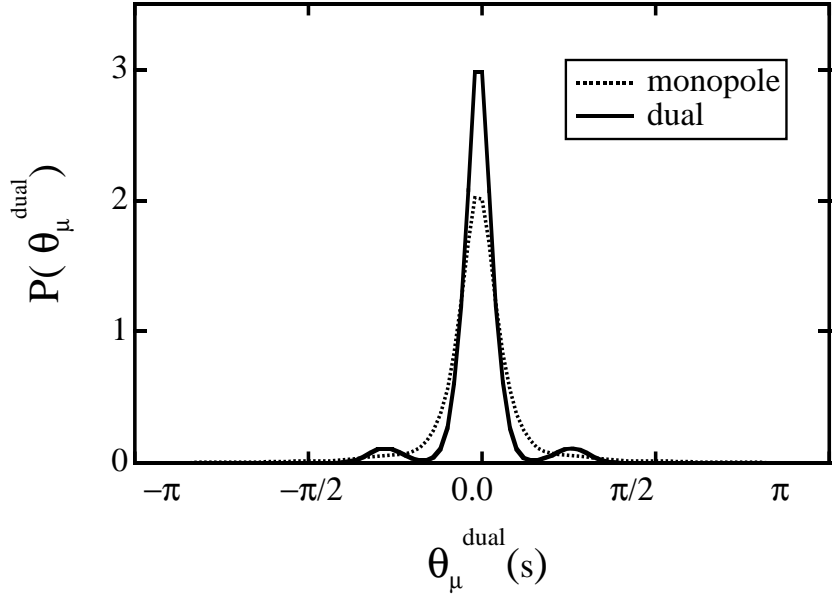}
\vspace{0cm}
\caption{   
The comparison of 
the probability distribution
$P(\theta_{\mu}^{\rm dual})$
with 
$P(\theta_{\mu}^{\rm Mo})$
in the monopole sector in the MA gauge.
The dual gauge variable $\theta^{\rm dual}_\mu$ is defined in the dual Landau
gauge $\partial_\mu \theta^{\rm dual}_\mu =0$.
The angle variable $\theta_\mu^{\rm Mo}$ in the monopole sector 
satisfies 
 the Landau gauge condition $\partial_\mu \theta_\mu^{\rm Mo} =0 $.
}
\label{dual1}
\vspace{0cm}
\end{figure}

\begin{figure}[bt]
\epsfxsize = 11 cm
\centering \leavevmode
\vspace{0cm}
\epsfbox{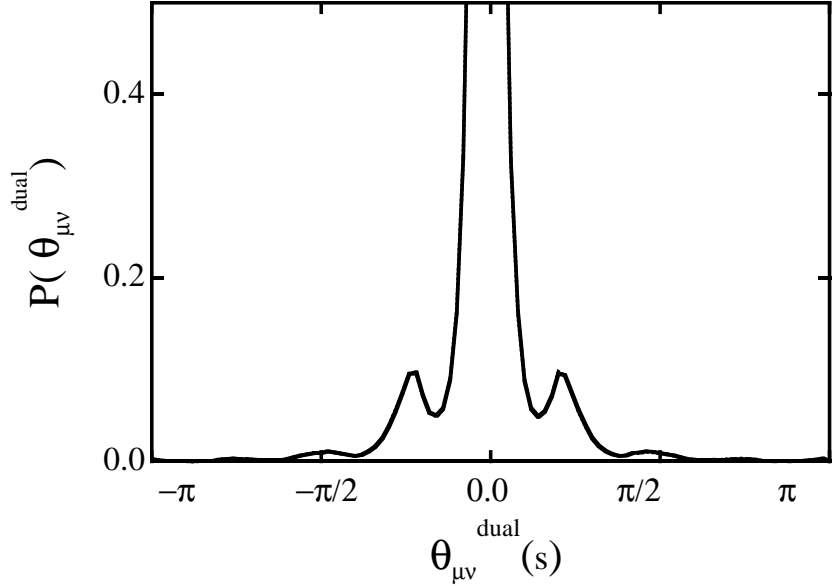}
\caption{   
The probability distribution 
$P( \bar \theta_{\mu\nu}^{\rm dual}) $ of the dual abelian 
field-strength $ \bar \theta_{\mu\nu}^{\rm dual}$
in the monopole sector in the MA gauge.
Here, $P( \bar \theta_{\mu\nu}^{\rm dual}) $ exactly coincides with 
 $P( \bar \theta_{\mu\nu}^{\rm Mo}) $.
}
\label{dual2}
\vspace{0cm}
\end{figure}

In Fig.\ref{dual1}, we compare
the probability distribution
$P(\theta_{\mu}^{\rm dual})$ of dual gauge variable 
$\theta_{\mu}^{\rm dual}$ and 
$P(\theta_{\mu}^{\rm Mo})$ of the gauge variable 
$\theta_{\mu}^{\rm Mo}$
in the monopole sector in the MA gauge. 
Here, $\theta^{\rm dual}_\mu$ is defined
in the dual Landau gauge $\partial _{\mu} \theta_\mu^{\rm dual} = 0$,
and $\theta^{\rm Mo}_\mu$ is defined in Eq.(\ref{eq:monopart}) satisfies  
the Landau gauge $\partial_\mu \theta_\mu ^{\rm Mo}=0$.
These two distributions seem to have almost the same widths.
This is because the Dirac sheet $n_{\mu\nu}$ is strongly suppressed 
in the Landau gauge, while there is no dual Dirac sheets 
with $n_{\mu\nu}^{\rm dual} \ne 0$ in the dual Landau gauge owing to the absence 
of $j_{\mu}$. In fact, only the essential minimal fluctuation remains 
both for the monopole variable $\theta_{\mu}^{\rm Mo}$ 
in the Landau gauge and for the dual variable $\theta_{\mu}^{\rm dual}$
in the dual Landau gauge.
Different from $\theta_{\mu}^{\rm Mo}$, however, $\theta_{\mu}^{\rm dual}$ 
has two  bumps around $\theta_\mu^{\rm dual} = \pm \frac{\pi}{4}$
originated from the monopole current.

We show in Fig.\ref{dual2} the probability distribution 
$P( \bar \theta_{\mu\nu}^{\rm dual}) $ of the dual abelian 
field-strength $ \bar \theta_{\mu\nu}^{\rm dual}$. There appear also 
some bumps originated from the quantization of the  monopole current
and lattice discretization.
Here, $P( \bar \theta_{\mu\nu}^{\rm dual})$ is the same as 
$P( \bar \theta_{\mu\nu}^{\rm Mo})$ 
because of $\bar \theta^{\rm dual}_{\mu\nu} 
= {}^{*}\bar \theta^{\rm Mo}_{\mu\nu} $ in the continuum limit.
However, the two-forms 
$\theta_{\mu\nu}^{\rm dual} \equiv (\partial \wedge \theta^{\rm 
dual})_{\mu\nu}$ differs from  $ {}^{*}\theta_{\mu\nu}^{\rm Mo} \equiv 
^{*}(\partial \wedge \theta^{\rm Mo})_{\mu\nu}$
according to the difference between $n_{\mu\nu}^{\rm dual}$ and
${}^* n_{\mu\nu}^{\rm Mo}$.
In the description by $\theta^{\rm Mo}_{\mu\nu}$, 
there inevitably appears the singularity as the Dirac sheet with 
$n_{\mu\nu} \ne 0$ to generate the monopole 
current through the breaking of the Bianchi identity,
although the appearance of the Dirac sheet is strongly suppressed in 
the Landau gauge.
In the dual formalism, on the other hand, the monopole sector becomes 
regular and does not include the dual Dirac sheet, i.e. 
$n_{\mu\nu}^{\rm dual}=0$,
in the dual Landau gauge owing to the dual Bianchi identity.
Actually in the monopole sector in the dual Landau gauge,
$\theta_{\mu\nu}^{\rm dual} \equiv (\partial  \wedge 
\theta^{\rm dual})_{\mu\nu}$ distributes only in the region of 
$-\pi < \theta^{\rm dual}_{\mu\nu} < \pi$, and hence one fines
$\bar \theta^{\rm dual}_{\mu\nu} \equiv \mbox{mod}_{2\pi} (\partial 
\wedge \theta^{\rm dual})_{\mu\nu} = \theta^{\rm dual}_{\mu\nu} $ and
$n_{\mu\nu}^{\rm dual} = 0$ everywhere on the lattice. 
Thus, the monopole sector in the lattice QCD is described by the dual 
gauge field without the singularity as the Dirac string in the dual 
Landau gauge.

Using the dual gauge variable  $\theta_{\mu}^{\rm dual}$,
the static potential between the 
monopole and the anti-monopole  is obtained from  the dual Wilson loop
${\rm tr} \prod_{i=1}^{\rm L}  u _{\mu_i}^{\rm dual}(s_i)$ with
$u_{\mu}^{\rm dual}(s) \equiv {\rm exp}\{ i \tau_{3}
\theta_{\mu}^{\rm dual}(s) \}$ \cite{atanaka}.
The large dual Wilson loop obeys the perimeter law,
and the inter-monopole potential can be fitted by the Yukawa potential
including the monopole size effect  \cite{atanaka}.
Thus, the dual gauge field $B_{\mu}$ seems to acquire its mass 
in the infrared region, which would be a direct evidence on the
dual Higgs mechanism by monopole condensation occurring in the QCD vacuum.

\section{Scaling Properties on Monopole Current and Dirac Sheet}
\label{subsec:SPMCDS}

In this section, 
we study the $\beta $-dependence of the gluon field and
the field strength in the abelian, monopole and photon sectors 
in the MA gauge with the U(1)$_3$ Landau gauge. 
We study also the monopole-current density $\rho_{\rm MC}$ 
and the Dirac-sheet density $\rho_{\rm DS}$ 
using the lattice QCD simulation with various $\beta $. 
We argue these quantities 
both in the lattice unit and in the physical unit, and 
examine their scaling property.

\begin{figure}[p]
\epsfxsize = 11 cm
\centering \leavevmode
\epsfbox{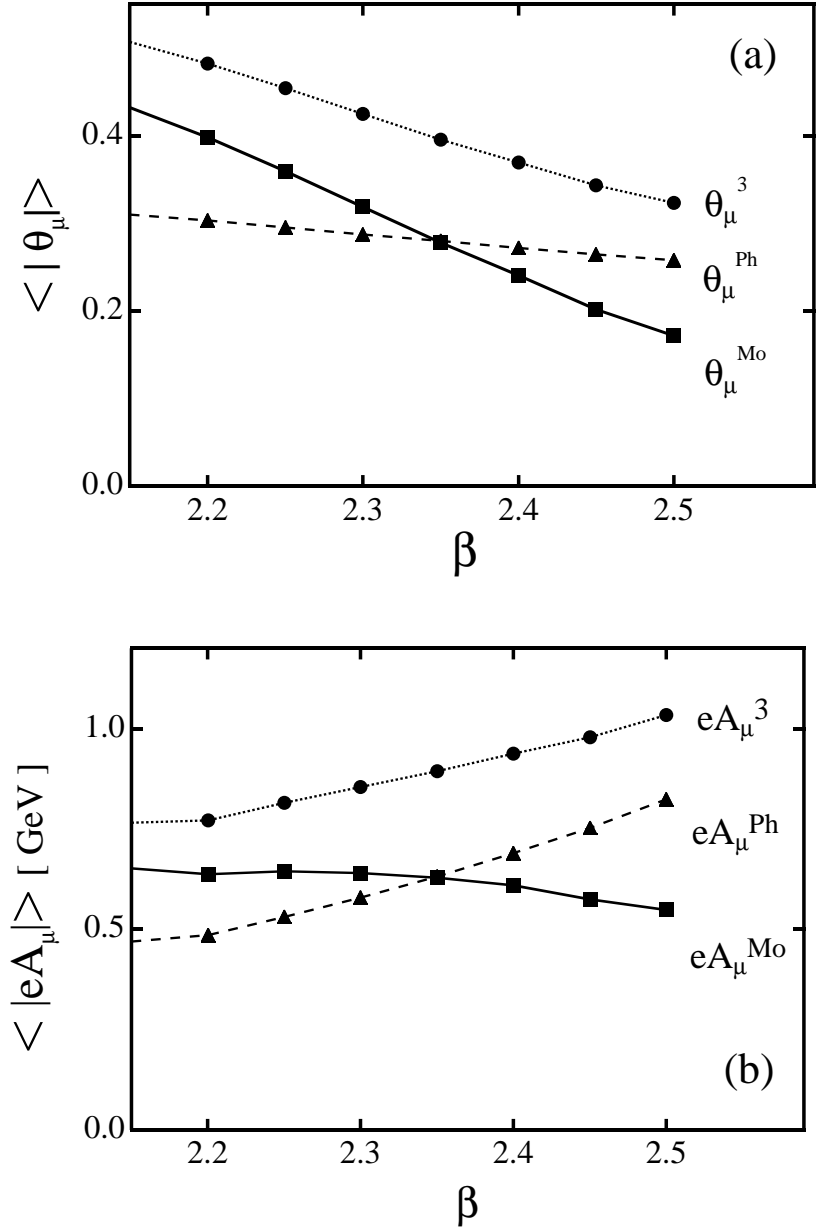}
\vspace{0cm}
\caption{   
(a)
The absolute value of the lattice angle variables
$ \theta^3_\mu(s)$, $ \theta^{\rm Mo}_\mu(s)$,
$ \theta^{\rm Ph}_\mu(s)$  in
the MA gauge with U(1)$_3$ Landau gauge fixing  on $16^4$ lattice
as the function of $\beta$.
The solid, dotted and dashed lines denote 
$|\theta^3_\mu(s)|$, $ |\theta^{\rm Mo}_\mu(s)|$ and
$|\theta^{\rm Ph}_\mu(s)|$, respectively.
(b)
The absolute value of the continuum abelian gauge fields 
$|eA^3_\mu|$, $|eA^{\rm Mo}_\mu|$  and $|eA^{\rm Ph}_\mu|$
in the MA gauge on $16^4$ lattice as the function of $\beta$. 
These are obtained from $\theta_\mu \equiv A_\mu \cdot \frac{ae}{2}$
with lattice spacing $a$. The lattice spacing $a$ is defined so as to
produce the string tension $\sigma =$0.89 GeV/fm.
}
\label{gfig14}
\vspace{0cm}
\end{figure}

%

In the MA gauge with the U(1)$_3$ Landau gauge, 
the probability distribution of 
abelian angle variables, 
$\theta _\mu ^3$, $\theta _\mu ^{\rm Mo}$ and $\theta _\mu ^{\rm Ph}$, 
also exhibit a peak around $\theta _\mu =0$ as was shown in 
Fig.\ref{gfig12}. 
We show in Fig.\ref{gfig14}(a) the $\beta $-dependence of 
$\langle | \theta_{\mu}^3| \rangle$, 
$\langle | \theta_{\mu}^{\rm Ph}| \rangle$ and 
$\langle | \theta_{\mu}^{\rm Mo}| \rangle$ 
in the MA gauge with the U(1)$_3$ Landau gauge, 
and find that the fluctuation of the abelian angle variables 
decreases with $\beta $ on the average. 
On the continuum abelian fields,
$eA_{\mu} \equiv \frac{2}{a}  \theta _{\mu}^3$,
$eA_{\mu}^{\rm Ph} \equiv \frac{2}{a}  \theta ^{\rm Ph} _{\mu}$ and
$eA_{\mu}^{\rm Mo} \equiv \frac{2}{a}  \theta ^{\rm Mo} _{\mu}$, 
we show in Fig.\ref{gfig14}(b) the $\beta $-dependence on 
$\langle | eA_{\mu}| \rangle$, 
$\langle | eA_{\mu}^{\rm Ph}| \rangle$ and 
$\langle | eA_{\mu}^{\rm Mo}| \rangle$. 
While there appears a $\beta$-dependence on $eA_{\mu}$ and 
$eA_{\mu}^{\rm Ph}$, 
the continuum field $eA_{\mu}^{\rm Mo}$ 
in the monopole sector is also found to be less $\beta $-independent.

\begin{figure}[p]
\epsfxsize = 11 cm
\centering \leavevmode
\epsfbox{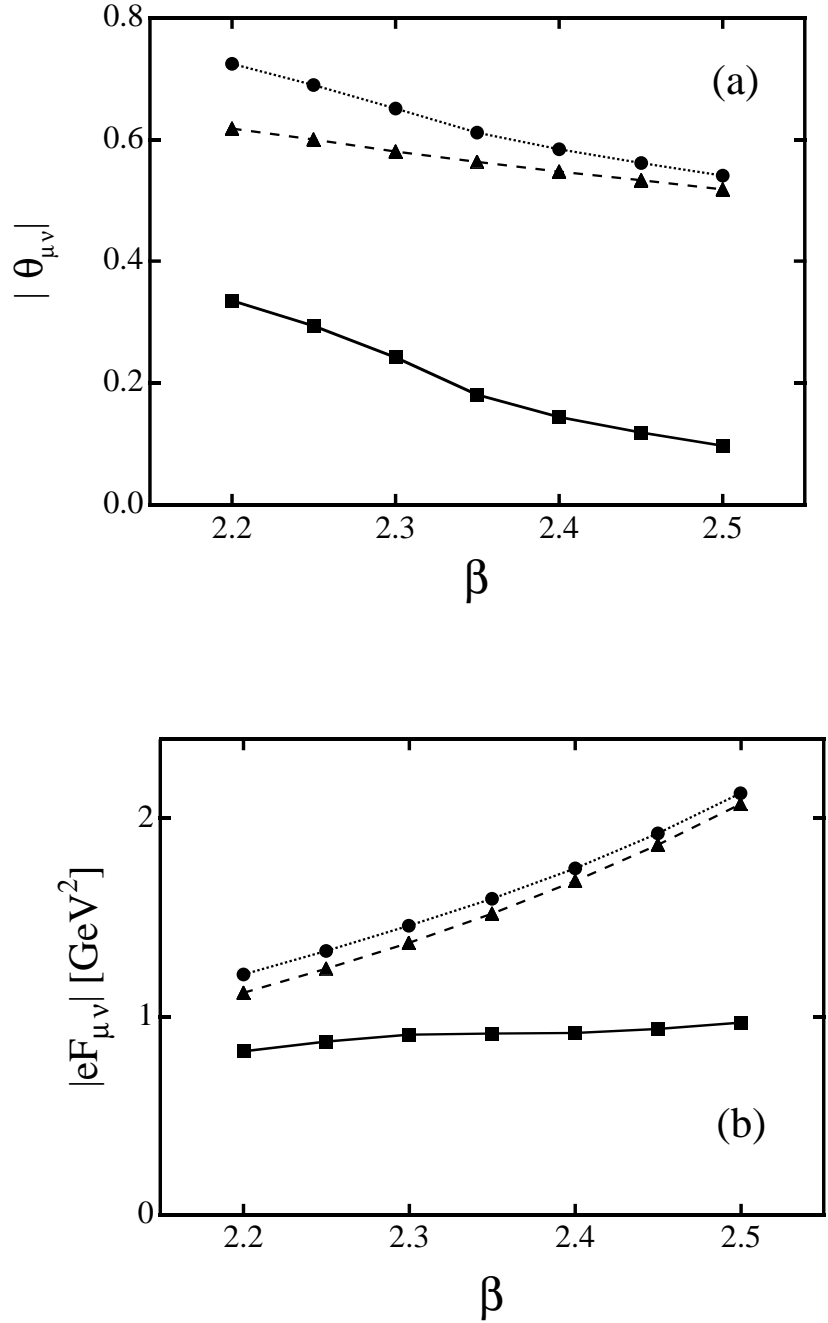}
\vspace{0cm}
\caption{ 
(a)  The absolute value of the abelian field strength
$|\bar \theta_{\mu\nu}|$ in the MA gauge as the function of $\beta$.
The circle, square and triangle denote  $|\bar \theta_{\mu\nu}|$ in the
abelian, monopole  and photon sectors, respectively. 
(b) The absolute value of the continuum abelian field strength
$ |eF_{\mu\nu}| \equiv |\bar \theta_{\mu\nu}| \cdot \frac{2}{a^2}$
in the physical scale as the function of $\beta$.
The circle, square and triangle denote  $|eF_{\mu\nu}|$ in the 
abelian, monopole  and photon sectors, respectively.
}
\label{gfig15}
\vspace{0cm}
\end{figure}



In the MA gauge, the probability distribution of 
the abelian plaquette variable 
$\bar \theta _{\mu \nu }$, 
has a peak around 
$\bar \theta _{\mu \nu }=0$ 
as was shown in Fig.\ref{gfig5c}, 
and hence the abelian plaquette action seems to control 
the abelian system like the SU(2) total action. 
Such a tendency on the gathering around $\bar \theta _{\mu \nu }=0$ 
becomes clearer as $\beta$ increases.
Quantitatively, we show in Fig.\ref{gfig15}(a) the $\beta $-dependence of 
$\langle |\bar \theta _{\mu\nu}| \rangle$, 
$\langle |\bar \theta_{\mu\nu}^{\rm Ph}| \rangle$ and 
$\langle |\bar \theta_{\mu\nu}^{\rm Mo}| \rangle$ 
in the MA gauge, and find decreasing of 
all the 
plaquette angles with $\beta $.
From these lattice data, 
we try to extract the continuum abelian field strength,
$eF_{\mu\nu} \equiv \frac{2}{a^2} \bar \theta _{\mu\nu}$,
$eF_{\mu\nu}^{\rm Ph} \equiv \frac{2}{a^2}\bar \theta ^{\rm Ph} _{\mu\nu} 
$ and
$eF_{\mu\nu}^{\rm Mo} \equiv \frac{2}{a^2}\bar \theta ^{\rm Mo} _{\mu\nu} 
$.
We show the $\beta $-dependence on 
$\langle | eF_{\mu\nu}| \rangle$, 
$\langle | eF_{\mu\nu}^{\rm Ph}| \rangle$ and 
$\langle | eF_{\mu\nu}^{\rm Mo}| \rangle$ 
in Fig.\ref{gfig15}(b), where the lattice constant $a$ used is determined 
so as to reproduce the string tension $\sigma $=0.89GeV/fm 
in the SU(2) sector. 
While there appears a $\beta$-dependence on $eF_{\mu\nu}$ and 
$eF_{\mu\nu}^{\rm Ph}$, the continuum field $eF_{\mu\nu}^{\rm Mo}$ 
in the monopole sector is almost $\beta$-independent.
In other words, we find a good scaling property even for the local 
variable as $eF_{\mu\nu}^{\rm Mo}$ in the monopole sector.

In order to understand the good scaling property 
on the abelian continuum variables in the monopole sector, 
we examine the $\beta $-dependence of 
the monopole-current density $\rho _{\rm MC}$ and 
the Dirac-sheet density $\rho_{\rm DS}$ in the MA gauge with the 
U(1)$_3$ Landau gauge.
In the physical unit, 
the monopole-current density $\rho _{\rm MC}$ is expressed as 
the ratio of the total monopole-current length $L_{\rm tot}^{phys}$ 
to 4-dimensional physical volume $V_{\rm tot}^{phys}$ of the system, 
\begin{eqnarray}
\rho _{\rm MC}^{phys} \equiv \frac{\rho _{\rm MC}^{lat}} {a^3}
={\displaystyle \sum_{s,\mu } a |k_\mu (s)| \over \displaystyle  
\sum_{s,\mu }a^4}
={L_{\rm tot}^{phys} \over V_{\rm tot}^{phys}}.
\end{eqnarray}
Similarly, the Dirac-sheet density $\rho_{\rm DS}$ 
in the physical unit is expressed as 
the ratio of the total Dirac-sheet area $S_{\rm tot}^{phys}$ 
to 4-dimensional physical volume $V_{\rm tot}^{phys}$ of the system, 
\begin{eqnarray}
\rho_{\rm DS}^{phys} \equiv \frac{\rho_{\rm DS}^{lat}}{a^2}
={\displaystyle \sum_{s,\mu ,\nu } a^2 |n_{\mu \nu }(s)| \over 
\displaystyle 
 \sum_{s,\mu ,\nu } a^4}
={S_{\rm tot}^{phys} \over V_{\rm tot}^{phys}}.
\end{eqnarray}

\begin{figure}[bt]
\epsfxsize = 11 cm
\centering \leavevmode
\epsfbox{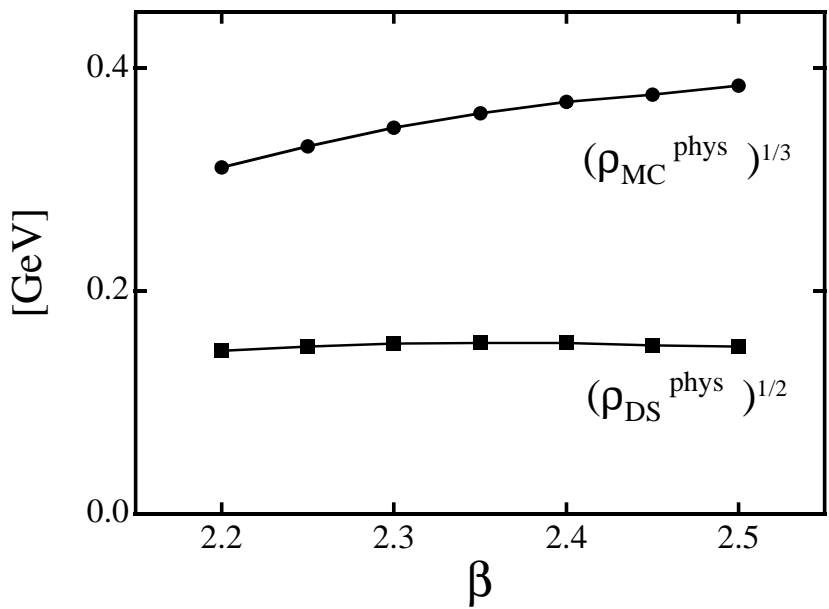}
\vspace{0cm}
\caption{  
The monopole-current density  $ ( \rho^{phys}_{\rm MC} )^\frac13 $
and the Dirac-sheet density
$ ( \rho_{\rm DS}^{phys} )^\frac12 $
as the function of $\beta$ in the MA gauge on $16^4$ lattice.
The area of the Dirac sheets is $\beta$-independent, while
the monopole-current length slightly increases with $\beta$.
}
\label{gfig16}
\vspace{0cm}
\end{figure}

If the monopole-current length $L_{\rm tot}^{phys}$ 
and the Dirac-sheet area $S_{\rm tot}^{phys}$ 
are physical values in the continuum limit, 
$\rho _{\rm MC}^{phys}$ and $\rho_{\rm DS}^{phys}$ 
are to be almost $\beta $-independent as long as the mesh $a$ 
is small enough. 
We show in Fig.\ref{gfig16} the $\beta $-dependence of 
$\rho _{\rm MC}^{phys}$ and $\rho_{\rm DS}^{phys}$ 
in the MA gauge with the U(1)$_3$ Landau gauge. 
As for the monopole density $\rho _{\rm MC}^{phys}$, 
there is a $\beta $-dependence to some extent. 
On the other hand, the Dirac-sheet density $\rho_{\rm DS}^{phys}$ 
is almost $\beta $-independent, and exhibit a clear scaling property. 

Thus, $n_{\mu \nu }^{phys}(s) \equiv n_{\mu \nu }(s)/a^2$ 
is expected to be almost $\beta $-independent on the average over the 
whole system. 
Near the continuum limit, we can then expect 
$a$-independence of $eA_\mu^{\rm Mo}$ as 
\begin{eqnarray}
eA_\mu ^{\rm Mo}(s)
= \frac2a  \theta _\mu (s)^{\rm Mo} 
= \frac{2}{a} \sum_{s'} D^{lat}(s-s') 
\partial_\alpha ^{lat} n_{\alpha \mu }(s')
\simeq 
\int d^4x' D(s-x') \partial_\alpha  n_{\alpha \mu }^{phys}(x'), 
\nonumber 
\end{eqnarray}
where the last expression is to be $a$-independent.
Thus, the good scaling property of the Dirac-sheet density is 
considered as the origin of the approximate $\beta $-independence of 
the abelian continuum variables 
$eA_\mu ^{\rm Mo}$ and $eF_{\mu \nu }^{\rm Mo}$ in the monopole sector. 

It is noted that the Dirac-sheet area $S_{\rm tot}^{phys}$ 
seems to have a continuum limit 
in a fixed 4-dimensional physical volume $V_{\rm tot}^{phys}$, 
although the monopole-current length $L_{\rm tot}^{phys}$ 
tends to increase as the mesh $a$ decreases. 
Here, let us consider the reason of the good scaling for 
the Dirac-sheet area in the U(1)$_3$ Landau gauge. 
In the U(1)$_3$ Landau gauge, the U(1)$_3$ gauge variable $\theta^3 _\mu 
(s)$ is 
forced to be mostly continuous as $\theta^3 _\mu  \simeq 0$ on the 
lattice, which suppresses 
the large $\theta _{\mu\nu }$ fluctuation 
which generates $n_{\mu \nu }\ne 0$ as was 
 shown in Fig.\ref{gfig5b}. 
Thus, the generation of the Dirac sheet is strongly 
suppressed in the U(1)$_3$ Landau gauge, although 
the Dirac sheet induced by the monopole current 
must appear.
Geometrically, 
the area of the Dirac sheet is minimized in the U(1)$_3$ Landau gauge, 
within the constraint that the boundary of the Dirac sheet is 
fixed as the monopole current.
This situation resembles the ``soap bubble'' with the boundary fixed.
Thus, the Dirac-sheet area is insensitive to the small fluctuation of 
its boundary or the monopole current due to the 
minimizing condition of the area in the U(1)$_3$ Landau gauge.

On the lattice, as the mesh $a$ becomes small, 
the monopole current 
may be fractal, and the length of the monopole current 
may become large with small $a$. 
Nevertheless, the Dirac-sheet area remains almost unchanged 
in the U(1)$_3$ Landau gauge owing to the minimizing condition of the 
area, 
even when its boundary or the monopole current becomes fractal. 
This would be the advanced point of the use of 
the Dirac sheet in the U(1)$_3$ Landau gauge 
for the scaling property rather than the monopole current.

\section{Estimation of the Wilson Loop from the Monopole}

\begin{figure}[bt]
\centering \leavevmode
{\tt    \setlength{\unitlength}{0.92pt}
\begin{picture}(377,200)
\thinlines    \put(36,120){\vector(-1,-2){0}}
              \put(120,134){\vector(-1,0){0}}
              \put(148,95){\vector(1,2){0}}
              \put(50,80){\vector(1,0){0}}
              \put(276,176){\line(0,-1){142}}
              \put(9,25){\begin{picture}(223,338)
                  \put(24,26){\begin{picture}(120,120)
\thinlines            \put(60,60){\vector(1,2){28}}
                      \put(60,60){\circle*{12}}
                      \put(60,60){\vector(-1,2){28}}
                      \put(60,60){\vector(-2,1){52}}
                      \put(60,60){\vector(-1,-2){28}}
                      \put(60,60){\vector(-2,-1){52}}
                      \put(60,60){\vector(1,-2){28}}
                      \put(60,60){\vector(2,-1){52}}
                      \put(60,60){\vector(2,1){52}}
                      \put(60,60){\vector(0,-1){60}}
                      \put(60,60){\vector(0,1){60}}
                      \put(60,60){\vector(-1,0){60}}
                      \put(60,60){\vector(1,0){60}}
                      \end{picture}}
\thinlines        \put(277,175){\line(-1,1){1}}
                  \end{picture}}
              \put(10,78){\begin{picture}(164,58)
\thinlines        \put(162,57){\line(-3,-5){33}}
                  \put(3,2){\line(1,0){126}}
                  \put(35,57){\line(-3,-5){33}}
                  \put(35,56){\line(1,0){126}}
                  \end{picture}}
\thinlines    \put(365,132){\line(-3,-5){33}}
              \put(206,77){\line(1,0){126}}
              \put(238,132){\line(-3,-5){33}}
              \put(238,131){\line(1,0){126}}
              \put(344,97){\vector(2,3){0}}
              \put(226,112){\vector(-2,-3){0}}
              \put(316,131){\vector(-1,0){0}}
              \put(252,76){\vector(1,0){0}}
              \put(270,176){\line(0,-1){142}}
              \put(282,176){\line(0,-1){142}}
              \put(277,17){\circle{14}}
              \put(276,189){\circle*{14}}
              \put(227,43){\circle*{0}}
\end{picture}}
\caption{   
(a) The monopole in the plaquette.
In the lattice unite, the
 magnetic field surrounding the monopole is $\theta^{\rm total}_{\mu\nu} = 2 \pi$, 
and therefore the magnetic field penetrating the north or south half 
sphere is
$\theta_{\mu\nu} = \pi$.  
The contribution of the monopole to the Wilson loop is
$e^{ia^2 e f_{\mu\nu} \frac{\tau_3}{2}} =e^{i\theta_{\mu\nu} \tau_3}
= e^{i\pi \tau_3} =-1$.
(b) The magnetic flux, which has positive and negative charges at the ends, 
penetrated 
in the plaquette.
The quantization condition of the flux requires to  $\theta_{\mu\nu} = 2 \pi n$ 
with $n \in {\bf Z}$.
The contribution of this flux to the Wilson loop is 
$e^{ia^2 e f_{\mu\nu} \frac{\tau_3}{2}} =e^{i\theta_{\mu\nu} \tau_3}
= e^{i2\pi n \tau_3} =1$.
}
\label{gfig19}
\end{figure}
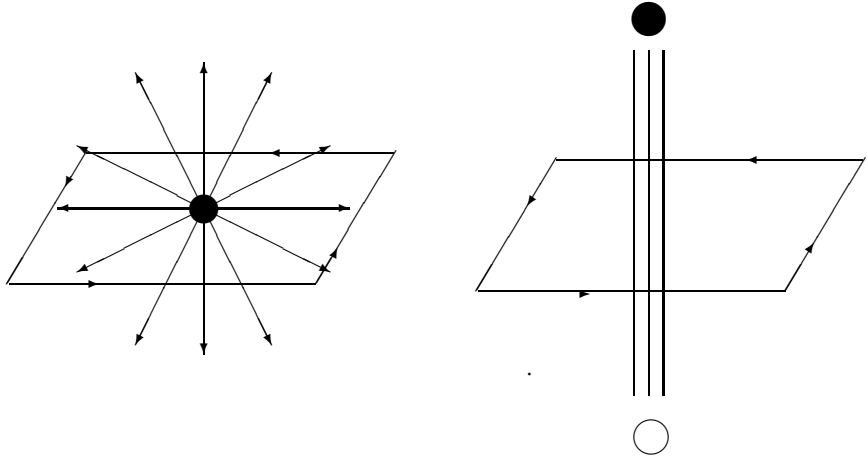

Up to now, we have found that the monopoles appearing in the MA gauge
brings the large fluctuation of the abelian field variable around 
them, which would be the origin of the string tension.
In this section, we consider the relation between the monopole density
$\rho_{\rm Mo}$ and the string tension $\sigma$ in order to clarify 
the role of monopole for the confinement  force in the analytical 
manner.
Here, we idealizes the large field fluctuation created by the 
monopole, and pay attention to the area-law reduction of 
the Wilson loop $\langle W \rangle$.
For the system with the monopole density $\rho_{\rm Mo}$, 
one (anti-) monopole is expected to occupy the area of the cube with the 
length $d_{\rm Mo} \equiv \rho_{\rm 
Mo}^{-1/3}$,
which would be a typical scale of the confinement physics \cite{ichie3}.
Here, $r_{\rm Mo} \equiv \frac{d_{\rm Mo}}{2}$ corresponds to the size of 
the QCD monopole, and is estimated as $r_{\rm Mo} \equiv 0.25 \sim0.3$fm
from the lattice QCD with $\beta=2.4$ as shown in Fig.\ref{gfig11}.
As an important feature, the unit-charge monopole provides a minus factor 
to the Wilson.
Let us consider the static monopole with the unite charge and the
``spatial'' Wilson loop surrounding the monopole as shown in 
Fig.\ref{gfig19} (a).
Taking the plane sheet $S$ as the surface area inside the loop, one 
can use the simple estimation of the monopole effect
as  a ``minus factor'', 
\begin{eqnarray}
{\rm exp} \{ i e\oint ds_\mu A_\mu^3 \frac{\tau^3}{2} \} 
= 
{\rm exp} \{ i e \! \int  \!\!\! \int \!\! d \sigma_{\mu\nu}
F_{\mu\nu}^3 \frac{\tau_3}{2} \} 
= {\rm exp} \{ i2\pi \!\! \times \!\!\frac{\tau^3}{2} \} = 
{\rm exp} \{  i\pi \tau^3 \} = -1,
\nonumber 
\end{eqnarray}
which corresponds to the  {\it half} of the quantized magnetic flux 
between the monopole and anti-monopole as shown in Fig.\ref{gfig19}(b).
In general, unit-charge monopole current which perpendicularly penetrates 
to the Wilson loop $W$ provides the ``minus factor'' to $W$.

On the other hand, the effect of a monopole apart from the Wilson  loop 
is expected largely canceled by the nearest anti-monopole, and is not expected to 
provide a large contribution to the Wilson loop in the dense monopole 
system like the AP-QCD vacuum.
Such a screening effect is also suggested by recent lattice simulations 
on the inter-monopole potential \cite{atanaka}.
Here, we try to estimate the contribution of monopoles to the Wilson 
loop of the $R \times T$ rectangular, where $R$  and $T$ are physical 
lengths. Defining the plane area $S$ inside the loop,
we expect that only nearest monopoles to  $S$ can 
contribute to  the Wilson loop without the screening effect by other 
monopoles.
Then, only monopoles inside $R \times T\times d_{\rm Mo}$ plate area near 
$S$ are provided the  minus factor to the Wilson loop.
The total monopole number in the  $R \times T\times d_{\rm Mo}$ plate
area is estimated as
\begin{eqnarray}
N_{\rm Mo}  \equiv (R \times T\times d_{\rm Mo}) \times \rho_{\rm Mo}
= \frac{ R \times T}{d_{\rm Mo}^2}.
\end{eqnarray}

Introducing the lattice with a fine mesh $a \ll  d_{\rm Mo}$, we try
to estimate the Wilson loop as the function of $d_{\rm Mo}$.
Since the lattice plaquette number on $S$ is $N^{lat} \equiv
\frac{R \times T}{a^2}$,
 we divide the plate area $V$ into $N^{lat}$
``cells'' as $a \times a \times d_{\rm Mo}$ boxes.
When $a$ is taken small enough, each cell includes at most one (anti-)monopole, 
and the existence probability $q$ of the monopole inside one cell is
obtained by $ q = \frac{N_{\rm Mo}}{N^{lat}} = \frac{a^2}{ d^2_{\rm Mo}}
 \ll 1$.
 Next, we apply  the statistical consideration for the estimation of 
 the Wilson loop $\langle W \rangle$. 
 The probability that $n$ cells include the monopole among $N^{lat}$
 cells is simply given ${}_{N^{lat}}C_n q^n (1-q)^{N^{lat} -n}$
 in a statistical source.
Since $n$ monopoles provide $(-)^n$ factor to the Wilson loop $W$,
the expectation value $\langle W \rangle$ is estimated as
\begin{eqnarray}
\langle W \rangle & = & 
\frac{ \displaystyle
\sum_{n=0}^{N^{lat}} {}_{N^{lat}}C_n q^n (1-q)^{N^{lat} -n} 
\times (-)^n 
}{\displaystyle
\sum_{n=0}^{N^{lat}} {}_{N^{lat}}C_n q^n (1-q)^{N^{lat} -n} 
} 
 =  (1-2q)^{N^{lat}} \nonumber \\
&  =  & {\rm exp} \{ N^{lat} {\rm ln}(1-2q)  \} =
{\rm exp} \{ {\frac{R \times T}{a^2} {\rm ln}(1-2q) } \}.
\end{eqnarray}
Therefore, the Wilson loop obeys the area law, and the string tension 
$\sigma$
is obtained as 
\begin{eqnarray}
\sigma & = & - \frac{1}{R \times T} {\rm ln} \langle W \rangle 
= -\frac{1}{a^2} {\rm ln} (1-2q)  
  =  -\frac{1}{a^2} {\rm ln} (1-2\frac{a^2}{d_{\rm Mo}^2}) 
\simeq \frac{2}{d_{\rm Mo}^2},
\end{eqnarray}
 because of $q = \frac{a^2}{d_{\rm Mo}^2} \ll 1$.
 Thus, we derive a simple relation between the string tension $\sigma$
 and the inter-monopole distance $d_{\rm Mo}$, 
\begin{eqnarray}
\sqrt{\sigma} \simeq \sqrt{2} \: d^{-1}_{\rm Mo}. 
\end{eqnarray}
As shown in Fig.\ref{gfig16}, we have found
 $d_{\rm Mo} \equiv \rho^{-1/3}_{\rm Mo} =$ 
0.5 $\sim$ 0.6 fm from
the lattice
result with $\beta = 2.2 \sim 2.5.$
Then, the string tension is evaluated as 
$\sqrt{\sigma} \simeq
\sqrt{2} d^{-1}_{\rm Mo} = (0.42 \simeq 0.57)$ GeV from the monopole 
density  $\rho$  using this estimation. 





\chapter{Monopole Current Dynamics}
\label{sec:MCD}

In the infrared region, color confinement can be understood by the 
dual Higgs theory with monopole condensation.
The string tension is almost reproduced by the monopole sector of the 
abelian gluon component, 
which is called as monopole dominance for the color confinement force.
Thus, the monopole degrees of freedom would be the key variable for 
the confinement physics.
In the dual Higgs theory, the monopole is assumed to be condensed, 
which has been suggested by the formation of global network of the 
monopole current in the lattice QCD.
In this chapter, we examine the realization of monopole condensation 
in the QCD-vacuum at the infrared scale.
However, QCD is described by the gluon field not by the monopole 
current, and therefore it is difficult to clarify  monopole 
condensation only with the lattice QCD simulation.
To this end, we generate the monopole-current system on the lattice 
using a simple monopole current action and study the role of the 
monopole current to the color confinement\cite{ichie2}.
Comparing the QCD vacuum with the monopole current system, we try to 
clarify the monopole condensation occurred in the non-perturbative QCD.

\section{Monopole Current Dynamics and Kosterlitz-Thouless-type Transition}
\label{sec:MCDKTTT}

To begin with, we consider the monopole-current action $S[k_\mu(s)]$
which simulates the monopole-current system observed in the lattice QCD
in the MA gauge. In general, $S[k_\mu(s)]$ includes the nonlocal
Coulomb interaction as $S_C=\int d^4x d^4y k_\mu(x) D(x-y) k_\mu(y)$
with the Coulomb propagator $D(x)$. In the dual Higgs phase, however,
the effective interaction between the monopole currents would be
short-range due to the {\it screening effect} by the dual Higgs mechanism
\cite{ezawa,atanaka} similar to the Debye screening
\cite{stack2}.
Then, the partition functional of the monopole current
$k_\mu(x) \equiv k_\mu^3(x) \cdot \tau^3/2$ would be written as
\begin{equation}
Z = \int {\it D}k_\mu {\rm exp} \{ -\alpha \int_a d^4x  {\rm tr}  k^2_\mu (x) \}
\delta(\partial^\mu k_\mu),
\label{eq:part}
\end{equation}
where $\alpha$ is the energy per unit length of the monopole current.
Here, the $\delta$-function is necessary to
ensure the current conservation $\partial^\mu k_\mu=0$,
and $a$ is an ultraviolet cutoff larger than the screening length.
To perform the path-integral (\ref{eq:part}), 
we put the system on the 4-dimensional lattice with the lattice spacing 
$a$.
In the lattice formalism, the monopole currents are defined on the dual 
lattice, 
\begin{equation}
k^{lat}_\mu (s) \equiv e /(4\pi) \cdot a^3 k^3_\mu(s).
\end{equation}
The partition function is given as
\begin{equation}
Z = \sum_{k^{lat}_\mu} {\rm exp} \{ -\alpha^{lat} 
\sum_s  k_\mu^{lat}(s) ^2 \} 
\delta (\partial_\mu k_\mu^{lat}(s)),
\label{eq:monoaction}
\end{equation}
where $\alpha^{lat} \equiv \alpha / 2 \cdot (4\pi/ea)^2 $.

\begin{figure}[p]
\vspace{-1cm}
\epsfxsize = 12 cm
\centering \leavevmode
\epsfbox{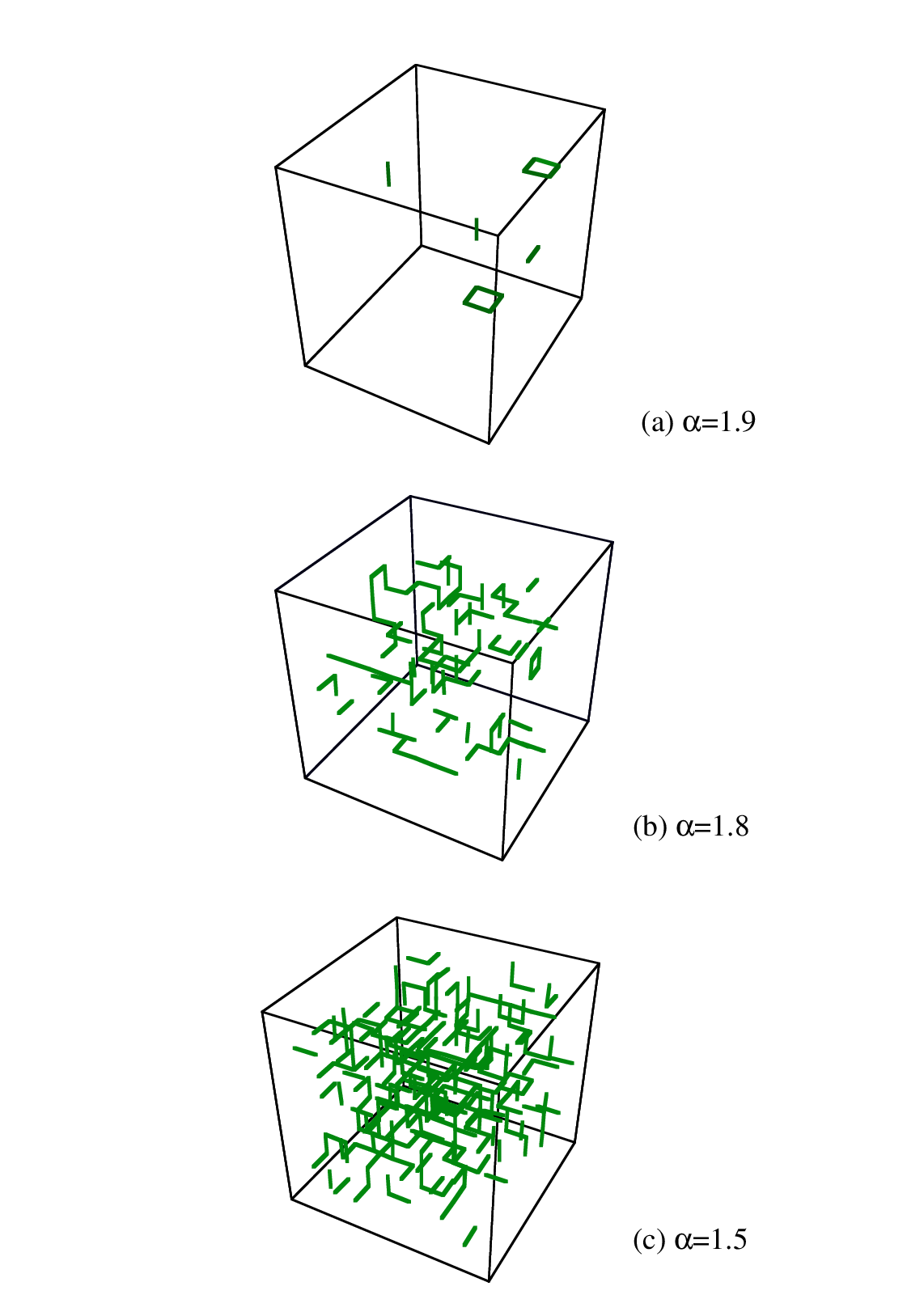}
\vspace{0cm}
\caption{  
The monopole current in the 
monopole-current system in ${\bf R}^3$ at a fixed time
:
(a) the ``critical phase'' ($\alpha=1.9 \simeq \ln 7$),
(b) the ``monopole condensed phase'' near the critical point ($\alpha=1.8 $)
and 
(c) the ``monopole condensed phase'' ($\alpha=1.5$).
}
\label{fig:mcfig1}
\vspace{0cm}
\end{figure}

\begin{figure}[htbp]
\epsfxsize = 7 cm
\centering \leavevmode
\epsfbox{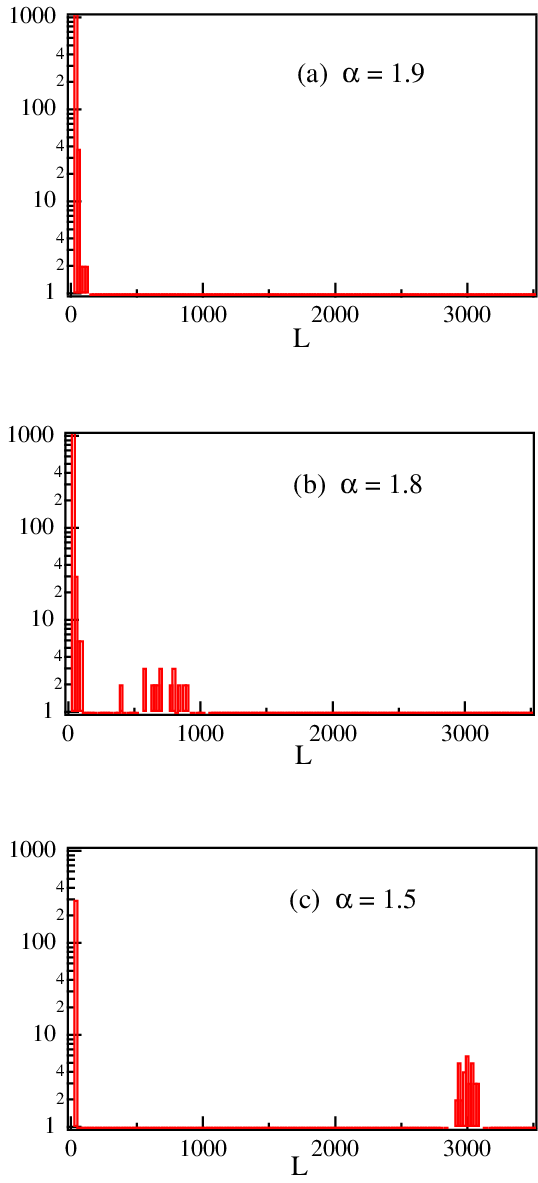}
\vspace{0cm}
\caption{    
The histograms of the monopole-current length $L$
of each monopole cluster in the monopole-current system.
The data at each $\alpha$ are taken 
from 40 current configurations.
(a) In the ``critical phase'', 
only short monopole loops appear.
(b) In the `` monopole condensed phase'' near the critical point,
monopole currents become in a cluster.
(c) In the ``confinement phase'', 
there appears one large monopole cluster in each current configuration.}
\label{fig:mcfig2}
\vspace{0cm}
\end{figure}

The lattice QCD simulation shows that 
one long monopole loop and many short loops appear in 
the confinement phase \cite{brandstater}.
Only the long loop becomes important 
for the properties of the QCD vacuum,
while many short loops are originated 
from the fluctuation in the ultraviolet region and therefore 
can be neglected. 
In this system, the partition function can be approximated 
as the single monopole loop ensemble with the length $L$, 
\begin{equation}
Z=\sum_{L}\rho(L)e^{-\alpha L},
\end{equation}
where $L$ and $\rho(L)$ are length of the monopole loop and its 
configuration number, respectively. 
The monopole current with length $L$ is regarded as the $L$ step
self-avoiding random walk, where $2D-1=7$ direction is possible 
in each step in the $D$ dimensional space-time.
Therefore, $\rho(L)$ is roughly estimated as $(2D-1)^L=7^L$.
Using this partition function, the expectation value of the monopole 
current length is found to be 
\begin{equation}  
\langle L \rangle =\frac{1}{Z}\sum_{L}\rho(L) L e^{-\alpha L} =
  \left\{ \begin{array}{ll}
  \{ \alpha-\ln(2D-1) \}^{-1}  & \mbox{if $\alpha > \ln(2D-1)$} \\
  \infty & \mbox{if $\alpha < \ln(2D-1)$}.
       \end{array}
  \right. 
\end{equation}
When the energy $\alpha$ is larger 
than the entropy $\ln(2D-1)$, the monopole loop length is finite. 
However, when $\alpha$ is smaller than the entropy, 
the monopole loop length 
becomes infinite, which corresponds to monopole condensation 
in the current representation \cite{ezawa}.
Here, the critical value on 
monopole condensation $\alpha_{c} \simeq {\rm ln}(2D-1) \simeq {\rm 
ln}7 = 1.945$,
which corresponds to the ``entropy'' of the self-avoiding random 
walk. Such a transition is quite similar to the Kosterlitz-Thouless
type transition in (1+2)-dimensional superconductor \cite{kosterlitz},
where vortex condensation plays an important role to the transition.

\begin{figure}[tb]
\epsfxsize = 10 cm
\centering \leavevmode
\epsfbox{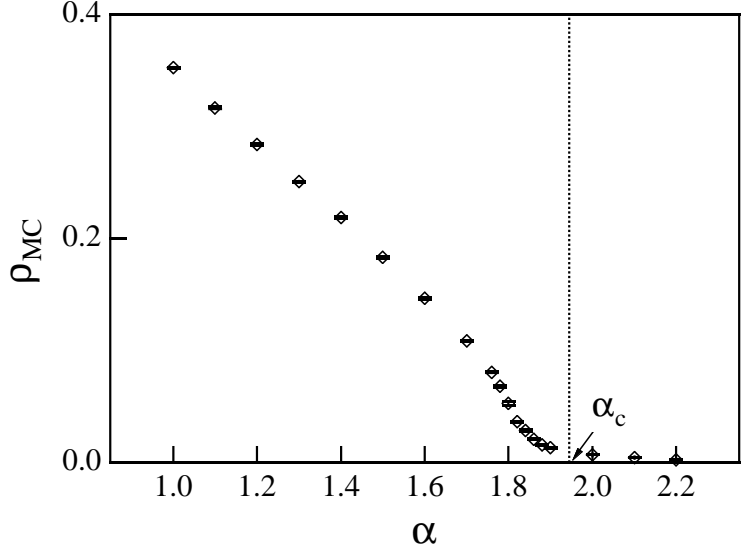}
\vspace{0cm}
\caption{    
The monopole density $\rho_{MC}$  as the function of $\alpha$.
The monopole density becomes 
small for large $\alpha$ and almost vanishes for $\alpha > 
\alpha_c  =
 {\rm ln}7  = 1.945$.
}
\label{fig:mcfig3}
\vspace{0cm}
\end{figure}
\begin{figure}[tb]
\epsfxsize = 10 cm
\centering \leavevmode
\epsfbox{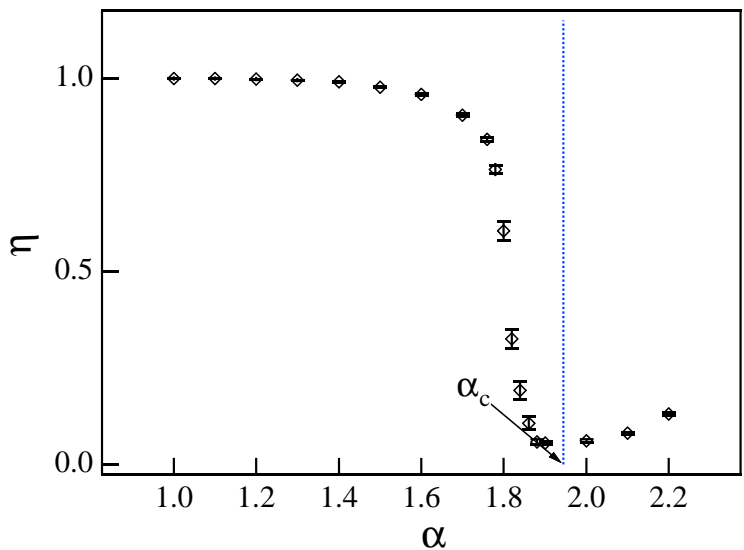}
\vspace{0cm}
\caption{   
The clustering parameter $\eta \equiv \sum_i L_i^2/(\sum_i L_i)^2$
as the function of $\alpha$. Near the critical point
$\alpha_c = \ln 7 = 1.945$, $\eta$ is drastically changed from unity to zero.
}
\label{fig:mcfig4}
\vspace{0cm}
\end{figure}

In performing the simulation,
we construct the monopole current
as the sum of  plaquettes of the monopole current
because of the current conservation condition, $\partial_\mu k_{\mu}$ = 0.
As the initial current configuration, we prepare a random monopole 
current system (hot start) or no monopole current system (cold start). 
Then, we update the {\it link} of the monopole current using the Metropolis 
method \cite{rothe}.

We generate the monopole current system 
using the monopole current action \ref{eq:monoaction}.  
Fig.\ref{fig:mcfig1} shows the monopole current 
on $8^4$ lattices for the typical cases,
($\alpha = 1.5,1.8,1.9$) at a fixed time. 
For these cases,  we show in Fig.\ref{fig:mcfig2}
the histograms of  monopole loop length \cite{bode}
 of each monopole cluster with 40 current configurations.
For large $\alpha$, only 
small loops  appear
and monopole density is small.
On the other hand, for small $\alpha$, 
there appears a global network of one large monopole cluster, 
and 
the monopole currents are complicated and dense.
The lattice monopole density
\begin{eqnarray} 
\rho_{MC} \equiv \frac{1}{4V}\sum_{s,\mu} |k_\mu(s)|
\end{eqnarray}
and the clustering parameter 
\begin{eqnarray}
\eta \equiv \frac{\sum_i L^2_i}{(\sum_i L_i)^2}
\end{eqnarray}
\cite{sasakidoc} are shown in Fig.\ref{fig:mcfig3}. and Fig.\ref{fig:mcfig4}, respectively.
Here, $L_i$ is the loop length of the $i$-th monopole  cluster.
For the extreme limit of $\eta = 1$, all the monopole loops
are linked in a cluster, while many monopole loops are isolated for 
small $\eta$.
As $\alpha$ decreases, the monopole density $\rho_{MC}$ becomes
larger gradually for  
$\alpha 
\stackrel{<}{\scriptstyle{\sim}} \alpha_c$.
However, the clustering parameter is drastically changed  
at $\alpha = 1.8$  closed to ${\rm ln7}$.
Thus, these monopole current simulations clearly show the 
Kosterlitz-Thouless-type transition around $\alpha_{c} = {\rm ln} 7$
in agreement with the theoretical consideration on monopole (-current)
condensation.

\section{Role of Monopoles for Confinement}
\label{sec:RMC}
In this section, we study how these monopole currents 
contribute to the color confinement properties.
Quark confinement is characterized by the linear inter-quark potential,
which can be obtained from the area-law behavior of the Wilson 
loop,
$\langle W \rangle 
= \langle P {\rm exp}({ie\oint A_\mu dx_\mu}) \rangle$.
Therefore, it is desired to extract the gauge variable from the monopole 
current $k_\mu$.
We now derive the abelian gauge variable in stead of the non-abelian gauge 
field, because the lattice QCD results show that
the color confinement phenomena can be discussed only with abelian part 
to some extent.
However,
in the presence of the magnetic monopoles, the ordinary abelian gauge 
field $A_\mu(x)$  inevitably includes the singularity as the Dirac string, 
which leads to some difficulties in the field theoretical treatment.
Instead, the dual field formalism is much useful to describe the monopole
current system, because the dual gauge field $B_\mu(x)$ can be introduced 
without the singularity for such a system with $k_\mu \neq 0$ and $j_\mu=0$.
This is the dual version of the ordinary gauge theory with $A_\mu(x)$
for the QED system with $j_\mu \neq 0$ and $k_\mu=0$.

\begin{figure}[tb]
\epsfxsize = 10 cm
\centering \leavevmode
\epsfbox{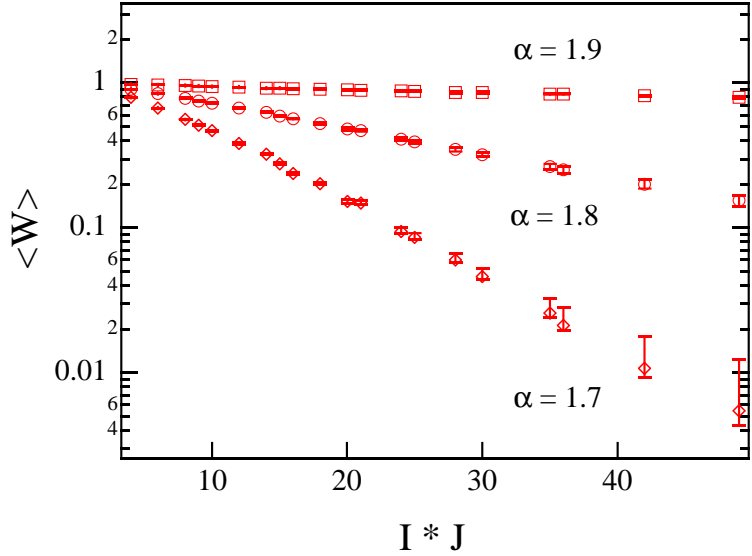}
\vspace{0cm}
\caption{    
The expectation value of the Wilson loop
$\langle W (I \times J) \rangle$ on the fine mesh $c=a/2$
for $\alpha=1.7,1.8,1.9$. The area-law behavior of
$\langle W \rangle$ indicates the linear inter-quark potential,
although the string tension is zero for $\alpha=1.9> \alpha_c$.}
\label{fig:mcfig5}
\vspace{0cm}
\end{figure}

Now, we apply the dual field formalism to the monopole current 
system discussed in Section \ref{sec:MCDKTTT}. 
Since the monopole current $k^{lat}_\mu(s)$ is generated 
on the lattice with the mesh $a$, 
the continuous dual field $B_\mu(x)$ is derived from 
$k_\mu(x)$ 
using Eq.(\ref{eq:dual}), in principle. 
In estimating the integral in Eq.(\ref{eq:dual}) numerically, we use 
a fine lattice with a small mesh $c$.
{\it To extract} $B_\mu(x)$ {\it correctly, the mesh }$c$
{\it is to be taken enough small}.  
However, too fine mesh is not necessary because 
the original current $k^{lat}_\mu(s)$ includes the error 
in the order of $a$.
Numerical analyses show that the use of $c \simeq a$ is too 
crude for the correct estimation of the integral in Eq.(\ref{eq:dual}).
Instead, the calculation with $c \le a/2$ is good enough for the 
estimation of $B_\mu(x)$, and hence we take $c=a/2$ hereafter.

\begin{figure}[tb]
\epsfxsize = 10 cm
\centering \leavevmode
\epsfbox{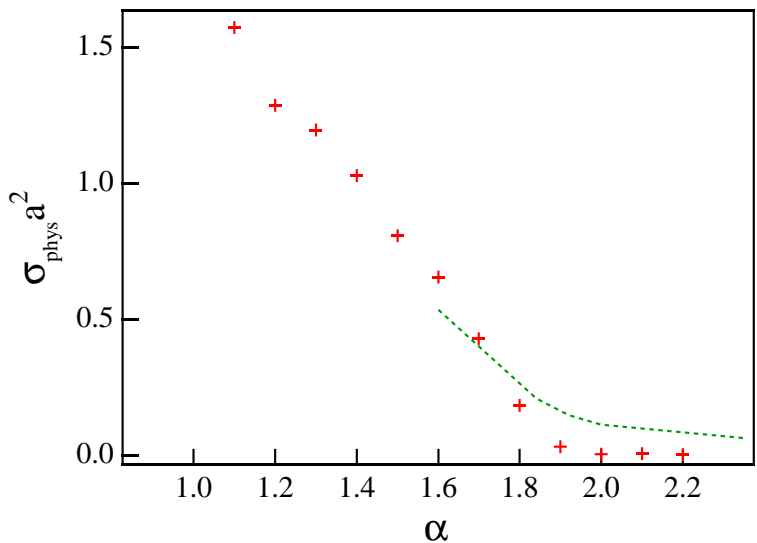}
\vspace{0cm}
\caption{    
 The Creutz ratio as the function of $\alpha$ in the monopole current system.
The dotted line denotes the Creutz ratio as the function of $\beta = 
1.25\alpha$
in the lattice QCD.
}
\label{fig:mcfig6}
\vspace{0cm}
\end{figure}

The expectation value of the Wilson loop $\langle W \rangle$ is shown in
Fig.\ref{fig:mcfig5}.
The Wilson loop exhibits the area-law and the linear confinement 
potential:
${\rm ln} \langle W \rangle$ decreases linearly with the inter-quark 
distance, where its slope corresponds to the string tension.
Quantitatively, the string tension is measured by the Creutz ratio,
and we show in Fig.\ref{fig:mcfig6} $\chi(3,3)$ as a typical example.
For the monopole condensed phase as $\alpha < \alpha_c$, the string 
tension gets a finite value, while it vanishes for the non-condensed 
phase of monopole as $\alpha \le \alpha_c$.
Thus, the confinement phase directly corresponds to the monopole 
condensed phase and therefore monopole condensation is essential as 
 essence of the confinement mechanism.

\section{Monopole Condensation in the QCD Vacuum}

\begin{figure}[h]
\vspace{-1.5cm}
\epsfxsize = 15 cm
\centering \leavevmode
\epsfbox{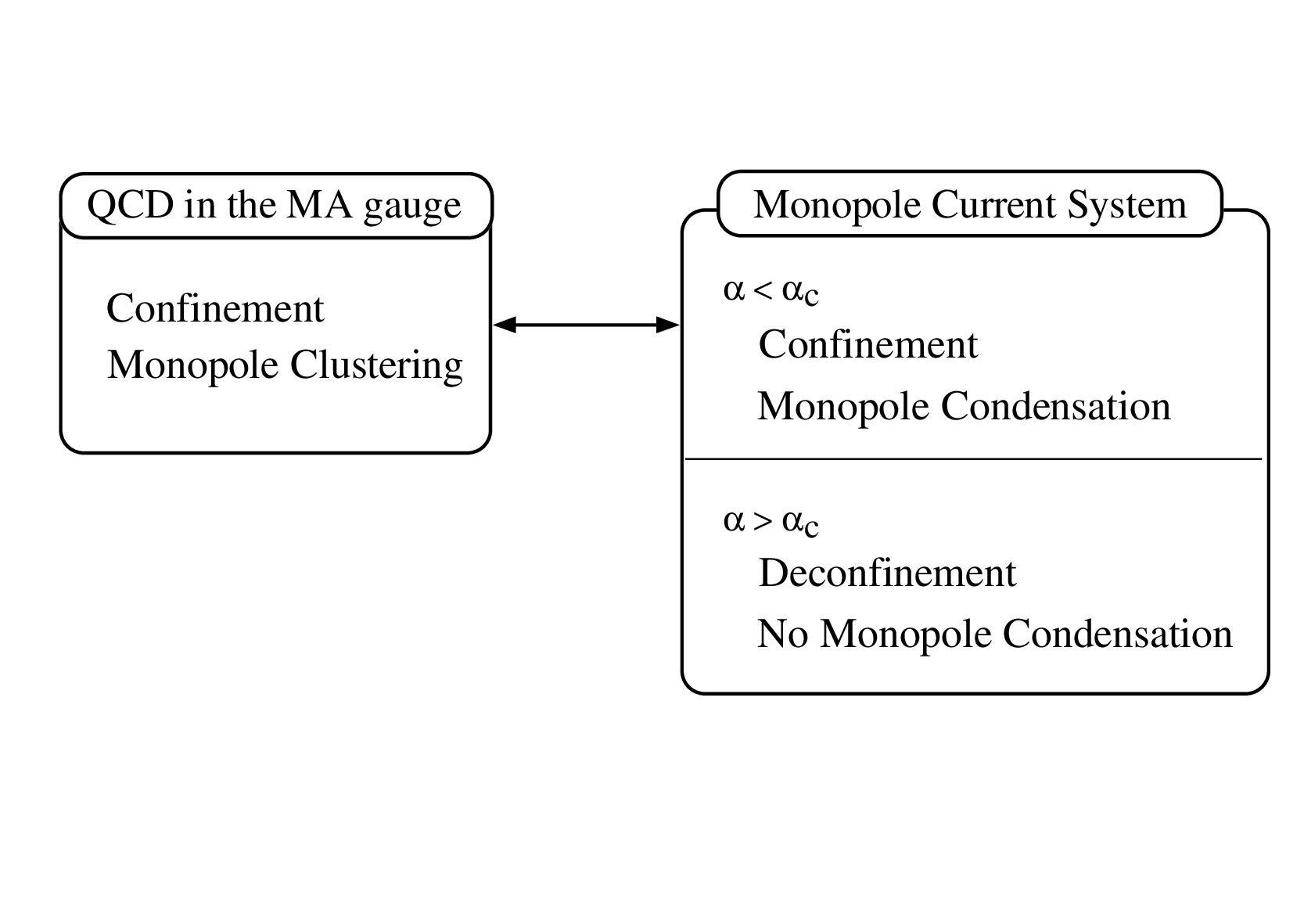}
\vspace{-2.5cm}
\vspace{0cm}
\caption{    
Comparison between QCD in the MA gauge and the monopole current system.
The QCD vacuum corresponds to the monopole-condensed phase in the monopole
current system because of the similarity on confinement and the 
monopole clustering.
}
\label{fig:mcfig8}
\end{figure}

In this section, we compare the lattice QCD with the monopole-current 
system in terms of  monopole condensation and confinement properties,
as shown in Fig.\ref{fig:mcfig8}.
The lattice QCD simulation shows that the QCD vacuum in the MA gauge 
holds the global network of monopole current as shown in 
Section \ref{monolattice}.
Considering the similarity on the monopole clustering, the QCD vacuum 
can be regarded as the monopole-condensed phase with $\alpha < \alpha_{c}$
in the monopole current system, as shown in Fig.\ref{fig:mcfig1}(c).
Such identification of the QCD vacuum with the monopole-condensed 
phase is also suggested 
in terms of the confinement properties, because the confinement phase 
corresponds to the monopole condensed phase as shown in Section \ref{sec:RMC}.

\vspace{1cm}

\section{Comparison with Vortex Condensation
 in 1+2 Superconductor}

In this section, we summarize 
in Table 7.1 the correspondence between 
monopole condensation in the QCD vacuum and vortex condensation in 
the 1+2 dimensional superconductor \cite{kosterlitz}
in terms of the topological object and the phase transition.
For these two systems, condensation of the line-like topological object 
plays the relevant role for the determination of the phase.
The phase structure is controlled by the balance of power 
between ``entropy'' (configuration number of the topological object)
and ``energy'' (self-energy of the topological object).

In the 1+2 dimensional superconductor, the Abrikosov vortex is the 
important topological object.
While the vortex scarcely appears at the low $T$,
such topological excitations frequently occur at the high $T$, which 
enhances the entropy factor in the free energy.
Then, at the critical temperature, vortex condensation occurs due to 
``entropy dominance'' and the system goes to the normal phase.
This phase transition is known as the Kosterlitz-Thouless transition.

Similarly in the QCD vacuum, the monopole current plays the relevant role 
as the line-like topological object in the 4-dimensional space-time.
In the monopole-current system, the control parameter is the self-energy 
$\alpha$ of the monopole current.
While only the local fluctuation of the monopole current appears for 
$\alpha > \alpha_c \simeq$ ln7, monopole condensation occurs for $\alpha < 
\alpha_c$ as the result of entropy dominance on the monopole-current 
configuration.
Moreover, this monopole condensation leads to the electric confinement.

\vspace{1cm}

\begin{tabular}{|c||c|c|} \hline
System 
& 
\begin{tabular}{c}
1+2 dim  \\
Superconductor
\end{tabular}
&
\begin{tabular}{c}
 4-dim  \\
 Monopole  system
\end{tabular}
\\  
\hline  \hline
\begin{tabular}{c}
Line-like  \\
topological object
\end{tabular}
 & Vortex & Monopole current \\ 
\hline  \hline
Control parameter & Temperature $T$ & Monopole self-energy $\alpha$ \\ 
  \hline \hline
\begin{tabular}{c}
Condensed phase  \\
(Entropy dominance) \\
\end{tabular}
&
\begin{tabular}{c}
$High$ $temperature$ \\
 Normal phase  \\
$\uparrow$  \\ 
 Vortex condensation \\ 
\end{tabular}
&
\begin{tabular}{c}
$ Small$ $\alpha$ \\
  Confinement phase  \\
$\uparrow$ \\
Monopole condensation \\
\end{tabular} \\
\hline 
\begin{tabular}{c}
Non-condensed phase \\
(Energy dominance) 
\end{tabular}
&
\begin{tabular}{c}
$Low$ $temperature$ \\
 Superconducting phase  
 \end{tabular}
&
\begin{tabular}{c}
$Large$ $\alpha$ \\
 Deconfinement phase 
\end{tabular} \\
\hline 
\end{tabular} 
\vspace{0.5cm}

\noindent {\small { Table 7.1} Correspondence 
between the monopole current system and the 1+2 dimensional superconductor}

\section{Monopole Size and Critical Scale in QCD}

Up to now, we have argued about the correspondence between the lattice 
QCD and the monopole current system described by (\ref{eq:part}) in terms of 
the qualitative aspects as monopole condensation and confinement.
In the final section of this chapter, we attempt further consideration 
on the correspondence between the coupling constants, $\alpha$ 
and $\beta \equiv \frac{2N_{c}}{e^{2}}$.
Here, we note also the limitation of the simple ``monopole current 
approach'' in the ultraviolet region and the critical scale of the 
dual superconductor picture of QCD.

To begin with, let us consider one magnetic monopole 
with an intrinsic radius $R$ in the multi-monopole system. 
In the static frame of the monopole, 
it creates a spherical magnetic field, 
%
%
$
{\bf H}(r)  =
  {g(r) \over 4\pi r^3} {\bf r}={{\bf r} \over e(r) r^3} 
$
for $ r \ge R $
and
$
{\bf H}(r)  =  {g(r) \over 4\pi R^3} {\bf r}={{\bf r} \over e(r) R^3} 
$
for $ r \le R $,
where the QCD running gauge coupling $e(r)$ is used to 
include the vacuum polarization effect. 
Here, the effective magnetic-charge distribution is assumed to be 
constant inside the monopole for the simplicity. 

Now, we consider the lattice formalism with a large mesh $a>R$. 
The electromagnetic energy observed on the lattice around 
a monopole is roughly estimated as 
\begin{eqnarray}
M(a) \simeq \int_a^\infty d^3x {1 \over 2}{\bf H}(r)^2 
\simeq {g^2(a) \over 8\pi a}
={2\pi \over e^2(a)a},
\label{eq:massa}
\end{eqnarray}
which is largely changed depending on the lattice mesh $a$.
%
%
This simple estimation neglects the possible reduction of $g(r)$ 
in the infrared region due to the asymptotic freedom of QCD. 
The screening effect of the magnetic field by other monopoles 
also reduces $g(r)$ effectively in a dense monopole system. 
However, $M(a)$ is modified by at most factor 2 
($M(a) = \frac{\pi}{ e^2(a)a }$)
even for the screening case as $g(r)=g(a) \cdot \theta(2a-r)$. 
Then, even in the multi-monopole system, 
$M(a)$ would provide an approximate value for
the electromagnetic energy on lattices 
created by one monopole, 
and we call $M(a)$ as  ``lattice monopole mass''.

For the large mesh $a>R$, the monopole contribution to the 
lattice action reads $S = M(a)a \cdot L$, 
where $L$ denotes the length of the monopole current 
measured in the lattice unit $a$. 
Therefore, $M(a)$ is closely related to the monopole-current coupling 
$\alpha^{lat}$ 
 and $\beta=2N_c/e^2$, 
$
\alpha^{lat} \simeq M(a)a  \simeq {2\pi \over e^2(a)}
={\pi \over 2} \cdot \beta_{\rm SU(2)}.
$
For the above screening case, this relation becomes 
$
\alpha^{lat} \simeq M(a)a  \simeq {\pi \over e^2(a)}
={\pi \over 4} \cdot \beta_{\rm SU(2)},
$
which is consistent with the numerical result, $\alpha = 0.8\beta$,
discussed in Section \ref{sec:RMC}.
Here, as long as the mesh is large as $a>R$, the lattice monopole 
action 
would not need modification by the monopole size effect, 
and the current coupling $\alpha^{lat}$ is proportional to $\beta$.
%
%
%
%
%
%
%
%
%
%
%
Quantum mechanically, 
there is the energy fluctuation about $a^{-1}$ at the scale $a$, 
and therefore monopole excitation occurs very often at the long-distance 
scale satisfying $M(a) 
\stackrel{<}{\scriptstyle{\sim}}
 a^{-1}$. 
Thus, $M(a)a  \simeq \alpha^{lat} 
$ 
is the control parameter for monopole excitation 
at the scale $a(>R)$, 
and we can obtain the quantitative criterion 
for ``monopole condensation'' as $M(a)a \stackrel{<}{\scriptstyle{\sim}}
 {\rm ln}(2D-1)$ 
from the analysis using the current dynamics in Section \ref{sec:MCDKTTT}.  

Second, we discuss the ultraviolet region with $a < R$.
%
In the current dynamics, 
there exists a critical coupling $\alpha_c \simeq {\rm ln} (2D-1)$ 
as shown in Section \ref{sec:MCDKTTT} and \ref{sec:RMC}. 
Above $\alpha_c$, the lattice current action 
provides 
no monopole condensation and no confinement, while 
$\beta \rightarrow \infty $ can be taken in the original QCD 
keeping the confinement property shown in Fig.\ref{fig:mcfig6}. 
%
%
Such a discrepancy between $\beta$ and $\alpha$ 
can be naturally interpreted by introducing the monopole size effect.
Obviously, the monopole-current theory should be drastically 
changed in the ultraviolet region as $a<R$, 
if the QCD-monopole has its peculiar size $R$.

%
%
Here, let us reconsider the relation between $a$ and 
$\alpha^{lat}$ in the lattice current action. 
Similarly in the lattice QCD, the action 
 has no definite 
scale except for the lattice mesh $a (>R)$, and therefore 
the scale unit $a$ would be determined so as to reproduce 
a suitable dimensional variable, 
e.g. the string tension $\sigma \simeq$ 1GeV/fm, 
in the monopole current dynamics. 
For instance, $a$ is determined as a function of $\alpha^{lat}$ 
using the Creutz ratio $\chi \simeq \sigma a^2$ in Fig.\ref{fig:mcfig6}. 
Therefore, $a$ should reach $R$ before realizing 
$\alpha^{lat} \rightarrow \alpha^{lat}_c$, 
and the framework of the current theory is to be largely 
modified due to the monopole size effect for $a<R$. 


In conclusion, the QCD-monopole size $R$ provides a critical scale $R_c$ 
for the description of QCD in terms of the dual Higgs 
mechanism. 
Quantitatively, the difference between the monopole current system
and QCD appear larger than $\beta = 1.25 \alpha$, which corresponds to
$R_c \simeq 0.25$fm.
In the infrared region as $a>R_c$, 
QCD can be approximated as a local monopole-current action \cite{ezawa}, 
and the QCD vacuum can be regarded as the dual superconductor. 
On the other hand, in the ultraviolet region as $a<R_c$, 
the monopole theory becomes nonlocal and complicated 
due to the monopole size effect, and the perturbative QCD 
would be applicable instead. 







\chapter{Dual Ginzburg-Landau Theory}
\label{sec:DGLT}

The lattice QCD simulation shows 
abelian dominance and monopole dominance for the string tension in the 
maximally abelian (MA) gauge \cite{kronfeld,schierholz,hioki}, 
and hence 
the monopole is considered to be the essential degrees of freedom for the 
confinement properties.
In the confinement phase in the MA gauge, there appears  the long and 
complicated monopole current covering the whole lattice space,
which suggests monopole condensation.
Thus, monopole condensation scheme is one of the realistic candidates of 
the physical interpretation for the confinement mechanism.
In this chapter,
we study the confinement phenomena
using the dual Ginzburg-Landau  (DGL) theory \cite{maedan,suganuma}, 
where the confinement mechanism is described by monopole 
condensation \cite{nambu,thoa,mandelstam}.

The DGL theory \cite{maedan,suganuma} is 
the infrared effective theory of QCD
 based on the dual superconductor picture.
In the ordinary Ginzburg-Landau theory,
condensation of the Cooper-pair field leads to exclusion or 
squeezing of the magnetic field in the superconductor. 
Here, the U(1)$_e$-gauge symmetry is spontaneously broken 
and the abelian gauge field, the photon, becomes massive as the result
of the Higgs mechanism.
The Cooper-pair field plays a role of the Higgs field.
In the QCD vacuum, condensation of the ``monopole field''
would provide the exclusion or squeezing of the  color-electric field
through the dual Meissner effect.
In the monopole-condensed system,
 the dual gauge symmetry is spontaneously broken and 
the ``dual gauge field'' becomes massive.

Let us formulate the DGL theory starting from QCD.
The QCD Lagrangian has the SU($N_c$) symmetry and is written 
by the quark field $q(x)$ and the gluon field 
$A_\mu(x)$ as
\begin{eqnarray}
{\cal L}_{\rm QCD}  = 
    - \frac 12{\rm tr}\left( G_{\mu \nu} G^{\mu \nu}\right)
    + \bar{q} \left( i \gamma_\mu D^\mu -  m \right) q, 
\end{eqnarray}
where $G_{\mu \nu}$ denotes the SU($N_c$) field strength,
\begin{eqnarray}
G_{\mu \nu} \equiv \frac{1}{ie}
([{\hat D_\mu},{\hat D_\nu}]- [\hat \partial_\mu, \hat \partial_\nu])
\hspace{1cm} \mbox{\rm with} \hspace{0.5cm}
\hat D_{\mu }  \equiv  \hat \partial_\mu + ieA_\mu. 
\end{eqnarray}
In terms of the Cartan decomposition, 
the gluon field $A_{\mu}(x) \equiv A_{\mu}^a(x)T^a$ $\in$ su($N_c$) 
with $N_c$=3 is decomposed into 
the diagonal component $\vec{A_{\mu}}=(A_{\mu}^{3},A_{\mu}^{8})$
 and the off-diagonal component $C^a_{\mu} \in {\bf C}$,
\begin{eqnarray}
A_\mu  =  \vec A_\mu \cdot \vec H+\sum\limits_{a =1}^3 
({C_\mu ^{*a }}E_a +C_\mu ^a E_{-a }).
\end{eqnarray}
Here, the Cartan subalgebra 
$\vec H$ is defined as
\begin{eqnarray}
{\vec H\equiv (H_1, H_2) \equiv (T_3, T_8),} 
\end{eqnarray}
and the raising and the lowering operators $E_{\pm a} (a=1,2,3)$ 
are defined as
\begin{eqnarray}
{E_{\pm 1}\equiv {1 \over {\sqrt 2}}(T_1\pm iT_2)}, \hspace{0.5cm}
{E_{\pm 2}\equiv {1 \over {\sqrt 2}}(T_4\mp iT_5)}, \hspace{0.5cm}
{E_{\pm 3}\equiv {1 \over {\sqrt 2}}(T_6\pm iT_7)}. 
\end{eqnarray}
One finds the relations
$
\left[ {\vec H, E_a} \right]=\vec \alpha _a
E_a$ and $
\left[ {\vec H, E_{-a} } \right]=-\vec \alpha _a
E_{-a} 
$,
where $\vec \alpha _a$ denotes the  
root vector of the SU(3) algebra:
${\vec \alpha _1=(1, 0)}$, 
${\vec \alpha _2=(-{1 \over 2}, -{{\sqrt 3} \over 2})}$
 and ${\vec \alpha _3=(-{1 \over 2}, {{\sqrt 3} \over 2})}$.
The off-diagonal gluon component $C_\mu^{a}$ is written as
\begin{eqnarray}
{C_\mu ^1={1 \over {\sqrt 2}}(A_\mu ^1+iA_\mu ^2)}, \hspace{0.5cm} 
{C_\mu ^2={1 \over {\sqrt 2}}(A_\mu ^4-iA_\mu ^5)}, \hspace{0.5cm} 
{C_\mu ^3={1 \over {\sqrt 2}}(A_\mu ^6+iA_\mu ^7)}. 
\end{eqnarray}
According to the Cartan decomposition,
the QCD Lagrangian is expressed as 
\begin{eqnarray}
{\cal L}_{\rm QCD} & =  &
{\cal L}^{\rm Abel} + {\cal L}^{\rm off},
 \label{eq:decomQCD}
\end{eqnarray}
where  ${\cal L}^{\rm Abel}$ is diagonal part of ${\cal L}_{\rm QCD}$,
\begin{eqnarray}
{\cal L}^{\rm Abel} & = & 
- \frac 14 
\left
( \partial_{\mu}{\vec{A_{\nu}}}
                         - \partial_{\nu}{\vec{A_{\mu}}} \right)^2
{+\bar q(i\gamma_\mu  \partial^\mu   - e \gamma_\mu \vec A^\mu
\cdot \vec H-m) q}, 
\end{eqnarray}
and
${\cal L}^{\rm off}$ is remaining part including the off-diagonal gluons, 
\begin{eqnarray}
{\cal L}^{\rm off}  
       & = &
  -{{1 \over 2}\sum\limits_{a=1}^3 {\left| {\left(
{(\partial _\mu-ie(\vec \alpha ^a\cdot \vec A_\mu ))
\wedge C^a} \right)_{\mu \nu }-{{ie} \over {\sqrt 2}}
\alpha _{abc }C_\mu ^{*b }C_\nu ^{*c }}
\right|}^2} \nonumber \\
       & &
{-{{ie} \over 2}\sum\limits_{a=1}^3 {(\vec f_{\mu \nu }
\cdot \vec \alpha _a)}(C^{*a}\wedge C^a)
^{\mu \nu }}
{+{e^2 \over 4}\left[ {\sum\limits_{a=1}^3 {\vec
\alpha _a(C^{*a}\wedge C^a)_{\mu \nu }}}
\right]^2}  \nonumber \\
       & &
 -{e\sum\limits_{a=1}^3 {\left[ {C^{*a}_\mu
(\bar q\gamma ^\mu E_aq)+C^a_\mu (\bar q\gamma ^\mu
E_{-a}q)} \right]}} 
\end{eqnarray}
with $(a \wedge b)_{\mu \nu} 
\equiv a _\mu 
b _\nu - b _\mu a _\nu $.
%

In the abelian gauge defined by diagonalizing a gauge dependent 
operator $\Phi$,
the SU(3) gauge theory is reduced into 
[ U(1)$_3$ $\times$ U(1)$_8$]$^{e}$ 
gluon gauge theory   \cite{maedan,suganuma}  and monopoles appear.
Particularly, in the MA gauge, 
off-diagonal components are forced to be small and behave as  massive 
fields,  
while 
diagonal gluon components  remain largely propagate over the long range
 \cite{amemiya}.
In fact, in the MA gauge, off-diagonal gluons are infrared-irrelevant 
and can be neglected at the long distance.
Therefore, after MA gauge fixing with $\Omega$, we construct the abelian 
projected QCD by dropping off the charged 
part and define the abelian gauge field ${\cal A}_\mu$ as
\begin{eqnarray}
{\cal A}_\mu \equiv \frac12 {\rm tr}(A_\mu \vec H) \cdot \vec H \equiv 
\vec A_\mu \cdot \vec H.
\end{eqnarray}
Accordingly, the SU(3) field strength 
\begin{eqnarray}
G_{\mu\nu}^\Omega 
 =  (\partial_\mu A_\nu -\partial_\nu A_\mu )
+ ie[A_\mu, A_\nu] +\frac{i}{e}
\Omega[\partial_\mu, \partial_\nu] \Omega^{\dagger} 
\end{eqnarray}
is modified as 
\begin{eqnarray}
 {\cal F}_{\mu\nu}  = 
\partial_\mu  {\cal A}_\nu -\partial_\nu {\cal A}_\mu - 
 {\cal F}_{\mu\nu}^{\rm sing},
\end{eqnarray}
where 
$ {\cal F}_{\mu\nu}^{\rm sing} 
= \frac{i}{e} \Omega[\partial_\mu, \partial_\nu]  \Omega^{\dagger}$.
This corresponds to Eq.(\ref{eq:twopars}) in the lattice formalism.
Here, the breaking of the abelian  Bianchi identity is brought by the appearance
of $  {\cal F}_{\mu\nu}^{\rm sing}$.
In the SU(3) case, there are three kinds of magnetic charges of the 
QCD-monopole 
corresponding to the subspace on 
degeneracy of the diagonalized operator $\Phi_{\rm diag}$;
for $\Phi_{diag} = {\rm diag}(\Phi_1,\Phi_2,\Phi_3)$, $\Phi_1 =\Phi_2$,
$\Phi_2 =\Phi_3$  and $\Phi_3 =\Phi_1$. For example,
the magnetic charge is proportional to the root vector $\vec \alpha_1$ 
in the case of the $\Phi_1 =\Phi_2$.
Thus, the magnetic charge of the monopole current is discretized as
\begin{eqnarray}
\vec{m} = g \sum\limits_{a=1}^3 \xi_a
\vec{\alpha_a}, \hspace{1cm}  \xi_a \in {\bf Z}.
\end{eqnarray}
For $N$ monopole system, 
the magnetic current $k_\mu  \equiv \vec k_\mu \cdot \vec H$ is 
expressed as
\begin{eqnarray}
\vec k^\mu(x) = -g\sum\limits_{l=1}^N 
{\vec m_l\int {d\tau _l}} 
\frac{ d \overline x_l^\mu (\tau _l)}{d \tau_{l}}
\delta^{4}(x-\bar x_{l}(\tau_{l})).
\end{eqnarray}
In the DGL theory, the monopole field is introduced by integrating over
all trajectories of the monopole particle as will be shown in the later section.

\section{Dual Gauge Field}

The extended Maxwell equation is described 
by the field strength $F_{\mu\nu}$ as
\begin{eqnarray}
\partial _\nu F^{\nu \mu }  =  
j^\mu , \hspace{2cm}
\partial _\nu ^*F^{\nu \mu }  =  k^\mu 
\end{eqnarray}
with the electric current $j_\mu$ and the magnetic current $k_\mu$,
and the electro-magnetic duality 
($F_{\mu \nu} \leftrightarrow {}^*F_{\mu \nu}$,
$j_\mu \leftrightarrow k_\mu$) is manifest in this formalism.
In the ordinary QED,
the description by the gauge field $A_\mu$ is useful,
since  QED includes only the electric current.
In this  description by $A_\mu$,
however, this electro-magnetic duality 
in the Maxwell equation is not manifest and
the introduction of the magnetic monopole leads to the singularity
in the gauge field $A_\mu$.
As shown in Section \ref{dff}, for  ``the dual system'' including only 
the magnetic current, it is rather useful to take the dual description 
with the dual gauge field $B_\mu$  instead of $A_\mu$.
Here, the dual gauge field $B_\mu$  is introduced so as to
satisfy
\begin{eqnarray}
^*F^{\mu \nu} = \frac 12 \varepsilon^{\mu \nu \alpha \beta}
  F_{\alpha \beta} 
\equiv \partial^\mu B^\nu - \partial^\nu B^\mu, 
\end{eqnarray}
or equivalently, 
the electro-magnetic field is written as
\begin{eqnarray}    
E^i & = & \frac 12 \varepsilon ^{ijk} {}^*F^{jk} 
= \varepsilon ^{ijk}\partial^{j}B^{k}, \\
H^i & = & -^*F^{0i}    = -\partial^{0}  B^{i} 
                       +\partial^{i}  B^{0}.
\end{eqnarray}

Now, we adopt here the Zwanziger formalism \cite{zwan} described both with 
$A_{\mu}$ and $B_{\mu}$ in order to keep the electro-magnetic duality 
in the extended Maxwell equation manifest. 
In the general case with $j_\mu \neq 0$ and $k_\mu \neq 0$,
we rewrite the field strength to keep the duality manifest
in the following way.
For an arbitrary anti-symmetric tensor $G_{\mu \nu}$ and
an arbitrary constant four-vector $n_\mu$, there is an identity
\begin{eqnarray}
G={1 \over {n^2}}\left\{ {
\left[ {n\wedge \left( {n\cdot G} \right)} \right]
-^*\left[ {n\wedge \left( {n\cdot ^*G} \right)
} \right]} \right\}
\end{eqnarray}
with $(n \cdot G)_\mu \equiv n_\alpha G^{\alpha \mu}$.
Substituting $G=F=(\partial \wedge A)$  and 
${}^*G={}^*F=(\partial \wedge B)$, we get
\begin{eqnarray}
F & = & {1 \over {n^2}}
(\left\{ { {n\wedge 
\left[ {n\cdot (\partial \wedge A)} \right]
}} \right\}
-^*\left\{ {n\wedge \left[ 
{n\cdot (\partial \wedge B)} \right]} \right\}), \\
^*F & = & {1 \over {n^2}}
({}^*\left\{ { {n\wedge 
\left[ {n\cdot (\partial \wedge A)} \right]
} } \right\}
+\left\{ {n\wedge \left[ 
{n\cdot (\partial \wedge B)} \right]} \right\}),
\end{eqnarray}
and then the extended Maxwell equations 
$\partial _\nu F^{\nu \mu }  =  
j^\mu$ and
$\partial _\nu ^*F^{\nu \mu }  =  k^\mu $
are written as
\begin{eqnarray}
\partial _\nu F^{\nu \mu } & = & 
{1 \over {n^2}} \Bigl( \left\{  {(n\cdot \partial )^2A^\mu 
-(n\cdot \partial )\partial ^\mu (n\cdot A)
-n^\mu (n\cdot \partial )(\partial \cdot A) 
+n^\mu \partial^2 (n\cdot A)}     \right\}  \nonumber \\
& & -{(n\cdot \partial )\varepsilon ^\mu {}_
{\nu \kappa \lambda } n^\nu \partial ^\kappa B^\lambda} \Bigr) 
\nonumber  =  j^\mu,  \\ 
\partial _\nu ^*F^{\nu \mu } & = & 
{1 \over {n^2}} \Bigl( \left\{ {(n\cdot \partial )^2 B^\mu 
-(n\cdot \partial ) \partial ^\mu (n\cdot B)
-n^\mu (n \cdot \partial ) (\partial \cdot B)
+n^\mu \partial^2 (n\cdot B)} \right\}  \nonumber
\\
& & +{(n\cdot \partial )\varepsilon ^\mu {}_
{\nu \kappa \lambda }n^\nu \partial ^\kappa A^\lambda}\Bigr) 
\nonumber 
 =  k^\mu. 
\end{eqnarray}
Therefore, the Lagrangian which leads to these field equations is
derived as
\begin{eqnarray}
{\cal L} & = & -{1 \over {2n^2}}
\left[ {n\cdot (\partial \wedge A)} \right]
\left[ {n\cdot ^*(\partial \wedge B)} \right]
+{1 \over {2n^2}}
\left[ {n\cdot (\partial \wedge B)} \right]
\left[ {n\cdot ^*(\partial \wedge A)} \right] 
\nonumber \\
& & -{1 \over {2n^2}}
\left[ {n\cdot (\partial \wedge A)} \right]^2 
 -{1 \over {2n^2}}
\left[ {n\cdot (\partial \wedge B)} \right]^2-j
_\mu A^\mu -k_\mu B^\mu. 
\label{eq:zwanlag}
\end{eqnarray}

In the Zwanziger formalism, 
the duality transformation 
$F_{\mu \nu} \leftrightarrow {}^*F_{\mu \nu}$ is given by the 
replacement $ A_\mu \leftrightarrow B_\mu$ and the electro-magnetic
duality is manifest, and  
the system can be described without 
singularities.
Particularly in the absence of $j_{\mu}$ and $k_{\mu}$,
the theory can be written using 
``only $A_\mu$'', ``only $B_\mu$'', or ``both $A_\mu$
and $B_\mu$'' in the Zwanziger formalism as
\begin{eqnarray}
\int {{\cal D} A e^{i\int {d^4x{\cal L}_{0}[A]}}}
=\int {{\cal D}B e^{i\int {d^4x{\cal L}_{0}[B]}}}
=\int {{\cal D}A {\cal D} B 
e^{i\int {d^4x{\cal L}_{0}[A,B]}}},
\end{eqnarray}
where
\begin{eqnarray}
{\cal L}_{0}[A ]& \equiv & -{1 \over 4}(\partial \wedge A)^2, \\
{\cal L}_{0}[A ,B] & \equiv & -{1 \over {2n^2}}
\left[ {n\cdot (\partial \wedge A)} \right]
\left[ {n\cdot ^*(\partial \wedge B)} \right]
+{1 \over {2n^2}}
\left[ {n\cdot (\partial \wedge B)} \right]
\left[ {n\cdot ^*(\partial \wedge A)} \right] \nonumber \\
& & -{1 \over {2n^2}}
\left[ {n\cdot (\partial \wedge A)} \right]^2 
-{1 \over {2n^2}}
\left[ {n\cdot (\partial \wedge B)} \right]^2, 
\label{eq:duallag}
\\
{\cal L}_{0}[B ] & \equiv & -{1 \over 4}(\partial \wedge B)^2.
\end{eqnarray}

In the Zwanziger formalism, 
the Lagrangian can be written with holding the duality between 
the gauge field $A_\mu$ and the
dual gauge field $B_\mu$.
However, occasionally  it is more useful to describe 
only with $A_\mu$ or $B_\mu$ for  practical calculations. 
Here, we try to describe the  Lagrangian with  only  $A_\mu$ or $B_\mu$
by the functional integration over $B_\mu$ or $A_\mu$ 
in the partition functional \cite{blago}.

Since the dual gauge field $B_\mu$ is bilinear in the Lagrangian
(\ref{eq:duallag}), $B_\mu$ can be analytically integrated in the 
partition functional in the Zwanziger formalism,
\begin{eqnarray}
  \int {\cal D} A  {\cal D} B e^{i \int d^{4} x {\cal L}[A,B]} 
= \int {\cal D} A   e^{i \int d^{4} x {\cal L}_{k}[A]} 
\end{eqnarray}
with 
\begin{eqnarray}
{\cal L}_k [A]=-{1 \over 4}
\{ (\partial \wedge A)_{\mu \nu } - F^{\rm sing}_{\mu\nu} \}^{2}.
\label{eq:aform}
\end{eqnarray}
Here, the singular term $ F_{\mu\nu}^{\rm sing} \equiv 
 \varepsilon _{\mu \nu \alpha \beta }
{1 \over {n\cdot \partial }}n^\alpha k^\beta $
appears due to existence of the magnetic current $k_\mu$,
and includes the nonlocal operator 
$\langle x | \frac{1}{n \cdot \partial} | y \rangle =
\theta(x_{n} - y_{n}) \delta^{3}(x_{\bot}-y_{\bot})$.

Thus,  the field strength $F_{\mu\nu}$ includes the nonlocality in this 
description only by $A_{\mu}$ in the presence of the magnetic current $k_\mu$.
Similarly, the theory can be described only with the dual gauge field $B_\mu$
after the functional integration  on $A_\mu$.  
In this case, the theory  is  replaced by $(B_{\mu},j_{\mu})$ in the above 
argument, and the Lagrangian reads
\begin{eqnarray}
{\cal L}_j [B]=-{1 \over 4}
\{ {}^*(\partial \wedge B)_{\mu \nu }
-\varepsilon _{\mu \nu \alpha \beta }
{1 \over {n\cdot \partial }}n^\alpha j^\beta \}^{2},
\label{eq:bform}
\end{eqnarray}
where the nonlocal part appears 
$^*F_{\mu \nu }$
reflecting  existence of the electric current $j_\mu$.

In this way, 
the system including both the electric and
the monopole currents can be described without any singularities using 
the Zwanziger 
formalism, where the dual field $B_{\mu}$ is introduced as well as the 
gauge field $A_{\mu}$.
On the other hand, the description 
only with  $A_\mu$ or $B_\mu$ leads to the appearance of the 
nonlocal operator in the field strength in the general case.
In constructing the DGL theory from  abelian projected QCD,
the above formalism can be applied by the simple replacement of 
$A_{\mu} \rightarrow  {\cal A}_\mu \equiv \vec A_\mu \cdot \vec H$,
$B_{\mu} \rightarrow  {\cal B}_\mu \equiv \vec B_\mu \cdot \vec H$
and so on.


\section{Monopole Field from the Monopole Particle}

In this section, we derive the Lagrangian for the monopole field from
the monopole-particle trajectories following  
Stone and Thomas \cite{thomas,samual,nagaosa}.
At the infrared scale 
in the MA gauge, the monopole current becomes dense
and complicated, and 
the monopole currents interact each other
only in the short distance due to the 
screening effect \cite{atanaka}
corresponding to monopole current.
The infrared effective action $k_\mu$ is expected to be described by the 
local action of $k_\mu$  
\begin{eqnarray}
Z = \int_a dk_\mu e ^{-\alpha k^{2}_\mu}.
\label{eq:cursys}
\end{eqnarray}
Here, $\alpha$  is the monopole self-energy and the 
``mesh'' $a$ is introduced as an ultraviolet cut-off larger 
than  the screening length \cite{ichie2}.
As shown Section \ref{appearance}, almost all monopoles have the unit charge 
in the abelian gauge, and therefore
the partition functional is approximately described by the  ensemble of 
the  monopole loops 
\begin{eqnarray}
Z = \sum_N \frac{Z^N_{\rm loop}}{N!}  \hspace{1cm} \mbox{with}
\hspace{1cm} Z_{\rm loop} \equiv 
\sum_{L=0}^{\infty} 
 \rho(L) e^{-\alpha L},
\end{eqnarray}
where $\rho(L)$ is the number of closed loop configurations with the length 
$L$.
Here, we have taken the lattice unit $a=1$.
Each monopole loop with the length $L$ can be approximated as $L$ step 
random walk \cite{thomas} because of the absence of the nonlocal 
interaction.
 
Let us consider the grand canonical ensemble for closed-loops
on the $D$-dimensional lattice  in the unit of the 
lattice spacing 1.
At the infrared scale, the partition function 
for a single monopole loop is written as
\begin{eqnarray}
Z_{\rm loop} = \int_0^\infty d \tau  \rho( \tau ) e^{-\alpha \tau}.  
\end{eqnarray}
The probability distribution of starting at $x$ and ending at $x'$
after a walk of $\tau$ steps is given as
\begin{eqnarray}
\rho(x,x',\tau) = \int_x^{x'} d[x(\cdot)] 
{\rm exp} \{ -\int_0^\tau { \frac{D}{2}\dot x(\tau')^2 d \tau' \}},
\end{eqnarray}
which satisfies
\begin{eqnarray}
\frac{\partial \rho(x,x',\tau)}{\partial \tau} 
= \frac{1}{2D} \frac{ \partial^2  \rho(x,x',\tau)}{\partial x_\mu^2}.
\end{eqnarray}
The total number of paths from $x$ to $x'$ with $\tau$ steps is estimated 
as
\begin{eqnarray}
\Gamma(x,x',\tau) & = & (2D)^\tau \rho(x,x',\tau) \nonumber \\
& = & N \int d[x(\cdot)] {\rm exp} \{ -\int_0^\tau d \tau' \frac{D}{2} \dot x^2
-{\rm ln}(2D)  \},
\end{eqnarray}
where the factor $2D$ is the configuration number for 
a walk of one step.   
Using the eigen-function $\Phi_m(x)$ satisfying
\begin{eqnarray}
\{ -\frac12 \partial_\mu^2-{\rm ln}(2D) \}  \Phi_m(x) = 
\omega_m \Phi_m \hspace{0.5cm}\mbox{and} \hspace{0.5cm}
\int d^D x \, \Phi_m^2(x) = 1,
\end{eqnarray}
$\Gamma(x,x',\tau)$ is written by 
\begin{eqnarray}
\Gamma(x,x',\tau) = \sum_n \Phi_n(x) \Phi_n(x') e^{- \omega_n \tau}. 
\end{eqnarray}
For the closed loop, all $\tau$ starting points on the loop define 
the same configuration, and hence $\Gamma(x,x',\tau)$  satisfies 
\begin{eqnarray}
\int d^D x \Gamma(x,x,\tau) = \tau \rho(\tau).
\end{eqnarray}
Then, we get
\begin{eqnarray}
-\frac{\partial Z_{\rm loop}}{\partial \alpha}  & = &
  \int_0^\infty d \tau  \tau \rho(\tau) e^{-\alpha \tau} 
= \int_0^\infty d \tau 
   \int d^D x \Gamma(x,x,\tau) 
e^{-\alpha \tau} = \int_0^\infty d \tau \sum_ne^{-(\omega_n+\alpha) \tau} 
\nonumber \\
& =&  \sum_n \frac{1}{\omega_n + \alpha} 
= {\rm Tr} \left(-\frac{1}{2D} \partial_\mu^2 - {\rm ln}(2D) + \alpha \right)^{-1}.
\end{eqnarray}
Integrating  on $\alpha$, we obtain
\begin{eqnarray}
Z_{\rm loop}&  =& -{\rm Tr} {\rm ln} (-\frac{1}{2D} \partial_\mu^2 - {\rm ln}(2D) 
+ \alpha) \nonumber \\
& = & - {\rm ln} \, {\rm Det} 
(-\frac{1}{2D}\partial_\mu^2 - {\rm ln}(2D) + \alpha).
\end{eqnarray}
Thus, 
the partition function for the whole system is expressed as 
\begin{eqnarray}
Z  & = & \sum_N \frac{Z^N_{\rm loop}}{N!} = {\rm exp}[Z^N_{\rm loop}] = 
{\rm Det^{-1}}(-\frac{1}{2D} \partial_\mu^2 +m^2) \nonumber \\
& & =  \int {\cal D} \varphi \int {\cal D}\varphi^*  
{\rm exp}[-\int d^D x \varphi^* (-\frac{1}{2D} \partial_\mu^2 +m^2)\varphi].
\label{eq:fieldrep}
\end{eqnarray}
Hence,  noninteracting loops can be written as a free-field 
theory of the complex scalar field with the 
mass $m^2 \equiv \alpha - {\rm ln}(2D)$ \cite{thomas}.

Here, considering the probability weight of $e^{-\alpha \tau}$,
the ``expectation value'' of the number of the walk which starts 
at $x$ and ends at $x'$ is proportional to 
\begin{eqnarray}
\lefteqn{ 
\int_0^\infty  d \tau \Gamma(x,x',\tau)e^{-\alpha \tau} } \nonumber \\ 
& = & \sum_n \frac{\Phi_n(x)\Phi_n(x')}{\omega_n + \alpha}
= \sum_n \langle x|n \rangle 
\langle n |  \frac{1}{-\frac{1}{2D} \partial^2_\mu + m^2} | n\rangle \langle
n | x' \rangle 
 \nonumber \\
& = & \frac{1}{Z} \int {\cal D} \varphi \int {\cal D} 
\varphi^* 
\varphi(x) \varphi ^* (x') 
{\rm exp} \{ -\int dx \varphi^* ({-\frac{1}{2D} \partial_\mu^2 + m^2})
 \varphi  \} 
 \nonumber \\
 & = & \langle \varphi(x) \varphi^*(x') \rangle. 
\end{eqnarray}
In particular for $|x-x'| \rightarrow \infty$, 
$\langle  \varphi(x) \varphi^*(x') \rangle $ 
indicates the density of loops whose length is infinity
and becomes $\langle  \varphi \rangle \langle \varphi^* \rangle
= |\langle  \varphi \rangle|^2$.
Thus, the expectation value 
$|\langle  \varphi \rangle|$ of the free massive scalar field 
is directly related to the loop density. 
When the self-energy $\alpha$ is larger than ``entropy'' ln$(2D)$, i.e.,
$m^2 > 0$,
the loop density is suppressed by the action factor and
the system is described by
the free scalar field theory. 
On the other hand, when $\alpha$ is less than ln$(2D)$, i.e., $m^2 < 0$,
infinite long loop appears and the
expectation value of the scalar field takes a non-zero value, i.e, 
the scalar field $\varphi$ condenses in the description of the scalar
field theory.
In order to stop the loop density being infinite, we need   $-\lambda 
|\varphi|^4$ with positive $\lambda$ in the Lagrangian,
corresponding to  a short-distance 
repulsion in Eq.(\ref{eq:fieldrep}).

In the monopole-current system (\ref{eq:cursys}) appearing in the QCD 
vacuum,
this complex scalar field $\varphi$ corresponds
to the ``monopole field''.
Here, the monopole current system can be regarded as a {\it self-avoiding} 
random walk, 
and then the factor ``$2D$'' in this argument is expected to be replaced 
by ``$2D-1$''.
Hence, 
whether the monopole condenses or not is determined by the balance of the 
``self energy'' $\alpha$ and the ``entropy'' ln$(2D-1)$, as is already shown 
in the Chapter \ref{sec:MCD}. 
In the presence of the dual gauge field $B_\mu$,
the derivative $\partial_\mu$ is simply replaced by the dual covariant 
derivative $\partial_\mu+ igB_\mu$
with the dual gauge coupling constant $g$.
Finally, the Lagrangian for the monopole field $\chi \equiv 
\frac{1}{\sqrt{2D}} \varphi$ is written as
\begin{eqnarray}
{\cal L}^{\rm mon} = \left| {(\partial _\mu +i g B_\mu )
 \chi _{}} \right|^2
- \lambda (\left| { \chi } \right|^2- v^2)^2,
\end{eqnarray}
with $2 \lambda v^2 = {2D} m^2$.

In the SU(3) QCD in the abelian gauge,
there appear three kinds of monopoles with the unit charge $\vec m_{a} 
= \pm \vec \alpha_{a}$ $(a=1,2,3)$ corresponding to the nontrivial 
root vector \cite{ezawa}. Among the three monopole currents
$k_{\mu}^{a}$ $(a=1,2,3)$,  
only two monopole-current degrees of freedom are 
independent due to the relation $\sum_{a=1}^{3} \vec \alpha_{a} = 0$.
In fact, $k_{\mu}^{3}$ can be expressed as $-(k_{\mu}^{1}+k_{\mu}^{2})$.
However, the description by the two independent monopole currents 
makes the global Weyl symmetry unclear and becomes complicated 
in the presence of the interaction with $B_{\mu}= \vec B_\mu \cdot \vec H$.
Therefore, in order to treat $k_{\mu}^{a}$ $(a =1,2,3)$ as the 
independent variable,
we introduce the Lagrange-multiplier field $B_{\mu}^0$ and rewrite the 
sum over the monopole current as
\begin{eqnarray}
\sum_{k_{\mu}^{1},k_{\mu}^{2}} \rightarrow 
\sum_{k_{\mu}^{1},k_{\mu}^{2},k_{\mu}^{3}} 
\delta ( \sum_{a=1}^{3} k_{\mu}^{a})
=\sum_{k_{\mu}^{1},k_{\mu}^{2},k_{\mu}^{3}} 
\int {\cal D} B_{\mu}^0 e^{i \int d^4x \{ B^0_{\mu} \sum_{a=1}^{3} k_{\mu}^{a}\} }
\end{eqnarray}
in the partition function of the monopole current.
This prescription corresponds to the introduction of the unphysical 
U(1)$_{0}^{m}$ magnetic charge coupled with  $ B_{\mu}^0$.
Now, we can apply the above formalism for each  $k_{\mu}^{a}$ $(a =1,2,3)$
independently, and then we obtain the partition functional,
\begin{eqnarray}
Z & = & \int {\cal D} \vec B_{\mu} 
e^{i \int d^4x \{ -\frac14 (\partial \wedge \vec B)^2 \}  }
 \int {\cal D} B_{\mu}^0 [\prod_{a=1}^{3} {\cal D} \chi_{a}]
\nonumber \\
&& \cdot
 {\rm exp} \{
i\int d^4x\sum_{a=1}^3 \{ |(\partial _\mu +ig\vec \alpha _a 
\cdot \vec B_\mu + ig B^0_\mu)\chi_a |^2 - \lambda (|\chi_a|^2-v^2)^2 
\} \},
\label{eq:monol}
\end{eqnarray}
where 
$\vec \alpha_{a} $ denotes the non-trivial root vector; 
$\vec \alpha_1 \equiv (1,0)$,
$\vec \alpha_2 \equiv (-\frac12, -\frac{\sqrt{3}}{2})$ and
$\vec \alpha_3 \equiv (-\frac12,\frac{\sqrt{3}}{2})$.
Since the partition functional $Z$ is invariant under the 
transformation
\begin{eqnarray}
\chi _\alpha(x)  \to  e^{i \theta (x)}  \chi _a(x) ,
\hspace{1cm} B_{\mu}^0(x)  \rightarrow B_{\mu}^0 (x)  +\frac{1}{g}
\partial_{\mu} \theta(x) ,
\end{eqnarray}
the theory holds the extra dual gauge symmetry U(1)$^m_{0}$
as well as U(1)$^m_{3}$ $\times$
U(1)$^m_{8}$. 
Here, the global Weyl symmetry is kept manifest in $Z$.
Due to the extra local U(1)$_{0}^m$ symmetry, the overall phase 
degrees of freedom  of $\chi_{a}(x)$ in Eq.(\ref{eq:monol})
is absorbed into $B_{\mu}^{0}$ and becomes unphysical.
In fact, without loss of generality, we can set the local condition as
\begin{eqnarray}
\sum\limits_{a =1}^3 {\rm arg} \chi _a = 0,
\end{eqnarray}
by using a suitable U(1)$_{0}^m$ dual gauge transformation and the 
shift of the integral variable $B_{\mu}^{0}$.
In principle,  $B_{\mu}^{0}$ can be integrated out in $Z$. After the 
integration on $B_{\mu}^{0}$, there appear some interaction terms as the 
functional determinant, however, those contribution is assumed to be 
included into the self-interaction term of $\chi_{a}$.
Thus, the final expression of $Z$ is obtained as
\begin{eqnarray}
Z = \int D \vec B_{\mu} 
e^{i \int d^4x \{ -\frac14 (\partial \wedge 
\vec B)^2 \} }   [\prod_{a=1}^{3} D \chi_{a}]
\, \, {\rm exp} \{ i \int d^4x {\cal L}^{\rm mon}
\},
\end{eqnarray}
where ${\cal L}^{\rm mon}$ denotes the monopole part of the 
Lagrangian  in the SU(3) case,
\begin{eqnarray}
{\cal L}^{\rm mon} = \sum_{a=1}^3 \{ |(\partial _\mu +ig\vec \alpha _a 
\cdot \vec B_\mu )\chi_a |^2 - \lambda (|\chi_a|^2-v^2)^2 \},
\end{eqnarray}
with  the constraint 
$
\sum\limits_{a =1}^3  \mbox{{\rm arg}} \chi _a = 0
$.

In the Zwanziger form (\ref{eq:duallag}), we obtain the dual 
Ginzburg-Landau Lagrangian as
\begin{eqnarray}
{\cal L}_{DGL} & = & -{1 \over {2n^2}}
[n\cdot (\partial \wedge \vec A)]^\nu
 [n\cdot {}^*(\partial \wedge \vec B)]_\nu  
 +{1 \over {2n^2}}[n\cdot (\partial \wedge \vec B)]^\nu 
[n\cdot {}^*(\partial \wedge \vec A)]_\nu  \nonumber \\
& & -{1 \over {2n^2}}[n\cdot (\partial \wedge \vec A)]^2
-{1 \over {2n^2}}[n\cdot (\partial \wedge \vec B)]^2 
{+\bar q(i\gamma_\mu  \partial^\mu  - e \gamma_\mu \vec A^\mu
\cdot \vec H-m)q}  
\nonumber \\
& & +\sum\limits_{a =1}^3 {\left[ {\left| 
{(i\partial _\mu -g\vec \alpha _a \cdot \vec B_\mu )
\chi _a } \right|^2-\lambda (\left| \chi_a  \right|^2-v^2)^2} 
\right]},
\label{eq:saishuu}
\end{eqnarray}
with  $\sum\limits_{a =1}^3 {\rm {\rm arg}} \chi_a = 0 $.

In terms of the Cartan decomposition,
this DGL Lagrangian is also expressed as
\begin{eqnarray}
{\cal L}_{DGL}   = & {\rm tr}  \{   &
-{1 \over {n^2}}[n\cdot (\partial \wedge {\cal A}]^\nu
 [n\cdot {}^*(\partial \wedge {\cal B})]_\nu  
 +{1 \over {n^2}}[n\cdot (\partial \wedge { \cal B})]^\nu 
[n\cdot {}^*(\partial \wedge {\cal A})]_\nu  \nonumber \\
& & -{1 \over {n^2}}[n\cdot (\partial \wedge {\cal A})]^2
-{1 \over {n^2}}[n\cdot (\partial \wedge {\cal B})]^2\} 
{+\bar q(i\gamma_\mu  \partial^\mu  - e \gamma_\mu {\cal A}^\mu
-m)q}
\nonumber \\
& & 
 +  [\hat{D}^{\rm dual}_\mu, \chi]^{\dagger}[\hat{D}^{\rm dual}_\mu, \chi]
-\lambda ( \chi^{\dagger} \chi -v^2)^2 \},
\label{eq:saishuu2}
\end{eqnarray}
where ${\cal A_\mu}$, ${\cal B_\mu}$, $\hat D^{\rm dual}_\mu $ and $\chi$ denote
${\cal A}_\mu \equiv \vec A_\mu \! \cdot \! \vec H$,
${\cal B}_\mu \equiv \vec B_\mu \! \cdot \! \vec H$,
$\hat D^{\rm dual}_\mu \equiv \hat \partial_\mu + ig{\cal B}_\mu$ and $\chi \equiv \sum_{a=1}^3
\sqrt{2} \chi_a E_a$, respectively.
The kinetic term of the monopole field leads to the original expression 
in Eq.(\ref{eq:saishuu})
\begin{eqnarray}
\lefteqn{ {\rm tr}([\hat{D}_\mu^{\rm dual}, \chi]^{\dag}[\hat{D}^{\rm dual}_\mu, \chi]) }
\nonumber \\
&=&
{\rm tr}([\hat{\partial}_\mu +ig{\cal B}_\mu, \chi^{\dag}]
[\hat{\partial}_\mu +ig{\cal B}_\mu, 
\chi]) \nonumber\\ 
&=&2{\rm tr}\{
(\partial_\mu \chi^*_a E_{-a} +i g\vec B_\mu \chi^*_a [\vec{H}, E_{-a}])
(\partial_\mu \chi_b E_b + i g\vec B_\mu \chi_b [\vec{H}, E_b])\}
\nonumber \\ 
&=&\left|(\partial_\mu+ig\vec \alpha_a \cdot \vec B_\mu )
\chi _a  \right|^2.
\end{eqnarray}
Here, the DGL Lagrangian (\ref{eq:saishuu}) has 
$[{\rm U(1)}_e]^2 \equiv {\rm U(1)}^3_e \times {\rm U(1)}^8_e$
gauge symmetry and 
$[{\rm U(1)}_m]^2 \equiv
{\rm U(1)}^3_m \times {\rm U(1)}^8_m$
dual gauge symmetry,  since the Lagrangian is invariant under the 
gauge transformation by
$\Omega_{e} \equiv e^{-i\theta_{e}} \equiv e^{-i \vec \theta_e \cdot \vec H}
\in $ U(1)$_{e}^3$ $\times$ U(1)$_{e}^8$
\begin{eqnarray}
 q  \rightarrow   \Omega_{e}(x) q, \hspace{0.5cm} 
{\cal A}_{\mu}   \rightarrow   
\Omega_{e} (x) \left( {\cal A}_{\mu}- \frac{i}{e} \partial_{\mu} \right) 
 \Omega_{e} (x)^{\dagger} = {\cal A}_\mu + \frac{1}{e} \partial_\mu \theta_e,
\end{eqnarray}
and also invariant under the dual gauge transformation by 
$\Omega_{m} \equiv e^{-i \theta_{m}} \equiv  e^{-i \vec \theta_{m} 
\cdot \vec H}\in $ U(1)$_{m}^3$ $\times$ U(1)$_{m}^8$ 
\begin{eqnarray}
 { \cal B}_{\mu}  & \rightarrow &   
\Omega_{m} (x) \left( {\cal B}_{\mu}- \frac{i}{ g}    \partial_{\mu} \right) 
 \Omega_{m} (x)^{\dagger} = {\cal B}_\mu + \frac{1}{e} \partial_\mu 
 \theta_m,
  \\
\chi   & \rightarrow &    \Omega_{m} \chi  \Omega_{m}^\dagger
=\sum_a \chi_a e^{i \vec \alpha_a \cdot \vec \theta_{m} } E_a.
\end{eqnarray}
In terms of the Weyl symmetry, it is useful to rewrite the field 
variable,
\begin{eqnarray} 
{\cal A}_\mu = \vec A_\mu \cdot \vec H =
\left( {\matrix{{ \vec A_\mu \cdot \vec \omega_1  }&{0}&{0}\cr
{0}&{ \vec A_\mu \cdot \vec \omega_2 }&{0}\cr
{0}&{0}&{\vec A_\mu \cdot \vec \omega_3 }\cr
}} \right)
\equiv
\left( {\matrix{{ A^{\rm R}_\mu } & {0} & {0}\cr
{0}&{ A_\mu^{\rm B}}&{0}\cr
{0}&{0}&{ A_\mu^{\rm G}}\cr
}} \right),
\end{eqnarray}
where $\vec \omega_{a}$($a = $1,2,3) denotes the SU(3) weight vector.
Here, $A^{\rm R}_{\mu}$, $A^{\rm B}_{\mu}$ and $A^{\rm G}_{\mu}$
satisfies the relations, 
\begin{eqnarray}
\sum_{a \in \{ R,B,G \}} A^a_{\mu}(x)  & =  &0,
\label{rel1} \\
(A^{\rm R}_\mu)^2 + (A^{\rm B}_\mu)^2 + (A^{\rm G}_\mu)^2
& = & \frac12 \{ (A^{\rm 3}_\mu)^2 + (A^{\rm 8}_\mu)^2 \} 
\label{rel2}
\end{eqnarray}
and
\begin{eqnarray}
(\partial \wedge \vec A)^2_{\mu\nu}  & = & 
(\partial \wedge  A^3)^2_{\mu\nu}  + (\partial \wedge A^8)^2_{\mu\nu} 
\nonumber \\ 
& = & 2 \{ (\partial \wedge  A^{\rm R})^2_{\mu\nu} 
+(\partial \wedge A^{\rm B})^2_{\mu\nu} +
(\partial \wedge  A^{\rm G})^2_{\mu\nu} \}.
\label{rel3}
\end{eqnarray}
Similarly, the dual gauge field ${\cal B}_\mu$ and the quark field $q$ are written as
\begin{eqnarray} 
{\cal B}_\mu  \equiv\left( {\matrix{ B^{\rm R}_\mu &{0}&{0}\cr
{0}&{ B_\mu^{\rm B}}&{0}\cr
{0}&{0}&{ B_\mu^{\rm G}}\cr
}} \right),
\hspace{1cm} q \equiv 
\left( {\matrix{q^R\cr
q^B\cr
q^G\cr
}} \right).
\end{eqnarray}
Here, $B_{\mu}^a$ also satisfies the similar relations to Eqs. 
(\ref{rel1})-(\ref{rel3}).
The DGL Lagrangian in the gauge  sector is  expressed as
\begin{eqnarray} 
{\cal L}^{\rm gluon} 
 & =  &\sum_{a \in \{ R,B,G \}} {\cal L}_a^{\rm gluon}  \nonumber \\
{\cal L}_a ^{\rm gluon}   & =   &-{1 \over {n^2}}[n\cdot (\partial \wedge  A)_a ]^\nu
 [n\cdot {}^*(\partial \wedge B_a )]_\nu  
 +{1 \over {n^2}}[n\cdot (\partial \wedge  B)]^\nu 
[n\cdot {}^*(\partial \wedge  A_a )]_\nu  \nonumber \\
& & -{1 \over {n^2}}[n\cdot (\partial \wedge  A_a )]^2
-{1 \over {n^2}}[n\cdot (\partial \wedge B_a )]^2. 
\end{eqnarray}
The quark sector in the DGL theory is written as 
\begin{eqnarray}
{\cal L}^{\rm quark} 
 & =  &\sum_{a \in \{ R,B,G \}} {\cal L}_a^{\rm quark}  
 \hspace{0.2cm} \mbox{with}  \hspace{0.2cm}
{\cal L}_a^{\rm quark}   =
 \bar q_a (i\gamma_\mu  \partial^\mu  - e \gamma_\mu  A^\mu_a
-m)q_a.
\end{eqnarray}
In this expression, the U(1)$_{\rm R} \times$ U(1)$_{\rm B} \times$ 
U(1)$_{\rm G}$
symmetry seems to  hold, however, one of U(1) symmetries is fixed by
the constraint as  (\ref{rel1}).
Under the Weyl transformation as
$A_\mu^a \leftrightarrow A_\mu^b$, 
$B_\mu^a \leftrightarrow B_\mu^b$ and 
$q_a \leftrightarrow q_b$ with
$a$, $b \in$ \{R,B,G\}, the Lagrangian $\sum_{a \in \{ R,B,G \}} {\cal L}_a$
is manifestly invariant.
Since the Lagrangian including the monopole field $\chi$ is also invariant under
the Weyl transformation as
$\chi \rightarrow \chi' = W \chi W^\dagger$  with  $W$ $\in$ Weyl,
the DGL Lagrangian has the Weyl global symmetry as the relic of the 
SU(3) symmetry.
In this way, the dual Ginzburg-Landau theory has  the 
[U(1)$_3$ $\times$ U(1)$_8$]$^{local}_{e}$
 $\times$
[U(1)$_3$ $\times$ U(1)$_8$]$^{local}_{m}$ $\times$
Weyl symmetry.

\section{Dual Meissner Effect  by   Monopole Condensation}

Based on the dual superconductor picture,
 monopole condensation leads to the color confinement,
which is brought as the result that the color-electric field is excluded from the 
QCD vacuum through dual Meissner effect.
In this section, we show the dual Meissner effect caused by
 monopole condensation in term of the DGL theory.
 
We separate the monopole field $\chi_a$ into the mean field 
$\langle \chi_{\alpha} \rangle = v$ and  its fluctuation $\tilde \chi_a$,
\begin{eqnarray}
\chi _a =(v +\tilde \chi_a )e^{i\xi _a },
\label{eq:mean}
\end{eqnarray}
where the monopole condensate does not depend on $a$ because of
the Weyl symmetry.
Due to the constraint $\sum_{a=1}^3$ arg$\chi_{a}=0$, there are only 
two independent degrees of freedom among the three phase variables
$\xi_{a}$ ($a=$1,2,3).
Using this expression (\ref{eq:mean}), 
the Lagrangian (\ref{eq:saishuu}) is expressed as
\begin{eqnarray}
{\cal L}_{DGL}& = & -{1 \over {2n^2}}
[n\cdot (\partial \wedge \vec A)]^\nu 
[n\cdot {}^* (\partial \wedge \vec B)]_\nu 
+{1 \over {2n^2}}[n\cdot (\partial \wedge \vec B)]^\nu 
[n\cdot {}^*(\partial \wedge \vec A)]_\nu  \nonumber \\
& & -{1 \over {2n^2}}[n\cdot (\partial \wedge \vec A)]^2
-{1 \over {2n^2}}[n\cdot (\partial \wedge \vec B)]^2 \nonumber \\
& & 
{+\bar q(i\gamma_\mu  \partial^\mu  - e \gamma_\mu \vec A^\mu
\cdot \vec H-m)q}
+{1 \over 2}m_B^2\vec B^2+\sum\limits_{a =1}^3 
{\left[ {(\partial _\mu \tilde \chi _a )^2-
m_\chi ^2 \tilde \chi_a^2} \right]}  \nonumber \\
& & +\sum\limits_{a =1}^3 
{\left[ 
{g^2(\vec \alpha _a \cdot \vec B_\mu )^2
(\tilde \chi _a ^2+2\tilde \chi _a v)^2
-\lambda (4v\tilde \chi _a ^3
+\tilde \chi _a ^4)} \right]},
\end{eqnarray}
where $m_{\rm B}=\sqrt 3gv$ and  $m_\chi =2\sqrt \lambda v$
denote  masses of the dual gauge field ${\cal B}_\mu$ 
and  the monopole $\chi_a$, respectively.
Thus, monopole condensation makes the  dual gauge field massive and 
the two independent phase variables of the monopole field  are changed into the
longitudinal degrees of freedom of the dual gauge field,
which is the dual Higgs mechanism.
Here, the color electric field ${\vec {\bf E}}$ cannot propagate
over the longer distance 
than $1/m_B$ and the dual Meissner effect occurs.
In this way, the color-electric flux is confined into the QCD vacuum
by the similar mechanism on the magnetic flux in the superconductivity.

\section{Color-flux-tube in the Monopole-Condensed Vacuum}

\begin{figure}[tb]
\epsfxsize = 10 cm
\centering \leavevmode
\epsfbox{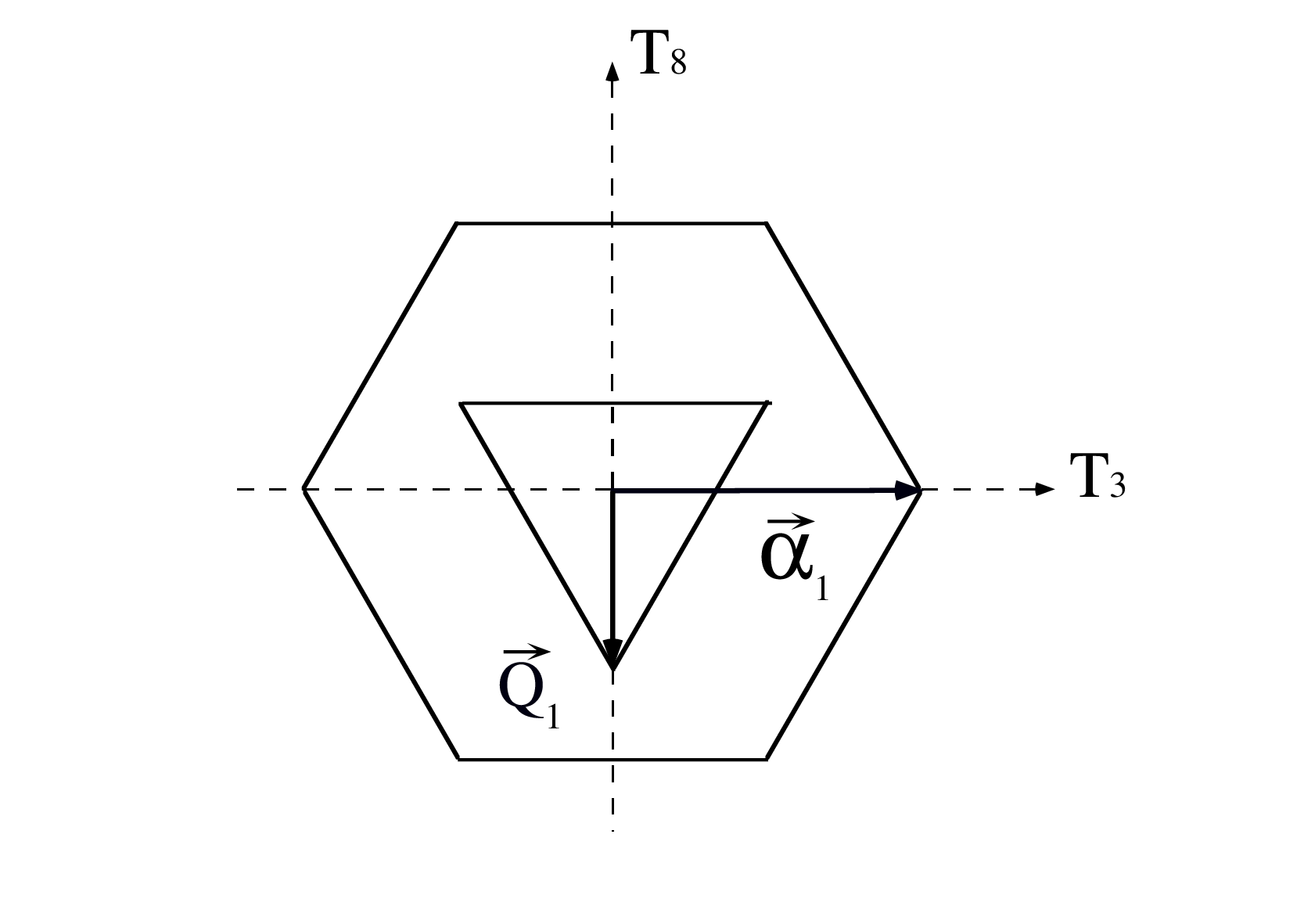}
\vspace{0cm}
\caption{ 
The electric charge of quarks and the magnetic charge
of monopoles. Since the quark color $\vec Q_1$ is vertical to
the magnetic charge $g \vec \alpha_1$ of the monopole field $\chi_1$,
condensation of $\chi_1$ does not contribute to  confinement of the color
$\vec Q_1$.
}
\label{fig:3E}
\end{figure}

As a result of the dual Meissner effect, 
the color electric flux is squeezed 
like a 
tube 
in the monopole condensed vacuum. 
In this section, 
we now investigate the structure of 
flux-tube with a quark and an antiquark at the both ends. 
In the standard notation,
 \cite{suganuma,kerson}, 
the quark charges  are $\vec Q_a \equiv e\vec w_a$,
where $\vec w_a(a=1,2,3)$ are the weight vectors of the SU(3) algebra,
$\vec w_1 = (0,- \frac{1}{\sqrt{3}})$,
$\vec w_2 = (-\frac{1}{2}, \frac{1}{2\sqrt{3}})$,
$\vec w_3 = (\frac{1}{2}, \frac{1}{2\sqrt{3}})$,
for the three color 
states \cite{kerson},
red($R$), blue($B$) and green($G$), respectively.
Using the Gauss law, one finds the color electric field $\vec{\bf{ E}}$
and then the dual gauge field $\vec B_\mu$, obeying 
$\vec F_{\mu\nu}={}^*(\partial \wedge \vec B)_{\mu\nu}$,
are proportional to  the color charge $\vec Q$.
The monopole $\chi_a$ couples with the quark charge $\vec Q_b$
in the form of $\vec \alpha_a \cdot \vec Q_b \chi_a$.
We note the algebraic relation between the root vector and the weight 
vector as,
\begin{eqnarray}
2 \vec w_a \cdot \vec\alpha_b= \sum_{c=1}^3 \varepsilon_{abc}
 \in\{-1,0,1\}, 
\end{eqnarray}
and therefore one kind of monopole $\chi_a$ couples with
not three but  two of the quark charges as is shown in Fig.\ref{fig:3E}.

For the example, in the case of ($R$-$\bar R$) system,
$|\chi_1|$ is never affected; $|\chi_1| = v$.
For this case, we can rewrite Eq.(\ref{eq:saishuu}) as, 
\begin{eqnarray}
{\cal L}_{DGL}=-{1 \over 3}\cdot{1 \over 4}
(\partial _\mu B^{\rm R}_\nu -\partial _\nu B^{\rm R}_\mu )^2
+2\left| {(\partial _\mu +\frac{i}{2} gB^{\rm R}_\mu )\chi^{\rm R}} \right|^2
-2\lambda (\left| \chi ^{\rm R}  \right|^2-v^2)^2,
\end{eqnarray}
where $\vec B_\mu \equiv \vec w_1 B_\mu ^{\rm R} $ and 
$|\chi_1| = v$, $\chi_2=\chi^{{\rm R}}$ and $\chi_3=\chi^{\rm R*}$, because 
the system is invariant under the transformation,
$\chi_2 \leftrightarrow                \chi_3$. 
In this case, we can rewrite the DGL lagrangian in the simple GL form;
\begin{eqnarray}
{\cal L}_{DGL}=-{1 \over 4}
(\partial _\mu \hat B_\nu -\partial _\nu \hat B_\mu )^2
+\left| {(\partial _\mu +i\hat g\hat B_\mu )
\hat \chi _{}} \right|^2
-\hat \lambda (\left| {\hat \chi } \right|^2-\hat v^2)^2,
\label{eq:GL1}
\end{eqnarray}
where the field variables and coupling constants are redefined as
\begin{eqnarray}
\hat B_\mu\equiv {1 \over {\sqrt 3}}B^{\rm R}_\mu,
 \hspace{0.5cm}
\hat g\equiv {{\sqrt 3} \over 2}g,
 \hspace{0.5cm}
\hat \chi \equiv \sqrt 2\chi^{\rm R},
 \hspace{0.5cm}
\hat v\equiv \sqrt 2v,
 \hspace{0.5cm}
\hat \lambda \equiv {\lambda \over 2}.
\end{eqnarray}
We get the same expression for the other two cases ($B$-$\bar B$,
$G$-$\bar G$).
Hereafter, we will drop the notation $\hat{}$ since there is no confusion.
The field equations for $B_\mu$  and $\chi$ are derived by the 
extreme condition,
\begin{eqnarray}
\partial ^2\chi +2igB^\mu (\partial _\mu \chi )
+ig(\partial _\mu B^\mu )\chi
-g^2B_\mu^2\chi +2\lambda (|\chi |^2-v^2)^2\chi =0,
\end{eqnarray}
\begin{eqnarray}
\partial _\mu (\partial ^\mu B^\nu
-\partial ^\nu B^\mu )
+ig\{(\partial ^\nu \chi ^*)\chi
-(\partial ^\nu \chi )\chi ^*\}
+2g^2B^\nu |\chi |^2=0.
\end{eqnarray}
For these equations, there is a static solution of vortex,
which is known as the Nielsen-Olesen vortex \cite{nielsen}.
Using the cylindrical coordinate($r,\theta,z$),  we consider the static solution satisfying
\begin{eqnarray}
{\bf B}  =  B_\theta {\bf e}_\theta, \hspace{2cm}
{\bf E}  =  \frac{1}{r} \frac{d}{dr}[rB(r)]
{\bf e}_z,  \hspace{2cm}
\chi  =   \bar \chi (r) e^{in\theta}
\end{eqnarray}
under the suitable boundary condition.

\begin{figure}[tb]
\epsfxsize = 10 cm
\centering \leavevmode
\epsfbox{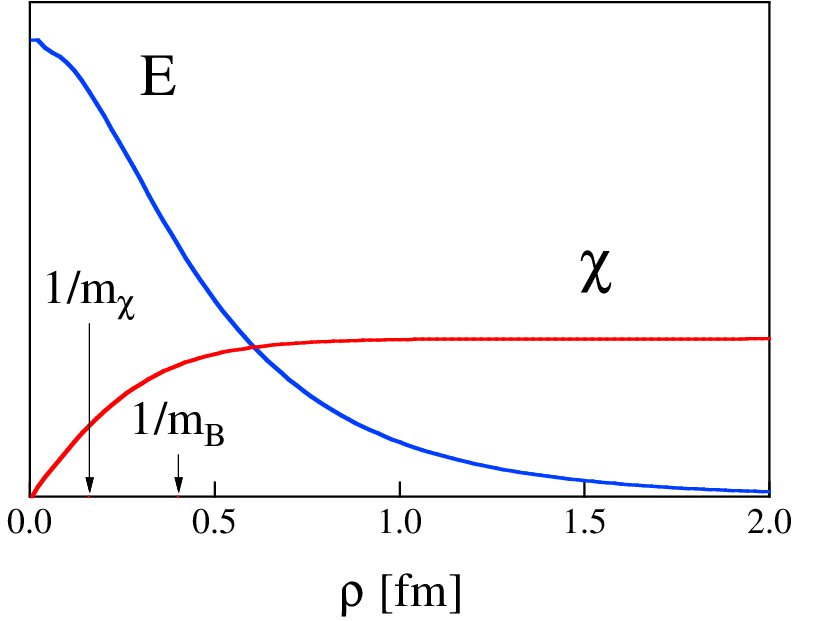}
\vspace{0cm}
\caption{    
The color-electric field $E(r)$ 
the monopole condensate $\bar \chi(r)$ 
are plotted as the function of $r$ in the cylindrical coordinate.
}
\label{fig:3A}
\end{figure}


We show in Fig.\ref{fig:3A} the profile of the color electric-field  $E(r)$ and 
the monopole field $\chi(r)$.
Apart from $r=0$, the monopole  condenses,  
while the monopole condensate disappears near the center of flux-tube.
Accordingly,
the color electric-field is
squeezed like a one-dimensional tube, whose radius is about $1/m_B$.
In this way, in the QCD vacuum, there are three types of the flux-tubes
according to the color of quarks, being different from the superconductivity.
However,   the DGL Lagrangian in the all color system becomes a simple GL type
and there is a solution where  the color-electric field is squeezed  like 
a one-dimensional tube as the Abrikosov vortex.
This result is consistent with the Regge trajectory of hadrons and 
the lattice QCD results,  which indicates that the QCD vacuum can be 
regarded as the dual version of the superconductor.

\section{Quark Confinement Potential}

In the DGL theory,
the color-electric flux between quarks is squeezed 
into {\it one-dimensional tube} as the result of  monopole condensation.
This indicates that the static potential between quark and anti-quark is
linear confinement potential.
Here, we investigate the interquark potential using the 
gluon propagator including the nonperturbative effect
 \cite{suganuma}.

The static potential between heavy quarks can be obtained from 
the vacuum energy where the static  quark and antiquark exist.
Here, we  take a quench approximation, i.e. we neglect the quantum effect of the
light quarks.  
As a first step, we  approximate the monopole filed $\chi_a$ as the mean field 
$|\chi_a| = v$ neglecting the effect of the quark sources on the 
condensed monopole.  
Later, we  include the correction  in order to eliminate the
divergence originated from this approximation. 
In this scheme, information on
 confinement is included in the gluon propagator, which leads to 
the strong interaction in the infrared region.

The vacuum energy $V(j) $ in the presence of the static quark sources $j$ 
is obtained from 
\begin{eqnarray}
Z  =  <0|e^{i \int\vec j_\mu \vec A^\mu d{^4}x}|0> 
 =  N\int {\cal D}\vec A_\mu {\cal D}\vec B_\mu 
e^{i\int {({\cal L}
+\vec j_\mu \vec A^\mu )d^4x}}  = e^{-iV(j)\int {dt}}.
\end{eqnarray}
The Lagrangian on the mean field level for the monopole field is written as
\begin{eqnarray}
{\cal L}_{DGL-MF} & = & -{1 \over {2n^2}}
[n\cdot (\partial \wedge \vec A)]^\nu 
[n\cdot {}^* (\partial \wedge \vec B)]_\nu 
+{1 \over {2n^2}}[n\cdot (\partial \wedge \vec B)]^\nu 
[n\cdot {}^*(\partial \wedge \vec A)]_\nu  \nonumber \\
& & -{1 \over {2n^2}}[n\cdot (\partial \wedge \vec A)]^2
-{1 \over {2n^2}}[n\cdot (\partial \wedge \vec B)]^2 \nonumber \\
& & 
{+\bar q(i\gamma_\mu  \partial^\mu   - e \gamma_\mu \vec A^\mu
\cdot \vec H-m)q}
+{1 \over 2}m_B^2\vec B_\mu^2.
\end{eqnarray}
Integrating out the dual gauge field $B_\mu$,  it becomes
\begin{equation}
{\cal L}_{DGL-MF} = -{1 \over 4}\vec f_{\mu \nu }
\vec f^{\mu \nu }+{1 \over 2}\vec A^\mu K_{\mu \nu }\vec A^\nu
{+\bar q(i\gamma_\mu  \partial^\mu   - e \gamma_\mu \vec A^\mu
\cdot \vec H-m)q}
\end{equation}
with
\begin{eqnarray}
K_{\mu \nu } & \equiv & {{n^2m_B^2} 
\over {(n\cdot \partial )^2+n^2m_B^2}}X_{\mu \nu } \\
X^{\mu \nu } & = & {1 \over {n^2}}
\epsilon _\lambda ^{\mu \alpha \beta }
\epsilon ^{\lambda \nu \gamma \delta }
n_\alpha n_\gamma \partial _\beta \partial _\delta \nonumber \\
& = & {1 \over {n^2}}[-n^2\partial ^2g^{\mu \nu }
+(n\cdot \partial )^2g^{\mu \nu } \nonumber 
+n^\mu n^\nu \partial ^2-(n\cdot \partial )
(n^\mu \partial ^\nu +n^\nu \partial ^\mu )
+n^2\partial ^\mu \partial ^\nu ],
\end{eqnarray}
where $X_{\mu\nu}$ satisfies the relation, 
$X_{\mu \nu} = X_{\nu \mu}$ and 
$X_{\mu \nu }\partial ^\nu =X_{\mu \nu }n^\nu =0$. 
When we introduce the external source $j$ in stead of dynamical quark part,
we get 
\begin{eqnarray}
{\cal L}_{DGL-MF} & = &
{1 \over 2}\vec A^\mu D^{-1}_{\mu \nu }\vec A^\nu 
+\vec j_\mu \vec A^\mu  \\
& = & {1 \over 2}(\vec A^\mu +\vec j_\alpha D^{\alpha \mu })
D_{\mu \nu }^{-1}(\vec A^\nu +D^{\nu \beta }\vec j_\beta )
-{1 \over 2}\vec j_\mu D^{\mu \nu }\vec j_\nu.
\end{eqnarray}
Integrating out $A_\mu$, we obtain the non-local current-current correlation as
\begin{eqnarray}
{\cal L}_j=-{1 \over 2}\vec j_\mu D^{\mu \nu }\vec j_\nu, 
\end{eqnarray}
where the propagator of the diagonal gluons is written as
\begin{eqnarray}
D_{\mu \nu }^{-1} &  = & \left( g_{\mu \nu }\partial ^2
-(1-{1 \over {\alpha _e}} )\partial _\mu \partial _\nu \right) 
+{{m_B^2n^2} \over {(n\cdot \partial )^2
+m_B^2n^2}}X_{\mu \nu } \label{eqn:dinv}\\
D_{\mu \nu } & = & {1 \over {\partial ^2}} \left( {g_{\mu \nu }
+(\alpha _e-1) 
{{\partial _\mu \partial _\nu } \over {\partial ^2}}} \right)
-{1 \over {\partial ^2}}{{m_B^2} \over 
{\partial ^2+m_B^2}}{{n^2} \over {
(n\cdot \partial )^2}}X_{\mu \nu }
\label{eqn:dmn}
\end{eqnarray}
in the Lorentz gauge $\partial_\mu A^\mu =0$.
If the monopole does not condense $v=0$ or $m_B=0$,  
the second term in (\ref{eqn:dmn}) disappears, and this propagator 
returns to familiar  propagator derived in the perturbative sense.
The nonperturbative effect is included in the 
second term. Indeed, the second term leads to the confinement potential
as is shown from now on.

The generating functional is given as 
\begin{eqnarray}
Z=e^{i\int {{\cal L}_j d^4x}}\equiv e^{iS_j}=e^{-iV(j)\int {dt}}.
\label{eqn:hankan}
\end{eqnarray}
Now we try to estimate
\begin{eqnarray}
S_j\equiv \int {{\cal L}_jd^4x}
=-{1 \over 2}\int {}\vec j_\mu 
D^{\mu \nu }\vec j_\nu d^4x.
\label{eq:sdef}
\end{eqnarray}
The static source of  the quark with charge $\vec Q_a$ located at $a$ and
the anti-quark $\vec Q_b$ at $b$  is given by 
\begin{eqnarray}
\vec j_\mu (x)=- g_{\mu 0}\{{\vec Q_a\delta ^3
( {\bf {x-a}}  )
+\vec Q_b\delta ^3({\bf{x-b}})} \}
\end{eqnarray}
and its Fourier transformation leads to
\begin{eqnarray}
\vec j_\mu (k)=-g_{\mu 0}( {\vec Q_a}e^
{-i \bf k \cdot \bf a}
+{\vec Q_b}e^{-i  \bf k \cdot \bf b}).
\label{eqn:sour}
\end{eqnarray}
Substituting the propagator (\ref{eqn:dmn}) and the source (\ref{eqn:sour})
into (\ref{eq:sdef}), we get using the unit vector $n$
\begin{eqnarray}
\lefteqn{ S_j   =  - {1 \over 2}\int {{{d^4k} \over {(2\pi )^4}}}
\vec j_\mu (-k)D^{\mu \nu }\vec j_\nu (k)  }\\
& = & {1 \over 2}\int {{{d^4k} \over {(2\pi )^4}}}
\vec j_\mu (-k)
[{1 \over {k^2-m_B^2}}g^{\mu \nu }
+{{-m_B^2} \over {k^2-m_B^2}}{{n^2} \over {(n\cdot k)^2}}
(g^{\mu \nu }-{{n^\mu n^\nu } \over {n^2}})]
\vec j_\nu (k) \nonumber \\
& = &-{1 \over 2}\int {{{d^3k} \over {(2\pi )^3}}}
(\vec Q_a+\vec Q_be^{i \mbox{\boldmath k$\cdot$r}})
(\vec Q_a+\vec Q_be^{-i \mbox{\boldmath k$\cdot$r}})
\times [{1 \over {\mbox{\boldmath k}^2+m_B^2}}+{{m_B^2} 
\over {\mbox{\boldmath k}^2+m_B^2}}
 {1 \over {(\mbox{\boldmath n$\cdot$k})^2}}]\int {dt} \nonumber ,
\end{eqnarray}
where {\bf r} $\equiv ${\bf b}$-${\bf a}
is relative vector between quark and anti-quark, and 
$2 \pi \delta (0) = \int {dt}$ is used.
Here, we take $n//{\bf r}$, because of the axial symmetry of the system 
and the energy minimum condition.
\begin{figure}[tb]
\epsfxsize = 10 cm
\centering \leavevmode
\epsfbox{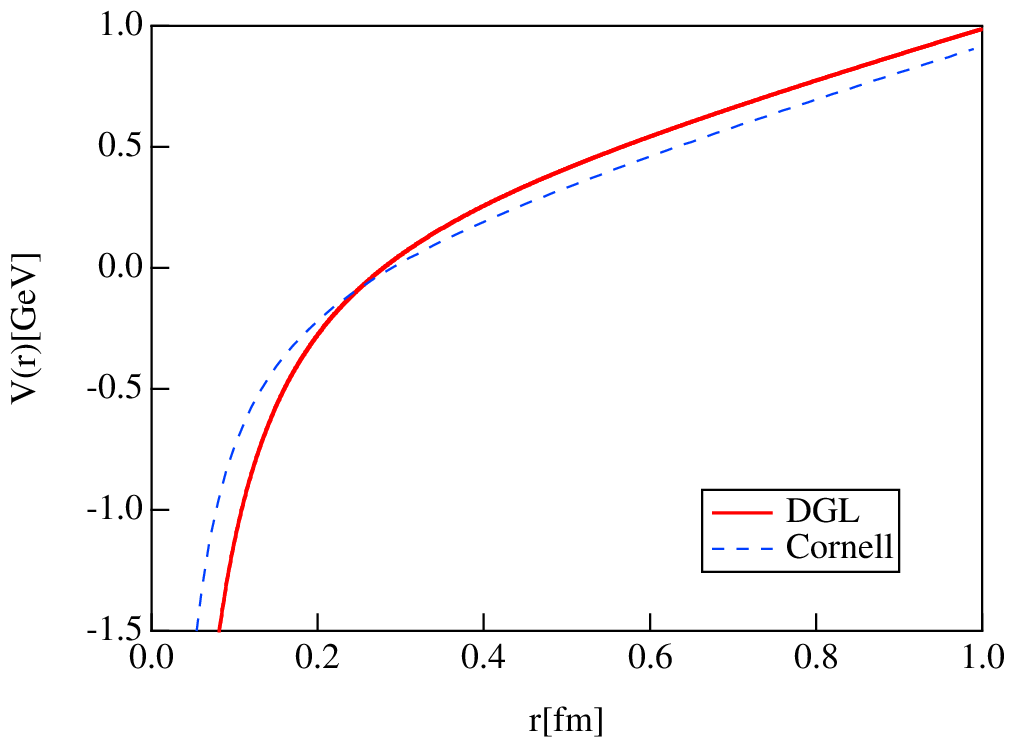}
\vspace{0cm}
\caption{  Confinement potential $V(r)$ 
as the function of the distance $r$ between
a static quark-antiquark.}
\label{fig:3D}
\vspace{0cm}
\end{figure}
The inter-quark potential  is obtained from (\ref{eqn:hankan})  as
\begin{eqnarray}
V(r;n) \equiv V_{\rm yukawa}(r)+V_{\rm linear}(r;n),
\end{eqnarray}
where  the yukawa term is 
\begin{eqnarray}
V_{\rm yukawa}(r) & = & {1 \over 2}\int {{{d^3k} \over 
{(2\pi )^3}}} 
(\vec Q_a+\vec Q_b e^{i {\mbox{\boldmath {k$\cdot$r}}}})
(\vec Q_a+\vec Q_be^{-i{\mbox{\boldmath k$\cdot$r}}} )  \nonumber
{1 \over {\mbox{\boldmath k}^2+m_B^2}}
 \nonumber \\
& = & \vec Q_a\cdot \vec Q_b\int {{{d^3k} \over 
{(2\pi )^3}}}{{e^{i{\mbox{\boldmath k$\cdot$r}}}} 
\over {\mbox{\boldmath k}^2+m_B^2}} 
 =  {{\vec Q_a\cdot \vec Q_b} \over {4\pi }}
{{e^{-m_Br}} \over r} 
\end{eqnarray}
and the linear term is 
\begin{eqnarray}
V_{\rm linear}(r;n) & = &{1 \over 2}\int {{{d^3k} 
\over {(2\pi )^3}}} (\vec Q_a+\vec Q_be^{i\mbox{\boldmath k$\cdot$r}})
(\vec Q_a+\vec Q_be^{-i{\mbox{\boldmath k$\cdot$r}}})
{{m_B^2} \over {\mbox{\boldmath k}^2+m_B^2}}{1 \over 
{\mbox{\boldmath (n$\cdot$k)}}^2} \nonumber \\  
& = & {1 \over 2}\int {{{d^3k} \over {(2\pi )^3}}}
\left(
{\vec Q_a^2+\vec Q^2_b+2\vec Q_a \cdot \vec Q_b
{\rm cos}( {\mbox{\boldmath k$\cdot$r}})}
\right)
{{m_B^2} \over {\mbox{\boldmath k}^2+m_B^2}}{1 \over
\mbox{\boldmath (n$\cdot$k)}^2}.
\end{eqnarray}
Introducing the ultraviolet cutoff $m_\chi$ corresponding to the core of the flux-tube,
we  get the expression,
\begin{eqnarray}
V_{\rm linear}(r)={1 \over {16\pi ^2}}
\int_{-\infty} ^\infty  {{{dk_r} \over {k_r^2}}}
({\vec Q_a^2+\vec Q^2_b+2\vec Q_a\cdot 
\vec Q_b\cos (k_rr)})
\ln ({{m_\chi ^2+k_r^2+m_B^2} \over {k_r^2+m_B^2}}),
\end{eqnarray}
where $k_r$ is parallel component of $k$ to the Dirac-string direction $n$.
Since  $\vec Q_a\cdot
\vec Q_b$  is $- \frac13 e^2$ for any color system, we finally get
\begin{eqnarray}
V_{}(r)=-{e^2 \over {12\pi }}
\cdot {{e^{-m_Br}} \over r}
+{{e^2m^2_B} \over {24\pi ^{}}}
\ln ({{m_\chi ^2+m_B^2} \over {m_B^2}}) r.
\end{eqnarray}
The string tension is given from the slope of the linear confinement potential
as 
\begin{eqnarray}
\sigma={{e^2m_B^2} \over {24\pi }} \ln 
\left( {{{m^2_B+m^2_\chi } \over {m^2_B}}} \right),
\label{eq:strten}
\end{eqnarray}
which corresponds to the vortex energy for the unit length in the typeII 
superconductor.
Here, the gauge field mass ${m_A^2}$ is neglected in the numerator
in the logarithm function, because it is much smaller than the mass of the Higgs field
(Cooper-pair field), ${m_A^2} \ll {m_\phi ^2}$.
Thus, the confinement potential arises from the second term
of the gluon propagator (\ref{eqn:dmn}).

The parameters in the dual Ginzburg-Landau theory is determined so as to
reproduce 
the interquark potential of heavy quarknium or  
results of the lattice QCD simulation;
$e=5.5$, monopole condensate $v=0.126$GeV and interaction strength between 
the monopoles $\lambda = $25.
These parameters lead to the unit magnetic charge $g=2.3$,
the mass of the dual gauge field  $m_B = 0.5$GeV, monopole field
mass $m_\chi = 1.26$GeV  and the string tension $\sigma = 1.0$GeV/fm.
Using these parameters,
we compare in Fig.\ref{fig:3D} the phenomelogical potential
like Cornell potential \cite{cor}
\begin{eqnarray}
V_{\rm Cornell}=-{{e^2_c} \over {3\pi }}{1 \over r}+\sigma_cr.
\end{eqnarray}
In the short range $r \le 0.2$fm= (1GeV)$^{-1}$, the Coulomb term  
dominates, while the linear potential dominates in the longer distance.

Thus, in the DGL theory, the gluon propagator includes the long range 
correlation between quarks 
through the non-local operator $\frac{1}{(n \cdot \partial)^2}$.
Such a long range correlation along  the direction of the Dirac string 
leads to the linear confinement potential.

\section{Asymptotic Behavior in DGL Theory}

Finally, we discuss the gauge coupling behavior in the DGL theory
in terms of the renormalization group \cite{suganuma5}.
In the abelian gauge fixing, the Dirac condition $eg=4\pi $ for the 
dual gauge coupling constant $g$ is derived  \cite{thooft,suganuma}. 
Since Dirac condition is a geometrical relation, 
it is renormalization group invariant. 
Therefore, 
``see-saw relation" between the gauge coupling constant $e$ and
the dual gauge coupling constant $g$ is expected;
a strong coupling system in one sector corresponds to a weak-coupling 
system in the other sector \cite{suganuma,suganumaB}. 

\begin{figure}[htbp]
\epsfxsize = 10 cm
\centering \leavevmode
\epsfbox{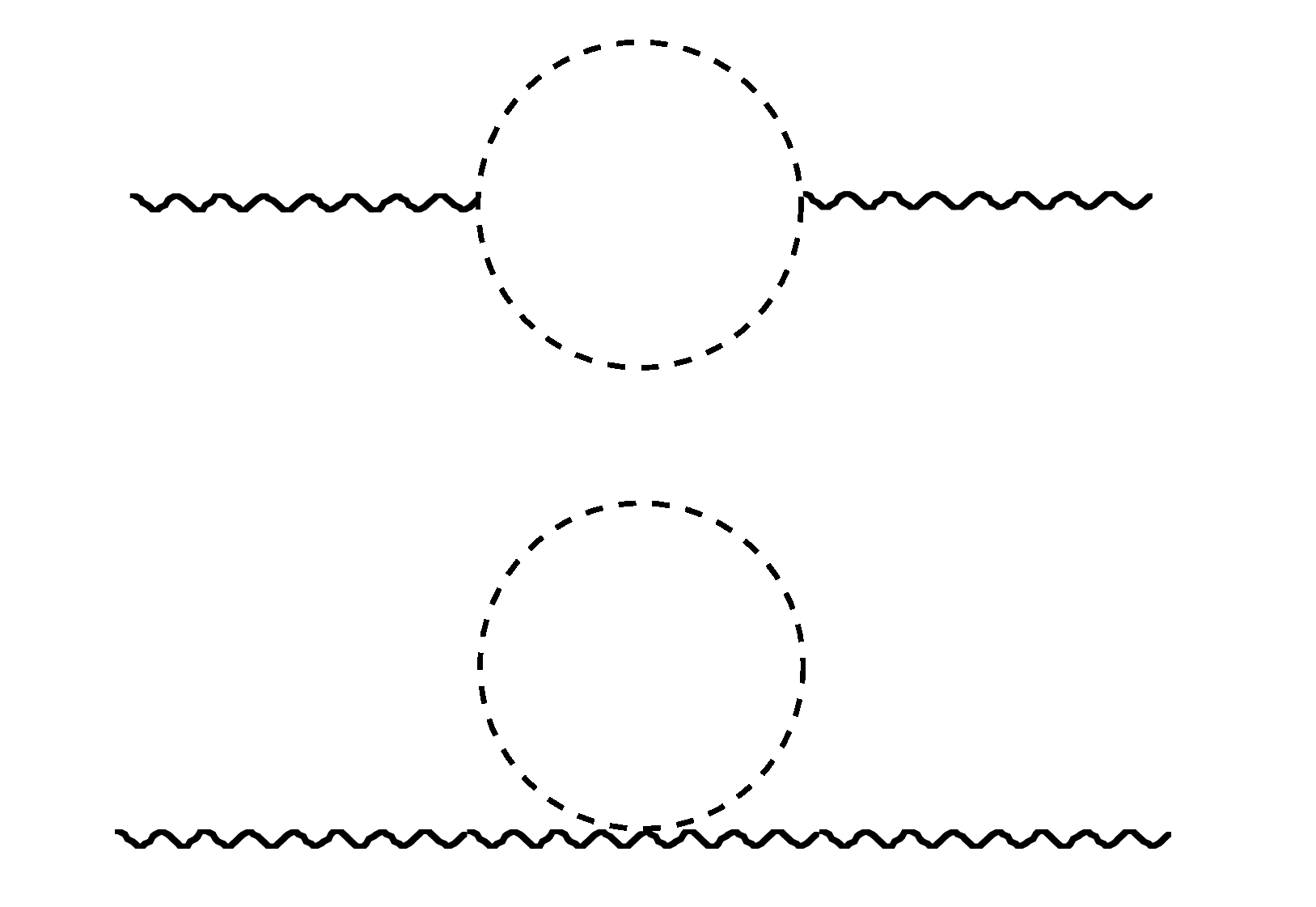}
\vspace{0cm}
\caption{    
The lowest-order polarization diagrams of the dual gauge field $B_\mu$, 
which is denoted by the wavy line, in the DGL theory. The dotted line 
denotes the monopole field.}
\label{fig:3C}
\vspace{0cm}
\end{figure}

The DGL theory in the pure gauge is renormalizable similar to
the abelian Higgs model \cite{itzykson},
and is not asymptotically 
free on $g$ in view of the renormalization group: 
$g(p^2)$ become large as  $p^2$ increases.
Hence, asymptotic freedom is expected for the QCD gauge coupling 
constant $e$ owing to the Dirac condition: 
$e(p^2)$ defined by $e(p^2)g(p^2)=4\pi$ become small
as $p^2$ increases.
Thus, the DGL theory qualitatively shows asymptotic freedom 
on the QCD gauge coupling $e$  \cite{suganuma,suganumaB}. 
This asymptotic behavior in the DGL theory is consistent 
with QCD qualitatively, 
and seems a desirable feature for an effective theory of QCD.

\begin{figure}[p]
\epsfxsize = 10 cm
\centering \leavevmode
\epsfbox{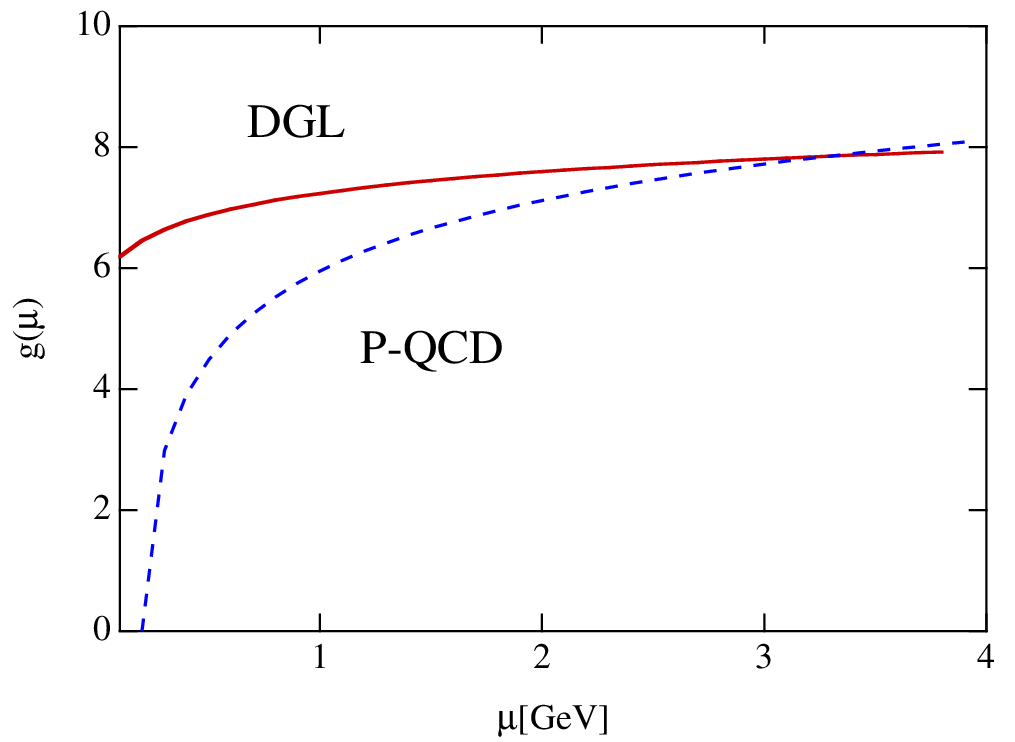}
\vspace{0cm}
\caption{    
 The running coupling constant $g(\mu)$ as the function of the
renormalization point $\mu$. The solid curve denotes the result in the
DGL theory, which is directly calculated.
The dashed curve denotes the leading-order perturbative QCD result,
where $g(\mu)$ is obtained using the Dirac condition $eg=4\pi$.
}
\label{fig:3B}
\vspace{2cm}
\epsfxsize = 10 cm
\centering \leavevmode
\epsfbox{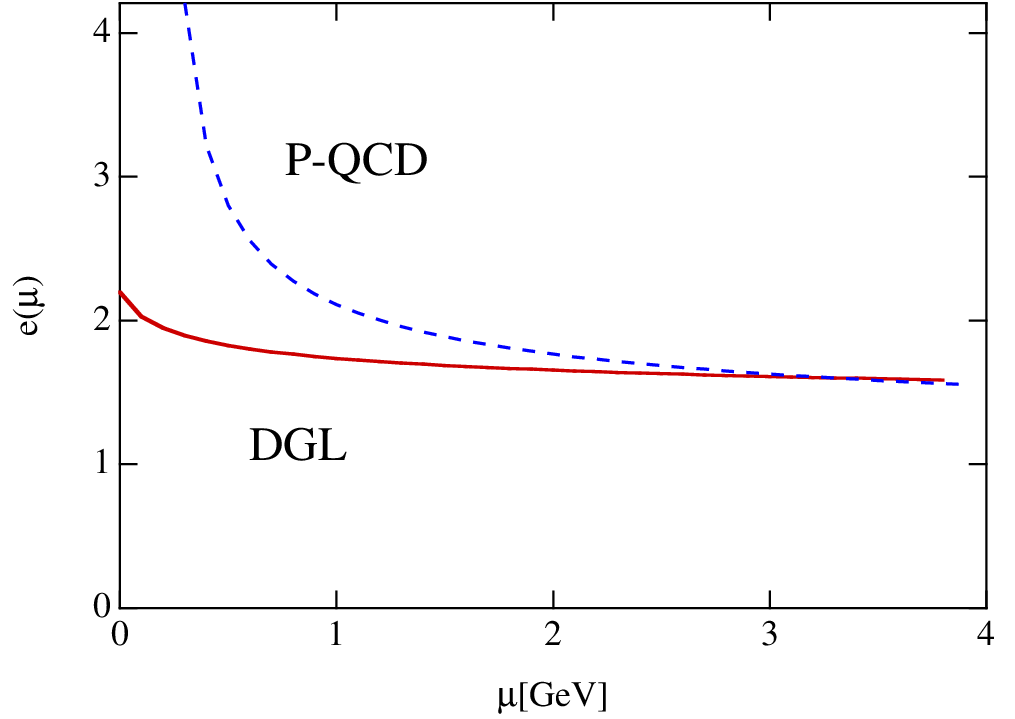}
\vspace{0cm}
\caption{   
The running coupling constant $e(\mu)$ as the function of the
renormalization point $\mu$. The solid curve denotes the result in the
DGL theory, where $e(\mu)$ is obtained using the Dirac condition $eg=4\pi$.
The dashed curve denotes the leading-order perturbative QCD result.}
\label{fig:3B2}
\vspace{0cm}
\end{figure}

Next, we attempt to calculate the $\beta$-function and 
the running coupling constant 
from the polarization tensor $\Pi _{\mu \nu}^{ab}(p)$ 
of the dual gauge field $B_\mu$ in the DGL theory. 
In particular, we are interested in the infrared region 
($p \le 1{\rm GeV}$), where the perturbative QCD calculation 
is not reliable. 

With the dimensional regularization  \cite{cheng,itzykson} 
by shifting the space-time dimension as $d=4-\epsilon $, 
we calculate the simplest nontrivial radiative correction 
from the monopole loop diagrams as shown in Fig.\ref{fig:3C}, 
\begin{eqnarray}
\Pi _{\mu \nu}^{ab}(p)=-{1 \over 32\pi^2}\delta ^{ab}(g\mu^{-\epsilon })^2 
(p^2g_{\mu\nu}-p_\mu p_\nu){1 \over \epsilon } + O(\epsilon ^0), 
\label{POLa}
\end{eqnarray}
where $g$ is the bare dual-gauge coupling and 
$\mu$ the renormalization point.
In the minimum subtraction scheme \cite{cheng}, 
the O(1/$\epsilon $) divergence is eliminated by the counter term contribution,
\begin{eqnarray}
\Pi _{\mu\nu}^{{\rm C} ab}(p)=-(Z_3-1)\delta ^{ab}(p^2 g_{\mu\nu} - p_\mu p_\nu),
\label{POLb}
\end{eqnarray}
where $Z_3$ is the wave-function renormalization constant \cite{itzykson} 
of the dual gauge field $B_\mu$, 
\begin{eqnarray}
Z_3(\mu)=1-{(g\mu^{-\epsilon })^2 \over 32 \pi^2}{1 \over \epsilon }.
\label{Zfactor}
\end{eqnarray}
Because of the Ward identity $Z_1=Z_2$ \cite{itzykson}, 
the renormalized coupling constant is simply given by 
$g(\mu) = Z_3(\mu)^{1/2} g$. 
The $\beta $-function  \cite{cheng,itzykson}
is then expressed as 
\begin{eqnarray}
\beta  \equiv \mu {d \over d \mu}g(\mu)
=\mu {d \over d \mu}\{Z_3(\mu)^{1/2}g\}
={1 \over 32\pi^2}g(\mu)^3+O[g(\mu)^5], 
\label{BETAf}
\end{eqnarray}
which determines the behavior of the running coupling $g(\mu)$ as  
\begin{eqnarray}
{1 \over g^2(\mu)}={1 \over g^2(\mu_0)}
-{1 \over 32\pi^2}\ln(\mu^2/\mu_0^2)
\label{RUNa}
\end{eqnarray}
within the leading order.

By summation of the multi-polarization diagrams, one obtains 
the final formula for the running coupling $g(\mu)$ 
including the higher order correction, 
\begin{eqnarray}
g^2(\mu)=g^2(\mu_0)-{1 \over 32\pi^2}\ln(\mu^2/\mu_0^2).
\label{RUNb}
\end{eqnarray}
In the DGL theory, the QCD gauge coupling $e(\mu)$ defined 
by $e(\mu)g(\mu)=4\pi$ is simply expressed as 
\begin{eqnarray}
e^2(\mu)=e^2(\mu_0) \left\{
1+{1 \over e^2(\mu_0)}\ln(\mu/\mu_0)
\right\}^{-1}.
\label{RUNc}
\end{eqnarray}

We show in Fig.\ref{fig:3B} the running coupling constants, 
$g(\mu)$ and $e(\mu)$, in the DGL theory 
with the parameter set in Ref. \cite{suganumaC}: $m_\chi$=1.67GeV.
Similarly in QED or the abelian Higgs model, 
we have imposed the renormalization condition as 
$g(\mu=2m_\chi)=7.9$ [ $e(\mu=2m_\chi)=1.59$ ], which 
is taken to be consistent with the parameter 
$g=6.28$ ($e=2.0$) in Ref. \cite{suganumaC} in the infrared region 
(see Fig.\ref{fig:3B}).
This result shows that
the gauge coupling $e(\mu)$ behaves as ``walking coupling constant", 
which means the coupling varies slowly, 
even in the infrared region as $\mu \le 1 {\rm GeV}$.

Thus, owing to the Dirac condition $eg=4\pi$,
the DGL theory has asymptotic freedom nature on the gauge coupling 
constant $e$, where the ``walking coupling constant'' is predicted for $e$
even in the infrared region.

\newpage

\chapter{QCD Phase Transition at Finite Temperature} 
\label{sec:QCDPTFT}

The QCD vacuum is non-trivial at zero temperature.
In this vacuum, quarks and gluons are confined in hadrons and the chiral 
symmetry is spontaneously broken.
As the temperature increases, however, the color
degrees of freedom in hadrons are defrozen. 
Above the critical temperature, the QCD vacuum is in the quark-gluon plasma
(QGP) phase, where quarks and gluons move almost freely.

The QCD phase transition is investigated in the various field of physics.
As for the lattice QCD, the QCD phase transition is one of the most important 
subject in the computer science  and studied using the Monte Carlo 
simulation.
The simulation demonstrates that such a phase transition 
happens at about 280 MeV for the pure gauge case  \cite{karsch}
and at about 100$\sim$200MeV for the full QCD case \cite{kanaya}. 
In the early Universe, 
the QCD phase transition is considered to have occurred, i.e, the system was 
changed from the QGP phase into the hadron phase around the
200-300 MeV as the temperature decreases.
As for the actual experience, in the recent years, 
some experimental groups are trying 
to create  QGP 
in the laboratory using high-energy heavy-ion collisions.
The RHIC (Relativistic Heavy Ion Collier)  project is aimed at forming 
QGP and at studying its properties.



In this chapter, we  investigate the behavior of the color-confinement at 
high temperature by studying the change of the properties in the QCD vacuum 
with temperatures in terms of the 
dual Ginzburg-Landau theory \cite{ichieB, ichieA}. 
To this end, we concentrate on the pure-gauge QCD case, where 
glueballs appear as the physical excitation.
Although such a pure gauge system is different from 
the real world, it is regarded as a proto-type of  real QCD 
and is well studied by using the lattice QCD simulation \cite{rothe}.
It is worth mentioning that 
our framework based on the DGL theory can be extended to include 
the dynamical quarks straightforwardly  \cite{maedan,suganuma} 
keeping the chiral symmetry of the system, 
which is explicitly broken in the color-dielectric model  \cite{nielsen2}
or in the lattice QCD with the Wilson fermion \cite{rothe}.

\section{Effective Potential Formalism at Finite Temperature}

In order to find the stable vacuum in the field theory, the 
effective potential formalism is useful \cite{kapusta}.
The effective potential indicates the vacuum energy at zero temperature
and corresponds to the thermodynamical potential at finite temperature.
Here, we investigate the effective potential in the path integral formalism
in the DGL theory \cite{ichieB}. 
The partition functional is written as 
\begin{eqnarray}
  Z[J] = \int {\cal D}{\chi_{\alpha}}{\cal D}{\vec{B}_{\mu}}
\exp{\left( i\int d^4x \{{\cal{L}}_{\rm DGL}
-J\sum_{\alpha=1}^3|\chi_\alpha|^2\} \right) },
\label{eq:Zj}
\end{eqnarray}
where we take the quadratic source term  \cite{ring}
instead of the standard linear source term  \cite{cheng,kapusta}. 
As is well-known in the $\phi^4$ theory  \cite{cheng,kapusta}, 
the use of the linear source term 
leads to an imaginary mass of the scalar field $\chi _\alpha $
in the negative-curvature region of the 
classical potential, and therefore 
the effective action cannot be obtained there due to 
the appearance of ``tachyons". 
In this respect, there is an extremely advanced point in 
the use of the quadratic source term \cite{ring}, 
because the mass of the scalar field $\chi _\alpha $ is always real even 
in the negative-curvature region of the classical potential 
owing to the contribution of the source $J$
to the scalar mass (see Eq.(\ref{eq:Ma}). 
Then, one obtains the effective action for the 
whole region of the order parameter without any difficulty 
of the imaginary-mass problem. 
Moreover, the effective action with the quadratic source 
can be formulated keeping the symmetry of the classical potential.
Since this method with the quadratic source term is quite general, 
it is convenient to formulate the non-convex effective potential 
in the $\phi ^4$ theory, the linear $\sigma $ model 
or the Higgs sector in the unified theory \cite{cheng}. 

The vacuum expectation value of $\chi _\alpha $ ($\alpha $=1,2,3)
is the same value $\bar \chi $ due to the Weyl symmetry \cite{maedan}, 
and therefore we separate the monopole field $\chi_\alpha $ into 
its mean field $\bar \chi$ and 
its fluctuation $\tilde{\chi}_\alpha $ as
\begin{eqnarray}
  \chi_\alpha = \left( \bar \chi + \tilde{\chi_\alpha} \right) 
  \exp{\left( i\xi_\alpha \right)}. 
\label{eq:Chia}
\end{eqnarray}
Here, the phase variables $\xi_\alpha$ have a constraint, 
$\sum_{\alpha =1}^3 \xi _\alpha =0$, where two independent degrees 
of freedom remain corresponding to the dual gauge symmetry 
$[{\rm U(1)}_3\times {\rm U(1)}_8]_m$  \cite{maedan,suganuma}.
When monopoles condense, the phase variables $\xi_\alpha $ 
turn into the longitudinal degrees of 
the dual gauge field $\vec B_\mu $, which is the dual Higgs mechanism.

Since we are interested in the translational-invariant system 
as the QCD vacuum, 
we consider the $x$-independent constant source $J$, 
which leads to a homogeneous monopole condensate. 
In the unitary gauge, 
the Lagrangian with the source term is rewritten as
\begin{eqnarray}
     {\cal L}_{\rm DGL}&-J\sum_{\alpha =1}^3 |\chi _\alpha |^2 
 = {\cal L}_{\rm cl}(\bar \chi ) - 3 J\bar \chi ^2 
 - 2 \bar \chi  [ 2\lambda  (\bar \chi ^2-v^2) + J ] 
  \sum_{\alpha =1}^3  \tilde \chi _\alpha  \nonumber \\
 &- {1 \over 4} (\partial_{\mu}\vec{B}_{\nu}
            - \partial_{\nu}\vec{B}_{\mu} )^2 
            + {1 \over 2} m_B^2 \vec{B}_{\mu}^2  
 +  \sum_{\alpha=1}^3 [ (\partial_{\mu} \tilde{\chi}_\alpha)^2 
      - m_{\chi}^2  \tilde{\chi}_{\alpha}^2 ] \\
 &+ \sum_{\alpha=1}^3  \{ 
    g^2(\vec \epsilon_\alpha  \cdot \vec B_\mu )^2 
    (\tilde \chi _\alpha ^2+2\bar \chi  \tilde \chi _\alpha )
   - \lambda (4 \bar \chi \tilde{\chi}_{\alpha}^3 + \tilde{\chi}
   _{\alpha}^4) \},  
\label{eq:Lb} \nonumber 
\end{eqnarray}
where ${\cal L}_{\rm cl}(\bar \chi )$ is the classical part, 
\begin{eqnarray}
{\cal L}_{\rm cl}(\bar \chi ) = -3\lambda (\bar \chi ^2-v^2)^2. 
\label{eq:Lcl}
\end{eqnarray}
Here, the masses of $\tilde \chi _\alpha $ and $\vec B_\mu $ are given by 
\begin{eqnarray}
   m_\chi ^2 = 2\lambda (3 \bar \chi ^2-v^2) + J = 4\lambda \bar \chi ^2, 
   \hbox{\quad} m_B^2 = 3 g^2 \bar \chi ^2, 
\label{eq:Ma}
\end{eqnarray}
where we have used the relation between the mean field 
$\bar \chi$ and the source $J$,
\begin{eqnarray}
J= -2\lambda (\bar \chi ^2-v^2). 
\label{eq:JCHI}
\end{eqnarray}
This relation is obtained by the condition that the linear 
term of $\tilde{\chi}_{\alpha}$ vanishes. 
It is remarkable that the scalar mass $m_\chi $ is always 
real owing to the source $J$ even in the negative-curvature 
region of the classical potential, $\bar \chi  < v / \sqrt{3}$. 

Integrating over $\vec B_\mu $ and $\tilde \chi _\alpha $
by neglecting the higher order terms of the fluctuations, 
we obtain the partition functional, 
\begin{eqnarray}
Z[J] = \exp{\left( i \int d^4x 
       \{{\cal L}_{\rm cl}(\bar \chi ) - 3J\bar \chi^2 \} \right) }
       [\hbox{Det} (iD_B^{-1})]^{-1}
       [\hbox{Det} (iD_{\chi}^{-1})]^{-3/2}, 
\label{eq:Za}
\end{eqnarray}
where the exponents, $-1$ and $-3/2$, originate from the 
numbers of the internal degrees of freedom.
Here, $D_B$ and $D_\chi $ are the propagators 
of $\vec B_\mu $ and $\tilde \chi _\alpha $ in the monopole 
condensed vacuum, 
\begin{eqnarray}
D_B =\left(g_{\mu \nu }-{ k_\mu k_\nu  \over m_B^2 } \right)
{i \over k^2- m_B^2 +i\epsilon }, \hbox{\quad} 
D_\chi ={-i \over k^2-m_\chi ^2+i\epsilon }
\label{eq:PRa}
\end{eqnarray}
in the momentum representation.
Hence, the effective action is given by the Legendre transformation \cite{cheng},
\begin{eqnarray}
\Gamma (\bar \chi ) & =  &-i \ln Z[J] + \int d^4x 3 J \bar \chi^2  
\nonumber \\
             &  =  & \int d^4x {\cal L}_{\rm cl}(\bar \chi ) 
                     +i \ln {\rm Det} (i D_B^{-1}) 
                     + {3 \over 2} i \ln {\rm Det} (i D_\chi ^{-1}). 
\label{eq:Ga}
\end{eqnarray}
The functional determinants are easily calculable in the momentum space, 
and we obtain the formal expression of the effective potential \cite{cheng}, 
\begin{eqnarray}
  V_{\rm eff}(\bar \chi) =
 {-\Gamma (\bar \chi ) / \int d^4x}
    = 3 \lambda (\bar \chi^2 - v^2 )^2 
    &+ 3 \int {d^4k \over i(2\pi )^4 } 
    \ln(m_B^2-k^2-i\epsilon ) \nonumber \\
    &+ {3 \over 2} \int {d^4k \over i(2\pi )^4 } 
    \ln(m_\chi ^2-k^2-i\epsilon ).
\label{eq:Vv}
\end{eqnarray}

In the finite-temperature system \cite{kapusta}, 
the partition functional $Z$ is described by 
the Euclidean variables; $x_{0}=-i\tau$, and the 
upper bound of the $\tau$ integration is $\beta=1/T$
with $T$ being the temperature. 
Then, the $k_0$-integration in the functional determinant 
becomes the infinite sum over the Matsubara frequency \cite{kapusta}.
The effective potential at finite temperatures 
physically corresponds to the thermodynamical potential, 
and is given by 
\begin{eqnarray}
 V_{\rm eff}(\bar \chi;T) = 3 \lambda (\bar \chi^2 - v^2 )^2 
    &+ 3 T \sum_{n=-\infty }^\infty  \int {d^3k \over (2\pi)^3 } 
    \ln\{(2n\pi T)^2 + k^2 + m_B^2\} \nonumber \\
    &+ {3 \over 2} T \sum_{n=-\infty }^\infty  \int {d^3k \over (2\pi)^3 } 
    \ln\{(2n\pi T)^2 + k^2 + m_\chi ^2\} 
\label{eq:Va}
\end{eqnarray}
in the DGL theory.
Performing the summation over $n$ and the angular integration, 
we obtain the final expression of the effective potential at 
finite temperatures, 
\begin{eqnarray}
 V_{\rm eff}(\bar \chi ;T) =   3 \lambda ( \bar \chi^2 - v^2 )^2 
          &+ 3 {T \over \pi^2} \int_0^\infty  dk k^2 \ln{
           \left(  1 - e^{ - \sqrt{ k^2 + m_B^2}/T }
           \right)  } \nonumber \\
          &+ {3 \over 2} {T \over \pi^2}\int_0^\infty  dk k^2 \ln{
           \left(  1 - e^{ - \sqrt{ k^2 + m_{\chi}^2}/T }
           \right)  }, 
\label{eq:Vb}
\end{eqnarray}
where $m_B$ and $m_\chi $ are functions of $\bar \chi $ 
as shown in Eq.(\ref{eq:Ma}).
Here, we have dropped the $T$-independent part 
(quantum fluctuation), because 
we are interested in the 
thermal contribution to the QCD vacuum \cite{monden}.

Before the numerical calculation, we examine the outline of the
phase transition using the high-temperature (high-$T$) expansion.
The effective potential is well approximated by
the high-$T$ expansion, when the particle mass
is much smaller than the temperature.
Since the particle masses are almost zero
for $|\bar \chi| \simeq 0$ as shown in Eq.(\ref{eq:Ma}),
the high-$T$ expansion is applicable for the effective potential
around $|\bar \chi|=0$.


The temperature-dependent part of the effective potential
for bosons is expressed as
\begin{eqnarray}
V_T(\bar \chi ;T) & \equiv &
{T \over {\pi ^2}}\int_0^\infty 
{dkk^2\ln (1-e^{-\sqrt {k^2+m^2}/ T})} \nonumber \\
 & = & {{T^4} \over {\pi ^2}}\int
 {dyy^2\ln (1-e^{-\sqrt {y^2+a^2}})},
\label{eq:kouon}
\end{eqnarray}
with $y^2 \equiv \frac{k^2}{T^2}$ and $a^2 \equiv \frac{m^2}{T^2}$. 
In the high-$T$ expansion,
$V_T(\bar \chi;T)$ is expanded in powers of $a$,
\begin{eqnarray}
V_T(\bar \chi ;T) & = & V(a^2(\bar \chi ))\left| {_{a^2=0}
+{{\partial V} \over {\partial a^2}}} 
\right.\left| {_{a^2=0}a^2} \right.+... \nonumber \\
& \approx & {{T^4} \over {\pi ^2}}
\int  _0^\infty {dyy^2\ln (1-e^{-y})}
+{{T^4} \over {2\pi ^2}}
\int _0^\infty {dyy^2{1 \over {\sqrt {y^2+a^2}}}
{1 \over {e^{\sqrt {y^2+a^2}}-1}}}
\left| {_{a^2=0}a^2} \right. \nonumber \\ 
& = &{{T^4} \over {\pi ^2}}(-{{\pi ^4} \over {45}})
+{{T^4} \over {2\pi ^2}}{{\pi ^2} \over 6}
({m \over T})^2
=-{{\pi^2 T^4} \over {45}}
+{{T^4} \over {12}}({m \over T})^2.
\end{eqnarray}
Thus, $V_{\rm eff}$ in Eq.(\ref{eq:Vb}) is expressed as
\begin{eqnarray}
V_{\rm eff}(\bar \chi ;T)
 \simeq 3\lambda (\bar \chi ^2-v^2)^2
-{{\pi ^2T^4} \over {10}}
+{{T^2} \over {12}}({3 \over 2}m_\chi ^2+3m_B^2)
\label{eq:kouonten}
\end{eqnarray}
in the high-$T$ expansion.
As the temperature increases,
the effective potential at $|\bar \chi| =0$ is changed from the
minimum point to maximum point
at the lower  critical temperature 
\begin{eqnarray}
T_{low} = 2 v( \frac {6 \lambda}{2 \lambda + 3 g^2}),
^{\frac 12}
\end{eqnarray}
which satisfies the flat curvature condition as 
\begin{eqnarray}
\frac{\partial^2 V}{\partial \bar \chi^2} |_
{\bar \chi=0} = 0.
\end{eqnarray}
In fact, $V_{\rm eff}(\bar \chi;T)$ at $\bar \chi=0$ is a
local maximum for $T<T_{low}$, while it becomes a (local) minimum
and the system is (meta-) stable for $T>T_{low}$.
For the 2nd-order phase transition,
$V_{\rm eff}$ has only one absolute minimum
at each $T$ because of continuous variation of the order parameter,
and therefore $T_{low}$ coincides the thermodynamical critical
temperature $T_c$. On the other hand,
for the 1st-order phase transition with a jump of the order parameter,
there are two absolute minima at $T_c$, and
the second minimum at $\bar \chi=0$ appears at the lower critical
temperature $T_{low}$, so that $T_{low}$ is lower than $T_c$.
In any case,
using the high-$T$ expansion,
we derive the analytical expression for $T_{low}$, which
provides a lower bound for the critical temperature.

\section{Numerical Results on QCD Phase Transition}

Here, we show the results on the QCD phase transition which obtained from the
numerical calculation \cite{ichieB}.

\begin{figure}[tb]
\epsfxsize = 8 cm
\centering \leavevmode
\epsfbox{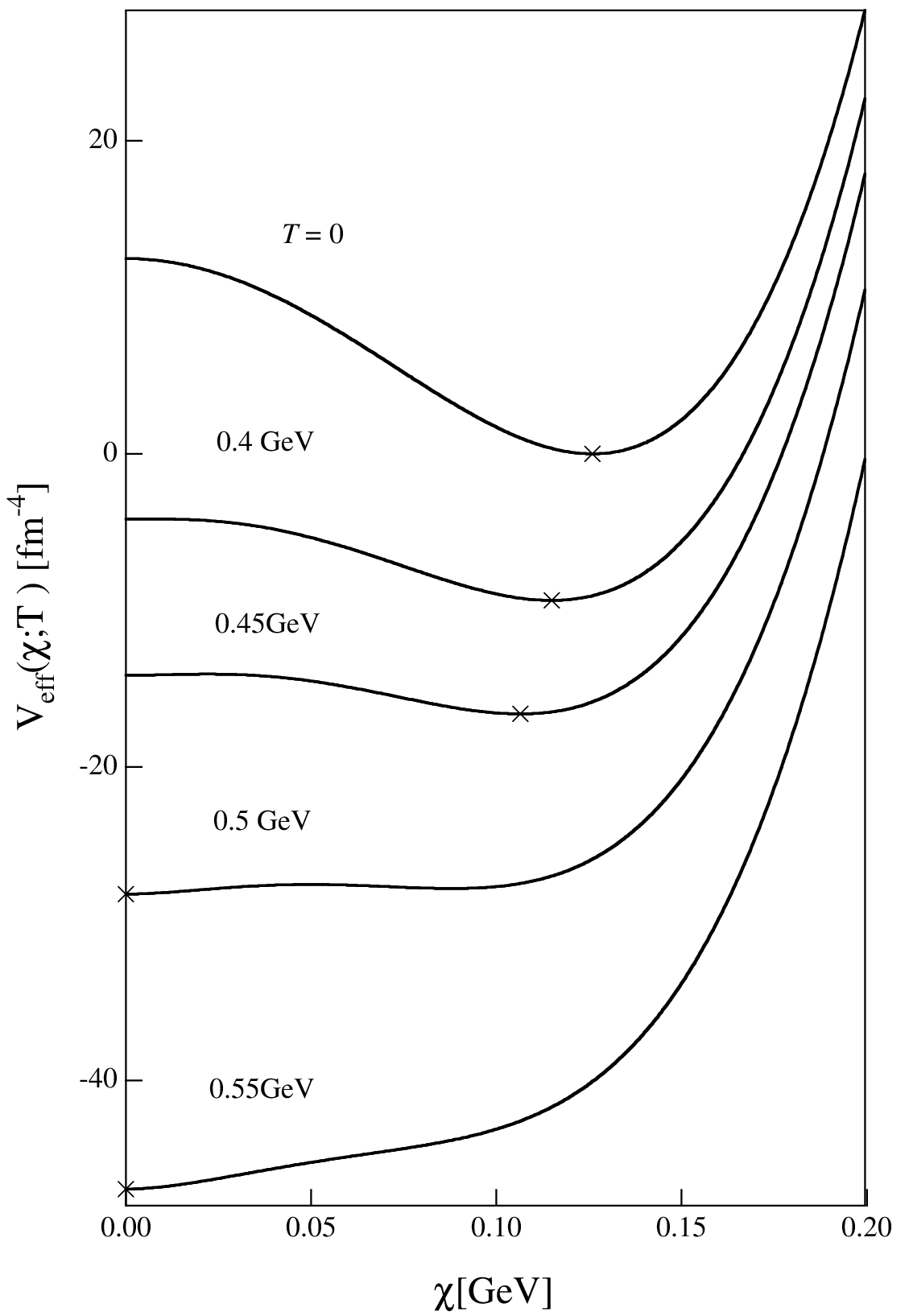}
\vspace{0cm}
\caption{     
The effective potentials at various temperatures as
          functions of the monopole condensate $\bar \chi$. 
          The numbers beside each curve are the temperatures. 
          The absolute minimum points of the effective potentials are 
          shown by crosses.
}
\label{fig:efpo1}
\end{figure}

We first show in Fig.\ref{fig:efpo1} the effective potential at various temperatures 
(thermodynamical potential), $V_{\rm eff}(\bar \chi ;T)$, 
as a function of the monopole condensate $\bar \chi $, 
an order parameter for the color confinement. 
The parameters, $\lambda$ = 25, $v$ = 0.126GeV 
and $g$ = 2.3, are extracted by fitting the static potential
in the DGL theory to the Cornell potential
\footnote{
We examined several possible parameter sets, and  
found a small parameter dependence on our results shown 
in this paper. 
}.
These values provide $m_B$ = 0.5GeV and $m_{\chi}$ = 1.26GeV 
at $T$ = 0.
The (local-)minimum point of $V_{\rm eff}(\bar \chi ;T)$ 
corresponds to the physical (meta-)stable vacuum state.
As the temperature increases, 
the broken dual gauge symmetry tends to be restored, 
and the monopole condensate in the physical vacuum, 
$\bar \chi _{\rm phys}(T)$, is decreased.
A first order phase transition is found at the thermodynamical 
critical temperature, $T_C \simeq 0.49$ GeV, 
and the QCD vacuum becomes trivial, $\bar \chi _{\rm phys}(T)=0$, 
for $T \ge T_C$. 
This phase transition is regarded as the 
deconfinement phase transition, 
because there is no confining force among colored particles 
in the QCD vacuum with $\bar \chi _{\rm phys}(T)=0$.

\begin{figure}[tb]
\epsfxsize = 10 cm
\centering \leavevmode
\epsfbox{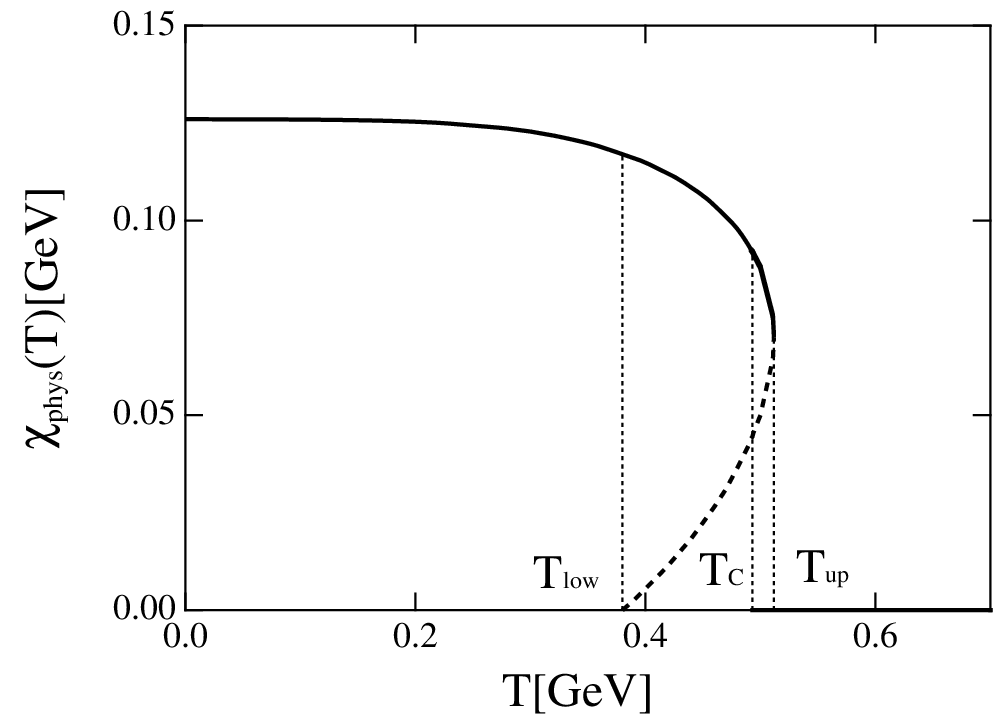}
\vspace{0cm}
\caption{ 
The monopole condensate $\bar \chi _{\rm phys}(T)$ 
 at minima of the effective potential 
 as a function of the temperature $T$. 
 The solid curve denotes $\bar \chi _{\rm phys}(T)$ corresponding 
 to the confinement phase, 
 which is the absolute minimum up to 
 $T_C = $0.49 GeV and becomes a local minimum up to 
 $T_{\rm up} = $0.51 GeV.
 A minimum appears at $\bar \chi=0$ 
 above $T_{\rm low} = $0.38 GeV and becomes the
 absolute minimum above $T_C = $0.49 GeV. 
 The dot-dashed curve denotes
 the value of $\bar \chi$ at the local maximum.
}
\label{fig:efpo2}
\end{figure}

We show the behavior of the monopole condensate 
in the physical vacuum, $\bar \chi _{\rm phys}(T)$,
as a function of the temperature $T$ in Fig.\ref{fig:efpo2}. 
One finds $\bar \chi _{\rm phys}$= 0.126 GeV at $T=0$, 
and the monopole condensate decreases monotonously up to 
$\bar \chi _{\rm phys}$ = 0.07 GeV at 
the upper critical temperature $T_{\rm up} = $ 0.51 GeV, 
where the minimum at finite $\bar \chi $ 
disappears in $V_{\rm eff}(\bar \chi ;T)$.
On the other hand, the local minimum is developed at $\bar \chi $ = 0 
in $V_{\rm eff}(\bar \chi ;T)$ above the lower 
critical temperature $T_{\rm low} = $ 0.38 GeV,
which is analytically obtained by 
using the high-temperature expansion in previous section 
\cite{kapusta,monden}.
The minimum value of $V_{\rm eff}(\bar \chi ;T)$ 
at $\bar \chi =0$ becomes deeper than that 
at finite $\bar \chi $ above the thermodynamical critical temperature
$T_C$ = 0.49 GeV.
Here, we get the first-order phase transition because we have 
considered full orders in $\bar \chi ^2$ as shown in Eq.(\ref{eq:Vb}).
On the other hand, Monden et al \cite{monden} 
did not get the first-order phase transition due to the 
use of only the lowest order in $\bar \chi ^2$
in the high-temperature expansion \cite{kapusta},
and therefore 
they had to introduce the cubic term in $\chi _\alpha $ in 
the Lagrangian.

\begin{figure}[tb]
\epsfxsize = 10 cm
\centering \leavevmode
\epsfbox{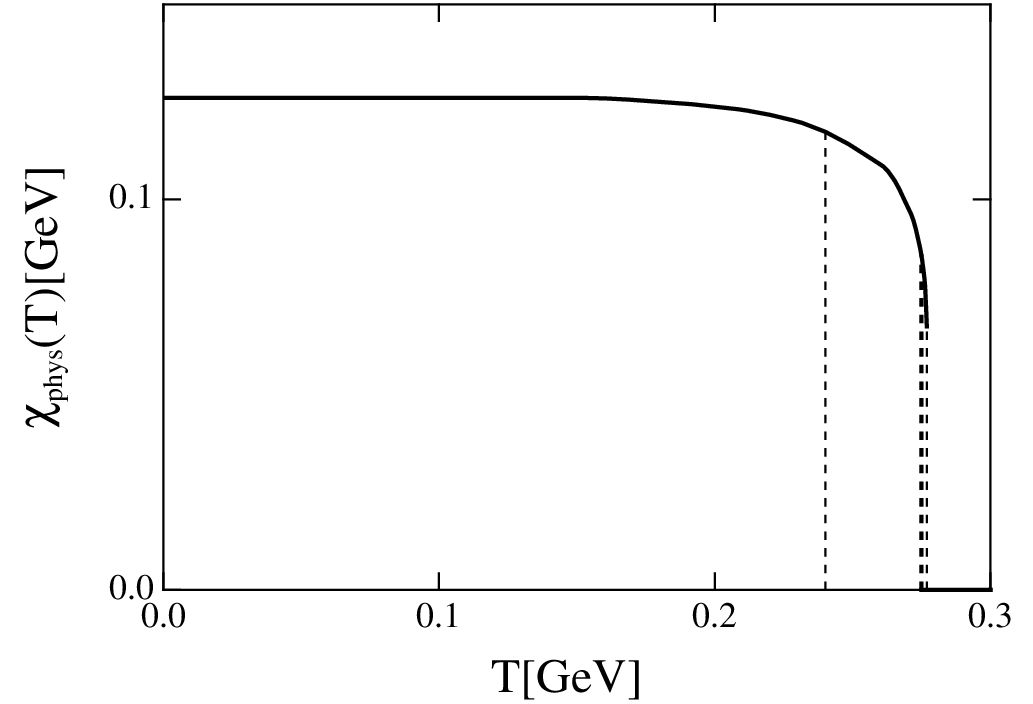}
\vspace{0cm}
\caption{   
 The monopole condensate $\bar \chi _{\rm phys}(T)$ at minima of the 
 effective potential as a function of the temperature 
 in the case of variable $\lambda (T)$
 so as to reproduce $T_C = $0.28GeV.
 The meanings of the curves are the same as in Fig.\ref{fig:efpo2}.
}
\label{fig:efpo3}
\end{figure}

Here, we consider the possibility of the temperature dependence 
on the parameters ($\lambda $,$v$) in the DGL theory.
The critical temperature, $T_C$ = 0.49 GeV, seems much larger 
than one of the recent lattice QCD prediction, 
$T_C = 0.26 \sim 0.28$GeV,
which is, for instance, estimated from the relation, 
$T_C/\sqrt{\sigma} \simeq 0.62$  \cite{karsch2}
 and the string tension $\sigma=0.89 \sim 1.0$GeV/fm. 
However, we should remember that the self-interaction term of $\chi _\alpha $ 
has been introduced phenomenologically in the DGL Lagrangian. 
In particular, the asymptotic freedom behavior of QCD leads to a 
possible reduction of the self-interaction among monopoles 
at high temperatures. 
Hence, we use a simple ansatz for the temperature dependence on $\lambda $, 
\begin{eqnarray}
\lambda (T) = \lambda  \left( {T_C - aT \over T_C}  \right), 
\label{eq:Lam}
\end{eqnarray}
keeping the other parameter $v$ constant.
Here, the constant $a$ is determined as $a=0.89$
so as to reproduce $T_C=0.28$GeV. (We take $\lambda (T)=0$ for $T>T_C/a$.) 
The results for the monopole condensate 
$\bar \chi _{\rm phys}(T)$ are shown in Fig.\ref{fig:efpo3}. 
The qualitative behavior is the same as in the 
above argument with a constant $\lambda $.
We find a weak first-order phase transition in this case also. 
Here, we find a large reduction of the 
self-interaction of the monopoles near the critical 
temperature $T_C$: $\lambda (T \simeq T_C) \simeq 2.7$ 
is considerably smaller than $\lambda (T=0)=25$. 

\begin{figure}[tb]
\epsfxsize = 10 cm
\centering \leavevmode
\epsfbox{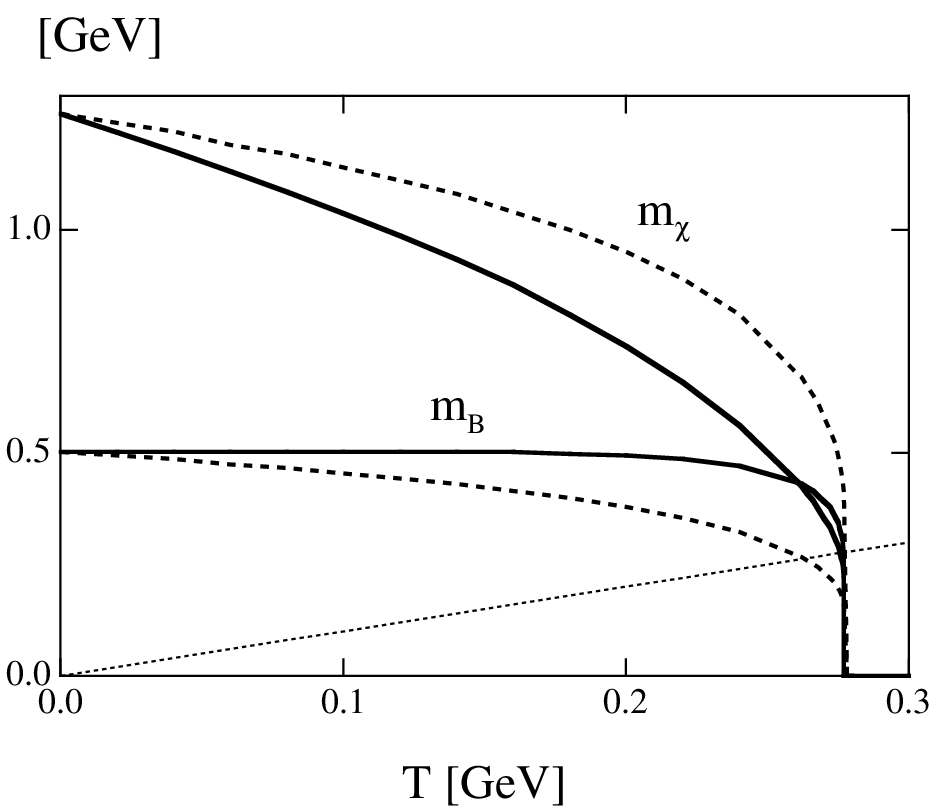}
\vspace{0cm}
\caption{     
The dual gauge field mass $m_B(T)$ and the monopole field mass
 $m_\chi (T)$.
The solid  lines denote the case of variable $\lambda (T)$ 
with a constant $v$. The dashed lines denote the case of 
variable $v(T)$ with a constant $\lambda $. 
A large reduction of these masses is found near the critical 
temperature. The dotted line denotes $m=T$.
The phase transition occurs at the temperature satisfying 
$m_B, m_\chi  \simeq T$.
}
\label{fig:efpo4}
\end{figure}

\begin{figure}[htbp]
\epsfxsize = 10 cm
\centering \leavevmode
\epsfbox{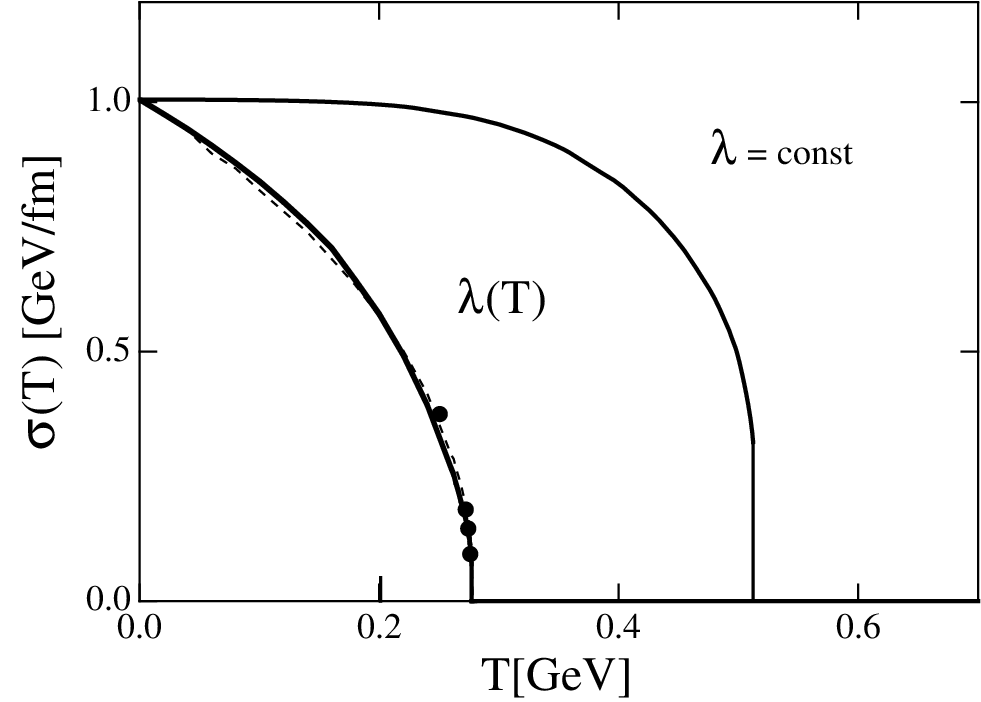}
\vspace{0cm}
\caption{   
The string tension $\sigma(T)$ as a function of the temperature $T$.
The solid and dashed lines correspond to the variable $\lambda (T)$ case 
with a constant $v$ and the variable $v(T)$ case with a constant $\lambda $, 
respectively. 
The constant ($\lambda $, $v$) case is also shown by the thin line.
The lattice QCD results in the pure gauge 
are shown by black dots near the critical temperature.
}
\label{fig:efpo5}
\end{figure}

Next, we investigate the variation of the masses of the 
dual gauge field $\vec B_\mu $ and the monopole field 
$\tilde \chi _\alpha $ at finite temperatures.
Here, $\tilde \chi _\alpha $ would appear as 
the color-singlet glueball field with  $0^+$ \cite{maedan,suganumaB}. 
These masses $m_B$ and $m_\chi $, 
at the finite temperature $T$ are given by 
\begin{eqnarray}
   m_B(T) = \sqrt{3} g \bar \chi _{\rm phys}(T), 
   \hbox{\quad} 
   m_\chi (T) = 2\sqrt{\lambda (T)}\bar \chi _{\rm phys}(T) 
\label{eq:GBM}
\end{eqnarray}
as shown in Eq.(\ref{eq:Ma}).
In Fig.\ref{fig:efpo4}, we show $m_B(T)$ and $m_\chi (T)$ as functions of the 
temperature $T$ using variable $\lambda (T)$ in Eq.(\ref{eq:Lam}). 
(Their behaviors are almost the same as the case of a 
constant $\lambda $ except for the difference of the value of $T_C$.)
It is worth mentioning that $m_B(T)$ and $m_\chi (T)$
drop down to $m_B, m_\chi  \sim T_C$(=0.28GeV)
from $m_B, m_\chi  \sim$ 1 GeV near the critical temperature $T_C$. 
In other words, 
the QCD phase transition occurs at the temperature satisfying 
$m_B, m_\chi  \simeq T$, 
which seems quite natural because the thermodynamical factor 
$
1 /\{ \exp(\sqrt{k^2+m^2}/T) \pm 1 \}
$
becomes relevant only for $m \stackrel{<}{\sim} T$.
Thus, our result predicts a large reduction 
of the dual gauge field mass $m_B$ and the monopole field mass $m_\chi $
near the 
critical temperature $T_C$. 
It is desirable to study the change of the scalar glueball mass 
at finite temperatures, especially near $T_C$, 
in the lattice QCD simulation with the larger lattice size and 
the higher accuracy.

We investigate the string tension $\sigma$ at finite temperatures, 
since $\sigma$ is one of the most important variables for 
the color confinement, 
and controls the hadron properties through the inter-quark potential.
We use the expression of the string tension $\sigma(T)$ 
provided by SST  \cite{suganuma}, 
\begin{eqnarray}
\sigma(T) = {e^2 m_B^2(T) \over 24 \pi} 
\ln \left(  {m_B^2(T) + m_\chi^2(T) \over m_B^2(T)} \right), 
\label{eq:STRa}
\end{eqnarray}
where the  masses $m_B(T)$ and $m_\chi(T)$ 
are given by Eq.(\ref{eq:GBM}).
The results are shown in Fig.\ref{fig:efpo5} as a function 
of the temperature $T$. 
In the case of constant $\lambda $, the string tension $\sigma(T)$ 
decreases very gradually up to the temperature, 
$T_{\rm up}$= 0.51 GeV.
On the other hand, in the case of variable
$\lambda (T)$, the string tension $\sigma(T)$ decreases rapidly with 
temperature, 
and $\sigma(T)$ drops down to zero around $T_C = $ 0.28 GeV. 
Hence, one expects a rapid change of 
the masses and the sizes of the quarkonia 
according to the large reduction of $\sigma(T)$ 
at high temperatures.
We plot also the pure-gauge lattice QCD results for the
temperature dependence of the string tension by black dots \cite{gao},
with $T_C=0.28$GeV \cite{karsch2}.
We find our results with variable $\lambda (T)$ agree with the lattice
QCD data.

We discuss further the temperature dependence 
of the parameters ($\lambda $,$v$) in the DGL theory.
Definitely, we should follow the lattice QCD data 
for this determination as the case of the Ginzburg-Landau 
theory of superconductors extracting the temperature dependence 
from experiments.
Since there exists the lattice QCD data on the string tension 
$\sigma$ \cite{gao},
 we try to reproduce $\sigma$ by taking a simple ansatz on 
$\lambda $ and $v$. 
We try the following ansatz,
\begin{eqnarray}
  B(T) \equiv 3\lambda (T)v^4(T)
  = 3\lambda v^4 \left( {T_C - aT \over T_C}  \right), 
\label{eq:LamZ}
\end{eqnarray}
where the constant $a$ is determined so as to reproduce $T_C=0.28$GeV. 
(We take $B(T)=0$ for $T>T_C/a$.) 
The variable $B(T)$ corresponds to the bag constant, the energy-density 
difference between the nonperturbative vacuum ($|\chi _\alpha | \ne 0$) 
and the perturbative vacuum ($|\chi _\alpha |=0$) in the DGL theory; 
see Eq.(\ref{eq:Vb}). 
The ansatz (\ref{eq:LamZ}) suggests the reduction of the bag constant 
at high temperatures, which provides the swelling of hot hadrons 
by way of the bag-model picture.
Since we have already examined a typical case for variable 
$\lambda (T)$ keeping $v$ constant, we show here another typical case 
for variable $v(T)$ keeping $\lambda $ constant.
The string tension $\sigma(T)$ in the variable $v(T)$ case with $a=0.91$
is shown by the dashed line in Fig.\ref{fig:efpo5}.
We find almost an identical result and find again a good agreement 
with the lattice QCD data.
Other combinations on $\lambda (T)$ and $v(T)$ under the relation 
(\ref{eq:LamZ})
also provide equally good results on $\sigma(T)$.

Finally, we investigate the relation between the scalar glueball mass 
$m_\chi (T)$ and the string tension $\sigma(T)$. 
For variable $\lambda (T)$ keeping $v$ constant, 
one finds from Eq.(\ref{eq:STRa}) an approximate relation, 
\begin{eqnarray}
{m_\chi (T) \over \sqrt{\sigma(T)}} \simeq {(24\pi )^{1/2} \over e} 
\simeq 1.6, 
\label{eq:GBMx}
\end{eqnarray}
near the critical temperature $T_C$.
On the other hand, for variable $v(T)$ keeping $\lambda $ constant, 
the scalar glueball mass at finite temperatures, 
$m_\chi (T)$, is shown by the dashed line in 
Fig.\ref{fig:efpo4}, 
and Eq.(\ref{eq:STRa}) leads to a simple relation, 
\begin{eqnarray}
{m_\chi (T) \over \sqrt{\sigma(T)}}
= \left( {2\lambda  \over \pi  \ln \{ (3g^2+4\lambda )/3g^2 \}} \right)^{1/2} 
\simeq 3.0, 
\label{eq:GBMy}
\end{eqnarray}
for the whole region of $T$. 
Thus, the DGL theory suggests a proportional relation between 
the scalar glueball mass and the square root of the string 
tension at least near $T_C$. 
It is worth mentioning that Engels et al. \cite{karsch2} 
obtained a similar relation, $m_{\rm GB}(T)= (1.7 \pm 0.5) \sqrt{\sigma(T)}$, 
for the lowest scalar glueball at finite temperatures 
from the thermodynamical study on the SU(2) lattice gauge theory.
Eqs.(\ref{eq:GBMx}) and (\ref{eq:GBMy}) can be examined from the 
thermodynamical study on the glueball mass in the lattice QCD, 
which may also reveal $T$-dependence on the parameters in the 
DGL theory.

\section{Hadron Bubble Formation in Early Universe}

The remaining part of this section,
we consider the application of the DGL theory to  big 
bang \cite{ichieA}.
As Witten proposed \cite{witten},
if the QCD phase transition is  of first order, 
the hadron and the QGP phase should coexist in early Universe.
Such a mixed phase may cause the inhomogeneity of the Universe in the 
baryon number distribution.
This inhomogeneity affects  the primordial nucleo-synthesis  \cite{kajino}.

\begin{figure}[p]
\epsfxsize = 10 cm
\centering \leavevmode
\epsfbox{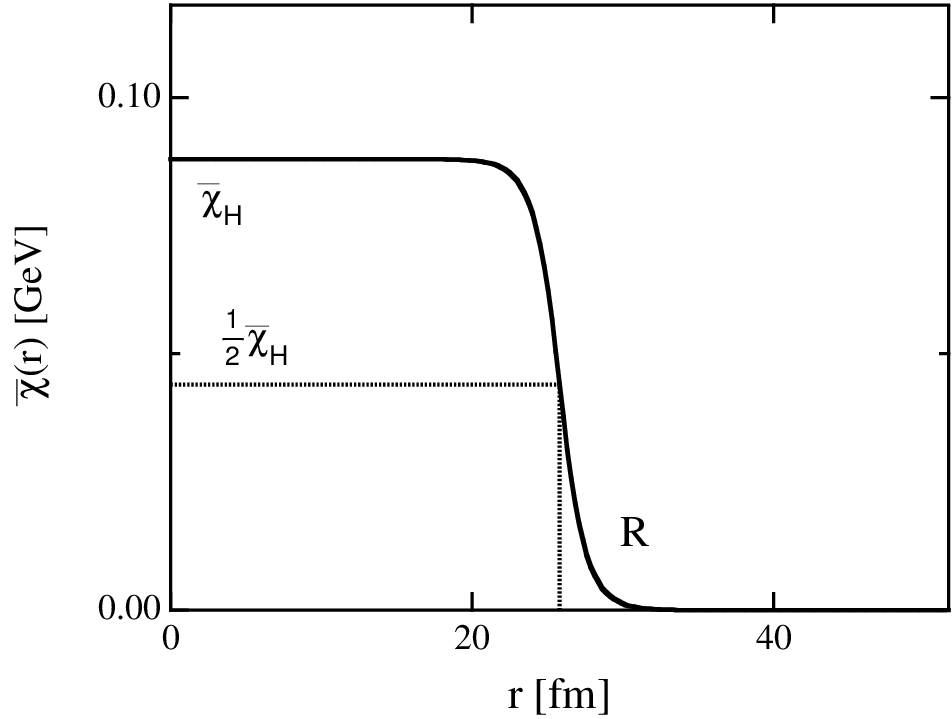}
\vspace{0cm}
\caption{    
The profile of the monopole condensate in the hadron bubble.
There is the QGP phase without monopole condensation outside the bubble,
while hadron phase with monopole condensation remains inside the bubble.}
\label{fig:efpo6}
\vspace{2cm}
\epsfxsize = 10 cm
\centering \leavevmode
\epsfbox{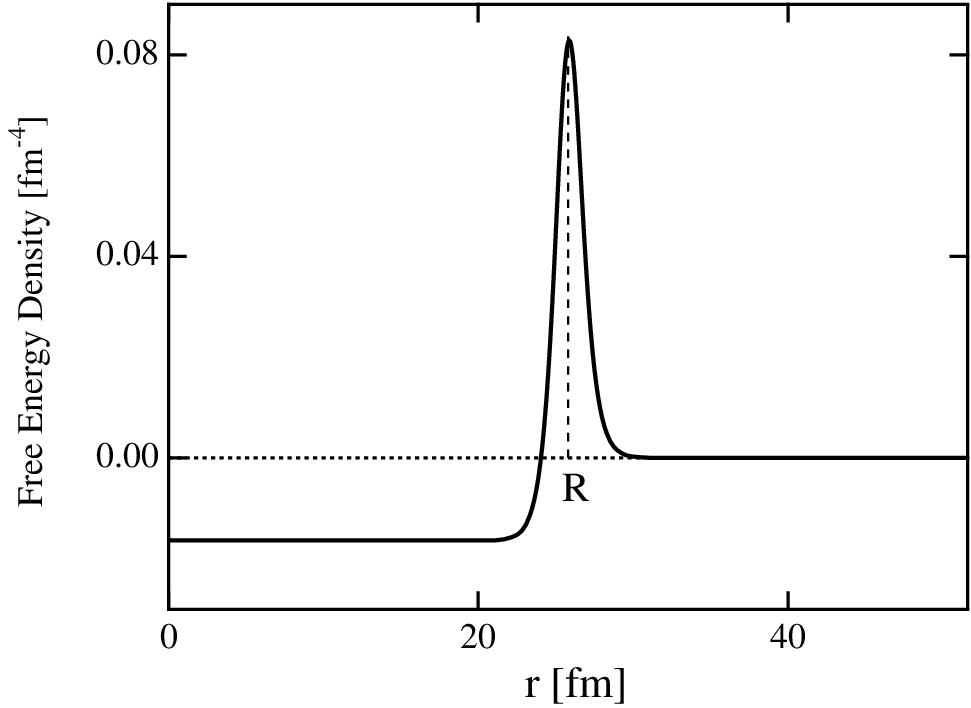}
\vspace{0cm}
\caption{    
The energy density of the hadron bubble.
It is negative inside and positive near the boundary surface.
}
\label{fig:efpo7}
\end{figure}

As a result of the 1st order phase transition, 
hadron bubbles appear in the QGP phase near the critical temperature.
We now consider how hadron bubbles are formed
in the DGL theory. 
In the supercooling system,
the free energy of the hadron bubble with radius $R$ profile
$\bar \chi(r;R)$ is written 
using the effective potential at finite temperature,
\begin{eqnarray}
E[\bar \chi(r;R)] = 4 \pi \int_0^{\infty}drr^2
\{ 3( \frac{d\bar \chi(r;R)}{dr})^2 + V_{\rm eff}(\bar \chi ;T)\}.
\end{eqnarray}
We use the sine-Gordon kink ansatz for the profile of 
the monopole condensate,
\begin{eqnarray}
\bar \chi (r;R)
=\bar \chi _H\tan ^{-1}e^{(R-r) / \delta }/ \tan ^{-1}e^{R / \delta },
\end{eqnarray}
where the thickness of the surface $\delta$ is 
determined by the free energy minimum conditions.
The result is shown in Fig.\ref{fig:efpo6}.
The monopole condensate $\bar \chi(r;R)$
is connected smoothly between inside and outside the bubble.
The energy density of the hadron bubble is shown in Fig.\ref{fig:efpo7}.
It is negative inside and positive near the boundary surface.
The total  energy is roughly estimated as the sum of the surface term
(corresponding to the positive region near the surface) and 
the volume term (corresponding to the negative region inside
the bubble).

\begin{figure}[p]
\epsfxsize = 10 cm
\centering \leavevmode
\epsfbox{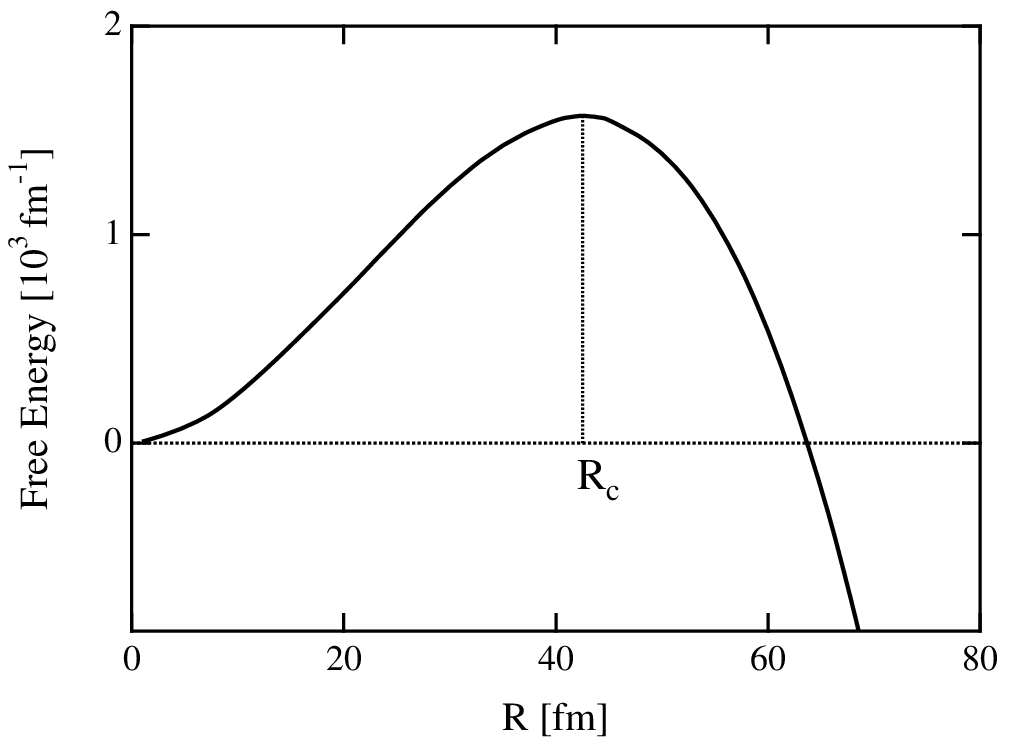}
\vspace{0cm}
\caption{   
The total energy of the  bubble is plotted as a
function of the hadron bubble radius$R$.
The total energy is roughly estimated as the sum of the 
volume term and the surface term.
}
\label{fig:efpo8}
\vspace{2cm}
\epsfxsize = 10 cm
\centering \leavevmode
\epsfbox{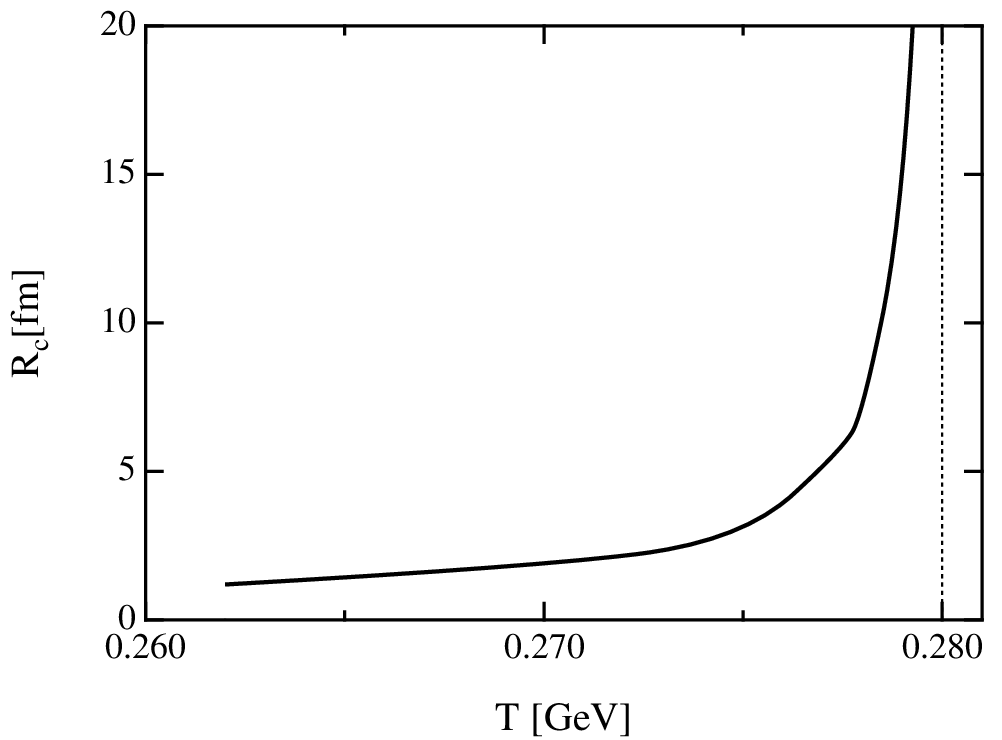}
\vspace{0cm}
\caption{    
The critical radius $R_c$, corresponding to maximum of the energy in 
Fig.\ref{fig:efpo8}, 
is plotted as a function of  temperature. 
}
\label{fig:efpo9}
\end{figure}

The energy of the hadron bubble with radius $R$ is shown in Fig.\ref{fig:efpo8}.
The bubble whose radius is smaller than critical radius $R_c$ collapses.
Only larger bubbles ( $R > R_c$) are found to grow up from 
the energetical argument.
However, the creation of large bubbles is suppressed because of formation
probability.
In the bubble formation process, there exists a large barrier height $h$
of the effective potential and therefore
the creation of large bubbles needs the large energy fluctuation above 
the barrier height.
Such a process is suppressed  because of the thermal dynamical factor
(proportional
to bubble formation rate), ${\rm exp}(-\frac43 \pi R_{\rm c}^3h/T )$.
Thus, the only small bubbles are created 
practically, although 
its radius should be larger than $R_{\rm c}$ energitically.
The temperature dependence of the critical radius
is shown in Fig.\ref{fig:efpo9}. 
In the temperature region of the supercooling state, i.e, 
$T_{\rm low} < T < T_{\rm c}$, the hadron bubbles are created.
As the temperature decreases, the size of hadron bubble becomes smaller,
however the bubble formation rate becomes larger.


\begin{figure}[tb]
\vspace{-1cm}
\epsfxsize = 14cm
\centering \leavevmode
\epsfbox{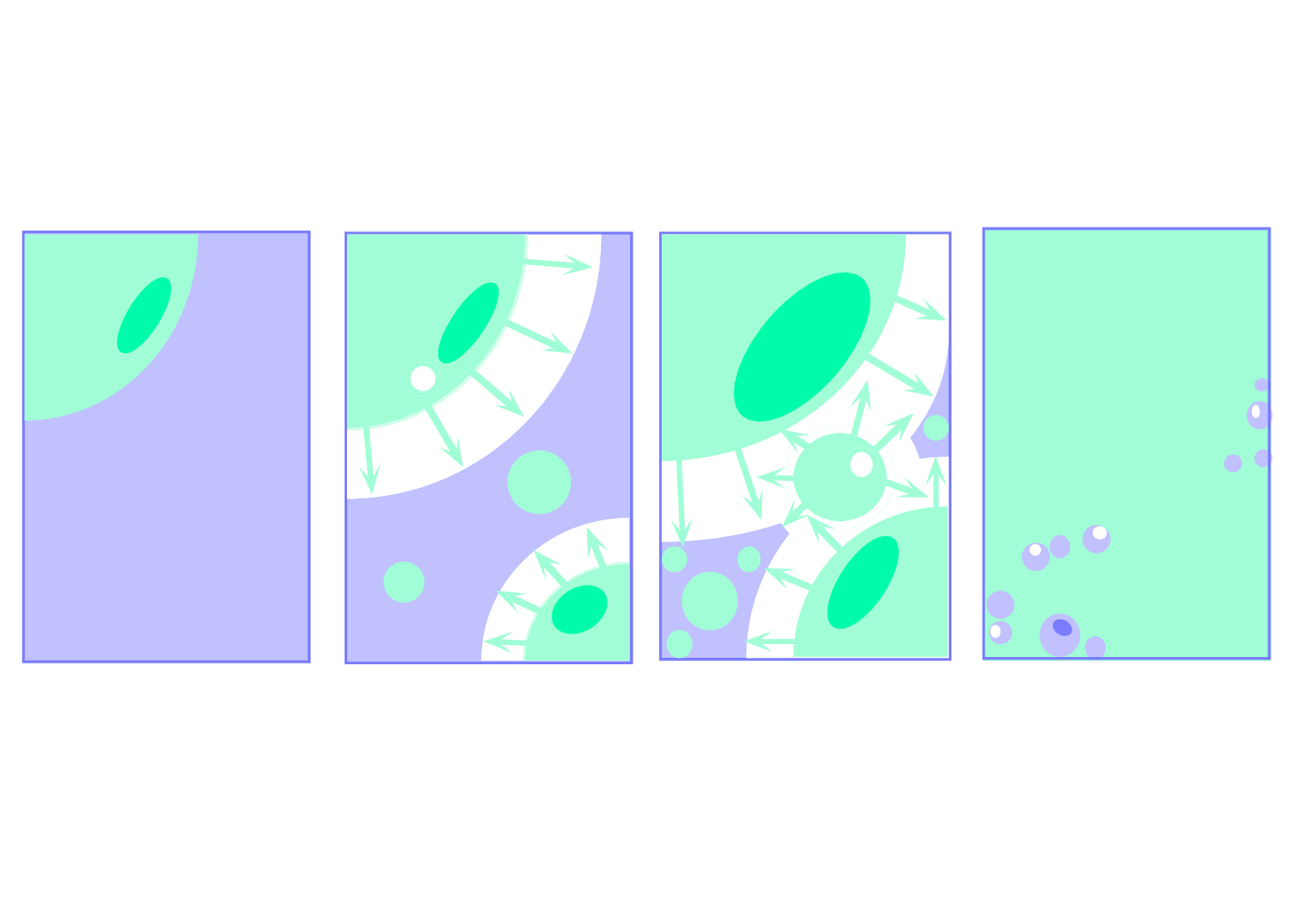}
\vspace{-2cm}
\caption{    
The scenario of the QCD phase transition in the early universe.
The shaded and white regions denote the hadron and QGP phases, respectively.
(a) Slightly below $T_{c}$, only large hadron bubbles appear with very 
small creation rate.  
(b) Hadron bubbles expand with radiating shock wave. 
(c) Near $T_{\rm low}$, many small hadron bubbles are created.
(d) The QGP phase pressured by the hadron phase is isolated. 
}
\label{fig:efpo10}
\vspace{1cm}
\end{figure}

From these results, we can imagine how the QCD phase transition happens in 
the big bang scenario \cite{fuller}, as shown in Fig.\ref{fig:efpo10}.
At the first stage slightly below $T_{\rm c}$, only large bubbles are 
created but  
its rate is quite small.
As temperature is lowered, smaller bubbles are  created with much 
formation rate.
During this process,
the created hadron bubbles expand with radiating shock wave which
reheats QGP phase \cite{fuller}.
Near $T_{\rm low}$ many small bubbles are violently created.
Finally QGP phase is isolated like the bubble \cite{fuller}.
Such an evolution of the hadron bubble 
can be obtained from the numerical simulation using the DGL theory.

Thus, using the quadratic source term instead of the 
linear source term, we have obtained the effective potential
in the DGL theory in the all $|\chi|$ region.
Thermal effects reduce the monopole condensate
and lead to deconfinement phase transition.
Since the critical temperature is found very large, $T_c \sim 0.5$GeV,
with the parameters unchanged at finite temperature,
we introduce a large reduction of the self-interaction between
monopoles so as to make the phase transition temperature about
0.28GeV, which is suggested in the lattice QCD. 
We find large reduction of string tension with the temperature
in accordance with the lattice QCD results.
We predict that the glueball mass decreases considerably 
near the critical temperature, 
which is to be checked by lattice QCD simulation and experiment.
Based on this effective potential at finite temperature,
we further investigated  properties of  hadron bubbles created
in the early Universe and
discussed the  hadron bubble formation process
from estimation of the size of hadron bubble
at various temperatures.

\newpage

\chapter{
Application to  Quark Gluon 
Plasma 
}
\label{sec:AQGP}

\begin{figure}[ht]
\vspace{-2cm}
\epsfxsize = 15 cm
\centering \leavevmode
\epsfbox{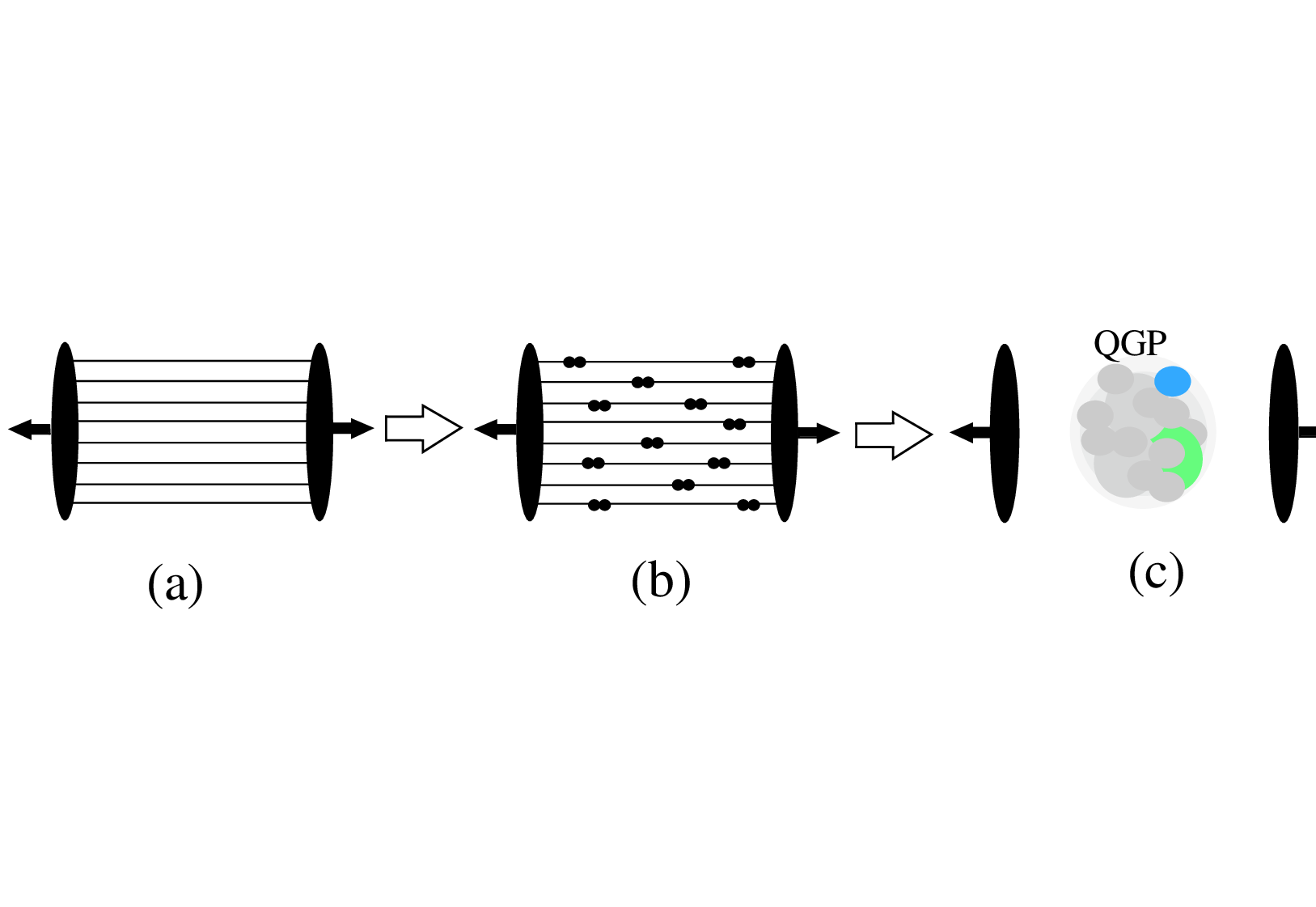}
\vspace{-3cm}
\caption{   
 The scenario of the QGP formation in high-energy heavy-ion collisions.
(a) There appear many color-flux-tubes between the projectile  nucleus and 
the target nucleus just after the collision.
(b) When the distance between the two nuclei becomes large,
 There appears 
the pair creation of quark and anti-quark.
(c) Many created quarks and anti-quarks make 
frequent collisions to form a thermal equilibrium and form QGP, when the 
energy density is larger than a critical value.
}
\label{fig:QGP1}
\end{figure}

In the recent years, some experimental groups are trying 
to create quark gluon plasma (QGP) as the new form of matter 
in the laboratory by high-energy heavy-ion collisions.
The RHIC (Relativistic Heavy Ion Collier)  project will  start
in the next year.
The scenario of producing QGP is based on Bjorken's picture \cite{bjorken}.
Just after heavy ions pass through each other,
many color-flux-tubes are produced between the projectile and  the target,
and pulled by them as shown in Fig.\ref{fig:QGP1}(a).
Usually, it is guessed that these flux-tubes are cut into several pieces 
through quark-antiquark pair creations, and
these short flux-tubes, which behave as excited 'mesons',
are thermalized by stochastic collisions among themselves.
If the energy deposition is larger than a critical value,
the thermalized system becomes the QGP phase, whereas if it is lower, 
the system remains to be the hadron phase.

The features of the  multi color--flux-tube system strongly depend 
on their density of the flux-tubes created by  
hard process in early stage.
When the flux-tube number density is low enough, the system is 
approximated as 
the incoherent sum of the individual flux-tube.
Its evolution is expected to  be superposition of 
random  multiple hadron creations of many color-flux-tubes produced 
between many nucleon-nucleon pairs. 
On the other hand, when the flux-tube number density is  sufficiently high, many flux-tubes overlap each other 
and would be melted  into a big flux-tube.
During this process, each flux-tube loses  
its individuality  and the whole system can be regarded as a huge 
flux-tube  between heavy-ions like a condenser \cite{matsui}.

In this chapter, we would like to study the properties of the 
multi-flux-tube system \cite{ichieC}.
QCD is very hard to deal with in the infrared
region analytically due to the breakdown of the perturbation technique.
Moreover,
for such a large scale system
there is a severe limit on the computational power even in the 
lattice QCD simulation.
For the study of the QGP formation,
the DGL theory would provide a useful method,
because it describes the properties
of the color-electric flux tube, which are important
in the pre-equillibrium system just after the
ultrarelativistic heavy-ion collisions.



\section{Formalism on Multi-flux-tube System}

In this section, we formulate the multi-flux-tube system in the DGL 
theory.
In this theory, the color-flux-tube is described as the dual version of
the Abrikosov vortex \cite{nielsen}. 
Different from the multi-vortex system in the 
superconductor, however,
there are some kinds of the color-flux-tubes corresponding to 
the kinds of color.
Furthermore, 
in the superconductor, 
the direction of the flux-tube is all same and
the system can be described only by the ground state such as triangle 
lattice system, while in the QCD vacuum
many color flux-tubes distribute randomly and the system includes the 
highly excited states. 
Here, we simplify such a complex system and discuss qualitatively.
For simplicity, we consider a single color charge system
using the dual Ginzburg-Landau Lagrangian Eq.(\ref{eq:GL1}).

\begin{figure}[htbp]
\epsfxsize = 13 cm
\centering \leavevmode
\epsfbox{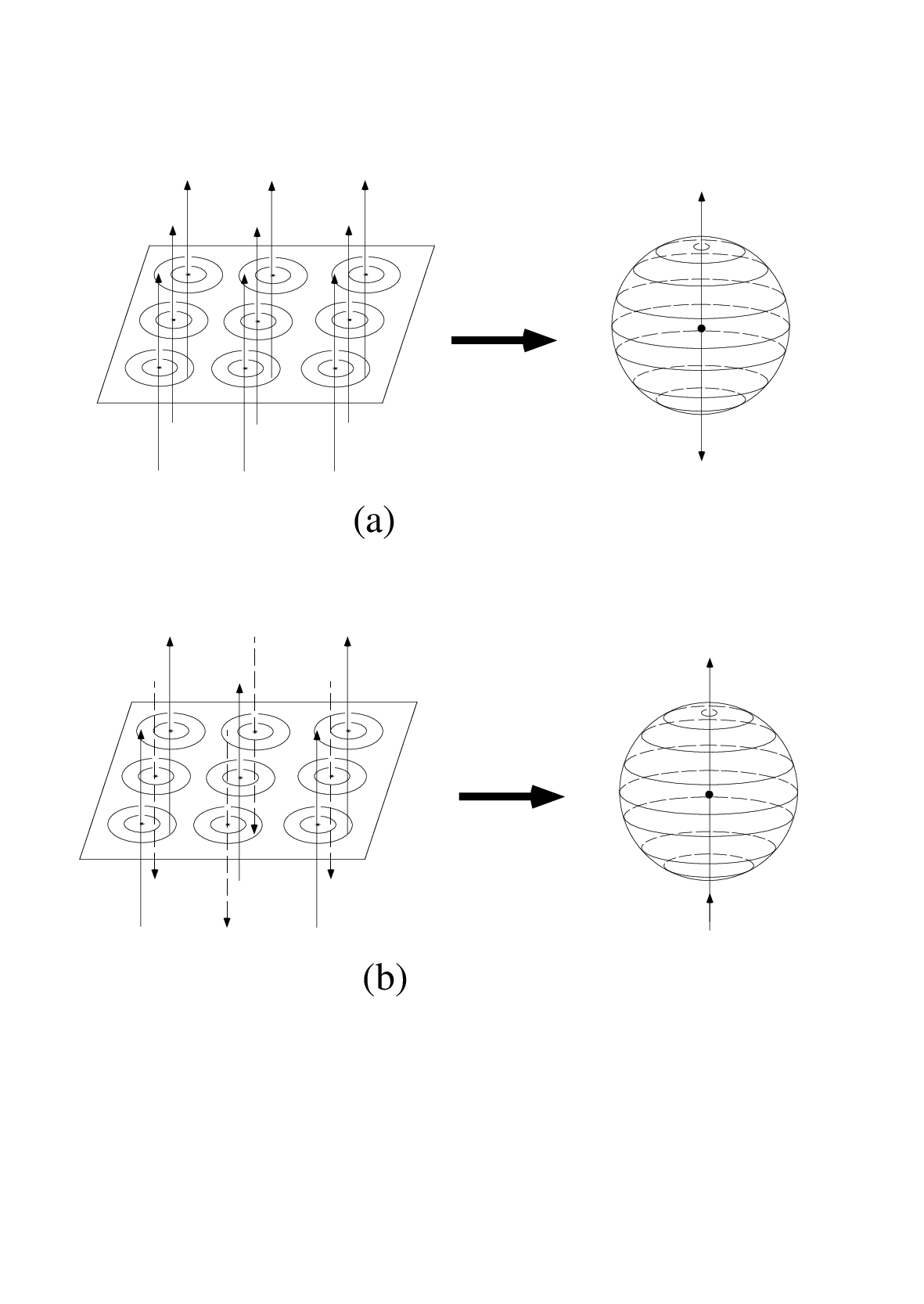}
\vspace{-2.5cm}
\caption{     
(a) A multi-flux-tube  system with the same direction of the 
flux-tubes is approximated by the two  color-flux-tubes going out 
from the north and the south poles on a sphere $S^2$.
(b) A multi-flux-tube system, where flux-tube direction is alternative, is 
approximated by the flux-tube and  'anti-flux-tube' system 
penetrating on $S^2$
with the color-flux going in from the south pole and leaving
out from the north pole.
}
\label{fig:QGP2}
\end{figure}

We consider two ideal cases of multi-flux-tube
penetrating on a {\it two dimensional  plane}.
The directions of all the color-flux-tubes are the 
same [Fig.\ref{fig:QGP2}(a)] in one 
case or alternative[Fig.\ref{fig:QGP2}(b)] in the other case.
When the flux-tubes are long enough,
the effect of  the flux edges is negligible.
Hence, taking the direction of the flux-tubes as the $z$-axis, the system 
is translationally invariant in the $z$-direction and 
is essentially described only with two spatial coordinates $(x,y)$.
For the periodic case in the $(x,y)$ coordinate, 
we can  regard the system as two flux-tubes going through two poles 
(north and south poles) of the $S^2$ {\it sphere}.
For the  system of  flux-tubes with all the directions being the same,
we take two flux-tubes passing through the two poles on $S^2$ sphere
as shown in Fig.\ref{fig:QGP2}(a) (which we call the two flux-tubes system).
For the alternative case, on the other hand,
we take a flux-tube coming in from the south pole and the other leaving 
out 
through the  north pole (which we call
flux-tube and anti-flux-tube system).
Such a prescription leads the exact solution for the periodic crystal of 
the sine-Gordon kinks, and also provides a simple but good description
for the finite density Skyrmion system studied by 
Manton \cite{skyrmion}.

The two color-flux-tube system on the sphere $S^2$  with 
radius $R$  corresponds to the multi-flux-tube system with the 
density $\rho=1/(2\pi R^2)$.
Introducing the polar coordinates $(R,\theta,\varphi)$ on $S^2$,
we consider the  static solution satisfying 
\begin{eqnarray}
B_0=0 , \hspace{2cm} {\bf B}=B(\theta){\bf e}_\varphi
\equiv \frac{\tilde B(\theta)}{R\sin \theta}  {\bf e}_\varphi ,
\hspace{2cm} \chi = \bar \chi(\theta) e^{in\varphi},
\end{eqnarray}
where we have used the axial symmetry of the system.
Here the electric field penetrates vertically on the {\it sphere} {surface,
${\bf E}//{\bf e}_r$;
\begin{eqnarray}
{\bf E}={\bf \nabla} \times {\bf B} = E{\bf e}_r,
\end{eqnarray}
which corresponds to the electric field penetrating vertically on the {\it 
plane}, ${\bf E}//{\bf e}_z$.
The field equations  are given by
\begin{eqnarray}
{1 \over {R^2\sin \theta }}
{d  \over {d \theta }}
(\sin \theta {{d \bar \chi }
\over {d \theta }})
-[{1 \over {R^2\sin ^2 \theta }}(n-g \tilde B(\theta))^2
+2\lambda (\bar \chi ^2-v^2)]\bar  \chi =0,
\label{eq:ren1}
\end{eqnarray}
\begin{eqnarray}
{d  \over {R^2d \theta }}
({1 \over {\sin \theta }}
{d  \over { d \theta }} \tilde B(\theta))
+\frac{2g}{\sin \theta}(n-g\tilde B(\theta))\bar \chi ^2=0.
\label{eq:ren2}
\end{eqnarray}
Consider the closed loop $C$ on $S^2$ with a constant  
 polar angle  $\theta = \alpha$ and $\phi \in [0,2\pi)$,
the electric flux penetrating the area surrounded by the loop $C$ is given 
by
\begin{eqnarray}
\Phi( \alpha ) = \int_{S} {\bf E} \cdot d{\bf S}
  = \int {\bf \nabla} \times {\bf B} \cdot
    d{\bf S}
  = \oint_c {\bf B} \cdot d{\bf l}
  = 2 \pi \tilde B(\alpha).
\end{eqnarray}
The boundary conditions for the two flux-tubes  system 
as shown in Fig.\ref{fig:QGP2}(a) are given as 
\begin{eqnarray}
\Phi( \alpha )=2 \pi \tilde B(\alpha) = 0,
  \hspace{2cm}              \bar \chi(\alpha) = 0   \hspace{2cm}
{\rm as}  \hspace{0.5cm} \alpha  \rightarrow 0,  
\label{eq:kashi1}
\end{eqnarray}
\begin{eqnarray}
\Phi( \alpha )=2 \pi  \tilde B(\alpha)
= \frac 12  \int _{S} E R^2 d\Omega 
= \pm \frac{2\pi n}{g}
\hspace{2cm} {\rm as} \hspace{0.5cm}
\alpha \rightarrow \frac{\pi}{2} \pm \epsilon,
\label{eq:kashi2}
\end{eqnarray}
Here, $n$ corresponds to the topological number of the flux-tube,
which appears also in the vortex solution in the superconductivity.
This boundary condition at $\alpha \rightarrow \frac{\pi}{2} \pm \epsilon$
has a discontinuity for the dual gauge field $\tilde B$.
Because the electric flux leaves out from the two poles,
there should be some sources to provide the electric flux.
In this case,
the Dirac-string like
 singularity appears on  the $\theta=\pi/2$ line, through which the electric
flux comes into the sphere from long distance. 
For the flux-tube and anti-flux-tube system as shown in Fig.\ref{fig:QGP2}(b), the 
boundary condition around $\theta = 0,\pi$ is given as  
\begin{eqnarray}
\Phi( \alpha )=2 \pi \tilde B(\alpha) = 0,
  \hspace{2cm}              \bar \chi(\alpha) = 0   \hspace{2cm}
{\rm as}  \hspace{0.5cm} \alpha  \rightarrow 0,\pi.  
\end{eqnarray}
In this case, there does not appear the Dirac-string like singularity,
since the electric flux is conserved.
The free energy for the unit length
of the color-flux-tube is written as
\begin{eqnarray}
F&=&\int { R^2 d\Omega}
\left[ {{1 \over 2}\left( {{1 \over {R^2\sin \theta }}
{d \over {d\theta }}\tilde B(\theta )} \right)^2
+\left( {{1 \over R}{{d\bar \chi } \over {d\theta }}} \right)^2} \right] 
\nonumber \\
&+&
\int { R^2 d\Omega}\left[ {{1 \over {R^2\sin ^2\theta }}
(n-g\tilde B(\theta ))^2\bar \chi ^2+\lambda
 \left( {\bar \chi ^2-v^2} \right)^2} \right] \nonumber\\
&=&\int_0^{\theta =\pi } 
{2\pi\sin }\theta d\theta \left[ 
{\left( {{{d\bar \chi } \over {d\theta }}} \right)^2
+{1 \over {\sin ^2\theta }}(n-g\tilde B(\theta ))\bar \chi ^2} 
\right]\nonumber\\
&+&\int_0^{\theta =\pi } {2\pi\sin }\theta d\theta \left[ 
{{1 \over 2}\left( {{1 \over {\sin \theta }}
{d \over {d\theta }}\tilde B(\theta )} \right)^2}
 \right]\cdot {1 \over {R^2}} \nonumber \\ 
 & + &\int_0^{\theta =\pi } 
 {2\pi\sin }\theta d\theta 
 \left[ {\lambda \left( {\bar \chi ^2-v^2} \right)^2} \right]\cdot R^2.
\end{eqnarray}
First, we consider a limit of $R \rightarrow \infty$, which corresponds
to the ordinary single vortex solution.
Introducing a new variable $\rho \equiv R \sin \theta$,
the free energy in  the limit $R \gg \rho$$(\theta \sim 0)$ is written as
\begin{eqnarray}
F&=&\int_{}^{} {2\pi \rho d\rho}\left[
 {{1 \over 2}\left( {{1 \over \rho }
 {d \over {d\rho }}\rho B(\rho )} \right)^2
 +\left( {{{d\bar \chi } \over {d\rho }}} \right)^2} \right] \nonumber \\
&+&\int_{}^{} {2\pi \rho d\rho}\left[ {{1 \over {\rho ^2}}
(n-g\rho B(\rho ))\bar \chi ^2
+\lambda \left( {\bar \chi ^2-v^2} \right)^2} \right],
\label{eq:enchu1}
\end{eqnarray}
and the  field equations are
\begin{eqnarray}
{1 \over \rho }{d \over {d\rho }}\rho \left( {{d \over {d\rho }}\bar  \chi } 
\right)
-{1 \over {\rho ^2}}(n-g\rho B(\rho ))^2
-2\lambda \left( {\bar \chi ^2-v^2} \right)\bar \chi =0,
\end{eqnarray}
\begin{eqnarray}
{d \over {d\rho }}{1 \over \rho }
{d \over {d\rho }}\left( {\rho B(\rho )} \right)
+{{2g} \over {\rho ^{}}}(n-g\rho B(\rho ))\bar \chi ^2=0,
\label{eq:enchu2}
\end{eqnarray}
with the boundary condition,
\begin{eqnarray}
\Phi = 2\pi\rho B(\rho)|^{\infty}_0=\frac{2\pi n}{g}
\hspace{0.5cm} \mbox{as} \hspace{0.5cm} \rho \rightarrow \infty.
\end{eqnarray}
Above equations, (\ref{eq:enchu1}-\ref{eq:enchu2}),
coincide exactly with those of ordinary  single vortex 
solution in the cylindrical coordinate.
Thus, we get the desired results.

One can analytically investigate the dependence of the  profile 
functions$(\tilde B(\theta),\chi(\theta))$ on the flux-tube number density.
For this purpose, we express the free energy as
\begin{eqnarray}
F \equiv  f_0+f_E\cdot {1 \over {R^2}}
+f_\chi \cdot R^2,
\end{eqnarray}  
where $f_0,f_E,$ and$f_\chi$ are $R$ independent functions and written as 
\begin{eqnarray}
f_0  & \equiv & 
\int_0^{\theta =\pi } 
{2\pi\sin }\theta d\theta \left[ {\left( {{{d\bar \chi } \over {d\theta }}} 
\right)^2
+{1 \over {\sin ^2\theta }}(n-g\tilde B(\theta ))^2\bar \chi ^2} \right], \\
f_E & \equiv &
\int_0^{\theta =\pi }
 {2\pi\sin }\theta d\theta  {{1 \over 2}
 \left( {{1 \over {\sin \theta }}{d \over {d\theta }}
 \tilde B(\theta )} \right)^2},
\\
 f_\chi & \equiv &
\int_0^{\theta =\pi }
 {2\pi\sin }\theta d\theta  {\lambda 
 \left( {\bar \chi ^2-v^2} \right)^2} .
\end{eqnarray}
In the large  $R$ case, which corresponds to the small 
color-flux-tube 
number density in the original multi-flux-tube system, the third term 
$f_\chi R^2$ is dominant.
Hence, the free energy $F$ is minimized as $\bar \chi \sim v$,
that is, the monopole tends to condense, and then the color electric 
field is localized only around $\theta=0$ (north pole) and $\theta=\pi$
(south pole) as shown in Fig.\ref{fig:QGP3}.
On the other hand, in the small  $R$ case,
the second term ${f_E}/{R^2} $ is dominant.
There is a constraint on the total flux penetrating on the upper sphere,
\begin{eqnarray}  
\Phi \equiv \int^{\frac{\pi}{2}}_0 E(\theta)2 \pi R^2 {\rm sin}
\theta d \theta = \frac{2 n\pi}{g},
\end{eqnarray}
that is,
\begin{eqnarray}
\int^1_0 E(t) dt= \frac{n}{gR^2} \equiv C \hspace{0.5cm}
 \mbox{with} \hspace{0.5cm} t = {\cos}\theta.  
\end{eqnarray}
Hence, one finds the equation,
\begin{eqnarray}
f_E \propto \int^{\frac{\pi}{2}}_{0}E(\theta)^2 {\rm \sin}\theta d \theta
=\int^1_0  E(t)^2 dt
=\int^1_0 \{(E(t)-C)^2\}dt +  C^2  
\geq C^2.
\end{eqnarray}
This condition leads to the  uniform color electric field $E = C$,
which provides the minimum of $f_E$.
Thus, the color electric field tends to spread over the space uniformly.

\begin{figure}[htbp]
\epsfxsize = 10 cm
\centering \leavevmode
\epsfbox{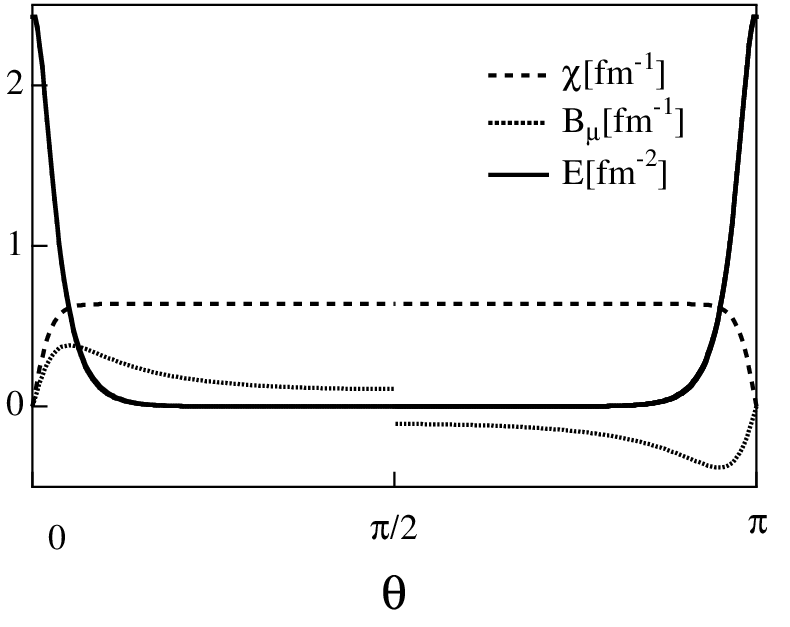}
\vspace{0cm}
\caption{    
The color-electric field $E(\theta)$ (solid curve), 
the monopole condensate $\bar \chi(\theta)$ (dashed curve)
and the dual gauge field $B(\theta)$ (dotted curve) 
are plotted as  functions of 
the polar angle $\theta$ for the low density case with $R=$4.0fm.
}
\label{fig:QGP3}
\end{figure}

Here, we consider the critical radius $R_c$ of the phase transition to 
the normal phase, where the monopole condensate disappears.
There are three useful inequalities on $f_E$, $f_\chi$, and $F$,
\begin{eqnarray}
f_E \geq 2\pi (\frac{ n}{g})^2
\label{eq:shikic}
\end{eqnarray}
\begin{eqnarray}
0 \leq f_\chi \leq\ 4 \pi \lambda v^4, \label{eq:shikia}
\end{eqnarray}
\begin{eqnarray}
F=f_0+f_E \cdot \frac{1}{R^2} + f_\chi \cdot R^2 \geq f_0 +2\sqrt{f_E 
f_\chi}.
\label{eq:shikib}
\end{eqnarray}
The equality is satisfied in Eq.(\ref{eq:shikib}),
\begin{eqnarray}
R^4 = \frac{f_E}{f_\chi} 
\end{eqnarray}
Using the inequalities equations(\ref{eq:shikic})-(\ref{eq:shikib}),
$R^4$ is larger than  the critical $R^4_c$;
\begin{eqnarray}
R^4 \geq R^4_c\equiv
\frac{2\pi n^2/g^2}{4 \pi \lambda v^4 }=
\large( \frac{n^2}{2\lambda g^2}\large)\frac{1}{v^4}.
\end{eqnarray}
For $R > R_c$, there exists a nontrivial inhomogeneous solution, which
differs from the normal phase.
For $R \leq R_c$, homogeneous normal phase provides the minimum of $F$.
Thus, $R_c $
 is the critical radius of the phase transition from the flux-tube
 phase to the normal one. 
In this case, the critical radius and the electric field are given by 
\begin{eqnarray}
\rho_c=\frac{1}{2 \pi R^2_c }=
\sqrt{\frac{\lambda}{2}}\frac{gv^2}{\pi n},
\label{eq:analy} 
\end{eqnarray}
\begin{eqnarray}
E_c
=\frac{n}{gR_c^2}=\sqrt{2\lambda}v^2, 
\end{eqnarray}
respectively.

\section{Numerical Results on Multi-flux-tube System}
We start with showing the low density case of two flux tubes system in 
Fig.\ref{fig:QGP3}.
The parameters of the DGL theory are fixed as $\lambda=$ 25,
$v=$ 0.126GeV, which lead the masses $m_B=$ 0.5GeV and 
$m_\chi=$ 1.26GeV \cite{suganuma}.
This parameter set provides the flux-tube radius $r_{FT}\sim$0.4fm and the 
suitable  interquark potential with 
the string tension as $\sigma=$1GeV/fm.
 The monopole condensate $\bar \chi(\theta)$,
the  dual gauge field $\tilde B(\theta)$ and
the color electric field  $E(\theta)$
are plotted as  functions of 
the polar angle $\theta$.
The electric field $E(\theta)$ is localized around 
the two poles($\theta$=0 and $\pi$)
and drops suddenly as $\theta$ deviates from the two poles.
The monopole condensate  vanishes at the two poles and 
becomes constant in the region away from these poles.
This behavior corresponds to the case of independent two vortices
in superconductivity.
Different from these physical quantities, the dual gauge field $\tilde 
B(\theta)$ is discontinuous at $\theta=\pi/2$.
There should be Dirac-string like  source to provide,
because the electric flux leaving out from the two poles.
It should be noted that the system has the reflection symmetry on
$\theta=\frac{\pi}{2}$ plane.

We show now in Fig.\ref{fig:QGP4} the number density dependence of the flux-tubes.
For the large radius of the sphere, ($R \geq $2fm), the color electric flux
is localized at $\theta=0$ and $\pi$ and there 
the  monopole condensate vanishes, 
while  
the $\bar \chi$  becomes constant $\bar \chi \simeq v$ around 
$\theta = \frac{\pi}{2}$.
As the radius $R$ decreases, the electric flux, localized at $\theta=0$,
$\theta=\pi$, starts to  overlap, and the value of the monopole 
condensate $\bar \chi$ becomes small.
The electric field $E(\theta)$ 
becomes constant and
$\bar \chi$ vanishes below  a critical radius $R_c
$.
We show in Fig.\ref{fig:QGP5} the monopole condensate at $\theta=\frac{\pi}{2}$
(the maximum value of the monopole condensate) as a function of the 
sphere radius $R$ (flux-tube number density $\rho=1/(2 \pi R^2)$).
The (first order) phase transition
occurs and the system becomes homogeneous normal 
phase above the critical value of the flux-tube number density.

\begin{figure}[htbp]
\epsfxsize = 8 cm
\centering \leavevmode
\epsfbox{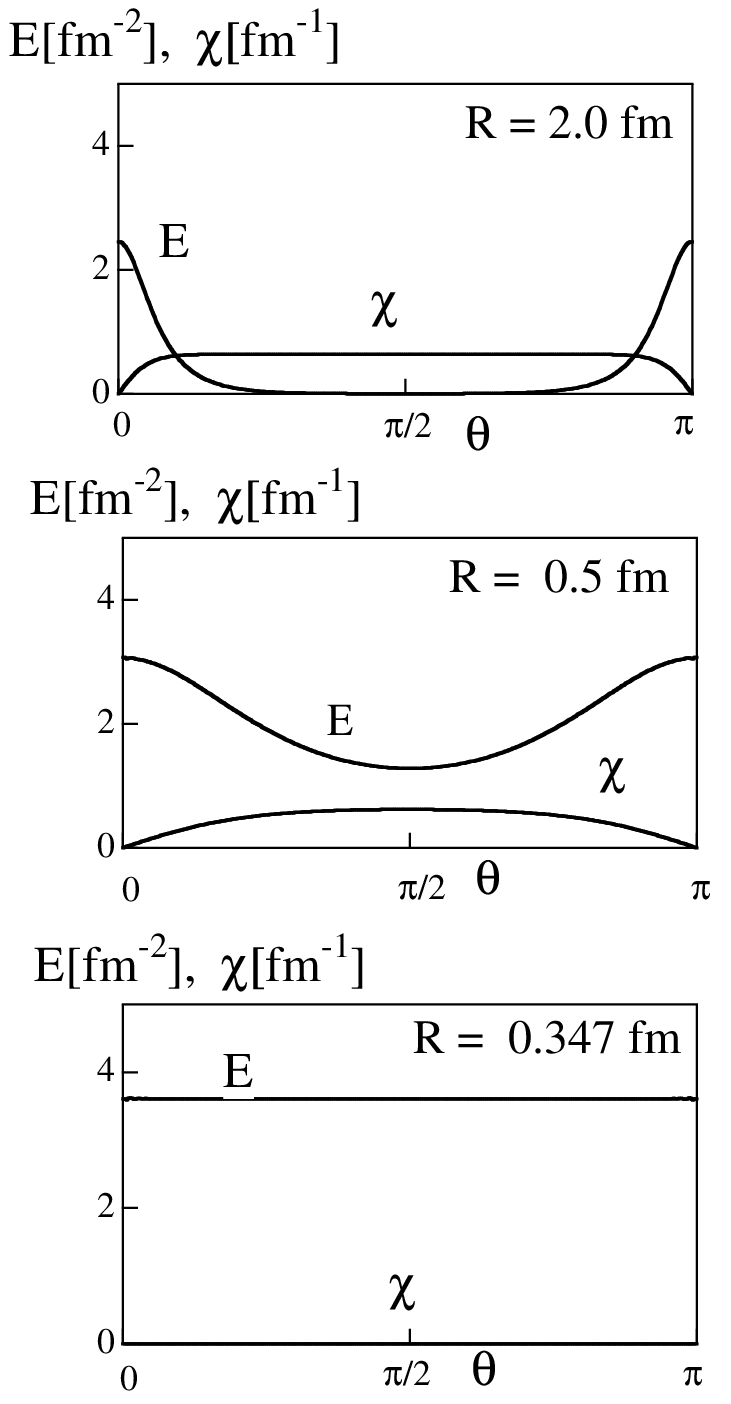}
\vspace{0cm}
\caption{     
The  flux-flux system on $S^2$.
The color-electric field $E(\theta)$ and the monopole condensate 
$\bar \chi(\theta)$
are depicted as  functions of the polar angle 
$\theta$ for the three radii; $R=$2.0fm, 0.5fm and 0.347fm.
Below the critical radius $R_c$=0.347fm, the color-electric field 
$E$  becomes constant and the monopole condensate $\bar \chi$ 
vanishes entirely.
}
\label{fig:QGP4}
\end{figure}

\begin{figure}[htbp]
\epsfxsize = 9 cm
\centering \leavevmode
\epsfbox{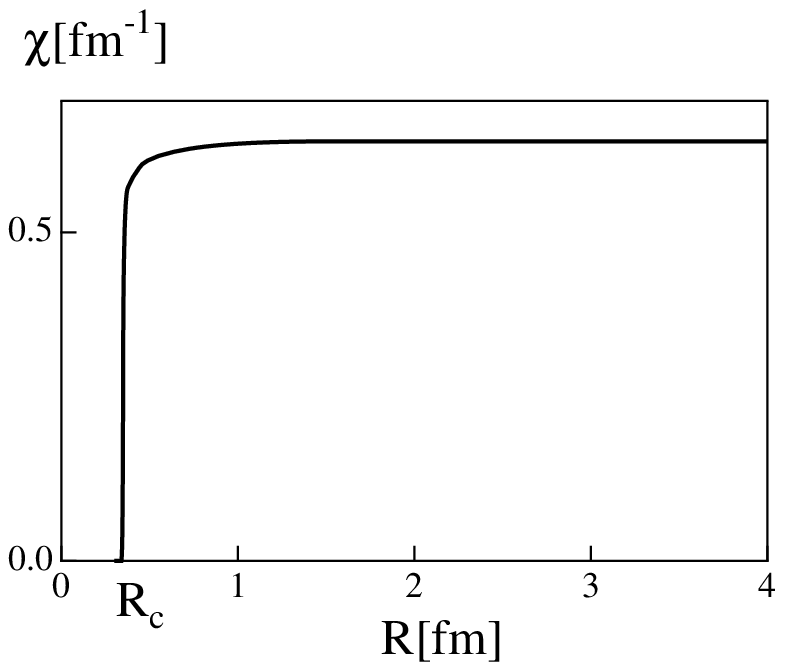}
\vspace{0cm}
\caption{     
The $R$ dependence
of the monopole condensate.
Here, the flux tube density is given by $1/(2\pi R^2)$.
The monopole condensate at $\theta=\frac{\pi}{2}$  decreases,
as the radius $R$ becomes smaller, and vanishes below $R=R_c$.}
\label{fig:QGP5}
\end{figure}

Here, we compare the free energy of two flux tube system with
that of inhomogeneous system in Fig.\ref{fig:QGP6}.
At large $R$, the former is smaller and the system favors the
existence of two flux tubes.
As  $R$ becomes smaller,
the energy difference  of the system becomes smaller.
Two flux-tubes melt and the electric field are changed to be 
homogeneous below the critical radius $R_c=0.35$fm, which   
 corresponds to the critical density $\rho_c=1/(2 \pi R_c^2)=
 1.3{\rm fm}^{-2}$. 
This critical density agrees  with the 
analytic estimation in Eq.(\ref{eq:analy}),
$\rho_c=
\sqrt{\frac{\lambda}{2}}\frac{gv^2}{\pi n}=
1.3{\rm fm}^{-2}$.  

\begin{figure}[htbp]
\epsfxsize = 9 cm
\centering \leavevmode
\epsfbox{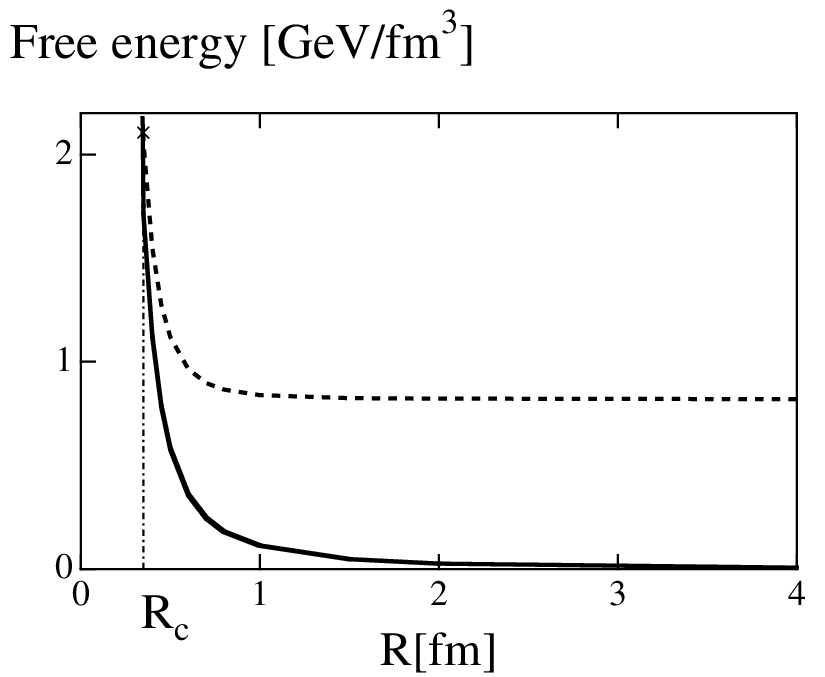}
\vspace{0cm}
\caption{    
The free energy in flux-flux system (solid curve) and
uniform system (dashed curve).
Because the lower free energy system is realized,
the inhomogeneous system  is changed into a homogeneous system
below the critical radius, $R_c=0.35{\rm fm}$.
Here, the critical flux-tube density reads 
$\rho_c
= 1.3{\rm fm}^{-2}$.
}
\label{fig:QGP6}
\end{figure}

We discuss now the system of flux-tube and anti-flux-tube with opposite 
direction placed at $\theta=$0 and $\theta=\pi$ respectively 
as shown in Fig.\ref{fig:QGP7}.
At low flux-tube number density ( $R \geq 2$fm),
the flux-tube  is localized at $\theta=0$, while the anti-flux-tube  at 
$\theta=\pi$.
The monopole condensate $\bar \chi(\theta)$ vanishes at the 
two poles ($\theta=0,\pi$) and becomes constant
away from these poles.
As  $R$ decreases, the electric flux starts to cancel each other 
and the monopole
condensate becomes small.
We also compare  the free-energy of the flux-tube and anti-flux-tube system
with the homogeneous system, in which both monopole condensate
and electric field are vanished. The critical radius, $R_c=$ 0.31fm,
is similar to the value of the two flux-tubes system.

\begin{figure}[htbp]
\epsfxsize = 8 cm
\centering \leavevmode
\epsfbox{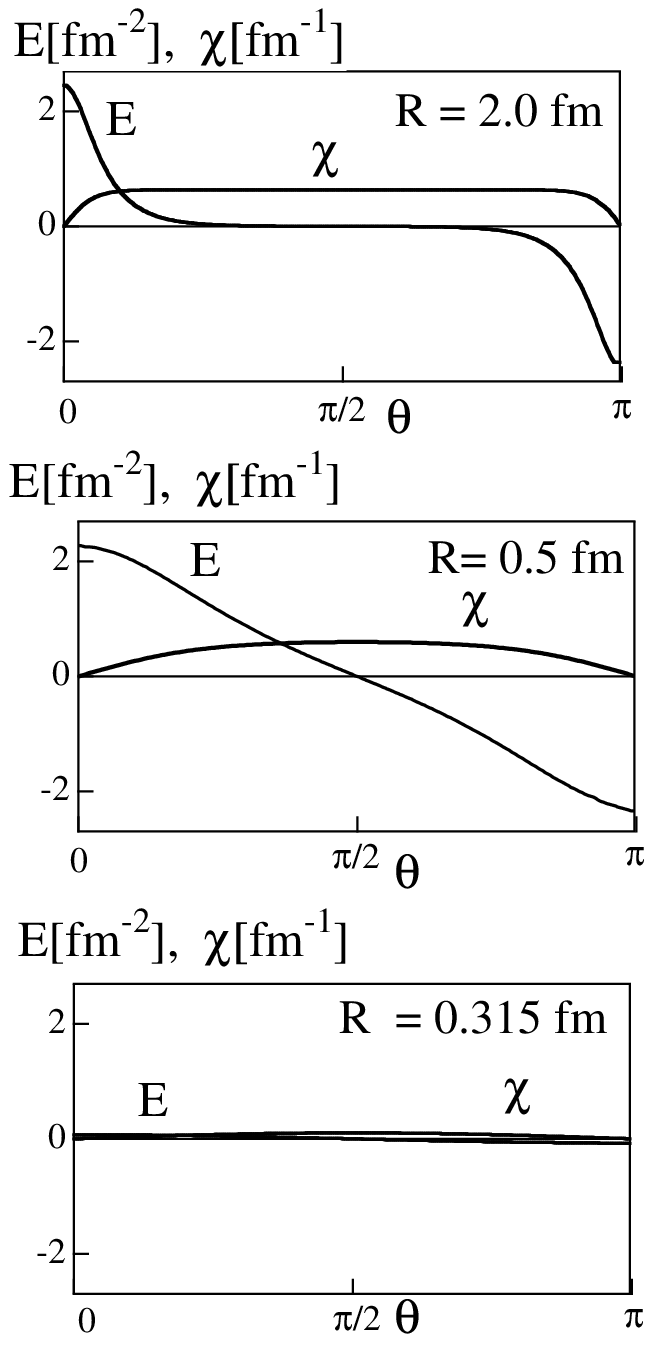}
\vspace{0cm}
\caption{   
The flux the and anti-flux-tube system on $S^2$. 
The color-electric field $E(\theta)$
and the monopole condensate  $\bar \chi(\theta)$ are depicted  as  functions 
of the polar angle $\theta$ for the three radii; $R=$2.0fm, 0.5fm, and  0.315fm.
Below $R=$0.315fm, both the color electric flux and the monopole 
condensate vanish entirely.
}
\label{fig:QGP7}
\end{figure}

%

Thus, we have found in both cases that the solution behaves as two independent
flux-tubes at a large $R$, i.e., a small number density $\rho$.
As the radius $R$ decreases, 
the monopole condensate decreases 
and eventually vanishes at a critical density,
where the color-electric field $E$ becomes uniform.
The critical density $\rho_c$ is found as 
$\rho_c=$1.3${\rm fm}^{-2}=(1.14{\rm fm})^{-2}$ 
for the two color-flux tube system;
$\rho_c=$1.7${\rm fm}^{-2}=(1.3{\rm fm})^{-2}$
for the  flux-tube and anti-flux-tube system.
Such similar values   in both cases suggest  that an actual 
flux-tube system would become uniform  around similar
density  to  $\rho_c \sim$1.5fm${}^{-2}$,
because realistic flux-tube system includes the flux-tubes and anti-flux 
tubes randomly, which  would correspond to an intermediate system between 
the above two ideal cases.  
Thus, the configuration of the color-electric field and the monopole field 
depend largely on the flux-tube density.

As discussed above, many flux-tubes are melted 
around $\rho_c \sim$ 1.5 fm${}^{-2}$.
Let us discuss here in  case of central collision between
heavy-ions with mass number $A$.
The nuclear radius $R$ is given by $R=R_0 A^{1/3}$, where
$R_0$=1.2fm corresponds to the nuclear radius,
and the normal baryon-number density is $\frac{A}{\frac{4}{3}\pi R^3}
=\frac{1}{\frac43 \pi R_0^3}$.
In the central region, one nucleon in the projectile makes hard
collisions with $\frac32 A^{1/3}$$(=2 \pi R_0^3 A^{1/3} \times \rho_0)$
nucleons in the target,
because the reaction volume is $\pi R_0^2 \times 2 R =2\pi R^3_0 A^{1/3}$.
Hence, nucleon-nucleon collision number is expected as 
 $(\frac32 A^{1/3})^2= \frac{9}{4} A^{2/3}$ 
per the single-nucleon area $\pi R_0^2$ between projectile and target 
nuclei.
Assuming one flux-tube formed in one nucleon-nucleon hard collision,
the flux-tube density is estimated as 
$\rho=\frac{ \frac{9}{4} A^{2/3}}{\pi R_0^2}=\frac{A^{2/3}}{(1.4{\rm fm})^2}$.
For instance,
$\rho$ would be 4.5, 5.8 and 17.5 fm${}^{-2}$ for $A=$ 27, 40 and 208.
This would indicate that the scenario of creating large sizes of flux tube 
becomes  much relevant for $A$-$A$ collision with larger $A$.

It would be important to reconsider  the process of the QGP formation 
in terms of the flux-tube number density.
When the flux-tube density is low enough, the flux-tubes are localized.
Each flux-tube evolution would be regarded as the multi-creation of hadrons
in the high energy p-p collision via the flux-tube breaking.
In this process, $q$-$\bar q$ pair creation plays an essential role 
on the QGP formation, which is the usual scenario.

On the other hand, for the dense flux-tube system,
neighboring flux-tubes are melted into a large cylindrical tube,  
where monopole condensate disappears.
Such a system, where the color electric field is made between heavy-ions,
becomes approximately homogeneous and is regarded as the 
'color condenser'.
In this case, large homogeneous QGP may be created in the central region.

In the actual case, however, the variations and directions 
of the color-flux-tubes are random.
For instance, in the peripheral region,
flux-tubes would be localized and are broken by 
quark-antiquark pair creations. 
In the central region,  dense flux-tubes are melted by
 annihilation or unification  \cite{ichieC}
 of flux-tubes, which will 
 be discussed in the next section.
 In this region, a huge system of dynamical gluons appears,
 because many dynamical gluons are created during this process.
Thermalization of such quarks and gluons leads to   QGP.  
Thus, the process  of QGP formation should  depend largely on the 
density of created flux-tubes, 
which is closely related to the incident energy, the impact parameter and 
the size of the projectile and the target nuclei.
 
\section{Flux-tube Interaction and QGP Formation Process}

Finally, we investigate the interaction among various color-flux-tubes,
which is considered to be important for the QGP formation process.
Each flux tube is characterized by the
color charge $\vec Q$  \cite{kerson,suganumax}
 at its one end.
To classify sorts of the flux tube,
we call the flux tube with a red quark ($R$) at its one end
as ``$R$-$\bar R$ flux tube'', and so on.
In this case, the ``direction'' of the color-electric flux
in the flux tube should be distinguished.
For instance, $\bar R$-$R$ flux tube is different
from $R$-$\bar R$ flux tube in terms of the flux direction.
We study the interaction between two color-electric flux tubes
using the DGL theory.
The color-electric charges at one end of the flux tubes are
denoted by $\vec Q_1$ and $\vec Q_2$.
We idealize the system as two sufficiently long flux tubes,
and neglect the effect of their ends.
We denote by $d$ as the distance between the two flux tubes.
For $d \gg m_\chi ^{-1}$,
the interaction energy per unit length in the two flux tube system
is estimated as
\begin{eqnarray}
E_{\rm int} \simeq {8\pi  \vec Q_1 \cdot \vec Q_2  \over e^2}
m_B^2 K_0(m_Bd),
\label{VVINT}
\end{eqnarray}
where $K_0(x)$ is the modified Bessel function.
Here, we have used the similar calculation
on the Abrikosov vortex in the type-II superconductor \cite{lifshitz}.

\begin{figure}[tb]
\vspace{-0.5cm}
\epsfxsize = 11 cm
\centering \leavevmode
\epsfbox{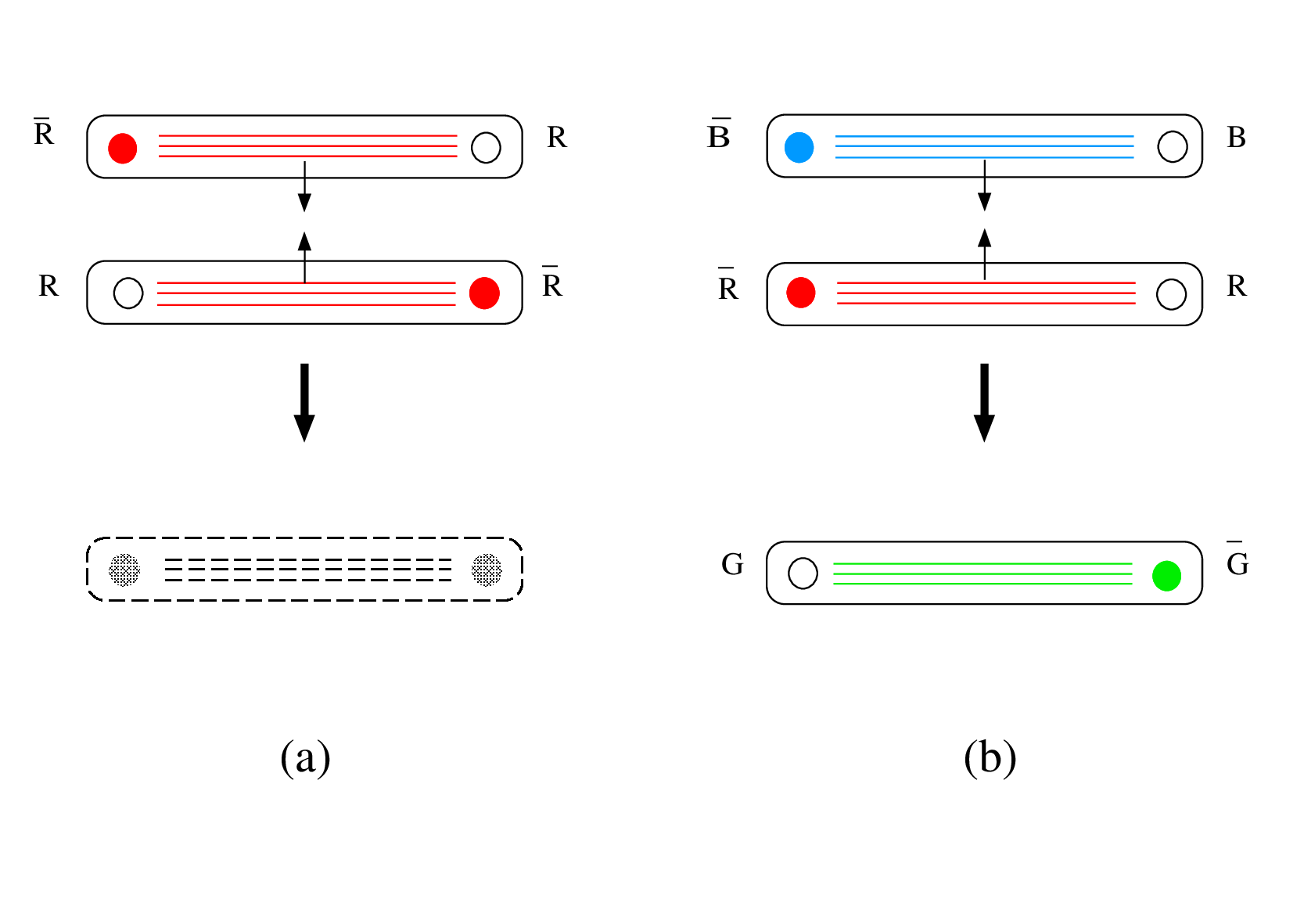}
\vspace{0cm}
\caption{    
The annihilation process of the color-electric flux tubes
during the QGP formation in ultrarelativistic
heavy-ion collisions.
(a) The same flux tubes with opposite flux direction are attracted
each other, and are annihilated into dynamical gluons.
(b) The different flux tubes (e.g. $R$-$\bar R$ and $B$-$\bar B$)
are attractive, and are unified into a single flux tube.
}
\label{fig:QGP8}
\end{figure}

As shown in Fig.\ref{fig:QGP8}, there are two interesting cases
on the interaction between two color-electric flux tubes.

\begin{enumerate}
\item (a)
For the same flux tubes with opposite flux
direction (e.g. $R$-$\bar R$ and $\bar R$-$R$),
one finds $\vec Q_1=-\vec Q_2$ i.e.
$\vec Q_1 \cdot \vec Q_2=-e^2/3$, so that
these flux tubes are attracted each other.
It should be noted that they would be annihilated into
dynamical gluons in this case.
\item (b)
For the different flux tubes satisfying
$\vec Q_1 \cdot \vec Q_2<0$
(e.g. $R$-$\bar R$ and $B$-$\bar B$),
one finds $\vec Q_1 \cdot \vec Q_2=-e^2/6$,
so that these flux tubes are attractive.
In this case, they would be unified into a single flux
tube (similar to $\bar G$-$ G$ flux tube).
\end{enumerate}

Based on the above calculation,
we propose a new scenario on the QGP formation via
the annihilation of the color-electric flux tubes.
When the flux tubes are sufficiently dense in the central region
just after ultrarelativistic heavy-ion collisions,
many flux tubes are annihilated or unified.
During the annihilation process of the flux tubes,
lots of dynamical gluons (and quarks) would be created.
Thus, the energy of the flux tubes turns into that of the
stochastic kinetic motion of gluons (and quarks).
The thermalization is achieved through the stochastic
gluon self-interaction, and finally the hot QGP would be created.
Here, the gluon self-interaction in QCD plays an
essential role to the thermalization process,
which is quite different from the photon system in QED.

In more realistic case, both the quark-pair creation and the flux-tube
annihilation would take place at the same time.
For instance, the flux tube breaking  \cite{matsui,gatoff,tatsumi}
would occur before the flux tube annihilation for the
dilute flux tube system.
On the contrary,
in case of the extremely high energy collisions,
these would be lots of flux tubes overlapping
in the central region between heavy ions,
and therefore the flux tube annihilation
should play the dominant role in the QGP formation.
In any case, the DGL theory would provide a calculable method
for dynamics of the color-electric flux tubes in the QGP formation.




\chapter{Summary and Concluding Remarks}

In the basis of the dual superconductor picture, we have systematically 
studied the confinement phenomena using the lattice QCD Monte-Carlo 
simulation, the monopole-current dynamics and the dual Ginzburg-Landau
(DGL) theory, an infrared effective theory of QCD. 
In the dual Higgs theory, color confinement is understood 
by  one-dimensional squeezing of the color-electric flux in the QCD 
vacuum through the dual Meissner effect caused by monopole condensation.
For the construction of the dual superconducting theory from QCD, 
there are two large gaps on speciality of the abelian sector and 
the appearance of monopoles, however, these gaps are expected to be 
fulfilled by taking the 't Hooft abelian gauge fixing, which is 
defined by the
diagonalization of a suitable gauge-dependent variable, $\Phi[A_{\mu}(x)]$. 
In the abelian gauge, the SU($N_c$) gauge theory is reduced into 
the U(1)$^{N_c-1}$ gauge theory including  color-magnetic 
monopoles,
which topologically appears corresponding to the nontrivial homotopy 
group, $\Pi_2($SU$(N_c)/$U$(1)^{N_c-1}) = {\bf Z}^{N_c-1}$.
In this gauge,
the diagonal and the off-diagonal gluon components behave as the 
abelian gauge field and the charged matter field, respectively, 
in terms of the residual gauge symmetry. As a remarkable fact, ``abelian 
dominance'', irrelevance of off-diagonal gluons, is numerically 
observed for the nonperturbative QCD phenomena like confinement and 
dynamical chiral-symmetry breaking in the lattice QCD simulation in 
the MA gauge, which is a sort of the abelian gauge.
Monopole condensation has 
been also suggested by the lattice QCD 
as the appearance of the global network of the 
monopole current in the MA gauge.


First, we have studied the origin of abelian dominance for the 
confinement force in MA gauge in terms of the gluon-field properties 
using the lattice QCD. 
In the MA gauge, the gluon-field fluctuation is maximally 
concentrated in the abelian sector.
As the remarkable feature in the MA gauge, we have found that 
the amplitude of the 
off-diagonal gluon is strongly suppressed, and therefore the phase 
variable of the off-diagonal (charged) gluon tends to be random, 
according to the weakness of the constraint from the QCD action.
Using the random-variable approximation for the charged-gluon phase 
variable, we have found the perimeter law of the charged-gluon contribution 
to the Wilson loop and have proved abelian dominance for the string  
tension in the semi-analytical manner.
These theoretical results have been  also numerically confirmed using the 
lattice QCD simulation.

Second, we have studied th QCD-monopole appearing in the abelian sector in 
the abelian gauge. The appearance of monopoles have been transparently 
formulated in terms of the gauge connection, and is originated from the 
singular nonabelian gauge transformation to realize the abelian gauge.
We have investigated the gluon field around the monopole in the lattice QCD. 
The QCD-monopole carries a large fluctuation of the gluon field and 
provides a large abelian action of QCD.
Nevertheless, QCD-monopoles can appear in QCD without large cost of 
the QCD action, due the large cancellation between the abelian and 
off-diagonal parts of the QCD action density around the monopole.
We have derived a simple relation between the confinement force and the 
monopole density by idealizing the monopole contribution to the 
Wilson loop.

Third, we have studied the monopole-current dynamics using the infrared 
monopole-current action defined on a lattice. We have adopted the local 
current action, considering the infrared screening of the 
inter-monopole interaction due to the dual Higgs mechanism.
When the monopole self-energy $\alpha$ is smaller then $\alpha_{c} = 
$ ln$ (2D-1 )$, monopole condensation can be analytically shown, and 
we have found this system being the confinement phase from the Wilson 
loop analysis.
By comparing the lattice QCD with the monopole-current system, the 
QCD vacuum has been found to corresponds to the monopole-condensed phase in 
the infrared scale.
We have considered the derivation of the DGL theory from the monopole 
ensemble, which would be essence of the QCD vacuum in the MA gauge 
because of abelian dominance and monopole dominance.

In the second half part of this paper, we have studied the 
deconfinement phase transition and the hadron flux-tube system
using the DGL theory. 
Deconfinement phase transition is one of the most interesting subject
in the QCD phase transition 
both for ultra-relativistic heavy-ion collisions and for 
early universe. 
We have formulated the effective potential in the  DGL theory,
introducing the quadratic source term
to study the QCD vacuum at finite temperatures.
We have found the reduction of the monopole 
condensate at finite temperatures, 
and found a first-order deconfinement phase transition 
at the critical temperature $T_C \simeq $ 0.49GeV 
using the temperature-independent parameters.
The monopole condensate vanishes and 
the broken dual gauge symmetry is restored above $T_C$. 
We have considered the temperature dependence of the monopole 
self-interaction noting $T_C=0.28$GeV 
as the lattice QCD simulation indicates. 
We have found a large reduction of the 
monopole self-interaction near the critical temperature. 
We have investigated the temperature dependence of the masses for
the relevant mode like $m_B$ and $m_\chi $, 
and have found their large reduction 
near the critical temperature $T_C$: $m_B, m_\chi  \sim T_C$.
We have calculated also the string tension at finite temperatures. 
The results agree with the lattice QCD data both in 
the variable $\lambda (T)$ and in the variable $v(T)$ cases.
In particular, the glueball mass reduction at high temperatures 
would be an important ingredient in the QCD phase transition.
In the pure gauge, there are only glueball excitations 
with the large mass ($\stackrel{>}{\sim}$ 1GeV) at low temperatures, 
and therefore it seems unnatural that the QCD phase transition 
takes place at a small critical temperature, $T_C \simeq$ 0.28GeV.
This problem would be explained by 
the large reduction of the glueball mass near the 
critical temperature as is demonstrated. 
In other words, this glueball-mass reduction may 
determine the magnitude of the critical temperature $T_C$ 
in the QCD phase transition.

Based on this effective potential at finite temperature,
we  have investigated  the properties of hadron bubbles created
in the early Universe and discussed the hadron bubble formation process.
From numerical results, we can imagine how the QCD phase transition
happens in the big bang scenario.
(a) Slightly below $T_{\rm c}$,
only large hadron bubbles appear, but the creation rate is quite small. 
(b) As temperature is lowered by the expansion of the Universe,
smaller bubbles are created with much formation rate.
During this process, the created hadron bubbles expand by
radiating shock wave, which reheats the QGP phase around them.
(c)  Near $T_{\rm low}$, many small hadron bubbles
are violently created in the unaffected region free from the shock wave.  
(d)  The QGP phase pressed by the hadron phase
is isolated as high-density QGP bubbles,
which provide the baryon density fluctuation.
Thus, the numerical simulation using the DGL theory would tell
how the hadron bubbles appear and evolve
quantitatively in the early Universe.

We have studied the multi-flux-tube system in terms of the DGL theory.
We have considered two ideal cases, where the directions of all
the color-flux-tubes are  the same in one case and alternative
in the other case for neighboring flux-tubes.
We have formulated the system of multi color-flux-tubes
by regarding it as the system  of two color-flux-tubes
penetrating through  a two dimensional  sphere surface.
We have found the multi flux-tube configuration becomes uniform above  some
critical flux-tube number density $\rho_c = 1.3 \sim 1.7 {\rm fm}^{-2}$.
On the other hand, the inhomogeneity on the color electric  distribution
appears
when the flux-tube density is smaller than $\rho_c$.
We have discussed the relation between the inhomogeneity in the color-electric
distribution
and the flux-tube number density in the multi-flux-tube system
created during the QGP formation process in the
ultra-relativistic heavy-ion collision.
When the flux-tube number density is low enough,  
the system can be approximated as the incoherent sum of the individual 
flux tube.
Its evolution would  lead to the creation of $q$-$\bar q$ pairs via 
flux-tube breaking. In this process, this $q$-$\bar q$ creation plays a 
essential role on the QGP formation.
On the other hand, for a dense flux-tube system, neighboring flux tubes 
are melted into a large cylindrical tube, which is approximately 
homogeneous in the whole space and is regarded as the ``color condenser''.
In this case, large homogeneous QGP may be created in the central region.
Thus, the process of QGP formation is expected depending largely on the 
density of created flux tubes.
We have finally investigated also the interaction of two color-flux-tubes in the DGL 
theory 
and showed that there is a possibility of 
annihilation or unification of two color-flux-tubes
when the flux tubes are sufficiently dense. 
This process would provide an enough energy to create
many dynamical gluons and to form QGP.

\newpage

{\bf Acknowledgment}

We are grateful for useful discussions and supports with 
Professor H. Toki and  all the members of the 
Research Center for Nuclear Physics theory group.
Simulations were performed on VPP500 at RIKEN and SX4 at RCNP.



\appendix



\chapter{Monte Carlo Method for Lattice QCD}

\section{Gauge Field on the Lattice}

Quantum Chromodynamics (QCD), a nonabelian gauge theory, is difficult
to solve analytically due to the large gauge
coupling in the low-energy region.
The Monte Carlo simulation based on the lattice gauge theory
is one of the promising methods for the direct
calculation of the QCD partition functional.
We review here the fundamentals of the lattice QCD and the numerical
simulation \cite{kerson}.

The Minskowski space-time is transformed into Euclidean space-time by 
replacing $x^0 = -ix^4$, where $x^0$ and $x^4$ are the Minskowski and
the Euclidean time,  respectively. 
A similar transformation is made in the time component of all four-vectors.
In the Euclidean space,  there is no distinction between upper and lower 
indices of four-vector, i.e, 
$x_{\mu}^{\rm E} = (x^1,x^2,x^3,x^4) = (x_1,x_2,x_3,x_4)$. 

In principle, to latticize the theory is performed by the replacement 
of derivatives by finite differences with the lattice spacing $a$.
This method, however, does not preserve gauge invariance, which is an 
essential attribute of the theory.
Wilson introduced a gauge-invariant lattice action using the 
path representation of the gauge group $G$.
The most elementary paths on the lattice are links out of 
which any path can be constructed. 
The link is identified by the site $s$ and the direction $\mu$.
With each link on the lattice, we associate a group element, called the 
``link variable'' as
\begin{eqnarray}
U_{\mu}(s) 
\equiv {\rm exp} \{ i a e A_\mu(s) \}
\equiv {\rm exp} \{i \theta_\mu(s) \} \hspace{0.5cm} \in G,
\end{eqnarray}
with the gauge coupling constant $e$.
The lattice angle variables $\theta_\mu$ defined as 
$ \theta_\mu \equiv \theta_\mu^a T^a \equiv aeA_\mu$ is dimensionless, 
and link
variable  $U_{\mu}(s)$ is transformed as
\begin{eqnarray}
U_\mu(s) \rightarrow V(s) U_\mu(s) V^\dagger(s+\hat \mu),
\end{eqnarray}
where $V(s)$ and $V(s+\hat \mu)$ are the gauge functions located at
the starting and the end points of the link $U_{\mu}(s)$.
Thus, the gauge transformation is described by simple multiplication 
of the group elements in the lattice formalism.

In the continuum, the field strength tensor 
$G_{\mu\nu} \equiv G_{\mu\nu}^a T^a \in {g}$ is defined in terms of an 
infinitesimal closed path.
By analogy, we defined it on the lattice in terms of
a ``plaquette'' $\Box_{\mu\nu}$,
a square bounded by four links.
The field tensor $ G_{\mu\nu}$, which is associated with  the oriented 
plaquette specified by the links $\{ \mu,\nu\}$ attached to site $s$, is 
defined through the relation 
\begin{eqnarray}
\Box_{\mu\nu}(s) \equiv 
U_\mu(s) U_\nu (s+\hat \mu) U ^\dagger_\mu(s+ \hat \nu) U_\nu^\dagger(s)
\equiv {\rm exp} (iea^2 G_{\mu\nu}(s))  \hspace{0.5cm} {\in G}.
\end{eqnarray}
For $a \rightarrow 0$, we recover the continuum field tensor,
\begin{eqnarray}
G_{\mu\nu}(s) = \frac1a [(A_\nu (s + \hat \mu) - A_\nu(s) ) - 
                       (A_\mu (s+ \hat \nu) - A_\mu(s) ) ]
                     + ie[A_\mu(s),A_\nu(s)] + O(a). \nonumber
\end{eqnarray}
The lattice action is constructed by using the relation,
\begin{eqnarray}
\sum_s \sum_{\mu,\nu} {\rm tr} \Box_{\mu\nu}(s)  =  
\sum_s \sum_{\mu,\nu} {\rm tr} [1-iea^2 G_{\mu\nu} -\frac12 e^2 a^4 
(G_{\mu\nu})^2+ \cdots],
\end{eqnarray}
where $\mu$ and $\nu$ are summed from 1 to 4. 
The first term on the right hand side is a constant,
and the second term vanishes because $G_{\mu\nu}$ is 
traceless. 
The sum over $s$,  $\mu$, $\nu$ on the left hand side is twice 
the sum over all plaquettes, so that one finds 
\begin{eqnarray}
\sum_s \sum_{\mu \ne \nu} {\rm tr} \Box_{\mu\nu} =
\sum_s \sum_{\mu > \nu} ( {\rm tr} \Box_{\mu\nu} + {\rm tr} 
\Box_{\mu\nu}^\dagger ) 
= 2 {\rm Re} \sum_s \sum_{\mu > \nu} {\rm tr} \Box_{\mu\nu}.
\end{eqnarray}
Because interchanging $\mu$ and $\nu$ changes the orientation of the 
designated plaquette. Thus, we have 
\begin{eqnarray}
\sum_s \sum_{\mu > \nu} {\rm Re} \,\,{\rm tr} \Box_{\mu\nu}
\rightarrow - \frac14 e^2 \int d^4x \sum_{\mu,\nu} 
{\rm tr} (G_{\mu\nu})^2 + {\rm const.},
\end{eqnarray}
as $a \rightarrow 0$. Here, the sum on the left hand side
extends over all plaquettes on the lattice.
 
Finally, we get the lattice  QCD action as
\begin{eqnarray}
S  & =& \int d^4x \sum_{\mu,\nu} \frac12 {\rm tr} (G_{\mu\nu})^2  =
\frac{2}{e^2} \sum_s \sum_{\mu > \nu} {\rm Re} \,{\rm tr} 
\left[1-\Box_{\mu\nu}(s) \right]
\nonumber \\
& =  &\beta  \sum_s \sum_{\mu > \nu} {\rm Re} \, \frac{1}{N_c}{\rm tr} 
\left[1- \Box_{\mu\nu}(s) \right],
\label{laction}
\end{eqnarray}
where $\beta \equiv \frac{2N_c}{e^2}$ is the control parameter 
relating to the lattice spacing $a$. 
The continuum limit $a \rightarrow 0$ corresponds to $e \rightarrow 0$
or $\beta \rightarrow \infty$ in the theory with the asymptotic 
freedom like QCD.
Thus, the QCD partition functional  in the lattice formalism is given as 
\begin{eqnarray}
Z = \int dU_{\mu} e^{-S[U_{\mu}(s)]} 
= \int [\Pi_{s,\mu} dU_\mu(s)] e^{-S[U_{\mu}(s)]}.
\end{eqnarray}
For SU(2), the link variable $U_{\mu}(s)$ is parameterized as 
$U_{\mu} \equiv U_{\mu}^0 + i \tau^a U_{\mu}^a$,
and the measure takes a form 
\begin{eqnarray}
dU_{\mu}(s) \equiv dU_{\mu}^0 dU_{\mu}^1 dU_{\mu}^2 dU_{\mu}^3
\delta ( U_{0}^2  + U_{1}^2 +U_{3}^2 +U_{3}^2 -1 ).
\end{eqnarray}
The expectation value of arbitrary operator $O$ is 
calculable as
\begin{eqnarray}
\langle O \rangle \equiv \frac{ \int d U_{\mu} O e^{-S[ U_{\mu}]}}
{ \int dU_{\mu} e^{-S[ U_{\mu}]}}
\end{eqnarray}
using the numerical simulation with the Monte Carlo method.

\vspace{1cm}

\section{Monte Carlo Method}

In this section, we briefly summarized the Monte Carlo method of the 
lattice QCD.
In the lattice QCD with the action (\ref{laction}),
the ensemble of gauge configuration is given by the 
canonical ensemble, characterized by 
Boltzmann distribution,
\begin{eqnarray}
 P(U) = \frac{1}{Z} e^{-\beta \hat S},
\end{eqnarray}
where we put $S[U]=\beta \hat S[U]$ and regard $\hat S[U]$ as a ``Hamiltonian''.
In principle, the lattice QCD partition function $Z$ can be calculated 
by the infinite number of Monte Carlo simulations using any random 
sampling of $U_{\mu}$.
In practical, however, almost all gauge configurations with large 
$\hat S$ do not contribute to the lattice QCD partition functional $Z$.
Therefore, it is necessary to sample the
``important gauge configuration'' with small $\hat S$ 
effectively in terms of the
numerical calculation with finite number of operations.
Such a ``important sampling'' can be realized by generating the random 
number with the weight $e^{-\beta \hat S[U]}$, and the Monte Carlo 
method enables us to calculate the lattice QCD partition function, 
numerically.

In the Monte Carlo method,  our goal is to generate a time sequence of
configurations such that the configuration $U$ 
occurs with probability $P(U)$ after a sufficiently long time.
Thus, the time average of any quantity 
 would be the same as its average over the canonical ensemble.

The time sequence is generated by a stochastic process.
A configuration $U$ is updated to $U'$  with a transition 
probability $T(U',U)$, which has the following general properties:

\begin{enumerate}

\item 
 $T(U',U) \ge 0$

\item
 $\int dU'  T(U',U) = \int dU T(U,U') = 1$,

\item
 $T(U',U) P(U) dU =  T(U,U') P(U') dU'$.

\end{enumerate}

The first two are just properties that any probability should have.
The last is known as the condition of detailed balance.
Integrating over $U'$ on both sides of the last condition, we obtain 
\begin{eqnarray}
 P(U)  = \int dU' T(U,U')  P(U'),
\end{eqnarray}
which states that the probability $P(U)$ is an eigenfunction of $T(U',U)$.
This means that transition probability preserves the equilibrium ensemble.
Thus, we get the gauge configuration ensemble obeying $P(U)$ as a ``time'' 
sequence of ``thermalization''. Using the obtained gauge configuration, 
the ensemble average of  arbitrary operator $O(U)$ is numerically 
estimated by
\begin{eqnarray}
\langle O \rangle = \int dU O(U) P(U).
\end{eqnarray}
On the actual numerical simulation, there are two cautions to be 
carefully checked.
One is the achievement of the thermo-equilibrium, the other is the 
vanishing of the cancellation among the sampling gauge configurations.

\chapter{Procedure of Maximally Abelian Gauge Fixing}

The maximally abelian (MA) gauge is the 
special abelian gauge  exhibiting infrared abelian dominance in the 
lattice QCD, and provides the theoretical basis of 
dual Higgs picture from QCD.
In the SU(2) lattice formalism, the MA gauge is defined so as to maximize
\begin{eqnarray}
R_{\rm MA}[U_\mu]  & \equiv & \frac12 \sum_{s}  R(s)  
\nonumber \\
 & \equiv & \sum_{s,\mu} {\rm tr} \{  U_\mu(s) \tau_3 U^{\dagger}_\mu(s) \tau_3 \}  
\nonumber \\
   & = & 
2\sum_{s,\mu}\{
U^0_\mu(s)^2+U^3_\mu(s)^2-U^1_\mu(s)^2-U^2_\mu(s)^2 \} \nonumber \\
   & = & 
2\sum_{s,\mu} \left[
1 - 2\{ U^1_\mu(s)^2+U^2_\mu(s)^2 \} \right]
\end{eqnarray}
by the gauge transformation.
Here, $R(s)$ is a local scalar variable defined as
\begin{eqnarray}
R(s) \equiv
\sum_{\pm \mu} {\rm tr} \{  U_{\pm \mu}(s) \tau_3 U^{\dagger}_{\pm \mu}(s)
 \tau_3 \} \hspace{1cm} \mbox{with} \hspace{1cm}  
U_{-\mu}(s) = U^{\dagger}_{\mu}(s-\mu).
\end{eqnarray}
Here, $R(s)$ is manifestly invariant under the lattice rotation and the 
reflection.
In the MA gauge, the operator 
\begin{eqnarray}
\Phi(s) \equiv \sum _{\mu,\pm } U_{\pm \mu}(s) \tau_3 U^{\dagger}_{\pm 
\mu}(s)
\end{eqnarray}
is diagonalized. 
In this appendix, we show the procedure of the MA gauge fixing on the 
lattice.

To begin with, we introduce a ``local'' gauge transformation,
whose gauge function $V_{s_0}(s)$ is not unity at the site $s_0$ only,
\begin{eqnarray}
\left\{
\begin{array}{llll}
V_{s_0}(s) & = &  V(s_0)  &  \hspace{1cm}   \mbox{for $s = s_0$} \\
V_{s_0}(s) & = &  1   & \hspace{1cm}   \mbox{for $s \ne s_0 $}.
\end{array}
\right.
\end{eqnarray} 
In order to maximize the value $R_{\rm MA}[U_{\mu}]$,
one may consider to maximize the local variables $R(s)$ at each site $s$ by
the local gauge transformation $V_s$.
However, 
since $R(s_0)$ at the site $s_0$
is changed not only by the gauge transformation  $V_{s_0}(s)$
but also by the gauge transformation $V_{s_0-\hat \mu}(s)$ 
with neighboring sites $s_0 - \hat \mu$, 
one cannot obtain the MA gauge configuration $U_\mu(s_0)^{\rm MA}$ 
only by simple local gauge transformation $V_{s_0}(s)$. 
After the local gauge transformation at all sites on the whole lattice,
one has to repeat this procedure until $R_{\rm MA}$ is maximized.
 
Now, let us derive the gauge transformation $V_{s_0}(s)$ 
to maximize $R(s_0)$.
After the gauge transformation by $V_{s_0}(s)$,
$R(s_0)$ is changed as
\begin{eqnarray}
R^{V} (s_0)
& = & \sum_{\mu} {\rm tr}
\{ V(s_0) U_\mu(s_0) \tau_3 U^{\dagger}_\mu(s_0) V^{\dagger}(s_0) \tau_3  
\nonumber \\
& &  + U_\mu(s_0- \hat \mu) V^{\dagger}(s_0) \tau_3 V(s_0) U^{\dagger} 
 (s_0- \hat \mu) \tau_3 \}  \nonumber \\
& = & {\rm tr} [  \sum_{\mu} 
\{  U_\mu(s_0) \tau_3 U^{\dagger} (s_0) 
  + U^{\dagger}_\mu(s_0-\hat \mu) \tau_3 U_\mu(s_0- \hat \mu) \}  \cdot
V^{\dagger}(s_0) \tau_3 V(s_0)  ]  \nonumber \\
& \equiv & {\rm tr} [\Phi(s_0) S(s_0) ].  
\label{eq:derivative}
\end{eqnarray}
Here, we define 
\begin{eqnarray}
S(s) \equiv  S^a(s) \tau^a \equiv \vec S \cdot \vec \tau \equiv 
 V^{\dagger}(s) \tau_3 V(s)  
\hspace{1cm}  \in \mbox{ su(2)},
\end{eqnarray}
\begin{eqnarray}
\Phi(s)  \equiv \Phi^a(s) \tau^a  \equiv 
\vec \Phi \cdot \vec \tau
 \equiv    \sum _{\pm \mu} U_{\pm \mu}(s) \tau_3 U^{\dagger}_{\pm \mu}(s)
 \hspace{1cm}  \in \mbox{su(2)},
\end{eqnarray}
which are both elements of Lie algebra  and 
satisfy relations  
tr$(\Phi) =$ tr$(S) = 0$ and $S^2 = 1$.
To maximize $R(s_0)$ by this gauge transformation,
$\vec S(s_0)$ is taken to be the same direction as $\vec \Phi(s_0)$
in the SU(2) $\simeq$ O(3) parameter space,
$\vec S // \vec \Phi$.
After this gauge transformation, $\Phi(s_0)$  is diagonalized as
$\Phi^{V}(s_0) = V(s_0) \Phi(s_0) V^{\dagger}(s_0) = \Phi^{V}_{3}(s_0) 
\tau_3$,
and $S(s_0)$ becomes $\tau_3$.  
Here, $\Phi$ plays a similar role as the Higgs field 
in the 't Hooft-Polyakov monopole.

\begin{figure}[bt]
\epsfxsize = 10 cm
\centering \leavevmode
\epsfbox{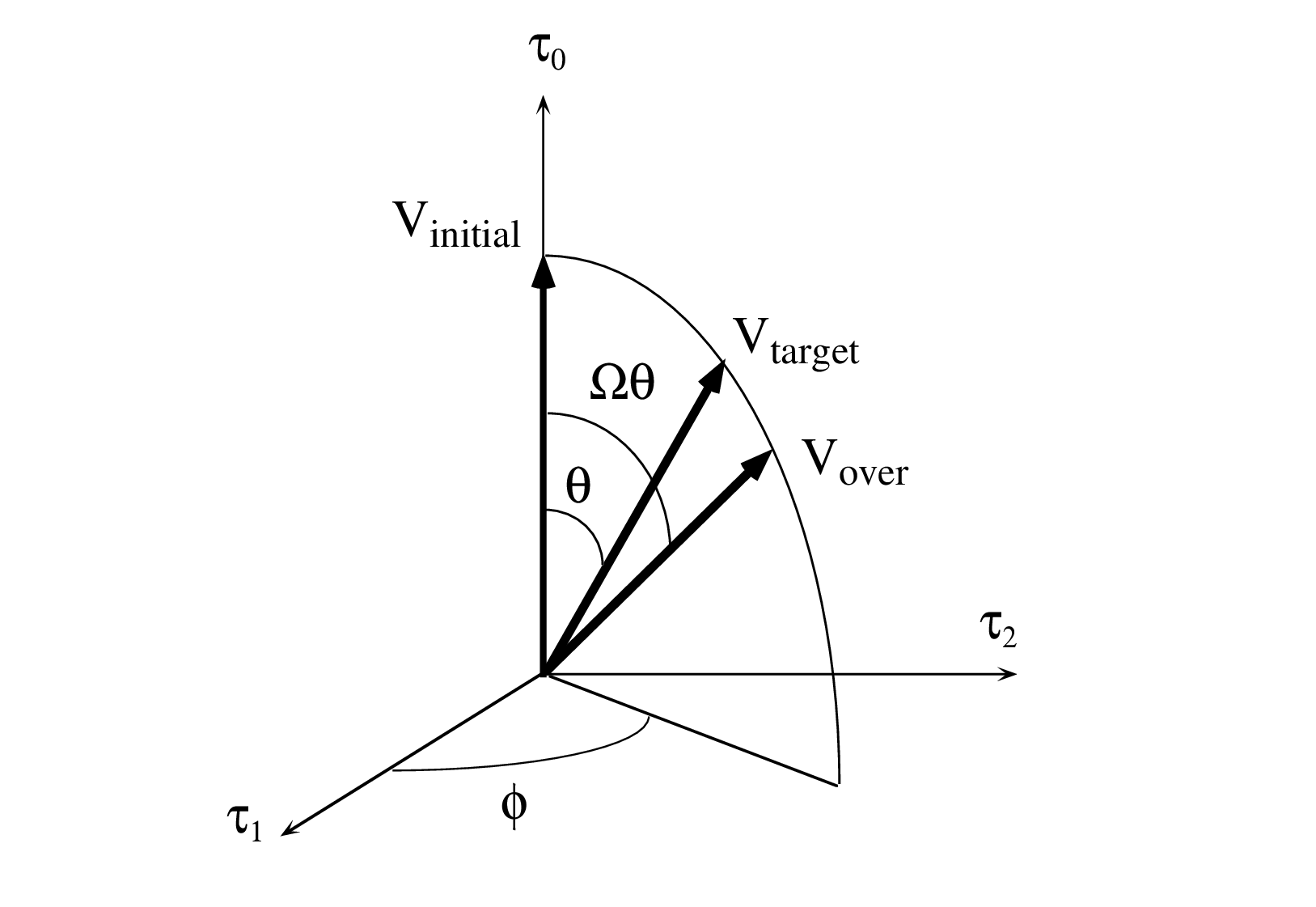}
\vspace{0cm}
\caption{
The gauge function $\vec V_{\rm over}$ used in the over-relaxation method.
The vector $\vec V_{\rm target}$ 
corresponds to the gauge function which maximizes $R(s_0)$.
}
\label{over}
\vspace{0cm}
\end{figure}

In the abelian gauge, 
the gauge function $V(s)$ is an element of the coset space 
SU(2)/U(1)$_3$ using the residual U(1)$_3$ degrees of freedom. 
We take 
the representative element of $V(s)$
so as to satisfy $V^3(s) = 0$, or 
\begin{eqnarray}
V(s) = V^0(s)  + i \{ V^1(s) \tau^1 + V^2(s) \tau^2 \}. 
\end{eqnarray}
Because of $( V^1)^2 + ( V^2)^2 + ( V^3)^2 = 1,$
we can parameterize $V(s)$ as
\begin{eqnarray}
\left\{
\begin{array}{lll}
V^0(s)  & = & \cos \theta(s)  \\
V^1(s)  & = & \sin \theta(s) \cos \phi(s)  \\
V^2(s)  & = & \sin \theta(s) \sin \phi(s), 
\label{eq:normal}
\end{array}
\right.
\end{eqnarray}
and then $S(s)$ is expressed as
\begin{eqnarray}
S \equiv S^a \tau^a \equiv V^{\dagger} \tau_3 V = \sin 2\theta \cos \phi \tau_1 
+ \sin 2\theta \sin \phi \tau_2 + \cos 2 \theta \tau_3. 
\end{eqnarray}
Since $\Phi(s_0) \equiv \Phi^a(s_0) \tau^a$ is obtained from the 
original gauge configuration $U_{\mu}(s)$, 
we get $V_{s_0} (s_0)$ as
\begin{eqnarray}
\left\{
\begin{array}{lll}
\tan^2 2 \theta(s_0)  & = &\frac{(S ^1)^2+(S^2)^2}{(S^3)^2} 
= \frac{(\Phi^1)^2+(\Phi^2)^2}{(\Phi^3)^2}  \\
\tan \phi (s_0) & = & \frac{S^2}{S^1} = \frac{\Phi^2}{\Phi^1}.
\end{array}
\right.
\end{eqnarray}
Thus, the gauge function $V(s)$ which maximizes $R(s_0)$ 
is obtained  so as to obey $\vec S(s_{0}) // \vec \Phi(s_{0})$.
This procedure makes $R(s_0)$ defined in  (\ref{eq:derivative}) maximum by 
$V_{s_0}(s)$.
This gauge transformation, however,  influences $R(s)$ of the neighboring 
sites, $s=s_0 \pm \hat \mu$, and in fact does not make them maximum.
Therefore, we have to perform this procedure to the neighboring sites.
By doing this, however, the original $R(s_0)$ gets some change and hence 
$R(s_0)$ is no  more in its maximum.
This fact forces us to repeat the local-gauge transformation many times.

To optimize the convergence, 
in the practical simulation,
we take an over-relaxation method. 
We show 
the vector 
$(\cos \theta, \sin \theta \cos \phi, \sin \theta \sin \phi)$
in Fig.\ref{over},  corresponding to the gauge function $V(s_0)$ in Eq.(\ref{eq:normal}). 
In the over-relaxation method,     
we take the angle value $\Omega \theta$
instead of $\theta$ obtained in Eq.(\ref{eq:normal}),
\begin{eqnarray}
\left\{
\begin{array}{lll}
V^0_{\rm over}(s_0)  & = & \cos (\Omega \theta)  \\
V^1_{\rm over}(s_0)  & = & \sin (\Omega \theta) \cos \phi  \\
V^2_{\rm over}(s_0)  & = & \sin (\Omega \theta) \sin \phi. 
\end{array}
\right.
\label{eq:overrelaxation}
\end{eqnarray}
This  overrelaxation parameter $\Omega$ is taken as $1 \sim 2$.


\end{document}